\documentclass[12pt]{article}

\usepackage{amsmath}
\usepackage{amssymb}
\usepackage{graphicx}
\usepackage{url}
\usepackage{multirow}
\usepackage{authblk}
\usepackage{mathtools, bbm}
\usepackage{float, subfig}
\usepackage{natbib}
\usepackage{nccmath}
\usepackage[margin=1in]{geometry}
\usepackage{longtable}
\usepackage{threeparttablex}
\usepackage{booktabs}
\usepackage{xcolor}
\usepackage{colortbl}
\usepackage{makecell}
\usepackage{titlesec}
\usepackage{adjustbox}

\usepackage{tikz}
\usepackage{tikzscale}
\usepackage{pgfplots}

\usetikzlibrary{arrows, automata, calc, shapes, arrows.meta, decorations.pathmorphing, bending, backgrounds, overlay-beamer-styles, tikzmark, positioning}

\pgfplotsset{compat=1.8}

\definecolor{shadegray}{rgb}{0.95,0.95,0.95}

\newcommand\numberthis{\addtocounter{equation}{1}\tag{\theequation}}

\newcommand{\bigCI}{\mathrel{\text{\scalebox{1.07}{$\perp\mkern-10mu\perp$}}}}

\newlength\imageheight

\pgfplotsset{compat=1.8}

\begin{document}

\title{A marginal structural model for normal tissue complication probability}

\author[1]{Thai-Son~Tang}
\author[1,2]{Zhihui (Amy) Liu}
\author[2]{Ali Hosni}
\author[2]{John Kim}
\author[1]{Olli~Saarela\thanks{Correspondence to: Olli Saarela, Dalla Lana School of Public Health, 155 College Street, Toronto, Ontario M5T 3M7, Canada. Email: \texttt{olli.saarela@utoronto.ca}}}

\affil[1]{Dalla Lana School of Public Health, University of Toronto}
\affil[2]{Princess Margaret Cancer Centre, University Health Network}

\maketitle

\begin{abstract}
The goal of radiation therapy for cancer is to deliver prescribed radiation dose to the tumor while minimizing dose to the surrounding healthy tissues. To evaluate treatment plans, the dose distribution to healthy organs is commonly summarized as dose-volume histograms (DVHs). Normal tissue complication probability (NTCP) modelling has centered around making patient-level risk predictions with features extracted from the DVHs, but few have considered adapting a causal framework to evaluate the safety of alternative treatment plans. We propose causal estimands for NTCP based on deterministic and stochastic interventions, as well as propose estimators based on marginal structural models that impose bivariable monotonicity between dose, volume, and toxicity risk. The properties of these estimators are studied through simulations, and their use is illustrated in the context of radiotherapy treatment of anal canal cancer patients.

\end{abstract}

\newpage




\section{Introduction}
The goal of radiation therapy for cancer is to deliver the prescribed amount of radiation to the tumor while minimizing the dose to the surrounding healthy tissues, also known as organs-at-risk (OARs). Exposure to radiation could cause damage to the OARs and lead to a range of side effects, such as skin fibrosis or changes to bowel function in treatment of cancer of the anal canal, which can interrupt treatment delivery and adversely affect a patient's quality of life. Guidelines for dose constraints to OARs are developed based on normal tissue complication probability (NTCP) models, which typically associate selected dosimetric features, as well as selected non-dosimetric factors, to an adverse outcome. For a general review of NTCP modeling and their use in clinical decision making, we refer to \citet{palma2019normal}.

In principle, radiation exposure to normal tissue is a 3-dimensional spatial function, and is recorded at voxel level. However, for analysis purposes this is commonly collapsed to a 1-dimensional dose distribution with a density function, which can be considered as functional data. The discretized histogram approximation of the density function is known as the dose-volume histogram (DVH); this loses the spatial information in the 3-D dose distribution, but simplifies modeling of the dose-response relationship. Alternatively, we can consider the complement of the cumulative distribution version of this, which consists of the proportional organ volumes receiving at least $d$ units. Biologically, a higher dose (to be defined later, but informally meaning a dose distribution shifted to the right) is expected to lead to a higher NTCP, and this monotonic relationship can be incorporated into statistical models. Empirically, the monotonic relationship between dose and NTCP has been demonstrated in many studies \citep{koper1999acute, choi2010stereotactic, brock2021more}.

Treatment planning may involve placing constraints on the OARs dose distributions to control NTCP \citep{kupchak2008experience, gulliford2010dose}. An example of such constraint could be to specify upper volume thresholds at a given radiation dose, such as restricting the volume of the bladder exposed to doses greater than 45 Gy to less than 80\% to lower the risk of radiation-induced genitourinary toxicity \citep{pederson2012late, jin2009dose, huang2019thyroid}. Causally, we may be interested in the effect of interventions corresponding to placing such constraints.

\citet{jackson2006atlas} proposed an ``atlas'' to visualize smoothed empirical NTCP estimates at different dose-volume thresholds as heatmaps or contour plots to identify dose-volume regions of high risk, which could be used to derive constraints for treatment plans \citep{wilkins2020derivation}. The contours, which are dose-volume combinations corresponding to the same NTCP, are referred to as iso-probability lines. Under monotonicity, the true underlying iso-probability lines are themselves monotonic decreasing with respect to dose and volume receiving it; however, \citet{jackson2006atlas} did not incorporate this restriction.

Statistical NTCP modelling has largely focused on making patient-level predictions and can be categorized into (i) regression models incorporating selected features of the dose distribution as predictors and (ii) functional regression models. The former are commonly logistic regression models regressing dichotomous toxicity outcome on selected summary statistics of the dose distribution (such as mean dose, maximum dose, or proportional OARs volume receiving a dose of at least $d$ units of radiation). Alternatively to parametric models, more flexible specification through for example smoothing splines have been used to improve prediction accuracy \citep{hansen2020prediction}.

In contrast to selected features of the dose distribution, scalar-on-function regression models make use of the information on the entire dose distribution by incorporating it as a functional covariate. \citet{benadjaoud2014functional} applied functional principal component regression dimension reduction purposes, while \citet{dean2016functional} compared functional principal component regression and functional partial least squares regression. The functional models proposed by \citet{schipper2007bayesian, schipper_2008} used splines and also enforced the monotonicity constraint. While this is biologically plausible, the constraint can also add stability to the estimation of flexible models without requiring very large sample sizes.

The previous modeling approaches do not explicitly consider NTCP modeling as a causal inference problem, which involves specifying the intervention of interest and accounting for confounding patient and tumor characteristics. Causal models with monotonicity constraints on a single exposure variable have been proposed in the literature \citep{qin2019using, westling2020causal, yuan2021enhanced}, as well as causal models for functional exposures, confounders and mediators \citep{lindquist2012functional, zhao2018functional, miao2020average}, but not in the context of NTCP modeling. \citet{nabi2022semiparametric} considered features of the dose distribution as a high dimensional exposure and outlined a semiparametric framework for estimating parameters of sufficient dimension reductions, with an application on parotid gland DVHs and weight loss in head and neck cancer patients. This approach requires specifying the dimension reduction a priori and does not enforce monotonicity. 

While the approaches based on dimension reduction may result in good predictive accuracy, due to the highly correlated nature of the DVH data, interpretation for the purpose of planning interventions is challenging. The same applies to the weight functions of the functional regression models of \citet{schipper2007bayesian, schipper_2008}; the individual coefficients forming the weight function cannot be directly interpreted causally, as the dose distribution cannot be intervened on pointwise \emph{ceteris paribus} (due to the restriction of it being a distribution). However, we can envision intervening on the dose distribution pointwise (e.g. on a dose at a given volume or volume at a given dose), and drawing the rest of the intervention dose distribution from a distribution, which induces a certain kind of marginal intervention effect. The distribution for drawing the corresponding stochastic intervention can be informed by the observed dose distributions, to ensure positivity. In the present manuscript, we aim to formulate a causal modeling framework and estimation methods for such effects. The marginal effects of interest relate closely to the marginal associations visualized by \citet{jackson2006atlas} and the iso-probability lines, but we will aim to also enforce the biologically plausible monotonicity constraint, leading us to consider models that are bivariable monotone with respect to dose and volume receiving this dose. Our proposed approach will also address the multiple comparison issues of estimating a large number of marginal effects of cumulative volumes at different doses, which we will achieve by fitting all of the effects using a single model.

The objective of this paper is to develop causal models for the NTCP under different treatment plans. The outline for this article is as follows. In Section 2, we adapt the Neyman-Rubin potential outcomes framework to formulate the causal estimand under multivariable exposures, and state the causal identifiability conditions required for estimation. We formulate causal contrasts of interest under stochastic interventions, and a bivariable monotone marginal structural model parametrizing the effects. In Section 3, we adapt the Bayesian non-parametric multiple monotone regression model proposed by \citeauthor{saarela2011method} (\citeyear{saarela2011method}) for estimation. In Section 4, we evaluate the performance of the proposed methods through simulation and in Section 5, apply them to data from a cohort of anal canal cancer patients to assess acute radiation-induced genitourinary toxicity. We conclude in Section 6 with a brief discussion.

\section{Proposed estimands}\label{section:framework}

\subsection{Notation}\label{section:notation}

\begin{figure}[H]
    \centering
    \subfloat[Differential DVH and dose density function]{\scalebox{0.85}{\pgfplotsset{every tick label/.append style={font=\scriptsize}}
\begin{tikzpicture}
\begin{axis}[
    axis lines = left,
    ymax = 0.11, 
    ymin = 0,
    scaled y ticks = false,
    yticklabel=\pgfkeys{/pgf/number format/.cd,fixed,precision=2,zerofill}\pgfmathprintnumber{\tick},
    xlabel = \text{Dose (in Grays)},
    ylabel = \text{Density},
    x label style = {yshift = -2em, font = \scriptsize},
    y label style = {yshift = 1em, font = \scriptsize},
    xtick = {0, 2, 4, 6, 8, 10, 12, 14, 16, 18, 20,  24},
    xticklabels = {0, 10, 20, 30, 40, 50, 60, 70, 80, 90, 100, $d_{\text{max}}$},
    extra x ticks = {1, 3, 5, 7, 9, 11, 13, 15, 17, 19, 23},
    extra x tick labels = {$1$, $2$, $3$, $4$, $5$, $6$, $7$, $8$, $9$, ${10}$,  $\mathcal{D}$},
    extra x tick style={color = brown, yshift=-3.5ex,
    major tick length=0pt}
]
\addplot+[ybar, color = white, bar width = 2, mark = no, samples at = {1, 3, 5, 7, 9, 11, 13, 15, 17, 19, 21, 23}] {x*e^(-(x/10)^2)/50 - 0.005} node[pos=0.0, scale=0.1, pin={[pin distance=3.5em]90:{\tiny $B_{1}$}}]{} 
node[pos=1/12, scale=0.1, pin={[pin distance=3em]90:{\tiny $B_{2}$}}]{} 
node[pos=2/12, scale=0.1, pin={[pin distance=2em]90:{\tiny $B_{3}$}}]{}
node[pos=3/12, scale=0.1, pin={[pin distance=1.25em]90:{\tiny $B_{4}$}}]{}
node[pos=4/12, scale=0.1, pin={[pin distance=1.25em]90:{\tiny $B_{5}$}}]{}
node[pos=5/12, scale=0.1, pin={[pin distance=1.25em]90:{\tiny $B_{6}$}}]{}
node[pos=6/12, scale=0.1, pin={[pin distance=1.25em]90:{\tiny $B_{7}$}}]{}
node[pos=8/12, scale=0.1, pin={[pin distance=1.25em]90:{\tiny $B_{8}$}}]{}
node[pos=9/12, scale=0.1, pin={[pin distance=1.25em]90:{\tiny $B_{9}$}}]{}
node[pos=10/12, scale=0.1, pin={[pin distance=1.25em]90:{\tiny $B_{10}$}}]{};
\addplot+[ybar, color = white, bar width = 2, mark = no, samples at = {23}] {x*e^(-(x/10)^2)/50 - 0.0015}
node[pos=1, scale=0.1, pin={[pin distance=1.25em]above:{\tiny $B_{\mathcal{D}}$}}]{};
\addplot[domain = 0:25, smooth] {x*e^(-(x/10)^2)/50};
\addplot+[ybar, color = brown, bar width = 2, mark = no, samples at = {1, 3, 5, 7, 9, 11, 13, 15, 17, 19, 21, 23}] {x*e^(-(x/10)^2)/50} node[pos=0.0, scale = 0.1, pin={[pin distance=3.5em]90:{\tiny $B_{1}$}}]{} 
node[pos=1/12, scale=0.1, pin={[pin distance=3em]90:{\tiny $B_{2}$}}]{} 
node[pos=2/12, scale=0.1, pin={[pin distance=2em]90:{\tiny $B_{3}$}}]{}
node[pos=3/12, scale=0.1, pin={[pin distance=1.25em]90:{\tiny $B_{4}$}}]{}
node[pos=4/12, scale=0.1, pin={[pin distance=1.25em]90:{\tiny $B_{5}$}}]{}
node[pos=5/12, scale=0.1, pin={[pin distance=1.25em]90:{\tiny $B_{6}$}}]{}
node[pos=6/12, scale=0.1, pin={[pin distance=1.25em]90:{\tiny $B_{7}$}}]{}
node[pos=8/12, scale=0.1, pin={[pin distance=1.25em]90:{\tiny $B_{8}$}}]{}
node[pos=9/12, scale=0.1, pin={[pin distance=1.25em]90:{\tiny $B_{9}$}}]{}
node[pos=10/12, scale=0.1, pin={[pin distance=1.25em]90:{\tiny $B_{10}$}}]{}
node[pos=1, scale=0.1, pin={[pin distance=1.25em]above:{\tiny $B_{\mathcal{D}}$}}]{};
\addplot+[ybar, fill opacity = 0.2, color = brown, fill = blue, bar width = 2, mark = no, samples at = {9, 11, 13, 15, 17, 19, 21, 23}] {x*e^(-(x/10)^2)/50};
\end{axis}
\end{tikzpicture}
}\label{fig:dvh-schematic-ddvh}}\hspace{0.4cm}
    \subfloat[Cumulative DVH and smoothened complementary cumulative distribution function]{\scalebox{0.85}{\pgfplotsset{every tick label/.append style={font=\scriptsize}}
\begin{tikzpicture}
\begin{axis}[
    axis lines = left,
    xmin = 0, xmax = 24,
    ymin = 0, ymax = 1.25,
    xlabel = \text{Dose (in Grays)},
    ylabel = \text{Volume of Organ-at-Risk (OAR)},
    x label style = {yshift = -2em, font = \scriptsize},
    y label style = {yshift = 1em, font = \scriptsize},
    xtick = {0, 2, 4, 6, 8, 10, 12, 14, 16, 18, 20, 24},
    xticklabels = {0, 10, 20, 30, 40, 50, 60, 70, 80, 90, 100, $d_{\text{max}}$},
    extra x ticks = {1, 3, 5, 7, 9, 11, 13, 15, 17, 19,  23},
    extra x tick labels = {$1$, $2$, $3$, $4$, $5$, $6$, $7$, $8$, $9$, ${10}$,  $\mathcal{D}$},
    extra x tick style={color = brown,
    yshift=-3.5ex,
    major tick length=0pt}
]
\addplot+[black, domain = 0:27, smooth, mark = none] {e^(-(x/10)^2)};
\addplot+[ybar, bar width = 2, mark = no, samples at = {1, 3, 5, 7, 9, 11, 13, 15, 17, 19, 21, 23, 25}] {e^(-((x-1)/10)^2)} node[pos=0.0, scale=0.1, pin={[pin distance=0.75em]60:{\scriptsize $G_{{1}}$}}]{} 
node[pos=1/12, scale=0.1, pin={[pin distance=0.75em]60:{\scriptsize $G_{{2}}$}}]{} 
node[pos=2/12, scale=0.1, pin={[pin distance=0.75em]60:{\scriptsize $G_{{3}}$}}]{}
node[pos=3/12, scale=0.1, pin={[pin distance=0.75em]60:{\scriptsize $G_{{4}}$}}]{}
node[pos=4/12, scale=0.1, pin={[pin distance=0.75em]60:{\scriptsize $G_{{5}}$}}]{}
node[pos=5/12, scale=0.1, pin={[pin distance=0.75em]60:{\scriptsize $G_{{6}}$}}]{}
node[pos=6/12, scale=0.1, pin={[pin distance=0.75em]60:{\scriptsize $G_{{7}}$}}]{}
node[pos=7/12, scale=0.1, pin={[pin distance=0.75em]60:{\scriptsize $G_{{8}}$}}]{}
node[pos=8/12, scale=0.1, pin={[pin distance=0.75em]60:{\scriptsize $G_{{9}}$}}]{}
node[pos=11/12, scale=0.1, pin={[pin distance=0.75em]above:{\scriptsize $G_{{{\mathcal{D}}}}$}}]{}
;

\draw[blue, dashed, thin] (90, 0) -- (90, 52);


\end{axis}
\end{tikzpicture}


}\label{fig:dvh-schematic-dvh}} \\
    \subfloat[DAG of the causal mechanism.]{\scalebox{0.75}{\begin{tikzpicture}[
                > = {Latex[length=2mm, width=2mm]}, 
                shorten > = 1pt, 
                auto,
                scale = 1.3,
                node distance = 1cm, 
                semithick 
            ]
    
            \tikzstyle{state}=[
                circle,
                draw = black,
                thick,
                minimum size = 15mm,
                inner sep = 6.5pt,
                fill = white,
                align = center,
                font = \Large
            ]
    
            \node[state] (Z) at (-1.5, -1.25) {$\boldsymbol{G}$};
            \node[state] (F) at (1.5, -1.25) {$Y$};
            \path[->, color = black] (Z) edge [color = black] node {}  coordinate[pos=0.5] (curve) (F);
            
            \node[state] (X1) at (0, 1) {$\boldsymbol{X}$};
            \node[state] (V) at (-2.5, 1) {$\boldsymbol{V}$};
            \path[->] (X1) edge node[midway] {} (V);
            \path[->] (X1) edge node[midway] {} (F);
            \path[->] (X1) edge node[midway] {} (Z);
\end{tikzpicture}}\label{figure:dag}} \\
    \caption{Panels (a) and (b): schematic diagrams of dose-volume histograms (DVHs) from a single OAR of a single patient. Discretized histogram bin indices are shown below the \emph{x}-axis. The area of the shaded histograms in the differential DVH plot (a) is equivalent to the height of the dashed line in the cumulative DVH plot (b). Panel (c): DAG. Here $\boldsymbol G$ represents the observed dose distribution, $\boldsymbol V$ the stochastic intervention dose distribution, $Y$ the toxicity outcome and $\boldsymbol X$ the covariates.}
    \label{fig:dvh-schematic}
\end{figure}

Conceptually, each observed dose distribution is a distribution of an underlying random variable, which can be defined as follows. For a given OAR, let $D^* \in [0, d_{\max}]$ denote an IID continuous random variable representing radiation dose of a voxel drawn randomly (uniformly) from the volume of the OAR, with density $f_{D^*}$. Let $D = \sum_{i=1}^{\mathcal D} \boldsymbol 1_{\{(i-1)/\mathcal D \le D^*/d_{\max} < i/\mathcal D \}} \in \{1, 2, ..., \mathcal{D}\}$ denote the discretized version of this using $\mathcal{D}$ equal partitions of $[0, d_{\max}]$ that are equivalent to histogram bins of size $\Delta d  = d_{\max}/\mathcal{D}$. In the example of Figure \ref{fig:dvh-schematic}, $\mathcal{D} = 12$ and $d_{\max} = 120$ Gy, a continuous dose realization of $D^* = 44$ Gy corresponds to the discrete dose interval $D = 5$ or $[40, 50)$ Gy histogram bin. The discrete radiation dose approximates the continuous radiation dose when the number of partitions is sufficiently large and the width of the histogram bins are sufficiently small.

We next outline notation for differential and cumulative DVHs as illustrated in Figure \ref{fig:dvh-schematic} \citep{kupchak2008experience}. Differential DVH, denoted by the random vector $\boldsymbol{B} = (B_{1}, \ldots, B_{\mathcal{D}})$, where the random variables $B_d = P(D = d) = P((d-1) \Delta d \le D^* < d \Delta d) =\int_{(d-1) \times \Delta d}^{d \times \Delta d} f_{D^*}(d') \,\textrm dd' \in [0, 1]$, $d \in \{1, 2, \ldots, \mathcal{D}\}$, characterize the discrete dose distribution, giving the histograms the interpretation of random probability mass functions with $\sum_{d = 1}^{\mathcal{D}} B_d = 1$. We note that conditional on $\boldsymbol{B}$, $D$ has a multinomial distribution. The random vector $\boldsymbol{B}$ is characterized by a joint density $f_{\boldsymbol{B}}$. When not needed, we suppress patient-level index $i = 1, \ldots, n$ to avoid cluttering the notation, but note that the sampling unit for each random variable/vector defined in this section is the individual, except for the random variable $D$, which reflects the dose bin of a randomly drawn voxel, and thus in principle would be indexed by individual and the random draw.
 
Cumulative DVH, denoted by $\boldsymbol{G} = (G_1, \ldots, G_{\mathcal{D}})$, where $G_d = P(D \ge d) = P(D^* \geq (d - 1)\Delta d) \in [0, 1]$, quantifies the proportion of OAR volume receiving dose greater than or equal to the $d$-th histogram bin lower bound, and can be considered random complementary cumulative distribution functions (CDFs) of the discretized radiation dose \citep{brady2013encyclopedia}. From here on for simplicity, we refer to `cumulative DVH' as `DVH', `differential DVH' as `dDVH', and `dose bin' as `dose'.  The relationship between DVH and dDVH for a particular dose $d \in \{1, 2, \ldots, \mathcal{D}\}$ can be expressed as
\begin{align}
    G_{d} &= P(D \ge d) = \sum_{d' = 1}^{\mathcal{D}} \boldsymbol 1_{\{d' \ge d\}} B_{d'} \equiv G_{d}(\boldsymbol{B})  \label{eq:dvh} \\
    B_{d} &= P(D = d) = \sum_{d' = 1}^{\mathcal{D}} \left[\boldsymbol 1_{\{d' = d - 1\}} - \boldsymbol 1_{\{d' = d\}}\right] G_{d'} \equiv B_{d}(\boldsymbol{G}).  \label{eq:ddvh}
\end{align}
We also introduce the vector notation $\boldsymbol{B}(\boldsymbol{G}) \equiv (B_1(\boldsymbol{G}), \ldots, B_{\mathcal D}(\boldsymbol{G}))$. We note again for clarity that random variable $D$ has a distribution over the voxels of an OAR of a given individual, while random vectors $\boldsymbol{B}$ and $\boldsymbol{G}$ have a distribution over the individuals in the patient population, representing the ``distribution of dose distributions''. In places, we use the notations $D(\boldsymbol g) \equiv D \mid \boldsymbol{G} = \boldsymbol g$ and $P(D = d \mid \boldsymbol{G} = \boldsymbol g)$ to emphasize this distinction.

Finally, we let $Y \in \{0, 1\}$ denote a binary outcome variable representing a particular kind of adverse event during a specified follow-up period, or a dichotomized ordinal toxicity grade. In addition, let $\boldsymbol{X} = (X_1, \ldots, X_p)$ be a $p$-dimensional vector of patient-level covariates, which include patient demographic, clinical and tumor characteristics.

\subsection{Intervening on the entire dose distribution}

Dose distributions to organs at risk can be manipulated as part of radiotherapy treatment planning, which motivates consideration of hypothetical interventions on these. How this happens in practice is through manipulation of radiotherapy delivery parameters, which control the number, shape and orientation of the beams. Intensity-modulated radiotherapy (IMRT) enables even more control as each beam is subdivided into hundreds of beamlets with an individually controllable intensity level \citep{taylor2004intensity}. IMRT uses computerized inverse planning, where the planner specifies the prescription dose and desired normal tissue dose constraints, and the software finds iteratively an optimal plan which satisfies the constraints. Different interventions may then correspond to for example different sets of constraints being placed on the dose distribution.

Thus, assuming that dose distributions can be intervened on, we outline our causal modeling framework where the exposure is represented as the random vector $\boldsymbol{G}$ introduced above. We denote $Y^{(\boldsymbol{g})} = Y^{\left(g_{1}, \ldots, g_{\mathcal{D}}\right)} \in \{0, 1\}$ as the potential outcome had a patient's DVH been set to $\boldsymbol{g} = \left(g_{1}, \ldots, g_{\mathcal{D}}\right)$. Assuming counterfactual consistency of the Stable Unit of Treatment Value Assumption (SUTVA), the observed and potential outcome intervened at the observed dose level are equal, $Y = Y^{\left(\boldsymbol{G}\right)}$. Under strong ignorability of the observed treatment assignment mechanism, we would have positivity, $ f_{\boldsymbol{G} \, \mid \boldsymbol{X}}(\boldsymbol{g} \mid \boldsymbol{X}) > 0$ and conditional exchangeability, $\boldsymbol{G} \bigCI Y^{(\boldsymbol{g})} \mid \boldsymbol{X}$ for all $g_d \in [0, 1], \; d \in \{1, 2, \ldots, \mathcal{D}\}$ and $\boldsymbol{X}$. As all dose configurations are unlikely to be possible, below we will introduce weaker versions of the positivity assumption, relating to specific hypothetical interventions.

The causal relationships are presented through a directed acyclic graph (DAG) in Figure \ref{figure:dag}. Under the aforementioned assumptions, the population average (causal) NTCP when simultaneously intervening on the entire dose distribution by setting it at the level $\boldsymbol{g}$ can be identified using the \emph{g}-formula as (Appendix \ref{apdx:ident_multexp})
\begin{align*}
    E\left[Y^{\left(\boldsymbol{g}\right)}\right] &= E_{\boldsymbol{X}}\left\{ E\left[ Y^{\left(\boldsymbol{B} \; = \; \boldsymbol{B}(\boldsymbol{g})\right)} \mid \boldsymbol{X}\right] \right\} = E_{\boldsymbol{X}}\left\{ E\left[ Y \mid \boldsymbol{B} = \boldsymbol{B}(\boldsymbol{g}), \boldsymbol{X}\right] \right\}.  \numberthis \label{eq:ident_multexp}
\end{align*}

For the estimand (\ref{eq:ident_multexp}), we define \emph{functional monotonicity} with respect to two stochastically ordered dose distributions. Consider manipulating the dose distribution from the level $\boldsymbol{g} = (g_{1}, \ldots, g_{{\mathcal{D}}})$ to $\boldsymbol{g}^* = (g_{1}^*, \ldots, g_{{\mathcal{D}}}^*)$, where $g_d^* \leq g_d$ for all $d \in \{1, 2, \ldots, \mathcal{D}\}$, which implies stochastic ordering of the random doses specified by two levels of intervention. We denote this by $D(\boldsymbol{g}^*) \preceq D(\boldsymbol{g})$. Visually, the dose distribution intervened at $\boldsymbol{g}^*$ is located below the one at $\boldsymbol{g}$ (Supplementary Figure \ref{fig:stochast-order-DVH-solid}) and exhibits stochastic ordering (Supplementary Figure \ref{fig:stochast-order-dDVH-solid}). Biologically it is then reasonable to assume that 
\begin{equation}
D(\boldsymbol{g}^*) \preceq D(\boldsymbol{g}) \Rightarrow Y^{(\boldsymbol{g}^*)} \le Y^{(\boldsymbol{g})}\label{equation:mono}
\end{equation}
at individual level. Conditional and population level monotonicity properties follow from the individual level property as $ Y^{(\boldsymbol{g})} \le Y^{(\boldsymbol{g}^*)} \Rightarrow 
E\left[ Y^{(\boldsymbol{g}^*)} \mid \boldsymbol{X}\right] 
\le E\left[ Y^{(\boldsymbol{g})} \mid \boldsymbol{X}\right] \Rightarrow
E\left[ Y^{(\boldsymbol{g}^*)}\right] \le E\left[ Y^{(\boldsymbol{g})}\right]$.

The inner conditional expectation of (\ref{eq:ident_multexp}) can be modeled by a causal version of the monotone functional regression model proposed by \citet{schipper2007bayesian, schipper_2008} model, which can be formulated as
\begin{align}
    \psi\left(E\left[Y^{(\boldsymbol{B} \; = \; \boldsymbol{b})} \mid \boldsymbol{X}; \boldsymbol{\theta} \right]\right) = \beta_0 + \sum_{j = 1}^{p} \beta_j X_j + \sum_{d = 1}^{\mathcal{D}} \alpha_d b_{d}, \qquad \alpha_d =  \sum_{k = 1}^d h\left\{ r((k-1) \Delta d)\right\} \Delta d \label{eq:schipper}
\end{align}
for $\boldsymbol{\theta} = (\beta_0, \ldots, \beta_p, \alpha_1, \ldots, \alpha_{\mathcal{D}})'$. 
The model parametrizes a non-decreasing set of dose-specific coefficients with an identifiability constraint on $\alpha_1$,
\begin{align}
0 = \alpha_1 \leq \ldots \leq \alpha_{\mathcal{D}} \label{eq:func-mono}
\end{align}
for some choice of monotone link function $\psi(\cdot)$, and dose bin size $\Delta d$, with $r(\cdot)$ specified flexibly using smoothing splines (e.g., B-splines). The non-negative function $h(\cdot)$ in the summand is needed to make the sum and thus the $\alpha_d$s non-decreasing. In Appendix \ref{apdx:func-mono}, we show that the conditional means specified by Model \eqref{eq:schipper} together with constraints \eqref{eq:func-mono} satisfy functional monotonicity as defined above.

\subsection{Stochastic interventions}\label{section:stochastic}

As mentioned before, considering arbitrary interventions on the entire dose distribution is problematic due to positivity issues; a given prescription dose for the tumor influences the dose to OARs, and variation in the treatment plans in an observational study may also be limited. Thus, in what follows, instead of fixing the entire dose distribution, we consider fully or partially stochastic interventions where the intervention is randomly drawn from a distribution of plausible dose distributions. These can be formulated as multivariable versions of the interventional effects considered for example by \citet{didelez2006direct} and \citet{vanderweele2014effect}.

Let $\boldsymbol{V} \in [0, 1]^{\mathcal{D}}$ denote a random draw from some pre-determined distribution of possible interventions. The intervention is defined such that $\boldsymbol{V} \bigCI \boldsymbol{G} \mid \boldsymbol{X}$ and $\boldsymbol{V} \bigCI Y^{(g)} \mid \boldsymbol{X}$ hold. The implied causal relationships are illustrated in the DAG of Figure \ref{figure:dag}. For identifiability of the stochastic intervention effects, in addition to previously assumed consistency and conditional exchangeability, we require positivity/absolute continuity of the observed and intervention dose distributions, expressed as $f_{\boldsymbol{V} \mid \boldsymbol{X}}(\boldsymbol{g} \mid \boldsymbol{X}) > 0 \Rightarrow f_{\boldsymbol{G} \mid \boldsymbol{X}}(\boldsymbol{g} \mid \boldsymbol{X}) > 0$ for all $\boldsymbol{g}$ and $\boldsymbol{X}$, meaning that the support of the intervention distribution is a subset of the support of the observed distribution \citep[cf.][]{dawid2010identifying}. This is a weaker assumption than assuming positivity $f_{\boldsymbol{G} \mid \boldsymbol{X}}(\boldsymbol{g} \mid \boldsymbol{X}) > 0$ for all $\boldsymbol{g}$ and $\boldsymbol{X}$.

Causal contrasts comparing the causal risk under the hypothetical and observed exposures can be specified for example in terms of risk differences $E\left[Y^{(\boldsymbol{V})}\right] - E\left[Y^{(\boldsymbol{G})}\right]$ or risk ratios $E\left[Y^{(\boldsymbol{V})}\right]/E\left[Y^{(\boldsymbol{G})}\right]$. Conditional average contrasts, conditional on selected features from $\boldsymbol X$, could be introduced similarly, and identified under the same assumptions. This would also allow estimation of individualized intervention effects.

While the above definitions are general, to aid interpretation and to further alleviate positivity issues, we introduce the split $\boldsymbol{V} = (V_d, \boldsymbol{V}_{-d})$, where $V_d$ denotes the volume receiving at least dose $d$ and $\boldsymbol{V}_{-d} = (V_1, ..., V_{d - 1}, V_{d + 1}, ..., V_{\mathcal{D}})$ the remaining dose distribution. In what follows we consider pointwise interventions defined such that only the volume $V_d$ receiving at least dose $d$ is directly intervened upon, while the rest of the dose distribution is drawn from the conditional density of the observed dose distributions, that is, $f_{\boldsymbol{V}_{-d} \mid V_d, \boldsymbol{X}}(\boldsymbol{g}_{-d} \mid g, \boldsymbol{X}) = f_{\boldsymbol{G}_{-d} \mid G_d, \boldsymbol{X}}(\boldsymbol{g}_{-d} \mid g, \boldsymbol{X})$ for all $g$, $\boldsymbol{g}_{-d}$ and $\boldsymbol{X}$. The intervention density $f_{V_{d} \mid \boldsymbol{X}}(g \mid \boldsymbol{x})$ can depend on the covariates $\boldsymbol{x}$, or can be specified independently of these. Some example interventions on $V_d$ could include truncating the distribution under a given dose or exponential tilting. Under the intervention, the causal NTCP under the hypothetical stochastic intervention can be identified as (Appendix \ref{apdx:stochast-ident}),
\begin{align*}
    E&\left[Y^{\left(V_d, \; \boldsymbol{V}_{-d}\right)}\right] = \int_{\boldsymbol{x}} \int_g E\left[Y \mid G_d = g, \boldsymbol{X} = \boldsymbol{x} \right]  \dfrac{f_{V_{d} \mid \boldsymbol{X}}(g \mid \boldsymbol{x})}{f_{G_{d} \mid \boldsymbol{X}}(g \mid \boldsymbol{x})} f_{G_{d}, \boldsymbol{X}}(g, \boldsymbol{x}) \, \mathrm{d}g \; \mathrm{d}\boldsymbol{x}, \numberthis \label{eq:stochast-int}
\end{align*}
that is, the marginal NTCP weighted by the ratio of transitioning from the observed dose to the intervention distribution at dose $d$ (importance sampling weights, cf. \citealp{saarela2015bayesian,liu2020estimation}). The same quantity can be alternatively identified in terms of the \emph{g}-formula,
\begin{align*}
    E&\left[Y^{(V_d, \; \boldsymbol{V}_{-d})}\right] \\
    &= \int_{\boldsymbol{x}} \int_{g} E_{\boldsymbol{G}_{-d} \mid G_d = g, \boldsymbol{X} = \boldsymbol{x}}\left\{ E\left[Y \mid G_d = g, \boldsymbol{G}_{-d}, \boldsymbol{X} = \boldsymbol{x} \right] \right\} f_{V_d \mid \boldsymbol{X}}(g \mid \boldsymbol{x}) f_{\boldsymbol{X}}(\boldsymbol{x}) \, \mathrm{d}g \, \mathrm{d}\boldsymbol{x}, \numberthis \label{equation:hypo_stand}
\end{align*}
and can be interpreted as the NTCP had patients from the original population been assigned treatment plans from the intervention distribution (Appendix \ref{apdx:stochast-ident}).

\subsection{Pointwise interventions}

To arrive at an interpretable model in the next subsection, we pay particular attention to a special case of the interventional effects introduced in Section \ref{section:stochastic} where the intervention at dose $d$ is deterministic such that $V_{d} = g$ with probability one, while still drawing the remaining dose distribution $\boldsymbol{V}_{-d}$ randomly. For example, limiting the volume receiving more than 80 Gy to exactly $10\%$ of the OAR, while drawing the rest of the intervention distribution $\boldsymbol{V}_{-d}$ from the conditional density $f_{\boldsymbol{V}_{-d} \mid V_d, \boldsymbol{X}}(\boldsymbol{g}_{-d} \mid g, \boldsymbol{X}) = f_{\boldsymbol{G}_{-d} \mid G_d, \boldsymbol{X}}(\boldsymbol{g}_{-d} \mid g, \boldsymbol{X})$. This ensures that the interventions are possible (positivity) as long as $V_{d} = g$ is supported by the observed data.

The ``pointwise'' causal NTCP, indexed with respect to dose $d$ and volume $g$, can be identified as (Appendix \ref{apdx:ptwise_ident}),
\begin{align}
     E\left[Y^{\left( g, \; \boldsymbol{V}_{-d} \right)} \right] &= E_{\boldsymbol{X}} \left\{ E_{\boldsymbol{G}_{-d} \mid G_{d} = g, \boldsymbol{X}}\left( E\left[Y \mid G_{d} = g, \boldsymbol{G}_{-d}, \boldsymbol{X} \right] \right) \right\}\nonumber \\ 
     &= E_{\boldsymbol{X}} \left\{ E\left[Y \mid G_{d} = g, \boldsymbol{X} \right] \right\},
     \label{eq:monotonicity_univariate}
\end{align}
where the innermost expectation on the right hand side of the first equality could be specified as in \eqref{eq:schipper}. However, we want to avoid modeling the conditional dose distribution in the middle expectation, as this would require specifying a non-standard high-dimensional nuisance model. Instead, in Section \ref{section:msm} we pursue parametrizing the marginal effect of the pointwise intervention directly, based on identification in terms of the second equality in \eqref{eq:monotonicity_univariate}.

Before that, we show that the pointwise causal NTCP exhibits a bivariable monotone relationship with respect to dose and volume assuming model (\ref{eq:schipper}) for the outcome and that the average dose distributions intervened at the dose-volume coordinates are stochastically ordered, that is, for volumes $g^{*} \le g$ at a fixed dose $d$ and doses $d^{*} \le d$ at a fixed volume $g$, we assume 
\begin{equation}
D_{d}(g^*) \preceq D_{d}(g) \quad\textrm{and}\quad D_{d^*}(g) \preceq D_{d}(g),\label{equation:so}
\end{equation}
where $D_{d}(g) \in \{1, 2, \ldots, \mathcal{D}\}$ is a random variable distributed by the probabilities
\begin{align*}
P(D_{d}(g) = d') = E_{\boldsymbol{G}_{-d} \mid G_{d}}\left[P(D = d' \mid G_{d} = g, \boldsymbol{G}_{-d})\right]. \numberthis \label{eq:cond-dose-relation}
\end{align*}
It is easy to see that the probabilities in \eqref{eq:cond-dose-relation} sum up to one. Assumption \eqref{equation:so} is illustrated in Supplementary Figure \ref{fig:stochast-order-average-DVH}. Under model \eqref{eq:schipper} and assumption \eqref{equation:so} we now have that (Appendix \ref{apdx:bivar-mono}),
\begin{align}
      E[Y^{(g^*, \; \boldsymbol{V}_{-d})}] \leq E[Y^{(g, \; \boldsymbol{V}_{-d})}] \label{eq:bivar-mono-volume}
\end{align}
and
\begin{align}
     E[Y^{(g, \; \boldsymbol{V}_{-d^*})}] \leq E[Y^{(g, \; \boldsymbol{V}_{-d})}]. \label{eq:bivar-mono-dose}
\end{align}
for all $g^* \leq g$ and $d^* \leq d$. From \eqref{eq:bivar-mono-volume} and \eqref{eq:bivar-mono-dose}, we finally obtain the bivariable monotonicity property
\begin{align*}
     D_{d^*}(g^*) &\preceq D_{d}(g)  \quad \Rightarrow \quad E[Y^{(g^*, \; \boldsymbol{V}_{-d^*})}] \leq E[Y^{(g, \; \boldsymbol{V}_{-d})}]. \numberthis \label{eq:bivar-mono}
\end{align*}
for all $g^* \leq g$ and $d^* \leq d$. 

Alternatively, without assuming model \eqref{eq:schipper}, the above monotonicity properties follow from the biological assumption \eqref{equation:mono} and the added regularity condition \eqref{equation:so} on the dose distributions.

\subsection{Marginal structural models}\label{section:msm}

In parametrizing the causal estimand in (\ref{eq:monotonicity_univariate}), a marginal structural model (MSM) for the conditional causal NTCP intervened at volume $g$ at $d$ could be specified as
\begin{align}
    \psi\left(E\left[Y_i^{\left(g, \; \boldsymbol{V}_{-d, i} \right)} \mathrel{\bigg|} \boldsymbol{x}_i;  \boldsymbol{\theta} \right] \right) =  \beta_{0d} + \sum_{j = 1}^{p} \beta_{jd} x_{ij} + \lambda_d(g)  \label{eq:MSMslice}
\end{align}
for some choice of monotone link function $\psi(\cdot)$ and a monotonic non-decreasing function $\lambda_d : [0,1] \rightarrow \mathbbm{R}$. However, such a model would have to be fitted separately at every $d \in \{1, \ldots, \mathcal{D}\}$, and the monotonicity property \eqref{eq:bivar-mono-dose} over the doses could not be enforced. Instead, we aim to fit the volume effects at every dose jointly and specify a model
\begin{align}
    \psi\left(E\left[Y_i^{\left(g, \; \boldsymbol{V}_{-d, i} \right)} \mathrel{\bigg|} \boldsymbol{x}_i;  \boldsymbol{\theta} \right] \right) =  \beta_0 + \sum_{j = 1}^{p} \beta_j x_{ij} + \lambda(d, g)  \label{eq:MSM}
\end{align}
with a bivariable monotone increasing function $\lambda: \{1, \ldots, \mathcal{D}\} \times [0,1] \rightarrow \mathbbm{R}$, where $\boldsymbol{\theta} = (\beta_0, \ldots, \beta_p, \lambda(\cdot, \cdot))$. Despite being conditional on covariates, we still refer to these models as ``marginal'' models since they parametrize the effect of a pointwise intervention, marginalized over the rest of the dose distribution. We focus on covariate conditional models, as individual-level predictions are usually the objective in NTCP modeling. However, in principle we could specify a model that is also marginal over the covariates as 
$$\psi\left(E\left[Y_i^{\left(g, \; \boldsymbol{V}_{-d, i} \right)} \mid \boldsymbol{\theta} \right] \right) =  \beta_0 + \lambda(d, g).$$ 
We discuss the estimation of such models briefly in Section \ref{section:discussion}. 

In Models \eqref{eq:MSMslice} and \eqref{eq:MSM} the covariate effects don't have to be assumed linear on the link function scale, generalized additive (GAM) type specifications are also possible. Models \eqref{eq:MSMslice} and \eqref{eq:MSM} can also be extended to parametrize effect modification; modeling a small number of effect modifiers non-parametrically would in principle be a straightforward extension. For example, if the potential confounder $p$ is also hypothesized to be an effect modifier, and takes values $x_{ip} \in \{0,1\}$, the right hand side of \eqref{eq:MSM} could be specified as $\beta_0 + \sum_{j = 1}^{p-1} \beta_j x_{ij} + \lambda(d, g, x_{ip})$, which would entail non-parametric estimation of two monotonic regression surfaces $\lambda(d, g, 0)$ and $\lambda(d, g, 1)$. If on the other hand $x_{ip} \in \mathbbm{R}$ but its effect on NTCP could be assumed monotonic, $\lambda(d, g, x_{ip})$ could be non-parametrically estimated as a non-decreasing function of all three arguments. Both of these cases could be implemented in the estimation approach proposed in Section \ref{section:msmestimation}. In the proposed estimation approach, the function values $\lambda_d(0)$ and $\lambda(0,0)$ also fix the absolute level of the linear predictor, so in practice in \eqref{eq:MSMslice} and \eqref{eq:MSM} we set $\beta_{0d} = 0$ and $\beta_0 = 0$ for identifiability.

\section{Proposed estimators}

In this section, we present estimators for the causal NTCP for the deterministic (\ref{eq:monotonicity_univariate}) and stochastic (\ref{eq:stochast-int}) interventions based on \emph{g}-computation and marginal structural models.

\subsection{\emph{g}-computation}

Under the pointwise intervention, the Monte Carlo integration-based estimator is
\begin{align*}
    E\left[Y_i^{(g, \; \boldsymbol{V}_{-d, i})}; \hat{\boldsymbol{\theta}} \right] &= \dfrac{1}{n} \sum_{i = 1}^n \left[  \dfrac{1}{m} \sum_{j = 1}^m E[Y_i \mid G_{d, i} = g, \boldsymbol{G}_{-d,i} = \boldsymbol{g}_{-d,ij}, \boldsymbol{x}_i; \; \hat{\boldsymbol{\theta}}] \right],
\end{align*}
where $\boldsymbol{g}_{-d, i1}, \ldots, \boldsymbol{g}_{-d, im}$ are i.i.d. draws from $f_{\boldsymbol{G}_{-d} \mid G_d, \boldsymbol{X}}(\boldsymbol{g}_{-d} \mid  g, \boldsymbol{X}_i)$. To incorporate monotonicity, the conditional mean is estimated using model (\ref{eq:schipper}) and stochastic ordering of the conditional dose distributions is assumed. Similarly, an estimator under the stochastic intervention is given by
{\begin{align*}
    E\left[Y_i^{(V_{d, i}, \; \boldsymbol{V}_{-d,i})}; \hat{\boldsymbol{\theta}} \right] = \dfrac{1}{n} \sum_{i = 1}^n \left[ \dfrac{1}{l} \sum_{k = 1}^l \left\{ \dfrac{1}{m} \sum_{j = 1}^m E[Y_i \mid G_{d, i} = v_{d, ik}, \boldsymbol{G}_{-d, i} = \boldsymbol{g}_{-d, ijk}, \boldsymbol{x}_i; \; \hat{\boldsymbol{\theta}}] \right\}\right],
\end{align*}}%
where $\boldsymbol{g}_{-d, i1k}, \ldots, \boldsymbol{g}_{-d, imk}$ are i.i.d. draws from $f_{\boldsymbol{G}_{-d} \mid G_d, \boldsymbol{X}}(\boldsymbol{g}_{-d} \mid v_{d, ik}, \boldsymbol{x}_i)$ and $v_{d,i1}, \ldots, v_{d,il}$ are i.i.d. draws from $f_{V_{d} \mid \boldsymbol{X}}(v_{d} \mid \boldsymbol{x}_i)$. The two estimators differ with the stochastic intervention requiring an additional averaging over the distribution of $V_d \mid \boldsymbol{X}$. Although the above estimators incorporate monotonicity, they require parametrization and estimation of the conditional dose distribution, unless this is fixed as part of the stochastic intervention. Also, as discussed before, the model estimates $\hat{\boldsymbol{\theta}}$ would not be directly interpretable as causal effects. Due to these reasons, instead of pursuing these estimators, we proceed to directly parametrize the marginal effects of interest.

\subsection{Marginal structural models}\label{section:msmestimation}

Models of type \eqref{eq:MSMslice} could be fitted using for example monotonic generalized additive models \citep[e.g.][]{pya2015shape}. However, instead of fitting separate models for the volume effects at each dose, we are primarily interested in model \eqref{eq:MSM} which parametrizes all of these at once. Fitting of such model is based on a replicated data set with $n \times \mathcal D$ rows, where the outcome of each individual is repeated $\mathcal D$ times, and the two variables for fitting the function $\lambda(d, g)$ are the grid of dose bins and the corresponding $\mathcal D$ observed cumulative volumes for each individual. For a binary toxicity outcome $Y_i$, the resulting objective function (quasi-log-likelihood) is given by
\begin{align}\label{equation:quasiloglik}
l_{n \mathcal D} (\boldsymbol\theta) &=
\sum_{i=1}^n \sum_{d=1}^{\mathcal D} Y_i \log \left[ \psi^{-1}\left(\sum_{j = 1}^p \beta_j x_{ij} + \lambda(d, g_{d,i})\right) \right] \nonumber \\
&\quad+ \sum_{i=1}^n \sum_{d=1}^{\mathcal D} (1 - Y_i) \log\left[ 1-\psi^{-1}\left(\sum_{j = 1}^p \beta_j x_{ij} + \lambda(d, g_{d,i}) \right)\right],
\end{align}
where $\psi^{-1}(a) = \text{expit}(a) = [1 + \exp(-a)]^{-1}$. In a different context, similar data replication has been considered by \citet{leisenring2000comparisons} and results in valid estimation of the mean structure in a generalized estimation equation framework under working independence correlation structure.

In principle, any model parametrizing bivariable monotonicity could be used to fit the function $\lambda$, as long as the resulting estimator satisfies the property $g^* \le g$ and $d^* \le d$ $\Rightarrow \hat\lambda(d^*,g^*) \le \hat\lambda(d,g)$. We also need a semi-parametric model as in \eqref{eq:MSM} to fit the covariate effects, as simultaneous non-parametric modeling of both the covariate effects and the dose-response relationship is difficult due to the curse of dimensionality. Here we will adapt the model of \citet{saarela2011method, saarela2015non, saarela2023bayesian} who proposed Bayesian multiple monotone regression based on a marked-point process construction for continuous, binary, ordinal and time-to-event outcomes. This model uses random configurations of points (covariate coordinates) and ordered marks (function levels) to construct piecewise constant regression function realizations, and the reversible jump Markov chain Monte Carlo (MCMC) sampling algorithm used for estimation makes local modifications to the functions without violating the monotonicity property. The resulting posterior mean surfaces averaged over the MCMC sample can approximate both smooth and non-smooth functions. The estimation procedure is implemented in R package \texttt{monoreg} \citet{monoreg}. While the objective function $l_{n \mathcal D} (\boldsymbol\theta)$ resulting from replicating the toxicity outcomes at each dose is not a true log-likelihood, it can still be combined with a prior $p(\boldsymbol \theta)$ and renormalized as
 \begin{align*}
 p_{n \mathcal D} (\boldsymbol\theta) = \frac{\exp\{l_{n \mathcal D} (\boldsymbol\theta)\} p(\boldsymbol\theta)}
 {\int_{\Theta}\exp\{l_{n \mathcal D}(\boldsymbol\theta)\} p(\boldsymbol\theta) \,\textrm d\boldsymbol\theta}
 \end{align*}
which defines a quasi-posterior distribution in the sense of \citep{chernozhukov2003mcmc}. This is a true density over $\Theta$, and can be sampled from using MCMC algorithms. The resulting quasi-posterior mean $\hat{\boldsymbol\theta} = \int_{\Theta} \boldsymbol\theta p_{n \mathcal D} (\boldsymbol\theta) \,\textrm d\boldsymbol\theta$ interpreted as a frequentist point estimate. Confidence bands can be produced via clustered bootstrap, resampling the individuals $i = 1,\ldots, n$ and reconstructing the replicate dataset of $n \times \mathcal D$ for each resample.
 
Under the pointwise interventions, the regression function estimate $\hat\lambda$ (quasi-posterior mean surface evaluated on a grid of $d$ and $g$ values) can be interpreted directly in terms of the covariate conditional intervention effects on the link function scale. Based on the identification result \eqref{eq:monotonicity_univariate}, population average effects on the original scale of the outcome may be calculated based on the fitted MSM as
\begin{align}
    &E_{\text{MSM}}\left[Y^{\left(g, \; \boldsymbol{V}_{-d}\right)}; \hat{\boldsymbol{\theta}} \right] = \dfrac{1}{n} \sum_{i = 1}^n \psi^{-1}\left( \sum_{j = 1}^p \hat{\beta}_j x_{ij} + \hat{\lambda}(d, g) \right) \label{eq:msm-pointwise}.
\end{align}
Based on the identification result \eqref{eq:stochast-int}, the effects of stochastic interventions can be obtained from taking an importance sampling weighted average of the pointwise effects as
\begin{align}
    E_{\text{MSM}}\left[Y^{(V_{d}, \; \boldsymbol{V}_{-d})}; \hat{\boldsymbol{\theta}} \right] = \dfrac{1}{n} \sum_{i = 1}^n \left[ \psi^{-1}\left( \sum_{j = 1}^p \hat{\beta}_j x_{ij} + \hat{\lambda}(d, g_{d, i}) \right) \dfrac{f_{V_{d} \mid \boldsymbol{X}}(g_{d,i} \mid \boldsymbol{x_i})}{f_{G_{d} \mid \boldsymbol{X}}(g_{d,i} \mid \boldsymbol{x_i})}\right]. \label{eq:msm-stochastic}
\end{align}
To estimate a causal contrast, such as risk ratio, \eqref{eq:msm-stochastic} can be compared to the observed counterpart $E\left[Y^{(G_{d}, \; \boldsymbol{G}_{-d})} \right]$, which can be calculated for example as the empirical risk $\frac{1}{n} \sum_{i = 1}^n Y_i$, or the model-based risk $\frac{1}{n} \sum_{i = 1}^n \psi^{-1}\left( \sum_{j = 1}^p \hat{\beta}_j x_{ij} + \hat{\lambda}(d, g_{d, i}) \right)$. The latter will ensure monotonicity of the causal contrasts.

To obtain the importance sampling weights, zero-and-one-inflated Beta regression models \citep{ospina2012general} for the distribution $f_{G_{d} \mid \boldsymbol{X}}$ could be considered, as this distribution is likely to be concentrated at the bounds. We note that unlike the \emph{g}-computation estimators, \eqref{eq:msm-pointwise} and \eqref{eq:msm-stochastic} do not require modeling the high-dimensional joint exposure distribution. The pointwise effects can be obtained entirely without modeling the exposure, while for the stochastic intervention effects, only the univariate distribution $f_{G_{d} \mid \boldsymbol{X}}$ is needed (with $f_{V_{d} \mid \boldsymbol{X}}$ known by definition). Large importance sampling weights may indicate practical violations of the positivity assumption $f_{V_d \mid \boldsymbol{X}}(g \mid \boldsymbol{x}) > 0 \Rightarrow f_{G_d \mid \boldsymbol{X}}(g \mid \boldsymbol{x}) > 0$, with truncation/trimming possible to control the bias/variance tradeoff.

\section{Simulation}\label{section:simulation}

\subsection{Simulation design}\label{section:simulation-design}

We carried out two simulation experiments to check the performance of the proposed estimation procedure. The objective of the first simulation was to demonstrate the ability of the flexible model formulations to capture the true dose-response relationship, compared to parametric specifications. The objective of the second simulation was to demonstrate the ability of the covariate adjustment to control for confounding.

In the first simulation, for $i = 1, \ldots, n = 100$ and $500$, we denote $X_i \sim \text{Bernoulli}(0.4)$ as a binary confounder. The dose distribution was discretized into $\mathcal{D} = 26$ histogram bins of size $\Delta d = \frac{20}{26}$ on the closed interval $[30, 50]$, resulting in the bins $\{[30, 30 + \frac{20}{26}), [30 + \frac{20}{26}, 30 + 2 \times\frac{20}{26}), ..., [30 + 25 \times \frac{20}{26}, 50]\}$. The corresponding volumes were simulated as complementary CDFs from a Normal distribution, $G_{d, i} = 1 - \Phi(30 + (d - 1) \Delta d \mid \mu_i(X_i), \sigma_i), \; d \in \{1, \ldots, 26\}$, where $\Phi(\cdot; \mu_i, \sigma_i)$ is the CDF of a $N(\mu_i, \sigma_i^2)$ random variable. The discretized dDVHs $B_{d, i}$ were calculated from $G_{d, i}$ using formula \eqref{eq:ddvh}. The patient-specific mean doses $\mu_i$ were simulated from a mixture of two domain-shifted Beta distributions dependent on the confounder, $\mu_i(X_i) = a_{\text{min}} + (a_{\text{max}} - a_{\text{min}}) \times \text{Beta}(\alpha_i(X_i), \beta_i(X_i))$, where $\alpha_i(X_i) = 2 \times \boldsymbol{1}_{\{X_i = 1\}} + \frac{4}{3} \times \boldsymbol{1}_{\{X_i = 0\}}$, $\beta_i(X_i) = 2 + \frac{4}{3} - \alpha_i(X_i)$,  $a_{\text{min}} = 35$ and $a_{\text{max}} = 45$ are the upper and lower limits of the domain of mean dose, while the standard deviations were simulated from $\sigma_i \sim U(1, 2)$. 

The binary normal tissue complication outcome was simulated using the mean dose and confounder, $Y_i \sim \mathrm{Bernoulli}(\text{expit}\{\gamma_0 + \gamma_1 \mu_i + \gamma_2 X_i\})$ with $(\gamma_0, \gamma_1, \gamma_2) = (-18, 0.45, 0.5)$. Note that $P(Y_i = 1 \mid \boldsymbol{G}_i, X_i) = P(Y_i = 1 \mid \boldsymbol{B}_i, X_i) = \text{expit}\{\gamma_0 + \gamma_1 \mu_i + \gamma_2 X_i\} = P(Y_i = 1 \mid X_i; \mu_i, \sigma_i)$ as the mean and standard deviation dose distribution parameters are sufficient statistics of the dose distribution. We define the stochastic intervention of interest as imposing an upper volume threshold of $q = 0.8$ at dose bin $d^* = 14$ corresponding to the histogram bin $[30 + 13 \times \frac{20}{26}, 30 + 14 \times \frac{20}{26})$, represented by the truncated distribution of the observed DVH volume at dose bin $d^*$,
\begin{align}
    f_{V_{d^*} \mid X}(g \mid X) = \dfrac{f_{G_{d^*}}(g \mid X) \cdot \boldsymbol{1}_{\{g \in [0, q]\}}}{F_{G_{d^*} \mid X}(q \mid X) - F_{G_{d^*} \mid X}(0 \mid X)} = \dfrac{f_{G_{d^*}}(g \mid X) \cdot \boldsymbol{1}_{\{g \in [0, q]\}}}{F_{G_{d^*} \mid X}(q \mid X)}
\end{align}
The data generating mechanism is illustrated in Supplementary Figure \ref{figure:dag_sim}.

Based on the above conditional data generating mechanism, the true marginal pointwise causal NTCP at dose-volume point ($d, g$) can be derived as (Appendix \ref{apdx:sim-estimands}),
\begin{align*}
    E&\left[Y^{\left(g, \; \boldsymbol{V}_{-d} \right)}\right] = \sum_{x = 0}^1 \left[ \int_{\mu_{\text{min}}(g)}^{\mu_{\text{max}}(g)} \text{expit}\{\gamma_0 + \gamma_1 \mu + \gamma_2 x\} \right. \\
    & \times \left. \dfrac{ (\mu - a_{\text{min}})^{\alpha(x) - 1} (a_{\text{max}} - \mu)^{\beta(x) - 1} |\text{\textbf{J}}|(\mu, g)}{\int_{\mu_{\text{min}}(g)}^{\mu_{\text{max}}(g)} (\mu - a_{\text{min}})^{\alpha(x) - 1} (a_{\text{max}} - \mu)^{\beta(x) - 1} |\text{\textbf{J}}|(\mu, g) \, \mathrm{d}\mu} \, \mathrm{d}\mu \right]  P(X = x) \numberthis, \label{eq:cond-risk}
\end{align*}
for which $\mu_{\text{min}}(g)$ and $\mu_{\text{max}}(g)$ are  minimum and maximum mean doses that solve for $\mu$ in $g = 1 - \Phi(d \mid x; \mu, \sigma)$ for a given value of $g$, and the Jacobian is given by
\begin{align}
    |\text{\textbf{J}}|(\mu, g) &= [\sigma(\mu, g)]^2 \left|\int_0^{d}   \phi\left(\frac{t - \mu}{\sigma(\mu, g)}\right) \left[ 1 - \left(\frac{t - \mu}{\sigma(\mu, g)}\right)^2 \right] \, \mathrm{d}t\right|^{-1}. \label{eq:jacobian}
\end{align}
The pointwise causal NTCP are calculated over a $26 \times 26$ grid of doses on $[30, 50]$, scaled to $[0, 1]$ for modeling purposes and relative volumes on $[0, 1]$. Similarly, the true causal NTCP under the stochastic intervention is (Appendix \ref{apdx:sim-estimands}),
\begin{align*}
    E&\left[Y^{(V_{d^*}, \boldsymbol{V}_{-d^*} )}\right] = \sum_{x = 0}^1 \left[ \int_0^q \int_{\mu_{\text{min}}(g)}^{\mu_{\text{max}}(g)} 
    \text{expit}\{\gamma_0 + \gamma_1 \mu + \gamma_2 x\}  \right. \\
    &\times \left. \dfrac{ (\mu - a_{\text{min}})^{\alpha_i(x) - 1} (a_{\text{max}} - \mu)^{\beta_i(x) - 1} |\text{\textbf{J}}|(\mu, g) \, }{\int_0^q \int_{\mu_{\text{min}}(g)}^{\mu_{\text{max}}(g)} (\mu - a_{\text{min}})^{\alpha_i(x) - 1} (a_{\text{max}} - \mu)^{\beta_i(x) - 1} |\text{\textbf{J}}|(\mu, g)  \, \mathrm{d}\mu  \, \mathrm{d}g} \, \mathrm{d}\mu
    \, \mathrm{d}g \right] P(X = x)
\end{align*}

We compare the performance of 4 logistic MSMs where the functional forms are parametrized by linear terms (\ref{eq:sim_mod_logi}),  polynomial terms (\ref{eq:sim_mod_polylog}), two additive flexible monotonic functions (\ref{eq:sim_mod_add}), and a bivariable flexible monotonic function (\ref{eq:sim_mod_bivar}): 
\begin{align}
    \text{logit}\left(E\left[Y^{(g, \; \boldsymbol{V}_{-d})} \mathrel{\big|} X; \boldsymbol{\beta}\right]\right) &= \beta_0 + \beta_1 X + \beta_2 d + \beta_3 g \label{eq:sim_mod_logi} \\
    \text{logit}\left(E\left[Y^{(g, \; \boldsymbol{V}_{-d})} \mathrel{\big|} X; \boldsymbol{\beta}\right]\right) &= \beta_0 + \beta_1 X + \beta_2 d + \beta_3 g + \beta_4 d^2 + \beta_5 g^2 \label{eq:sim_mod_polylog} \\
    \text{logit}\left(E\left[Y^{(g, \; \boldsymbol{V}_{-d})} \mathrel{\big|} X; \boldsymbol{\theta}\right]\right) &= \beta_0 + \beta_1 X + \lambda_1(d) + \lambda_2(g) \label{eq:sim_mod_add} \\
    \text{logit}\left(E\left[Y^{(g, \; \boldsymbol{V}_{-d})} \mathrel{\big|} X; \boldsymbol{\theta}\right]\right) &= \beta_0 + \beta_1 X + \lambda(d, g)  \label{eq:sim_mod_bivar}
\end{align}
All the models were fitted through the \texttt{monoreg} package \citet{monoreg} using 2000 iterations with a burn-in of 1000 iterations.

Using model (\ref{eq:sim_mod_bivar}) as the causal model, the estimator for the pointwise causal NTCP at ($d, g$) follows (\ref{eq:msm-pointwise}), while the NTCP intervened under the truncated distribution is given by
\begin{align*}
    E\left[Y^{(V_{d^*}, \; \boldsymbol{V}_{-d^*})}; \hat{\boldsymbol{\theta}} \right] &= \dfrac{1}{n} \sum_{i = 1}^n \left[ \psi^{-1}\left(\hat{\beta}_0 + \hat{\beta}_1 X_i + \hat{\lambda}(d^*, G_{d^*, i} ) \right) \cdot \dfrac{\mathbf{1}_{\{G_{d^*,i} \leq q\}}}{\hat{F}_{G_{d^*, i} \mid X_i}(q \mid X_i)} \right],
\end{align*}
where the dose distribution functions could be estimated empirically for the levels of the binary covariate. Absolute bias, Monte Carlo standard deviation and root mean squared error were calculated at every dose-volume grid point to evaluate the out-of-sample performance of the estimators, while the Brier score, deviance, the effective number of parameters and the Deviance Information Criteria were calculated to describe in-sample performance \citep{gelman1995bayesian}. The average of the out-of-sample metrics over the dose-volume domain were reported to provide a one-number summary. The design of the second simulation study is described in Appendix \ref{apdx:sims-scen}. The R code to reproduce results from both simulation studies are available at \url{https://github.com/thaisontang/marginalntcp}.

\subsection{Simulation results}

\begin{figure}[H]
    \centering
    \vspace{-1.5em}
     \subfloat[Logistic]{\includegraphics[width = 0.33\textwidth]{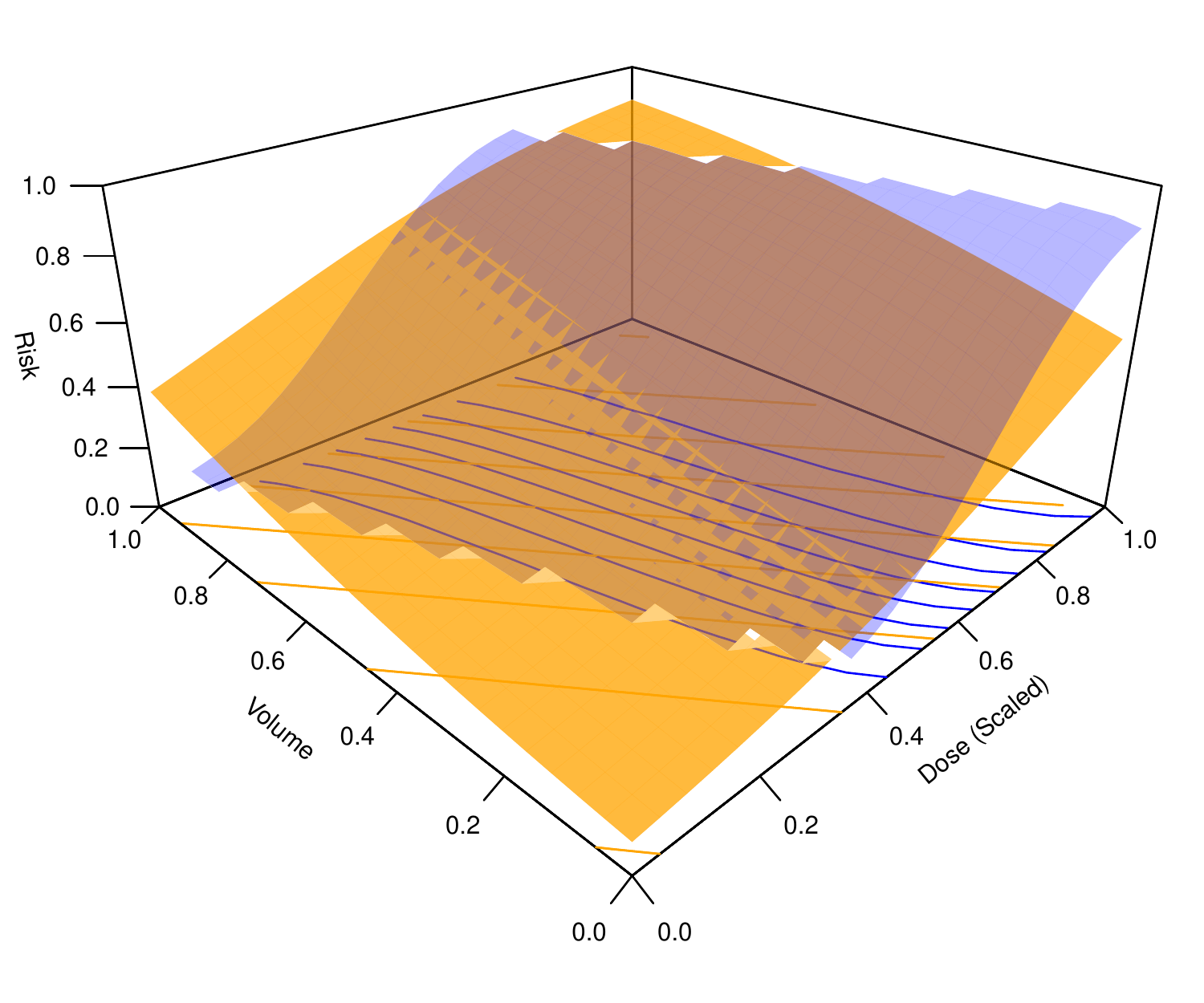} \label{figure:sim_a}} \hspace{0.6cm}
     \subfloat[Polynomial logistic]{\includegraphics[width = 0.33\textwidth]{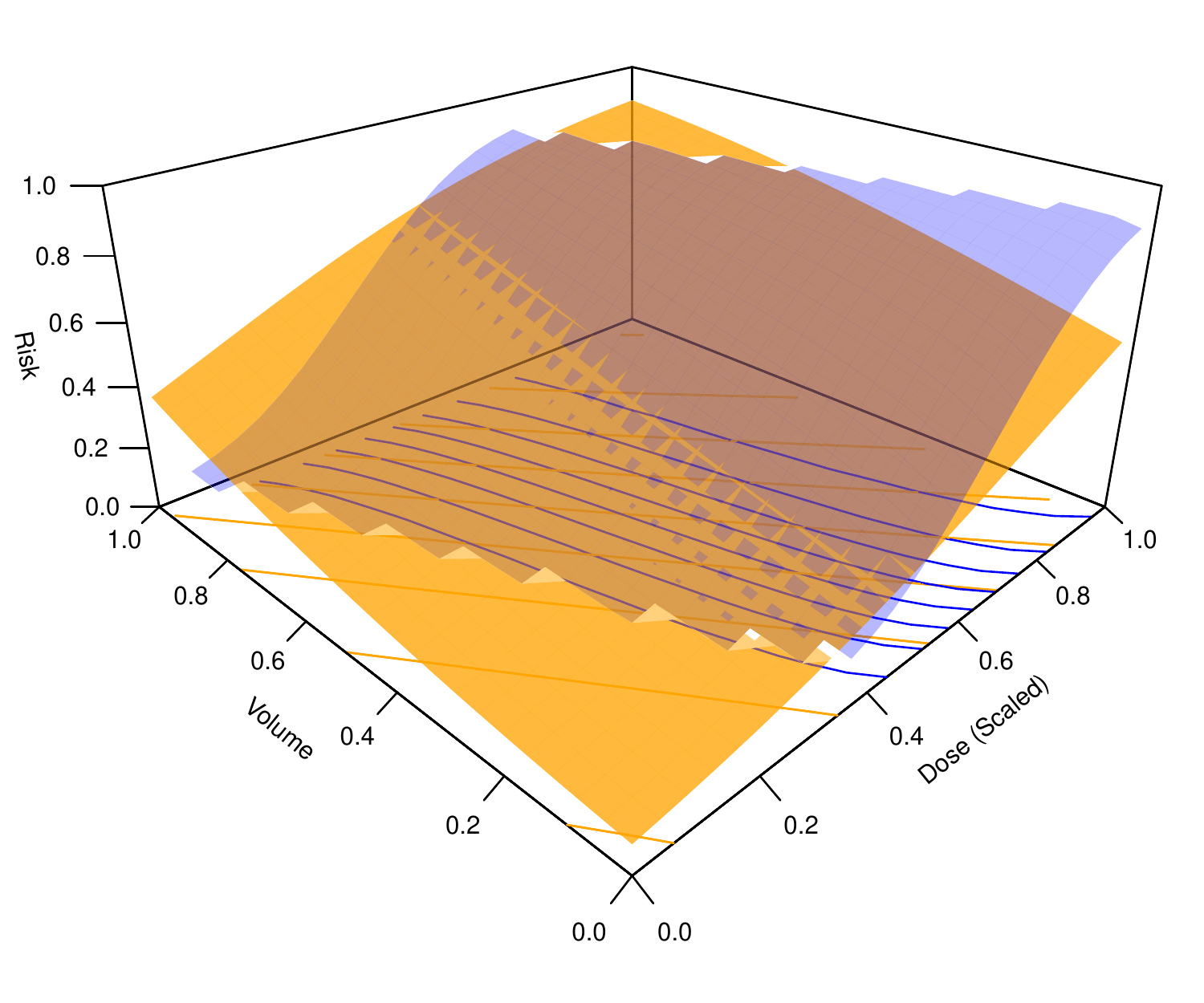} \label{figure:sim_b}}  \\[-0.5em]
    \subfloat[Additive]{\includegraphics[width = 0.33\textwidth]{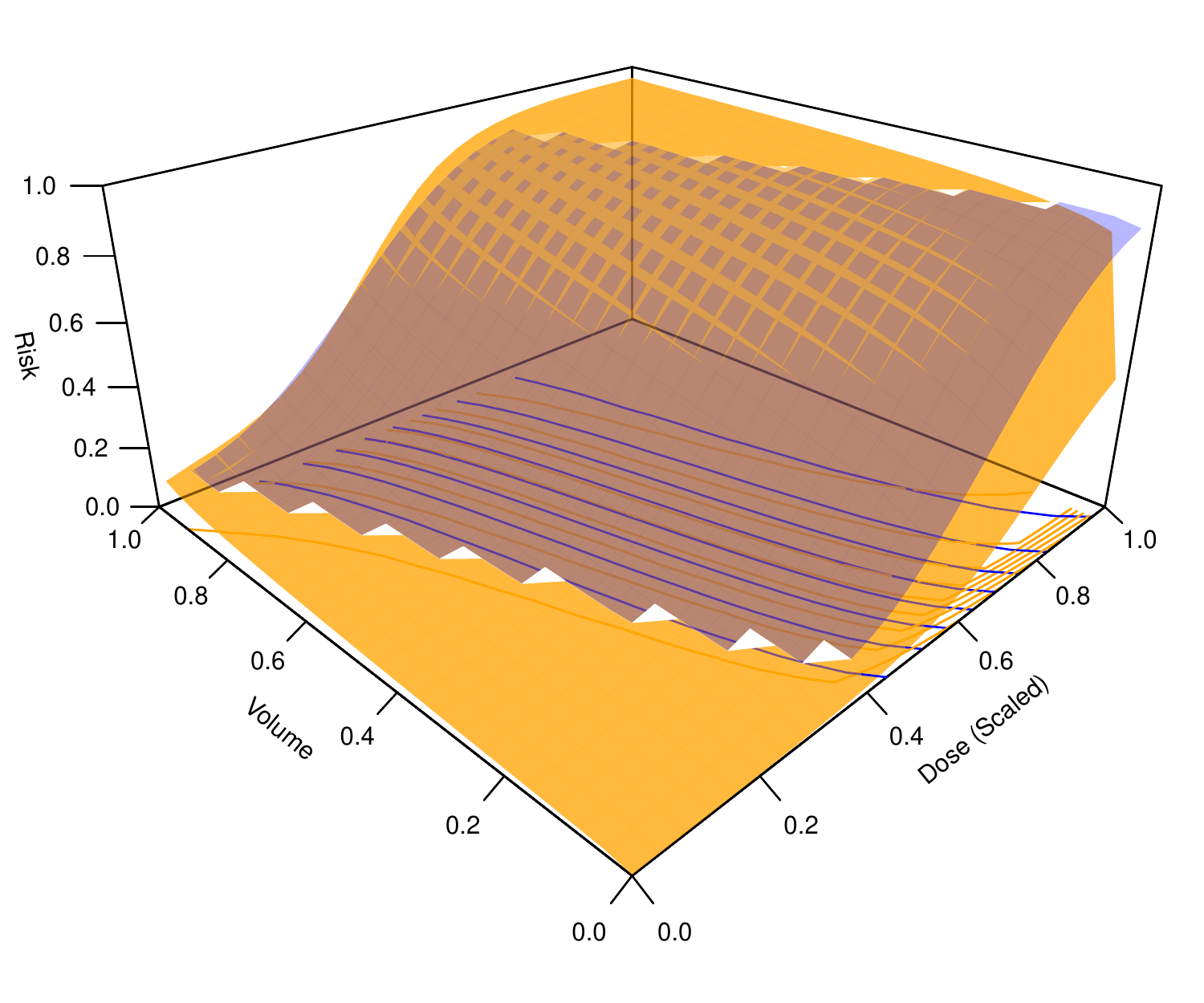} \label{figure:sim_c}} \hspace{0.6cm}
     \subfloat[Bivariable monotone]{\includegraphics[width = 0.33\textwidth]{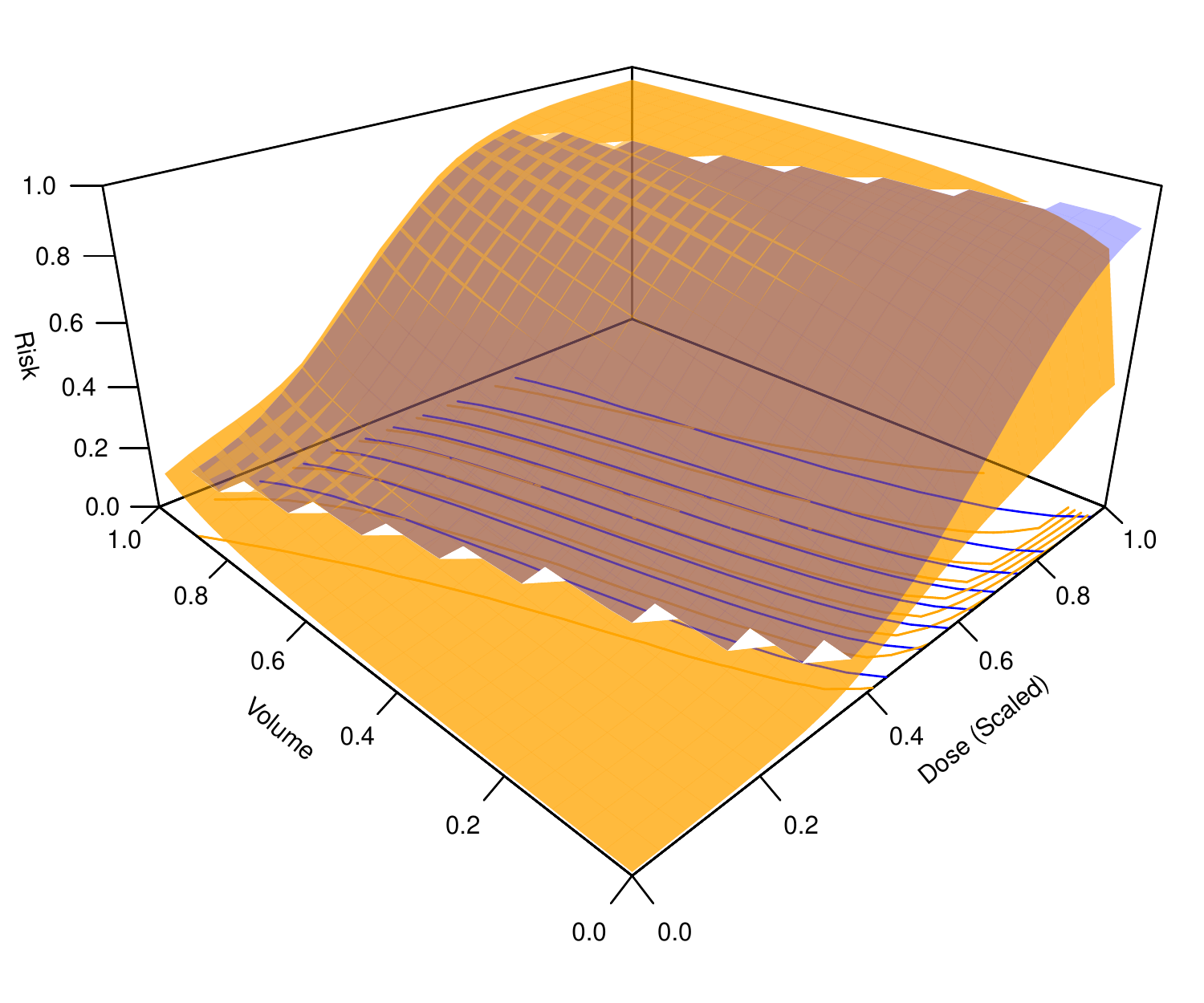} \label{figure:sim_d}}
\caption{Perspective plots of the model-based estimated (orange) and true (blue) pointwise causal NTCP by dose/volume coordinate ($n = 100$).}
    \label{figure:persp_sim}
\end{figure}

The true and average model-estimated pointwise causal NTCPs based on 104 replicates are illustrated by perspective plots in Figure \ref{figure:persp_sim} with the summarized performance metrics reported in Table \ref{tab:perf_mets}. Supplemental contour plots for a grid of estimates and performance metrics are presented in Appendix C. 

\begin{TableNotes}[flushleft]\item \linespread{1}\scriptsize $n$ = sample size, $|\mathrm{Bias}|$ = absolute bias, MCSD = Monte Carlo standard deviation, rMSE = root mean-square error, MCE = Monte Carlo error, $k$ = effective number of parameters, DIC = deviance information criterion \end{TableNotes}
    \newcolumntype{L}[1]{>{\raggedright\arraybackslash}m{#1}}
    \newcolumntype{C}[1]{>{\centering\arraybackslash}m{#1}}
    \newcolumntype{R}[1]{>{\raggedleft\arraybackslash}m{#1}}
    \renewcommand\theadgape{\Gape[2pt]}
    \renewcommand\cellgape{\Gape[0pt]}
     \renewcommand\theadfont{}
    {
    \small
    \newcommand{\colsize}{1.2cm}
    \renewcommand{\arraystretch}{1.1}
    \begin{longtable}{C{0.6cm}  C{2cm}  C{1.1cm} C{1.1cm}  C{1.1cm} C{1.1cm} C{1.1cm} C{1.5cm} C{1.1cm} C{1.5cm} }
    \caption{Summarized performance metrics averaged over a dose-volume grid of pointwise-causal risk estimates.} \\
    \toprule \multirow{2}{*}[-1em]{\textbf{\textit{n}}} &  \multirow{2}{*}[-1em]{\textbf{Model}} &  \multicolumn{4}{c}{\textbf{Out-of-Sample Metric}} & \multicolumn{4}{c}{\textbf{In-Sample Metric}} \\ \cmidrule(lr){3-6} \cmidrule(lr){7-10}
    & & $|\text{Bias}|$ & MCSD & rMSE & MCE & Brier & Deviance & $k$ & DIC  \\ 
    \midrule
   \multirow{4}{*}{100} &  Logistic & 0.166 & 0.055 & 0.180 & 0.005 & 21.55 & 3224.63 & 3.96 & 3228.59 \\ 
  & Polynomial & 0.165 & 0.059 & 0.181 & 0.006 & 21.54 & 3225.46 & 5.36 & 3230.82 \\ 
  & Additive & 0.018 & 0.062 & 0.065 & 0.006 & 19.12 & 2973.42 & 51.11 & 3024.53 \\ 
  & Bivariable & 0.025 & 0.078 & 0.084 & 0.008 & 18.90 & 2956.77 & 94.16 & 3050.93 \\ 
  \midrule
   \multirow{4}{*}{500} &  Logistic & 0.166 & 0.025 & 0.169 & 0.002 & 21.75 & 16236.82 & 3.90 & 16240.84 \\ 
  & Polynomial & 0.165 & 0.025 & 0.169 & 0.002 & 21.74 & 16235.71 & 5.22 & 16240.88 \\ 
  & Additive & 0.008 & 0.027 & 0.029 & 0.003 & 19.32 & 14888.30 & 133.56 & 15031.87 \\ 
  & Bivariable & 0.014 & 0.041 & 0.044 & 0.004 & 19.30 & 14920.23 & 260.29 & 15191.64 \\ 
    \bottomrule 
    \insertTableNotes
    \label{tab:perf_mets}
    \end{longtable}
    }

All models yielded bivariable monotone increasing estimates of NTCP surfaces with respect to dose and volume, evident by the noncrossing indifference curves generated by the contour lines (Figure \ref{figure:persp_sim}). The parametric logistic and polynomial logistic models were unable to capture the true surface in contrast to the flexibly specified additive and bivariable monotone functions. This was evident by the larger mean absolute bias and small Monte Carlo standard deviations of the logistic and polynomial logistic models as the parametric assumptions contributed to more stable but biased estimates  (Table \ref{tab:perf_mets}). From increasing the sample from 100 to 500, the stability of the additive and bivariable monotone models improved, while bias further reduced, pointing towards consistency of the non-parametric specifications. In contrast, the bias of the parametric specifications remained with increased sample size. In addition, the in-sample metrics favour the additive and bivariable monotone models with respect to model fit as they are flexible enough to approximate the true surfaces. All four models captured the true stochastic causal NTCP with low bias and similar levels of variability (Supplementary Figure 8). Compared to the bivariable monotone model, logistic, polynomial logistic and additive functional forms exhibited slightly larger bias but smaller 95\% confidence bands, with less variable estimates evident under larger sample sizes. The results from the second simulation study are presented Appendix D and indicate that the flexible models with covariate adjustment produced the lowest bias in two different confounding scenarios.

\section{Illustration}\label{section:illustration}

We illustrate the use of the marginal NTCP models using data from a cohort of patients with anal and perianal cancer treated at the Princess Margaret Cancer Centre between 2008-2013 \citep{hosni2018ongoing}. Radiation dose (27-36 Gy for elective target and 45-63 Gy for gross targets) was selected based on tumor clinicopathologic features and delivered over 5-7 weeks \citep{han2014prospective, hosni2018ongoing}. Acute toxicity and dose distribution of organs-at-risk were available and extracted from treatment planning system for 87 patients.

12 patients had Grade $\geq$2 acute genitourinary toxicity, with these patients, on average, having higher bladder radiation dose than those without toxicity (Supplementary Figure \ref{fig:pmh-dvh-gutox}). 35 patients with Grade $\geq$3 acute skin (defined as perianal, inguinal, or genital) toxicity also had, on average, higher skin radiation dose (Supplementary Figure \ref{fig:pmh-dvh-gitox}). There was greater variation in bladder dose distributions in comparison to skin dose distributions.

Our aim was to fit MSMs to estimate the pointwise causal NTCPs for genitourinary and skin toxicity and present these in contour plots. We also considered hypothetical stochastic interventions on the treatment plans where volume of the bladder exposed to greater than 40 Gy was restricted to less than or equal to 30\% ($G_{40 \text{ Gy}} \leq 0.3$) and the observed volume of the skin exposed to greater than 20 Gy restricted to less than or equal to 20\% ($G_{20 \text{ Gy}} \leq 0.2$). MSMs with linear, polynomial, additive and bivariable monotone specifications for dose and volume effects were fitted using the \texttt{monoreg} package. We considered age (under/over 65 years) as a potential confounder, as the presence of frailty could be associated both with less intensive treatment and increased toxicity. We did not consider indicators of disease progression (stage/tumor size) as potential confounders, because although they may be associated with the intensity of treatment, they are not expected to be associated with the toxicity outcomes considered here. More generally, as our outcomes are safety rather than effectiveness outcomes, we do not expect confounding by indication to be a major issue in our analyses. Bootstrap with 1000 resamples was used to generate confidence intervals.

\begin{figure}
    \centering
    \subfloat[Logistic]{\includegraphics[width = 0.45\textwidth]{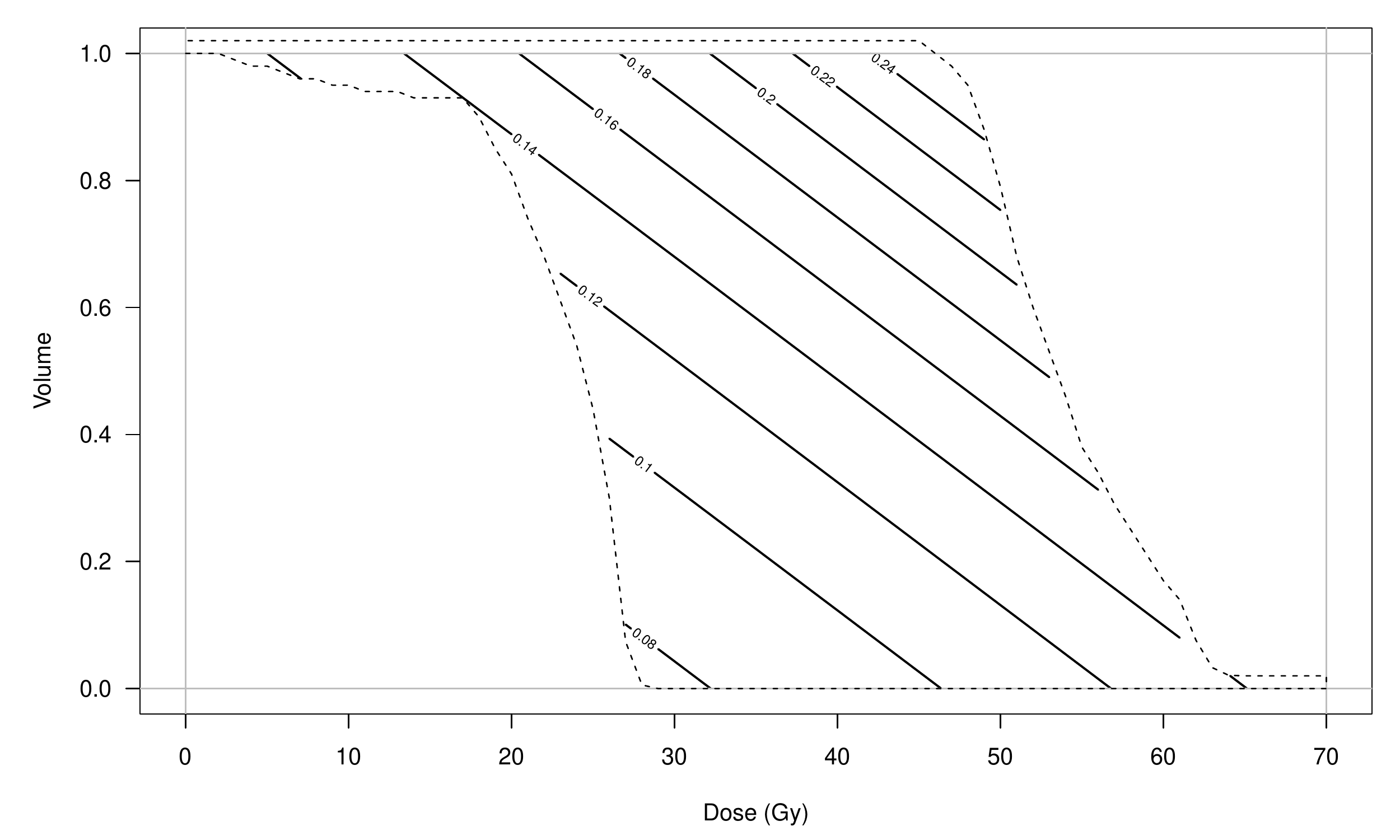}\label{fig:4a}} \hspace{0.6cm}
     \subfloat[Polynomial logistic]{\includegraphics[width = 0.45\textwidth]{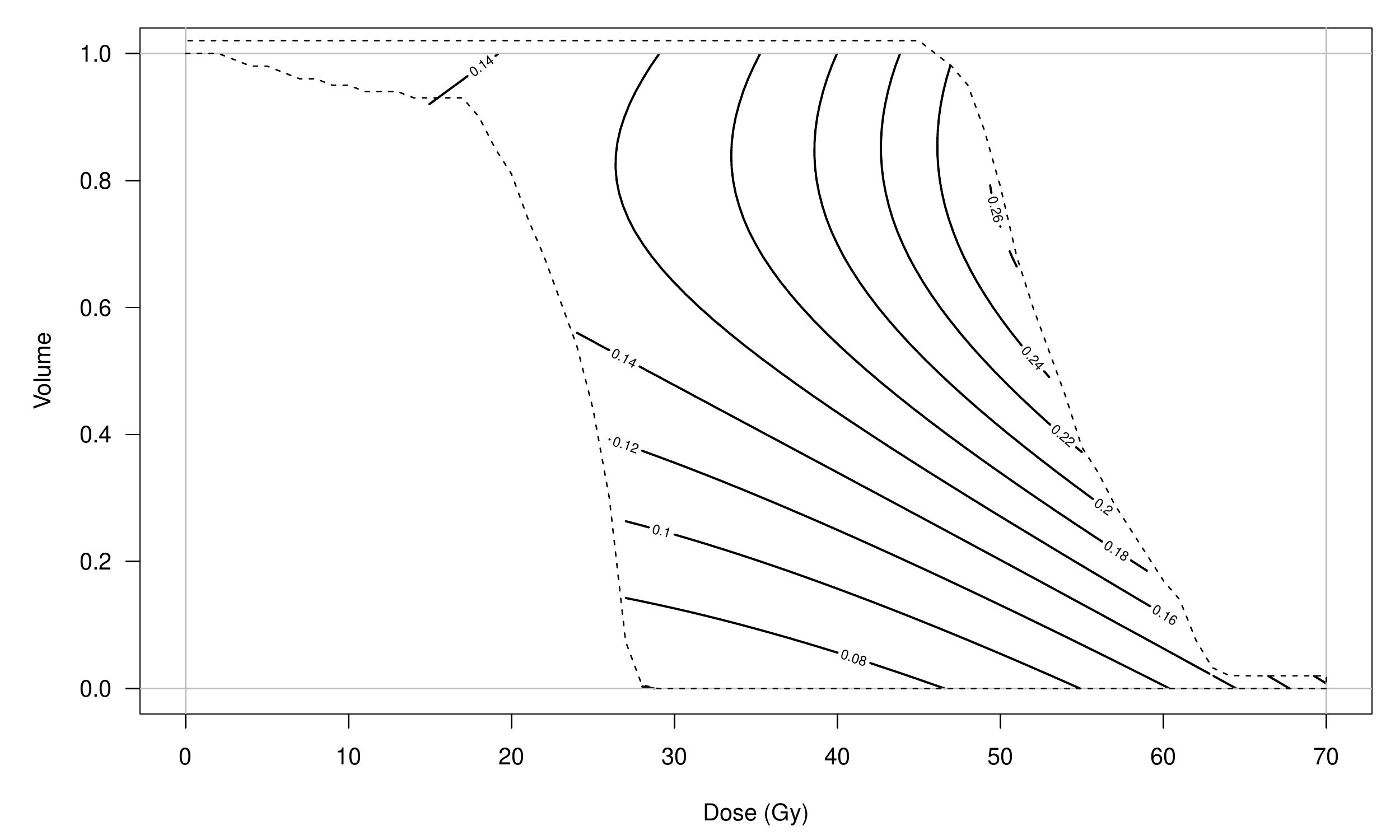}\label{fig:4b}} \\[-0.5em]
    \subfloat[Additive]{\includegraphics[width = 0.45\textwidth]{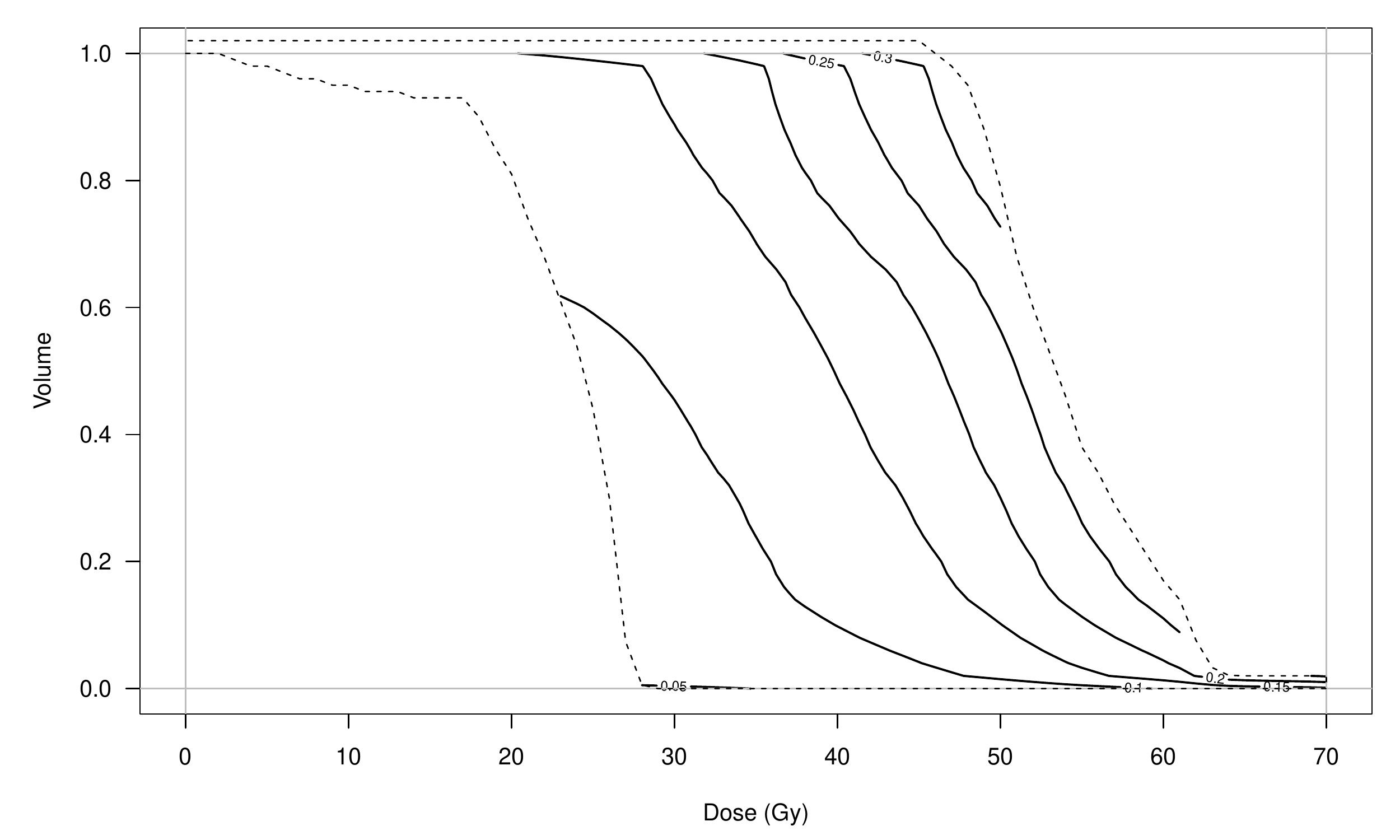}\label{fig:4c}} \hspace{0.6cm}
     \subfloat[Bivariable monotone]{\includegraphics[width = 0.45\textwidth]{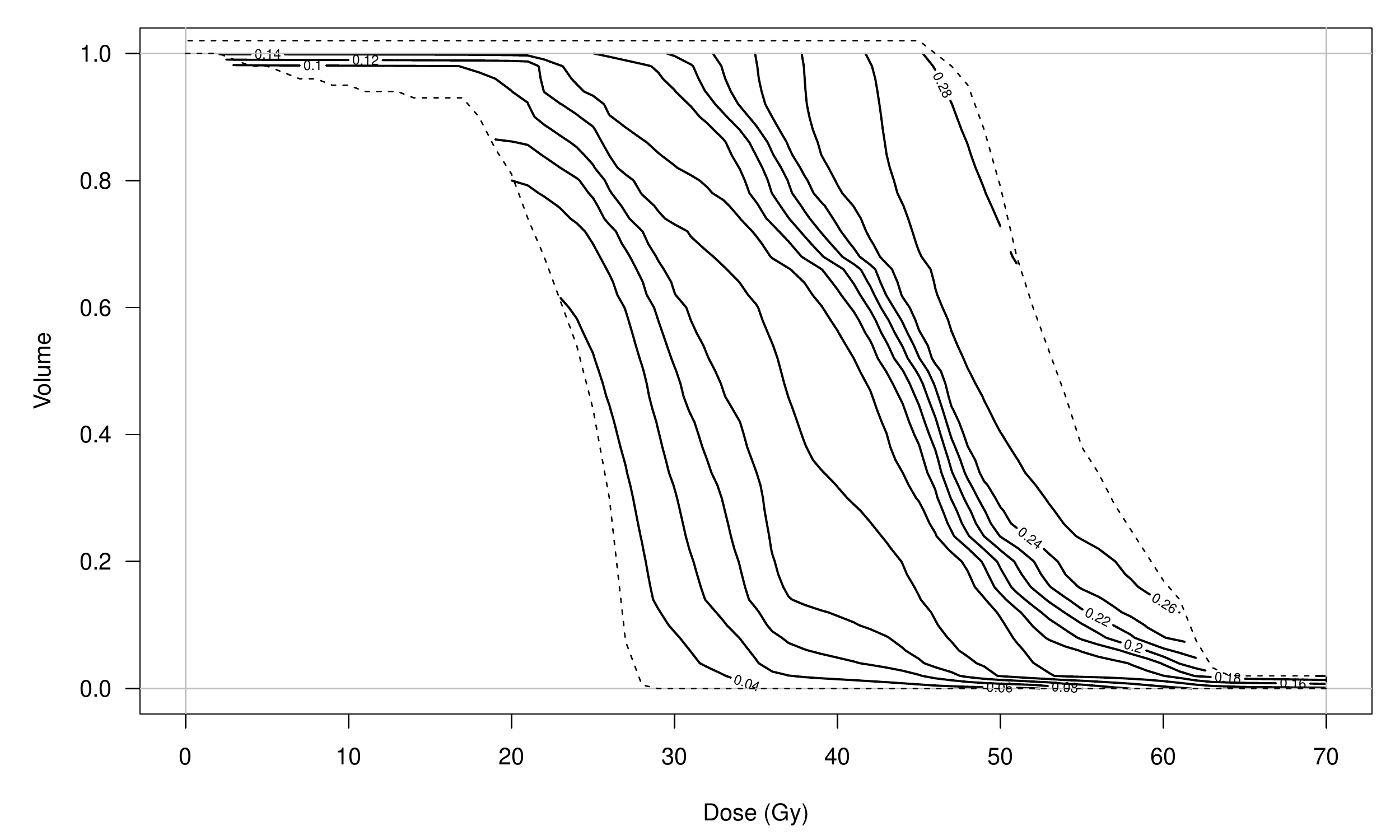}\label{fig:4d}}
\caption{Contour plots for the model-estimated pointwise causal NTCP for genitourinary toxicity at each dose-volume coordinate within the bladder DVH domains (outlined by dotted lines) in anal canal cancer patients.}
    \label{figure:mean_contour_gutox}
\end{figure}

The contour plots of model-based estimates of bladder and skin pointwise-causal NTCP are shown in Figure \ref{figure:mean_contour_gutox} and Supplementary Figure \ref{figure:mean_contour_gitox}, respectively. The contours reflect iso-probability curves similar to those presented in the empirically-estimated atlas of complication incidence \citep{jackson2006atlas} (these are not to be confused with the dose distributions in Supplementary Figure \ref{fig:pmh-dvh-gutox}). The NTCP estimates for the bladder exhibit bivariable monotone increasing surfaces under logistic, additive and bivariable monotone model specifications, suggesting that treatment plans reducing dose or volume pointwise yield a reduction in the NTCP (Figure \ref{figure:mean_contour_gutox}). Unlike the additive and bivariable monotone models, the polynomial logistic model does not enforce bivariable monotonicity and is not guaranteed to produce bivariable monotone NTCP estimates. The contours between the parametric specifications, additive and bivariable monotone models are visually quite different, suggesting that the true underlying surfaces may be better captured with increasing model flexibility. The bivariable monotone functional form resulted in the best model fit (lowest Brier score and DIC) (Table \ref{tab:stochastic-causal-ntcp}).

For the skin, all four model specifications produced bivariable monotone increasing NTCP estimates yielding visually fairly similar contour lines (Supplementary Figure \ref{figure:mean_contour_gitox}). However, the model fit statistics still indicated some differences between the models, with bivariable monotone model resulting in the best fit in terms of Brier score and DIC (Table \ref{tab:stochastic-causal-ntcp}). From contour plots produced by the bivariable monotone model, the NTCP had the dose distribution of all patients been intervened such that exactly 20\% of the bladder received at least 20 Gy of radiation is approximately 18\% (Figure \ref{figure:mean_contour_gutox}), while the NTCP had the dose distribution of all patients been intervened such that exactly 20\% of the skin received at least 30 Gy of radiation is approximately 65\% (Supplementary Figure \ref{figure:mean_contour_gitox}). Contours such as the ones presented could guide the placement of dose-volume constraints in lowering marginal NTCP during treatment planning.

\begin{TableNotes}[flushleft]\item \linespread{1}\scriptsize $k$ = effective number of MCMC parameters, DIC = Deviance Information Criterion \end{TableNotes}
    \renewcommand\theadgape{\Gape[2pt]}
    \renewcommand\cellgape{\Gape[0pt]}
     \renewcommand\theadfont{}
    {\small 
    \newcommand{\colsize}{1.2cm}
    \renewcommand{\arraystretch}{1.1}
    \begin{longtable}{ C{2.2cm} C{2.2cm}  C{3.5cm} C{1cm} C{1.5cm} C{1.2cm} C{1.2cm} } 
    \caption{Estimates of the risk ratio for the stochastic causal NTCP under the truncated (bladder volume $\leq$ 0.3  at $40$ Gy; skin volume $\leq$ 0.2 at $20$ Gy) and observed interventions of the DVH and associated in-sample performance metrics.} \\
    \toprule \multirow{2}{*}[-0.5em]{\textbf{DVH}} & \multirow{2}{*}[-0.5em]{\textbf{Model}} &  \textbf{Causal Risk Ratio} & \multicolumn{4}{c}{\textbf{In-Sample Metric}} \\ \cmidrule(lr){3-3} \cmidrule(lr){4-7}
   & & Estimate (95\% CI) & Brier & Deviance & $k$ & DIC  \\ 
    \midrule
  \multirow{4}{*}{Bladder} & Logistic & 0.71 (0.40, 1.12) & 11.48 & 4749.66 & 3.76 & 4753.42 \\ 
  & Polynomial & 0.65 (0.27, 1.14) & 11.43 & 4727.72 & 5.81 & 4733.53 \\ 
  & Additive & 0.67 (0.22, 1.00) & 11.34 & 4691.83 & 25.57 & 4717.39 \\ 
  & Bivariable  & 0.62 (0.18, 1.00) & 11.13 & 4599.30 & 72.43 & 4671.73 \\ 
  \midrule
  \multirow{4}{*}{Skin} & Logistic & 0.95 (0.89, 1.00) & 23.20 & 8099.83 & 3.87 & 8103.71 \\ 
  & Polynomial & 0.93 (0.85, 1.00) & 22.76 & 7984.43 & 131.32 & 8115.74 \\ 
  & Additive & 0.93 (0.82, 1.00) & 21.96 & 7802.06 & 89.52 & 7891.58 \\ 
  & Bivariable  & 0.92 (0.77, 1.02) & 21.37 & 7661.03 & 198.66 & 7859.69 \\ 
    \bottomrule 
    \insertTableNotes 
    \label{tab:stochastic-causal-ntcp}
    \end{longtable}
    }

The results for the stochastic intervention estimates are also presented in Table \ref{tab:stochastic-causal-ntcp} as causal risk ratios. Under the bivariable monotone model, the NTCP had the observed bladder volume of patients exposed to greater than 40 Gy been restricted to less than or equal to 30\% is 0.62 times (95\% CI: 0.18--1.00) the NTCP had the bladder dose distribution been unrestricted. Similarly, the NTCP had the observed skin volume of patients exposed to 20 Gy been restricted to less than or equal to 20\% is 0.92 times (95\% CI: 0.77--1.02) the NTCP had the skin dose distribution been unrestricted. Overall, the estimated stochastic causal risk ratios across all four models under both DVHs exhibited a reduction in causal NTCP from the unrestricted dose distribution. For the bladder, the width of the 95\% bootstrap confidence intervals was smaller under the additive and bivariable monotone models where monotonicity was parametrized. The enforced monotonicity generates a constraint in the risk ratio estimates and a modality around 1, while the logistic and polynomial logistic models were unconstrained. For the skin, the confidence intervals were larger under the additive and bivariable models with the higher variance under the latter producing bootstrap estimates above 1. Bivariable monotonicity is enforced under pointwise interventions but not between causal contrasts under different stochastic interventions. The R code implementing an illustration of the proposed methods on a simulated dataset is available at \url{https://github.com/thaisontang/marginalntcp}.

\section{Discussion}\label{section:discussion}

There were a few reasons why we did not pursue causal modeling of NTCP using functional regression, even though a functional model for DVH exposures had previously been proposed by \citet{schipper2007bayesian, schipper_2008}, and as we showed, can be formulated as a causal model \eqref{eq:schipper}. First, the individual coefficients $\alpha_d$ in \eqref{eq:schipper} are not themselves directly interpretable as causal contrasts, as a given volume $b_d$ cannot be intervened \emph{ceteris paribus}. 
Also, the identification of the model, even in the presence of the restriction $\alpha_1 = 0$ on the first coefficient, is fragile, due to the property $\sum_{d=1}^{\mathcal D} b_d = 1$. In particular, if $b_1 \approx 0$, then
\begin{align*}
    \psi\left(E\left[Y^{(\boldsymbol{B} = \boldsymbol{b})} \mid \boldsymbol{X}; \boldsymbol{\theta} \right]\right) &= \beta_0 + \sum_{j = 1}^{p} \beta_j X_j + \sum_{d = 1}^{\mathcal{D}} \alpha_d b_{d} \\
    &= \beta_0 + \sum_{j = 1}^{p} \beta_j X_j + \sum_{d = 1}^{\mathcal{D}} \alpha_d b_{d} - a + a \sum_{d=1}^{\mathcal D} b_d\\
&= (\beta_0 - a)+ \sum_{j = 1}^{p} \beta_j X_j + (\alpha_1 + a) b_{1} + \sum_{d = 2}^{\mathcal{D}} (\alpha_d + a) b_{d} \\
&\approx (\beta_0 - a) + \sum_{j = 1}^{p} \beta_j X_j + \sum_{d = 2}^{\mathcal{D}} (\alpha_d + a) b_{d},
\end{align*}
that is, the addition of an arbitrary constant to the coefficients does not change the likelihood. \citet[][]{schipper_2008} noted that estimating the quantities $\alpha_d - \beta_0$ is more stable than estimating the absolute level of $\alpha_d$. They also note that their model lacks identifiability when $\alpha_d$ is specified as a step function instead of smooth specifications through splines. We confirmed this by using the monotonic model of \citet{saarela2011method}, which is based on piecewise constant realizations, for the $\alpha_d$ coefficients, and could not achieve stable estimation. Our proposed marginal model does not have these limitations.

We also considered ``slice-based'' univariable marginal models \eqref{eq:MSMslice} fitted separately at every dose level. While a similar approach is sometimes used in practice to find regions of the dose distribution most strongly associated with the toxicity outcomes, this does not enforce monotonicity of the effects by dose, and is susceptible to multiple testing issues. Fitting all the ``slices'' together as a single bivariable model resolves these issues. It however raises another one; fitting the dose-specific models together requires replicating the outcomes of the individuals in a long format dataset, and as a result, the fitted objective function is no longer a likelihood. While the resulting objective function can still be used to fit the marginal mean structure, standard errors based on maximum likelihood estimation are not available. We used clustered bootstrap, resampling at individual level to obtain standard errors. In principle, generalized estimating equations with clustered robust/sandwich standard errors would also work, as long as the models are parametrically specified, similar to the replicated data setting of \citet{leisenring2000comparisons}. While we used the Bayesian non-parametric monotonic regression model of \citet{saarela2011method} to fit the marginal models, we interpreted the resulting point estimates (quasi-posterior means) in a frequentist sense, which can be justified asymptotically \citep{chernozhukov2003mcmc}. Bayesian interpretation of the results and estimation of posterior intervals when combining non-likelihood estimating functions with MCMC in the present setting is a topic for further research. While we used the monotonic regression model of \citet{saarela2011method}, in principle it can be substituted by any sufficiently flexible bivariable monotonic model specification.

To derive the monotonicity property of the marginal model, we assumed the monotonic functional regression model and stochastic ordering of the conditional average dose distributions being compared. This was done for the purpose of connecting the monotonicity of the marginal effects to the monotonicity encoded by the functional model. However, we note that while these conditions are sufficient to obtain the property \eqref{eq:bivar-mono}, they are not necessary. The property \eqref{eq:bivar-mono} could be assumed directly for the marginal model, or could potentially apply under weaker conditions on the dose distributions.

The monotonicity property itself is biologically plausible at least when the toxicity outcome is specific to the organ-at-risk. It may be questionable when the outcome is non-specific and can be caused by radiation exposure to different organs. This is a relevant reservation, as in treatment planning interventions to reduce the dose on one OAR may direct it somewhere else. While we have not addressed simultaneous modeling of multi-organ radiation exposured herein, we will address this limitation in future work. While we focused here on modeling of dichotomous toxicity outcomes, similar models can be specified to continuous, ordinal and time-to-event outcomes, and can be estimated using the functions available in the R package \texttt{monoreg} \citep{monoreg}.

Our proposed methods required the causal identifiability assumptions of conditional exchangeability, positivity, and counterfactual consistency. We argued that since our outcomes are of safety rather than effectiveness type, the usual issue of confounding by indication is of lesser importance in the current context. However, the presented research could be extended by studying further statistical methods to control for observed counfounders besides model adjustment, including inverse probability of treatment weighting, and doubly robust estimation. This was already suggested by the importance sampling ratio appearing in estimator \eqref{eq:msm-stochastic}, and similar weights could be added to the objective function \eqref{equation:quasiloglik}, which would also enable estimation of models that are marginal over the covariates. The present paper focused on covariate conditional models as individual-level predictions are usually sought in NTCP modeling. In Section \ref{section:msm} we outlined some ways to allow for effect modification in these models. If the hypothesized effect modifiers are dichotomous or their effect can be assumed jointly monotonic with the dose-volume, they can be incorporated using the existing functionality in the R package \texttt{monoreg}. As a caveat we note that due to the curse of dimensionality, only a limited number of covariates can be included in the estimation of the resulting multivariable non-parametric regression function. Also, a different approach would be needed for including effect modifiers with non-monotonic effects on the outcome.

Our consideration of localized interventions was aimed at reducing the positivity issues related to interventions on the entire dose distribution. As suggested in Section \ref{section:msmestimation}, any remaining positivity issues could be detected by studying the importance sampling weights. The counterfactual consistency assumption states that the potential outcome corresponding to the dose distribution intervened upon at the observed level $\boldsymbol G$ is equivalent to the observed outcome. The interpretation is that all the relevant information about the intervention is encoded in the DVH. Summarizing the dose distribution in terms of the DVH collapses over the spatial dose distribution, so this assumption could be violated if the spatial dimensions carry relevant information on the toxicity risk, and two different spatial dose distributions result in the same DVH. Such assumptions are usually implicit in NTCP modeling; the advantage of the causal modeling framework is making all the assumptions explicit, so that their plausibility in a specific application can be assessed.

\bibliographystyle{apalike}
\bibliography{ref}

\setcounter{section}{0}
\renewcommand\thesection{\Alph{section}}
\setcounter{figure}{0}
\setcounter{table}{0}
\titleformat{\section}[hang]{\bfseries}{Appendix \thesection:\ }{0pt}{}
\titleformat{\subsection}[hang]{\bfseries}{\thesubsection:\ }{0pt}{}
\renewcommand{\figurename}{Supplementary Figure}
\renewcommand{\tablename}{Supplementary Table}

\newpage

\section*{Supporting Information for ``A marginal structural model for normal tissue complication probability'' by Thai-Son Tang, Zhihui Liu, Ali Hosni, John Kim, and Olli Saarela}

\section{Mathematical Proofs}\label{apdx:math-sect}
\subsection{Identification of NTCP under the causal assumptions}\label{apdx:ident_multexp}

The identifiability of (\ref{eq:ident_multexp}) follows as
\begin{align*}
    E\left[Y^{\left(\boldsymbol{G} \; = \; \boldsymbol{g}\right)}\right] &=  E_{\boldsymbol{X}}\left\{ E\left[ Y^{\left(\boldsymbol{g}\right)} \mid \boldsymbol{X}\right] \right\} \\
    &= E_{\boldsymbol{X}}\left\{ E\left[ Y^{\left(\boldsymbol{g}\right)} \mid \boldsymbol{G} = \boldsymbol{g}, \boldsymbol{X}\right] \right\} &  \left(\boldsymbol{G} \bigCI Y^{(\boldsymbol{g})} \mid \boldsymbol{X} \right) 
    \\ &= E_{\boldsymbol{X}}\left\{ E\left[ Y \mid \boldsymbol{G} = \boldsymbol{g}, \boldsymbol{X}\right] \right\} & \left(Y = Y^{\left(\boldsymbol{G}\right)}\right) 
    \\
    &= E_{\boldsymbol{X}}\left\{ E\left[ Y \mid \boldsymbol{B} = \boldsymbol{B}(\boldsymbol{g}), \boldsymbol{X}\right] \right\}  & (\ref{eq:ddvh})
\end{align*}

\subsection{Functional monotonicity of NTCP when intervening on the entire dose distribution} \label{apdx:func-mono}

Under model (\ref{eq:schipper}) and assuming stochastic ordering of conditional dose distributions, intervening on the dose distribution from $\boldsymbol{g} = (g_{1}, \ldots, g_{{\mathcal{D}}})$ to $\boldsymbol{g}^{*} = (g_{1}^{*}, \ldots, g_{{\mathcal{D}}}^{*})$, where $g_d^* \leq g_d$ for all $d \in \{1, 2, \ldots, \mathcal{D}\}$ implies monotonicity with respect to the causal NTCPs, which we call functional monotonicity:
\begin{align*}
     g_d^* \leq g_d, \;\; &d \in \{1, 2, \ldots, \mathcal{D}\} \\
     \Rightarrow\quad & D(\boldsymbol{g}^*) \preceq D(\boldsymbol{g}) & (\text{stochastic ordering}) \\
    \quad \Rightarrow \quad & \textstyle \sum_{d = 1}^{\mathcal{D}} \alpha_d P(D = d \mid \boldsymbol{G} = \boldsymbol{g}^*)
     \leq \sum_{d = 1}^{\mathcal{D}} \alpha_d P(D = d \mid \boldsymbol{G} = \boldsymbol{g}) & (\text{non-decreasing } (\ref{eq:func-mono})) \\
     \quad \Rightarrow \quad & \textstyle \sum_{d = 1}^{\mathcal{D}} \alpha_d B_d(\boldsymbol{g}^*) \leq \sum_{d = 1}^{\mathcal{D}} \alpha_d B_d(\boldsymbol{g}) \; & (\ref{eq:ddvh}) \\
     \Rightarrow \quad &  \psi\left(E\left[ Y^{\left(\boldsymbol{B} \; = \; \boldsymbol{B}(\boldsymbol{g}^*)\right)} \mid \boldsymbol{X}\right]\right)  \leq \psi\left(E\left[ Y^{\left(\boldsymbol{B} \; = \; \boldsymbol{B}(\boldsymbol{g})\right)} \mid \boldsymbol{X}\right]\right)  & (\ref{eq:schipper}) \\ 
     \Rightarrow \quad  & E\left[ Y^{\left(\boldsymbol{B} \; = \; \boldsymbol{B}(\boldsymbol{g}^*)\right)} \mid \boldsymbol{X}\right]
     \leq E\left[ Y^{\left(\boldsymbol{B} \; = \; \boldsymbol{B}(\boldsymbol{g})\right)} \mid \boldsymbol{X}\right] & (\text{monotone } \psi(\cdot)) \\ 
     \Rightarrow \quad & E\left[Y^{\left(\boldsymbol{G} \; = \; \boldsymbol{g}^*\right)}\right] \leq E\left[Y^{\left(\boldsymbol{G} \; = \; \boldsymbol{g} \right)}\right] & (\ref{eq:dvh})
\end{align*}
where $B_d(\boldsymbol{g}) = P(D = d \mid \boldsymbol{G} = \boldsymbol{g})$ for $d \in \{1, 2, \ldots, \mathcal{D}\}$ and $\boldsymbol{B}(\boldsymbol{g}) = (B_{1}(\boldsymbol{g}), \ldots, B_{{\mathcal{D}}}(\boldsymbol{g}))$.

\subsection{Identification of NTCP under stochastic interventions}\label{apdx:stochast-ident}

The result (\ref{eq:stochast-int}) is obtained as follows:
\begin{align*}
    &E\left[Y^{\left(\boldsymbol{V}\right)}\right] = E_{\boldsymbol{X}}\left\{ E_{V_d \mid \boldsymbol{X}}\left[ E_{\boldsymbol{V}_{-d} \mid V_d, \boldsymbol{X}}\left( E\left[ Y^{\left( V_d, \; \boldsymbol{V}_{-d} \right)} \mathrel{\big|} V_d, \boldsymbol{V}_{-d}, \boldsymbol{X} \right] \right) \right] \right\}  \\
    &=  \int_{\boldsymbol{x}} \int_g \int_{\boldsymbol{g}_{-d}} E[Y^{(g, \; \boldsymbol{g}_{-d})} \mid V_{d} = g, \boldsymbol{V}_{-d} = \boldsymbol{g}_{-d}, \boldsymbol{X} = \boldsymbol{x}] \\
    &\qquad\qquad \times f_{\boldsymbol{V}_{-d} \mid V_d, \boldsymbol{X}}(\boldsymbol{g}_{-d} \mid g, \boldsymbol{x}) f_{V_{d} \mid \boldsymbol{X}}(g \mid \boldsymbol{x}) f_{\boldsymbol{X}}(\boldsymbol{x}) \, \mathrm{d}\boldsymbol{g}_{-d} \, \mathrm{d}g \; \mathrm{d}\boldsymbol{x} \\
    &=  \int_{\boldsymbol{x}} \int_g \int_{\boldsymbol{g}_{-d}} E[Y^{(g, \; \boldsymbol{g}_{-d})} \mid  \boldsymbol{X} = \boldsymbol{x}] f_{\boldsymbol{V}_{-d} \mid V_d, \boldsymbol{X}}(\boldsymbol{g}_{-d} \mid g, \boldsymbol{x}) f_{V_{d} \mid \boldsymbol{X}}(g \mid \boldsymbol{x})  \\
    &\qquad\qquad \times  f_{\boldsymbol{X}}(\boldsymbol{x}) \, \mathrm{d}\boldsymbol{g}_{-d} \, \mathrm{d}g \; \mathrm{d}\boldsymbol{x} & (\boldsymbol{V} \bigCI Y^{(\boldsymbol{g})} \mid \boldsymbol{X}) \\
    &= \int_{\boldsymbol{x}} \int_g \int_{\boldsymbol{g}_{-d}} E\left[Y^{(g, \; \boldsymbol{g}_{-d})} \mid \boldsymbol{X} = \boldsymbol{x} \right] \dfrac{f_{\boldsymbol{V}_{-d} \mid V_d, \boldsymbol{X}}(\boldsymbol{g}_{-d} \mid g, \boldsymbol{x})}{f_{\boldsymbol{G}_{-d} \mid G_d, \boldsymbol{X}}(\boldsymbol{g}_{-d} \mid g, \boldsymbol{x})} \dfrac{f_{V_{d} \mid \boldsymbol{X}}(g \mid \boldsymbol{x})}{f_{G_{d} \mid \boldsymbol{X}}(g \mid \boldsymbol{x})} \\
    &\qquad\qquad \times f_{\boldsymbol{G}_{-d} \mid G_{d}, \boldsymbol{X}}(\boldsymbol{g}_{-d} \mid g, \boldsymbol{x}) f_{G_{d} \mid \boldsymbol{X}}(g \mid \boldsymbol{x}) f_{\boldsymbol{X}}(\boldsymbol{x}) \, \mathrm{d}\boldsymbol{g}_{-d} \, \mathrm{d}g \; \mathrm{d}\boldsymbol{x} & (*) \\
    &= \int_{\boldsymbol{x}} \int_g \int_{\boldsymbol{g}_{-d}} E\left[Y^{(g, \; \boldsymbol{g}_{-d})} \mid \boldsymbol{X} = \boldsymbol{x} \right] \dfrac{f_{V_{d} \mid \boldsymbol{X}}(g \mid \boldsymbol{x})}{f_{G_{d} \mid \boldsymbol{X}}(g \mid \boldsymbol{x})}  \\
    &\qquad\qquad \times f_{\boldsymbol{G}_{-d} \mid G_{d}, \boldsymbol{X}}(\boldsymbol{g}_{-d} \mid g, \boldsymbol{x}) f_{G_{d} \mid \boldsymbol{X}}(g \mid \boldsymbol{x}) f_{\boldsymbol{X}}(\boldsymbol{x}) \, \mathrm{d}\boldsymbol{g}_{-d} \, \mathrm{d}g \; \mathrm{d}\boldsymbol{x} & (**) \\
    &= \int_{\boldsymbol{x}} \int_g \int_{\boldsymbol{g}_{-d}} E\left[Y^{(g, \; \boldsymbol{g}_{-d})} \mid G_{d} = g, \boldsymbol{G}_{-d} = \boldsymbol{g}_{-d},\boldsymbol{X} = \boldsymbol{x} \right] \dfrac{f_{V_{d} \mid \boldsymbol{X}}(g \mid \boldsymbol{x})}{f_{G_{d} \mid \boldsymbol{X}}(g \mid \boldsymbol{x})}  \\
    &\qquad\qquad \times f_{\boldsymbol{G}_{-d} \mid G_{d}, \boldsymbol{X}}(\boldsymbol{g}_{-d} \mid g, \boldsymbol{x}) f_{G_{d} \mid \boldsymbol{X}}(g \mid \boldsymbol{x}) f_{\boldsymbol{X}}(\boldsymbol{x}) \, \mathrm{d}\boldsymbol{g}_{-d} \, \mathrm{d}g \; \mathrm{d}\boldsymbol{x} & (\boldsymbol{G} \bigCI Y^{(\boldsymbol{g})} \mid \boldsymbol{X}) \\    
    &= \int_{\boldsymbol{x}} \int_g \int_{\boldsymbol{g}_{-d}} E\left[Y \mid G_{d} = g, \boldsymbol{G}_{-d} = \boldsymbol{g}_{-d},\boldsymbol{X} = \boldsymbol{x} \right] f_{\boldsymbol{G}_{-d} \mid G_{d}, \boldsymbol{X}}(\boldsymbol{g}_{-d} \mid g, \boldsymbol{x})  \\
    &\qquad\qquad \times  \dfrac{f_{V_{d} \mid \boldsymbol{X}}(g \mid \boldsymbol{x})}{f_{G_{d} \mid \boldsymbol{X}}(g \mid \boldsymbol{x})} f_{G_{d} \mid \boldsymbol{X}}(g \mid \boldsymbol{x}) f_{\boldsymbol{X}}(\boldsymbol{x}) \, \mathrm{d}\boldsymbol{g}_{-d} \, \mathrm{d}g \; \mathrm{d}\boldsymbol{x} & (Y^{(\boldsymbol{G})} = Y) \\
    &= \int_{\boldsymbol{x}} \int_g E\left[Y \mid G_{d} = g, \boldsymbol{X} = \boldsymbol{x} \right] \dfrac{f_{V_{d} \mid \boldsymbol{X}}(g \mid \boldsymbol{x})}{f_{G_{d} \mid \boldsymbol{X}}(g \mid \boldsymbol{x})} f_{G_{d} \mid \boldsymbol{X}}(g \mid \boldsymbol{x}) f_{\boldsymbol{X}}(\boldsymbol{x}) \, \mathrm{d}\boldsymbol{g}_{-d} \, \mathrm{d}g \; \mathrm{d}\boldsymbol{x},
\end{align*}
where ($*$) denotes assuming $f_{V_d \mid \boldsymbol{X}}(g \mid \boldsymbol{x}) > 0 \Rightarrow f_{G_d \mid \boldsymbol{X}}(g \mid \boldsymbol{x}) > 0$ and ($**$) denotes assuming $f_{\boldsymbol{V}_{-d} \mid V_d, \boldsymbol{X}}(\boldsymbol{g}_{-d} \mid g, \boldsymbol{x}) = f_{\boldsymbol{G}_{-d} \mid G_d, \boldsymbol{X}}(\boldsymbol{g}_{-d} \mid g, \boldsymbol{x})$ for all $g, \boldsymbol{g}_{-d}$ and $\boldsymbol{x}$.

\newpage

\noindent For the \emph{g}-formula based causal estimand (\ref{equation:hypo_stand}), we have
\begin{align*}
    &E\left[Y^{\left(V_d, \; \boldsymbol{V}_{-d}\right)}\right] = E_{\boldsymbol{X}}\left\{ E_{V_d \mid \boldsymbol{X}}\left[ E_{\boldsymbol{V}_{-d} \mid V_d, \boldsymbol{X}}\left( E\left[ Y^{\left( V_d, \; \boldsymbol{V}_{-d} \right)} \mathrel{\big|} V_d, \boldsymbol{V}_{-d}, \boldsymbol{X} \right] \right) \right] \right\}  \\
    &=  \int_{\boldsymbol{x}} \int_g \int_{\boldsymbol{g}_{-d}} E[Y^{(g, \; \boldsymbol{g}_{-d})} \mid V_{d} = g, \boldsymbol{V}_{-d} = \boldsymbol{g}_{-d}, \boldsymbol{X} = \boldsymbol{x}] \\
    &\qquad\qquad \times f_{\boldsymbol{V}_{-d} \mid V_d, \boldsymbol{X}}(\boldsymbol{g}_{-d} \mid g, \boldsymbol{x}) f_{V_{d} \mid \boldsymbol{X}}(g \mid \boldsymbol{x}) f_{\boldsymbol{X}}(\boldsymbol{x}) \, \mathrm{d}\boldsymbol{g}_{-d} \, \mathrm{d}g \; \mathrm{d}\boldsymbol{x} \\
    &=  \int_{\boldsymbol{x}} \int_g \int_{\boldsymbol{g}_{-d}} E[Y^{(g, \; \boldsymbol{g}_{-d})} \mid  \boldsymbol{X} = \boldsymbol{x}]  f_{\boldsymbol{V}_{-d} \mid V_d, \boldsymbol{X}}(\boldsymbol{g}_{-d} \mid g, \boldsymbol{x}) \\
    &\qquad\qquad \times f_{V_{d} \mid \boldsymbol{X}}(g \mid \boldsymbol{x}) f_{\boldsymbol{X}}(\boldsymbol{x}) \, \mathrm{d}\boldsymbol{g}_{-d} \, \mathrm{d}g \; \mathrm{d}\boldsymbol{x} & (\boldsymbol{V} \bigCI Y^{(\boldsymbol{g})} \mid \boldsymbol{X}) \\
    &= \int_{\boldsymbol{x}} \int_g \int_{\boldsymbol{g}_{-d}} E\left[Y^{(g, \; \boldsymbol{g}_{-d})} \mid \boldsymbol{X} = \boldsymbol{x} \right] \dfrac{f_{\boldsymbol{V}_{-d} \mid V_d, \boldsymbol{X}}(\boldsymbol{g}_{-d} \mid g, \boldsymbol{x})}{f_{\boldsymbol{G}_{-d} \mid G_d, \boldsymbol{X}}(\boldsymbol{g}_{-d} \mid g, \boldsymbol{x})}  \\
    &\qquad\qquad \times f_{\boldsymbol{G}_{-d} \mid G_{d}, \boldsymbol{X}}(\boldsymbol{g}_{-d} \mid g, \boldsymbol{x}) f_{V_{d} \mid \boldsymbol{X}}(g \mid \boldsymbol{x}) f_{\boldsymbol{X}}(\boldsymbol{x}) \, \mathrm{d}\boldsymbol{g}_{-d} \, \mathrm{d}g \; \mathrm{d}\boldsymbol{x} & (*) \\
    &= \int_{\boldsymbol{x}} \int_g \int_{\boldsymbol{g}_{-d}} E\left[Y^{(g, \; \boldsymbol{g}_{-d})} \mid \boldsymbol{X} = \boldsymbol{x} \right]  f_{\boldsymbol{G}_{-d} \mid G_{d}, \boldsymbol{X}}(\boldsymbol{g}_{-d} \mid g, \boldsymbol{x}) \\
    &\qquad\qquad \times f_{V_{d} \mid \boldsymbol{X}}(g \mid \boldsymbol{x}) f_{\boldsymbol{X}}(\boldsymbol{x}) \, \mathrm{d}\boldsymbol{g}_{-d} \, \mathrm{d}g \; \mathrm{d}\boldsymbol{x} & (**) \\
    &= \int_{\boldsymbol{x}} \int_g \int_{\boldsymbol{g}_{-d}} E\left[Y^{(g, \; \boldsymbol{g}_{-d})} \mid G_d = g, \boldsymbol{G}_{-d} = \boldsymbol{g}_{-d}, \boldsymbol{X} = \boldsymbol{x} \right]   \\
    &\qquad\qquad \times f_{\boldsymbol{G}_{-d} \mid G_{d}, \boldsymbol{X}}(\boldsymbol{g}_{-d} \mid g, \boldsymbol{x}) f_{V_{d} \mid \boldsymbol{X}}(g \mid \boldsymbol{x}) f_{\boldsymbol{X}}(\boldsymbol{x}) \, \mathrm{d}\boldsymbol{g}_{-d} \, \mathrm{d}g \; \mathrm{d}\boldsymbol{x} & (\boldsymbol{G} \bigCI Y^{(\boldsymbol{g})} \mid \boldsymbol{X}) \\
    &= \int_{\boldsymbol{x}} \int_g \int_{\boldsymbol{g}_{-d}} E\left[Y \mid G_d = g, \boldsymbol{G}_{-d} = \boldsymbol{g}_{-d}, \boldsymbol{X} = \boldsymbol{x} \right]   \\
    &\qquad\qquad \times f_{\boldsymbol{G}_{-d} \mid G_{d}, \boldsymbol{X}}(\boldsymbol{g}_{-d} \mid g, \boldsymbol{x}) f_{V_{d} \mid \boldsymbol{X}}(g \mid \boldsymbol{x}) f_{\boldsymbol{X}}(\boldsymbol{x}) \, \mathrm{d}\boldsymbol{g}_{-d} \, \mathrm{d}g \; \mathrm{d}\boldsymbol{x} & (Y^{(\boldsymbol{G})} = Y) \\
    &= \int_{\boldsymbol{x}} \int_{g}  E_{\boldsymbol{G}_{-d} \mid G_d = g, \boldsymbol{X} = \boldsymbol{x}}\left\{ E\left[Y \mid G_d = g, \boldsymbol{G}_{-d}, \boldsymbol{X} = \boldsymbol{x}\right] \right\} f_{V_d \mid \boldsymbol{X}}(g \mid \boldsymbol{x}) f_{\boldsymbol{X}}(\boldsymbol{x}) \, \mathrm{d}g \, \mathrm{d}\boldsymbol{x},
\end{align*}
where ($*$) denotes assuming $f_{V_d \mid \boldsymbol{X}}(g \mid \boldsymbol{x}) > 0 \Rightarrow f_{G_d \mid \boldsymbol{X}}(g \mid \boldsymbol{x}) > 0$ and ($**$) denotes assuming $f_{\boldsymbol{V}_{-d} \mid V_d, \boldsymbol{X}}(\boldsymbol{g}_{-d} \mid g, \boldsymbol{x}) = f_{\boldsymbol{G}_{-d} \mid G_d, \boldsymbol{X}}(\boldsymbol{g}_{-d} \mid g, \boldsymbol{x})$ for all $g, \boldsymbol{g}_{-d}$ and $\boldsymbol{x}$. 

\bigskip

\noindent For the special case of $f_{V_d \mid \boldsymbol{X}}(g \mid \boldsymbol{X}) = f_{G_d \mid \boldsymbol{X}}(g \mid \boldsymbol{X})$ for all $g$ and $\boldsymbol{X}$,
\begin{align*}
    E\left[Y^{\left(\boldsymbol{V}\right)}\right] &= \int_{\boldsymbol{x}} \int_{g}  E_{\boldsymbol{G}_{-d} \mid G_d = g, \boldsymbol{X} = \boldsymbol{x}}\left\{ E\left[Y \mid G_d = g, \boldsymbol{G}_{-d}, \boldsymbol{X} = \boldsymbol{x}\right] \right\} f_{G_d \mid \boldsymbol{X}}(g \mid \boldsymbol{x}) f_{\boldsymbol{X}}(\boldsymbol{x}) \, \mathrm{d}g \, \mathrm{d}\boldsymbol{x} \\
    &= E_{\boldsymbol{G}, \boldsymbol{X}}\left\{E[Y \mid G_d, \boldsymbol{G}_{-d}, \boldsymbol{X}] \right\} = E[Y].
\end{align*}

\newpage

\subsection{Identification of NTCP under pointwise interventions}\label{apdx:ptwise_ident}

The result (\ref{eq:monotonicity_univariate}) is obtained as follows for the special case of discrete random variable $V_d = g$ with probability 1, where $P(V_d = g \mid \boldsymbol{X}) = 1$ for all $\boldsymbol{X}$,
\begin{align*}
    E&[Y^{(g, \; \boldsymbol{V}_{-d})}] = E_{\boldsymbol{X}}\{E_{V_d \mid \boldsymbol{X}} ( E_{\boldsymbol{V}_{-d} \mid V_d, \boldsymbol{X}}[ E[Y^{(g, \; \boldsymbol{V}_{-d})} \mid V_d, \boldsymbol{V}_{-d}, \boldsymbol{X}] ] ) \} \\
        &= \int_{\boldsymbol{x}} \int_{g'} \int_{\boldsymbol{g}_{-d}} E[Y^{(g, \; \boldsymbol{g}_{-d})} \mid V_d = g', \boldsymbol{V}_{-d} = \boldsymbol{g}_{-d}, \boldsymbol{X}] \\
        & \qquad \times f_{\boldsymbol{V}_{-d} \mid V_d, \boldsymbol{X}}(\boldsymbol{g}_{-d} \mid g', \boldsymbol{x})  \delta(g' - g) f_{\boldsymbol{X}}(\boldsymbol{x}) \, \mathrm{d}\boldsymbol{g}_{-d} \, \mathrm{d}g' \, \mathrm{d}\boldsymbol{x}\\
        &= \int_{\boldsymbol{x}} \int_{\boldsymbol{g}_{-d}} E[Y^{(g, \; \boldsymbol{g}_{-d})} \mid V_d = g, \boldsymbol{V}_{-d} = \boldsymbol{g}_{-d}, \boldsymbol{X}] \\
        & \qquad \times f_{\boldsymbol{V}_{-d} \mid V_d, \boldsymbol{X}}(\boldsymbol{g}_{-d} \mid g, \boldsymbol{x})  f_{\boldsymbol{X}}(\boldsymbol{x}) \, \mathrm{d}\boldsymbol{g}_{-d} \, \mathrm{d}\boldsymbol{x} \\
        &= \int_{\boldsymbol{x}} \int_{\boldsymbol{g}_{-d}} E[Y^{(g, \; \boldsymbol{g}_{-d})} \mid \boldsymbol{X}] \cdot \dfrac{f_{\boldsymbol{V}_{-d} \mid V_d, \boldsymbol{X}}(\boldsymbol{g}_{-d} \mid g, \boldsymbol{x})}{f_{\boldsymbol{G}_{-d} \mid G_d, \boldsymbol{X}}(\boldsymbol{g}_{-d} \mid g, \boldsymbol{x})} \\
        & \qquad \times f_{\boldsymbol{G}_{-d} \mid G_d, \boldsymbol{X}}(\boldsymbol{g}_{-d} \mid g, \boldsymbol{x}) f_{\boldsymbol{X}}(\boldsymbol{x}) \, \mathrm{d}\boldsymbol{g}_{-d} \, \mathrm{d}\boldsymbol{x} & (Y^{(\boldsymbol{g})} \bigCI \boldsymbol{V} \mid \boldsymbol{X}) \\
        &= \int_{\boldsymbol{x}} \int_{\boldsymbol{g}_{-d}} E[Y^{(g, \; \boldsymbol{g}_{-d})} \mid \boldsymbol{X}] f_{\boldsymbol{G}_{-d} \mid G_d, \boldsymbol{X}}(\boldsymbol{g}_{-d} \mid g, \boldsymbol{x}) f_{\boldsymbol{X}}(\boldsymbol{x}) \, \mathrm{d}\boldsymbol{g}_{-d} \, \mathrm{d}\boldsymbol{x} & (**) \\
        &= \int_{\boldsymbol{x}} \int_{\boldsymbol{g}_{-d}} E[Y^{(g, \; \boldsymbol{g}_{-d})} \mid G_d = g, \boldsymbol{G}_{-d} = \boldsymbol{g}_{-d}, \boldsymbol{X}] \\
        & \qquad \times f_{\boldsymbol{G}_{-d} \mid G_d, \boldsymbol{X}}(\boldsymbol{g}_{-d} \mid g, \boldsymbol{x}) f_{\boldsymbol{X}}(\boldsymbol{x}) \, \mathrm{d}\boldsymbol{g}_{-d} \, \mathrm{d}\boldsymbol{x} & (Y^{(\boldsymbol{g})} \bigCI \boldsymbol{G} \mid \boldsymbol{X}) \\
        &= E_{\boldsymbol{X}} \left\{ E_{\boldsymbol{G}_{-d} \mid G_{d} = g, \boldsymbol{X}}\left( E\left[Y \mid G_{d} = g, \boldsymbol{G}_{-d}, \boldsymbol{X} \right] \right) \right\} & (Y^{(\boldsymbol{G})} = Y) \\
        &= E_{\boldsymbol{X}} \left\{ E\left[Y \mid G_{d} = g, \boldsymbol{X} \right] \right\},
\end{align*}
where ($**$) denotes assuming $f_{\boldsymbol{V}_{-d} \mid V_d, \boldsymbol{X}}(\boldsymbol{g}_{-d} \mid g, \boldsymbol{x}) = f_{\boldsymbol{G}_{-d} \mid G_d, \boldsymbol{X}}(\boldsymbol{g}_{-d} \mid g, \boldsymbol{x})$ for all $g, \boldsymbol{g}_{-d}$ and $\boldsymbol{x}$, $f_{V_d \mid \boldsymbol{X}} (g' \mid \boldsymbol{x}) = \delta(g' - g)$ for all $g'$, and the Dirac delta function is defined as
\begin{align*}
    \delta(a) = \begin{cases}
        \infty & a = 0 \\
        0 & \text{otherwise}
    \end{cases} \qquad \text{and} \qquad \int_{-\infty}^\infty \delta(a) \, \mathrm{d}a = 1.
\end{align*}

\newpage

\subsection{Bivariable monotonicity of NTCP under a pointwise intervention} \label{apdx:bivar-mono}

Under model (\ref{eq:schipper}) and assuming stochastic ordering of conditional dose distributions, intervening on dose from $d$ to $d^{*}$ at a fixed volume $g$ gives
\begin{align*}
      d^* \leq d \quad &\Rightarrow \quad D_{d^*}(g) \preceq D_{d}(g) & (\text{stochastic ordering})\\
     &\Rightarrow \quad \textstyle \sum_{d'  = 1}^{\mathcal{D}} \alpha_{d'} P(D = d' \mid G_{d^*} = g) 
     \leq \sum_{d' = 1}^{\mathcal{D}} \alpha_{d'} P(D = d' \mid G_{d} = g) & (\text{non-decreasing } (\ref{eq:func-mono})) \\
     &\Rightarrow \quad \textstyle \sum_{d' = 1}^{\mathcal{D}} \alpha_{d'} E_{\boldsymbol{G}_{-d^{*}} \mid G_{d^{*}}}\left[P(D = d' \mid G_{d^{*}} = g, \boldsymbol{G}_{-d^{*}})\right] \\
     &\quad\qquad \leq \textstyle \sum_{d' = 1}^{\mathcal{D}} \alpha_{d'} E_{\boldsymbol{G}_{-d} \mid G_{d}}\left[P(D = d' \mid G_{d} = g, \boldsymbol{G}_{-d})\right] & (\ref{eq:cond-dose-relation}) \\
     &\Rightarrow \quad \textstyle \sum_{d' = 1}^{\mathcal{D}} \alpha_{d'} E_{\boldsymbol{G}_{-d^*} \mid G_{d^*}}[B_{d'}(G_{d^*} = g, \boldsymbol{G}_{-d^*})] \\
     &\quad\qquad \leq \textstyle \sum_{d' = 1}^{\mathcal{D}} \alpha_{d'} E_{\boldsymbol{G}_{-d} \mid G_{d}}[B_{d'}(G_{d} = g, \boldsymbol{G}_{-d})] & (\ref{eq:ddvh})\\
     &\Rightarrow \quad E_{\boldsymbol{G}_{-d^*} \mid G_{d^*}}\left[\textstyle \sum_{d' = 1}^{\mathcal{D}} \alpha_{d'} B_{d'}(G_{d^*} = g, \boldsymbol{G}_{-d^*}) \right] \\
     &\quad\qquad \leq E_{\boldsymbol{G}_{-d} \mid G_{d}}\left[\textstyle \sum_{d' = 1}^{\mathcal{D}} \alpha_{d'} B_{d'}(G_{d} = g, \boldsymbol{G}_{-d}) \right] \\
     &\Rightarrow \quad  E_{\boldsymbol{G}_{-d^*} \mid G_{d^*}, \boldsymbol{X}}\left[\psi\left\{E( Y^{( \boldsymbol{B} \; = \; \boldsymbol{B}(G_{d^*} \; = \; g, \; \boldsymbol{G}_{-d^*} ) )} \mid \boldsymbol{X} )\right\}  \right] 
     \\
     &\quad\qquad \leq E_{\boldsymbol{G}_{-d} \mid G_{d}, \boldsymbol{X}}\left[\psi \left\{E\left( Y^{( \boldsymbol{B} \; = \; \boldsymbol{B}(G_{d} \; = \; g, \; \boldsymbol{G}_{-d}))} \mid \boldsymbol{X} \right)\right\}  \right] & (\ref{eq:schipper})\\
     &\Rightarrow \quad  E_{\boldsymbol{G}_{-d^*} \mid G_{d^*}, \boldsymbol{X}}\left[ E\left( Y^{( \boldsymbol{B} \; = \; \boldsymbol{B}(G_{d^*} \; = \; g, \; \boldsymbol{G}_{-d^*})
     )} \mid \boldsymbol{X} \right) \right] \\
     &\quad\qquad \leq E_{\boldsymbol{G}_{-d} \mid G_{d}, \boldsymbol{X}}\left[ E\left( Y^{(\boldsymbol{B} \; = \; \boldsymbol{B}(G_{d} \; = \; g, \; \boldsymbol{G}_{-d}))} \mid \boldsymbol{X} \right) \right] &  (\text{monotone } \psi(\cdot)) \\
     &\Rightarrow \quad  E_{\boldsymbol{G}_{-d^*} \mid G_{d^*}, \boldsymbol{X}}\left[ E\left( Y^{( g, \; \boldsymbol{G}_{-d^*} )} \mid \boldsymbol{X} \right) \right] \\
     &\quad\qquad \leq E_{\boldsymbol{G}_{-d} \mid G_{d}, \boldsymbol{X}}\left[ E\left( Y^{( g, \; \boldsymbol{G}_{-d})} \mid \boldsymbol{X} \right) \right]  & (\ref{eq:dvh}) \\
     &\Rightarrow \quad  E_{\boldsymbol{G}_{-d^*} \mid G_{d^*}, \boldsymbol{X}}\left[ E\left( Y^{( g, \; \boldsymbol{G}_{-d^*} )} \mid G_{d^*} = g, \boldsymbol{G}_{-d^*}, \boldsymbol{X} \right) \right] \\
     &\quad\qquad \leq E_{\boldsymbol{G}_{-d} \mid G_{d}, \boldsymbol{X}}\left[ E\left( Y^{( g, \; \boldsymbol{G}_{-d})} \mid G_d = g, \boldsymbol{G}_{-d}, \boldsymbol{X} \right) \right]  & (Y^{(\boldsymbol{g})} \bigCI \boldsymbol{G} \mid \boldsymbol{X}) \\
     &\Rightarrow \quad  E_{\boldsymbol{G}_{-d^*} \mid G_{d^*}, \boldsymbol{X}}\left[ E\left( Y \mid G_{d^*} = g, \boldsymbol{G}_{-d^*}, \boldsymbol{X} \right) \right] \\
     &\quad\qquad \leq E_{\boldsymbol{G}_{-d} \mid G_{d}, \boldsymbol{X}}\left[ E\left(  Y\mid G_d = g, \boldsymbol{G}_{-d}, \boldsymbol{X} \right) \right]  & (Y^{(\boldsymbol{G})} = Y) \\
     &\Rightarrow \quad E\left[Y^{(g, \; \boldsymbol{V}_{-d^*})}\right] \leq E\left[Y^{(g, \; \boldsymbol{V}_{-d})}\right] & (\ref{eq:monotonicity_univariate})
\end{align*}
where $B_{d'}(G_{d} = g, \boldsymbol{G}_{-d}) = P(D = d' \mid G_{d} = g, \boldsymbol{G}_{-d})$ for $d \in \{1, 2, \ldots, \mathcal{D}\}$ and $\boldsymbol{B}(G_{d} = g, \boldsymbol{G}_{-d}) = (B_{1}(G_{d} = g, \boldsymbol{G}_{-d}), \ldots, B_{\mathcal{D}}(G_{d} = g, \boldsymbol{G}_{-d}))$.

\newpage

\noindent Similarly, intervening on OAR volume from $g$ to $g^*$ at a fixed dose $d$ gives
\begin{align*}
      g^* \leq g \quad &\Rightarrow \quad D_{d}(g^*) \preceq D_{d}(g) & (\text{stochastic ordering}) \\
     &\Rightarrow \quad \textstyle \sum_{d' = 1}^{\mathcal{D}} \alpha_{d'} P(D = d' \mid G_{d} = g^*) \\
     &\quad\qquad\leq \textstyle \sum_{d' = 1}^{\mathcal{D}} \alpha_{d'} P(D = d' \mid G_{d} = g) & (\text{non-decreasing } (\ref{eq:func-mono})) \\
     &\Rightarrow \textstyle\quad \sum_{d' = 1}^{\mathcal{D}} \alpha_{d'} E_{\boldsymbol{G}_{-d} \mid G_{d}}\left[P(D = d' \mid G_{d} = g^*, \boldsymbol{G}_{-d})\right] \\
     &\quad\qquad\leq \textstyle \sum_{d' = 1}^{\mathcal{D}} \alpha_{d'} E_{\boldsymbol{G}_{-d} \mid G_{d}}\left[P(D = d' \mid G_{d} = g, \boldsymbol{G}_{-d})\right] \\
     &\Rightarrow \textstyle \quad \sum_{d' = 1}^{\mathcal{D}} \alpha_{d'} E_{\boldsymbol{G}_{-d} \mid G_{d}}[B_{d'}(G_{d} = g^*, \boldsymbol{G}_{-d})] \\
     &\quad\qquad\leq \textstyle \sum_{d' = 1}^{\mathcal{D}} \alpha_{d'} E_{\boldsymbol{G}_{-d} \mid G_{d}}[B_{d'}(G_{d} = g, \boldsymbol{G}_{-d})] & (\ref{eq:ddvh}) \\
     &\Rightarrow \quad  E_{\boldsymbol{G}_{-d} \mid G_{d}}\left[\textstyle \sum_{d' = 1}^{\mathcal{D}} \alpha_{d'} B_{d'}(G_{d} = g^*, \boldsymbol{G}_{-d}) \right] \\
     &\quad\qquad\leq E_{\boldsymbol{G}_{-d} \mid G_{d}}\left[\textstyle \sum_{d' = 1}^{\mathcal{D}} \alpha_{d'} B_{d'}(G_{d} = g, \boldsymbol{G}_{-d}) \right] \\
     &\Rightarrow \quad  E_{\boldsymbol{G}_{-d} \mid G_{d}, \boldsymbol{X}}\left[\psi\left\{E\left( Y^{(\boldsymbol{B} \; = \; \boldsymbol{B}(G_{d} \; = \; g^*, \; \boldsymbol{G}_{-d}))} \mid \boldsymbol{X} \right)\right\}  \right] 
     \\
     &\quad\qquad\leq E_{\boldsymbol{G}_{-d} \mid G_{d}, \boldsymbol{X}}\left[\psi\left\{E\left( Y^{(\boldsymbol{B} \; = \; \boldsymbol{B}(G_{d} \; = \; g, \; \boldsymbol{G}_{-d}))} \mid \boldsymbol{X} \right)\right\} \right]  & (\ref{eq:schipper}) \\
     &\Rightarrow \quad  E_{\boldsymbol{G}_{-d} \mid G_{d}, \boldsymbol{X}}\left[ E\left( Y^{(\boldsymbol{B} \; = \; \boldsymbol{B}(G_{d} \; = \; g^*, \; \boldsymbol{G}_{-d}))} \mid \boldsymbol{X} \right)   \right] 
     \\
     &\quad\qquad\leq E_{\boldsymbol{G}_{-d} \mid G_{d}, \boldsymbol{X}}\left[ E\left( Y^{(\boldsymbol{B} \; = \; \boldsymbol{B}(G_{d} \; = \; g, \; \boldsymbol{G}_{-d}))} \mid \boldsymbol{X} \right)  \right]  & (\text{monotone } \psi(\cdot)) \\
     &\Rightarrow \quad  E_{\boldsymbol{G}_{-d} \mid G_{d}, \boldsymbol{X}}\left[ E\left( Y^{(G_{d} \; = \; g^*, \; \boldsymbol{G}_{-d})} \mid \boldsymbol{X} \right)   \right] 
     \\
     &\quad\qquad\leq E_{\boldsymbol{G}_{-d} \mid G_{d}, \boldsymbol{X}}\left[ E\left( Y^{(G_{d} \; = \; g, \; \boldsymbol{G}_{-d})} \mid \boldsymbol{X} \right)  \right]  & (\ref{eq:dvh}) \\
     &\Rightarrow \quad  E_{\boldsymbol{G}_{-d} \mid G_{d}, \boldsymbol{X}}\left[ E\left( Y^{(G_{d} \; = \; g^*, \; \boldsymbol{G}_{-d})} \mid G_{d} = g^*, \boldsymbol{G}_{-d}, \boldsymbol{X} \right)   \right] 
     \\
     &\quad\qquad\leq E_{\boldsymbol{G}_{-d} \mid G_{d}, \boldsymbol{X}}\left[ E\left( Y^{(G_{d} \; = \; g, \; \boldsymbol{G}_{-d})} \mid G_{d} = g, \boldsymbol{G}_{-d}, \boldsymbol{X} \right)  \right]  & (Y^{(\boldsymbol{g})} \bigCI \boldsymbol{G} \mid \boldsymbol{X}) \\
     &\Rightarrow \quad  E_{\boldsymbol{G}_{-d} \mid G_{d}, \boldsymbol{X}}\left[E\left( Y \mid G_{d} = g^*, \boldsymbol{G}_{-d}, \boldsymbol{X} \right) \right] \\
     &\quad\qquad \leq E_{\boldsymbol{G}_{-d} \mid G_{d}, \boldsymbol{X}}\left[ E\left( Y \mid G_{d} = g, \boldsymbol{G}_{-d}, \boldsymbol{X} \right)\right] & (Y^{(\boldsymbol{G})} = Y) \\
     &\Rightarrow \quad E\left[Y^{(g^*, \; \boldsymbol{V}_{-d})}\right] \leq E\left[Y^{(g, \; \boldsymbol{V}_{-d})}\right], & (\ref{eq:monotonicity_univariate})
\end{align*}
where $B_{d'}(G_{d} = g, \boldsymbol{G}_{-d}) = P(D = d' \mid G_{d} = g, \boldsymbol{G}_{-d})$ for $d \in \{1, 2, \ldots, \mathcal{D}\}$ and $\boldsymbol{B}(G_{d} = g, \boldsymbol{G}_{-d}) = (B_{1}(G_{d} = g, \boldsymbol{G}_{-d}), \ldots, B_{{\mathcal{D}}}(G_{d} = g, \boldsymbol{G}_{-d}))$.

\newpage

\subsection{Derivation of the true causal estimand under the simulated data generating mechanism}\label{apdx:sim-estimands}

\begin{align*}
    E&\left[Y^{\left(g, \; \boldsymbol{V}_{-d} \right)}\right] = E_{X}\{E_{\boldsymbol{G}_{-d} \mid G_{d} = g, X}\left[E\left( Y \mid G_{d} = g, \boldsymbol{G}_{-d}, X \right) \right]\} \\
    &= \sum_{x = 0}^1 \left[ \int_{\boldsymbol{g}_{-d}} E[Y \mid G_d = g, \boldsymbol{G}_{-d} = \boldsymbol{g}_{-d}, X = x] f_{\boldsymbol{G}_{-d} \mid G_d, X}(\boldsymbol{g}_{-d} \mid g, x)  \, \mathrm{d}\boldsymbol{g}_{-d} \right] P(X = x) \\
    &= \sum_{x = 0}^1 \left[ \int_{\boldsymbol{g}_{-d}} \int_{\mu} \int_{\sigma} E[Y \mid G_d = g, \boldsymbol{G}_{-d} = \boldsymbol{g}_{-d}(\mu, \sigma(\mu, g)), \mu = \mu, \sigma = \sigma(\mu, g), X = x]  \right. \\ 
    &\qquad \left. \times f_{\boldsymbol{G}_{-d}, \mu, \sigma \mid G_d, X}(\boldsymbol{g}_{-d}(\mu, \sigma(\mu, g)), \mu, \sigma(\mu, g) \mid g, x)  \, \mathrm{d}\mu \, \mathrm{d}\sigma \, \mathrm{d}\boldsymbol{g}_{-d} \right] P(X = x) \\
    &= \sum_{x = 0}^1 \left[ \int_{\mu} E[Y \mid \mu = \mu, \sigma = \sigma(\mu, g), X = x] f_{\mu \mid G_d, X}(\mu \mid g, x)  \, \mathrm{d}\mu \right] P(X = x) \\
    &= \sum_{x = 0}^1 \left[ \int_{\mu} E[Y \mid \mu = \mu, \sigma = \sigma(\mu, g), X = x] \dfrac{f_{\mu, G_d \mid X}(\mu, g \mid x)}{\int_{\mu} f_{\mu, G_d \mid X}(\mu, g \mid x) \, \mathrm{d}\mu}  \, \mathrm{d}\mu \right] P(X = x) \\
    &= \sum_{x = 0}^1 \left[ \int_{\mu} \text{expit}\{\gamma_0 + \gamma_1 \mu + \gamma_2 x\} \cdot \dfrac{f_{\mu \mid X}(\mu \mid x) f_{\sigma}(\sigma(\mu, g)) |\text{\textbf{J}}|(\mu, g) }{\int_{\mu} f_{\mu \mid X}(\mu \mid x) f_{\sigma}(\sigma(\mu, g)) |\text{\textbf{J}}|(\mu, g) \, \mathrm{d}\mu}  \, \mathrm{d}\mu \right] P(X = x) \\
     &= \sum_{x = 0}^1 \left[ \int_{\mu_{\text{min}}(g)}^{\mu_{\text{max}}(g)}  \dfrac{\text{expit}\{\gamma_0 + \gamma_1 \mu + \gamma_2 x\} \cdot (\mu - a_{\text{min}})^{\alpha(x) - 1} (a_{\text{max}} - \mu)^{\beta(x) - 1} |\text{\textbf{J}}|(\mu, g)}{\int_{\mu_{\text{min}}(g)}^{\mu_{\text{max}}(g)} (\mu - a_{\text{min}})^{\alpha(x) - 1} (a_{\text{max}} - \mu)^{\beta(x) - 1} |\text{\textbf{J}}|(\mu, g) \, \mathrm{d}\mu}  \, \mathrm{d}\mu \right] P(X = x)
\end{align*}
where,
\begin{align*}
\alpha(x) &= 2 \cdot \mathbbm{1}(x = 1) + \frac{4}{3} \cdot \mathbbm{1}(x = 0) \\
\beta(x) &= 2 + \frac{4}{3} - \alpha(x) \\
\text{expit}\{\eta\} &= \dfrac{1}{1 + \exp\{-\eta\}} \\
        |\mathrm{\textbf{J}}| &= \begin{vmatrix} \frac{\partial \mu}{\partial \mu} & \frac{\partial \mu}{\partial \sigma} \\[1em]
        \frac{\partial g}{\partial \mu} & \frac{\partial g}{\partial \sigma}\end{vmatrix}^{-1} = \begin{vmatrix} 1 & 0 \\[1em]
        -\frac{1}{\sigma^2} \int_0^{d} \left(\frac{t - \mu}{\sigma}\right) \phi\left(\frac{t - \mu}{\sigma}\right) \, \mathrm{d}t & \frac{1}{\sigma^2} \int_0^{d}   \phi\left(\frac{t - \mu}{\sigma}\right) \left[ 1 - \left(\frac{t - \mu}{\sigma}\right)^2 \right] \, \mathrm{d}t \end{vmatrix}^{-1} \\
        &= [\sigma]^2 \left|\int_0^{d}   \phi\left(\frac{t - \mu}{\sigma}\right) \left[ 1 - \left(\frac{t - \mu}{\sigma}\right)^2 \right] \, \mathrm{d}t\right|^{-1}
    \end{align*}

\newpage

\noindent We note that the deterministic relationship between the DVHs, $\boldsymbol{g}$, and the DVH parameters, $(\mu, \sigma)$, are given by,
\begin{align*}
    g_1 &= 1 - \int_{30}^{30 + \Delta d} \dfrac{1}{\sigma} \cdot \phi\left(\dfrac{t - \mu}{\sigma}\right) \, \mathrm{d}t \\
    g_2 &= 1 - \int_{30}^{30 + 2 \Delta d} \dfrac{1}{\sigma} \cdot \phi\left(\dfrac{t - \mu}{\sigma}\right) \, \mathrm{d}t \\
    &\vdots \\
    g_\mathcal{D} &= 1 - \int_{30}^{30 + 26 \Delta d} \dfrac{1}{\sigma} \cdot \phi\left(\dfrac{t - \mu}{\sigma}\right) \, \mathrm{d}t
\end{align*}

\begin{align*}
    E&\left[Y^{(V_{d^*}, \; \boldsymbol{V}_{-d^*})}\right] \\
    &= \sum_{x = 0}^1 \left[ \int_{g} E_{\boldsymbol{G}_{-d^*} \mid G_{d^*} = g, X = x}\left[ E\left(Y \mid G_{d^*} = g, \boldsymbol{G}_{-d^*}, X = x \right) \right] f_{V_{d^*} \mid X}(g \mid x) \, \mathrm{d}g \right\} P(X = x) \\
    &= \sum_{x = 0}^1 \left[ \int_0^q E_{\boldsymbol{G}_{-d^*} \mid G_{d^*} = g, X = x}\left[ E\left(Y \mid G_{d^*} = g, \boldsymbol{G}_{-d^*}, X = x \right) \right] \dfrac{f_{G_{d^*}}(g \mid X = x) }{F_{G_{d^*} \mid X}(q \mid X = x) } \, \mathrm{d}g \right] P(X = x) \\
    &= \sum_{x = 0}^1 \left[ \int_0^q \int_{\mu_{\text{min}}(g)}^{\mu_{\text{max}}(g)} 
    \text{expit}\{\gamma_0 + \gamma_1 \mu + \gamma_2 x\}  \right. \\
    &\times \left. \dfrac{ (\mu - a_{\text{min}})^{\alpha(x) - 1} (a_{\text{max}} - \mu)^{\beta(x) - 1} |\text{\textbf{J}}|(\mu, g) \, }{\int_0^q \int_{\mu_{\text{min}}(g)}^{\mu_{\text{max}}(g)} (\mu - a_{\text{min}})^{\alpha(x) - 1} (a_{\text{max}} - \mu)^{\beta(x) - 1} |\text{\textbf{J}}|(\mu, g)  \, \mathrm{d}\mu  \, \mathrm{d}g} \, \mathrm{d}\mu
    \, \mathrm{d}g \right] P(X = x)
\end{align*}
substituting the expression with (\ref{eq:cond-risk}) and
\begin{align*}
    f_{G_{d^*} \mid X}(g \mid x) &= \int_{\mu}  f (\mu, \sigma(\mu, g) \mid x) \, \mathrm{d}\mu = \int_{\mu} f(\mu \mid x) f(\sigma(\mu, g)) |\text{\textbf{J}}|(\mu, g) \, \mathrm{d}\mu\\
    &= \int_{a_{\text{min}}}^{a_{\text{max}}} \dfrac{(\mu - a_{\text{min}})^{\alpha(x) - 1} (a_{\text{max}} - \mu)^{\beta(x) - 1}}{(a_{\text{max}} - a_{\text{min}})^{\alpha(x) + \beta(x) - 1} \cdot \mathcal{B}(2, \frac{4}{3})}  \cdot \dfrac{\mathbbm{1}^{(\sigma(\mu, g))}_{[1, 2]}}{b_{\text{max}} - b_{\text{min}}} \cdot |\text{\textbf{J}}|(\mu, g) \, \mathrm{d}\mu \\
    &= \kappa \int_{\mu_{\text{min}}(g)}^{\mu_{\text{max}}(g)} (\mu - a_{\text{min}})^{\alpha(x) - 1} (a_{\text{max}} - \mu)^{\beta(x) - 1} |\text{\textbf{J}}|(\mu, g) \, \mathrm{d}\mu  \\
    F_{G_{d^*} \mid X}(q \mid x) &= \int_0^q f_{G_{d^*} \mid X} (g \mid x) \, \mathrm{d}g, \qquad  F_{G_{d^*} \mid X}(0 \mid x) = \int_0^0 f_{G_{d^*} \mid X} (g \mid x)   \, \mathrm{d}g = 0
\end{align*}
where, $\kappa = \dfrac{[ \mathcal{B}(2, \frac{4}{3})]^{-1}}{(b_{\text{max}} - b_{\text{min}}) \cdot (a_{\text{max}} - a_{\text{min}})^{\frac{4}{3} + 2 - 1} }$.

\newpage

\section{Supporting figures}\label{apdx:figures-sect}


\begin{figure}[H]
    \centering
    \subfloat[DVHs]{\includegraphics[width = 0.5\textwidth]{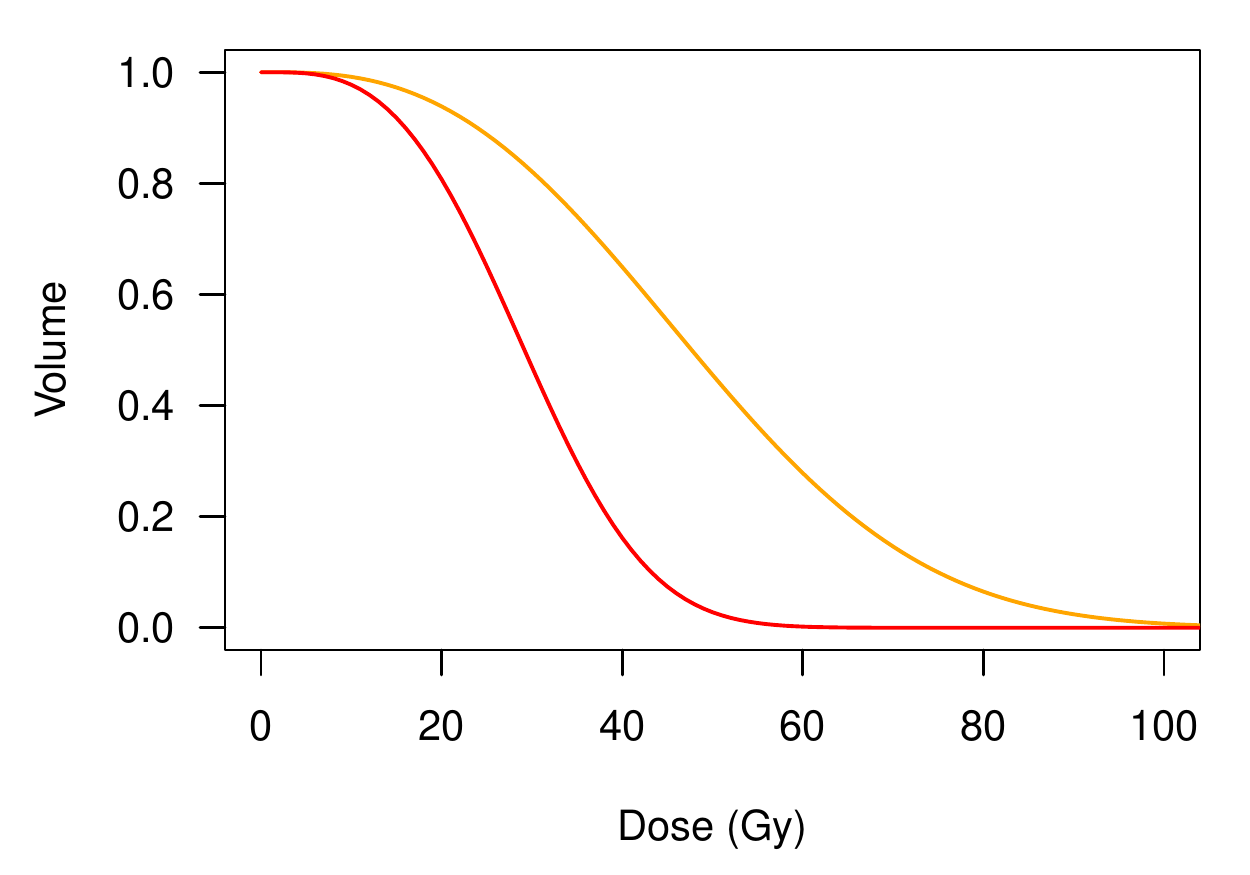}
    \label{fig:stochast-order-DVH-solid}
    }
    \subfloat[dDVHs]{\includegraphics[width = 0.5\textwidth]{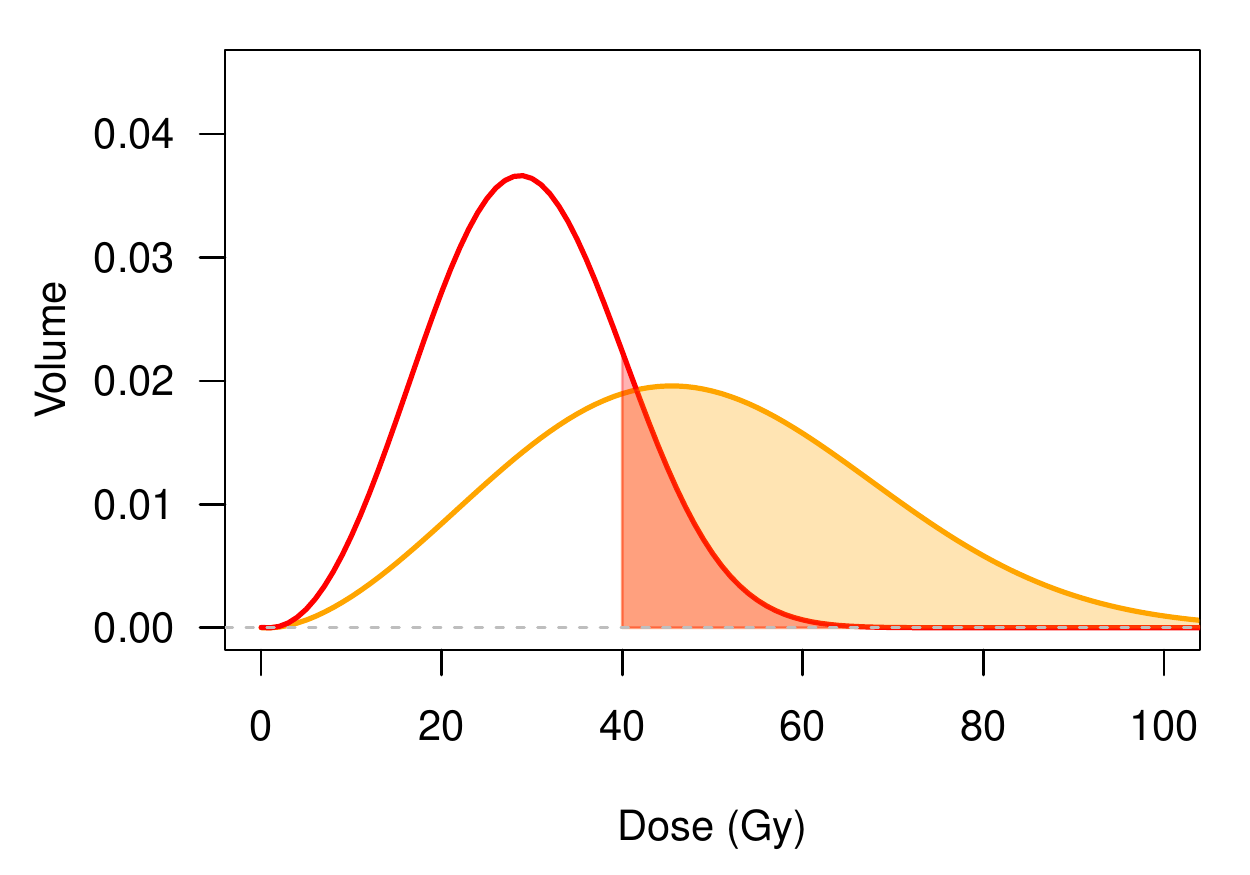}
    \label{fig:stochast-order-dDVH-solid}}

    \captionsetup{labelfont=bf}
    \caption{Stochastic ordering of (a) DVHs (represented as complementary CDFs) at $\boldsymbol{g}^*$(red) and $\boldsymbol{g}$ (orange) and their (b) corresponding density functions. Shaded regions reflect the complementary CDF evaluated at 40 Gy.}
\end{figure}

\begin{figure}[H]
    \centering
    \subfloat[Pointwise average DVHs]{\includegraphics[width = 0.5\textwidth]{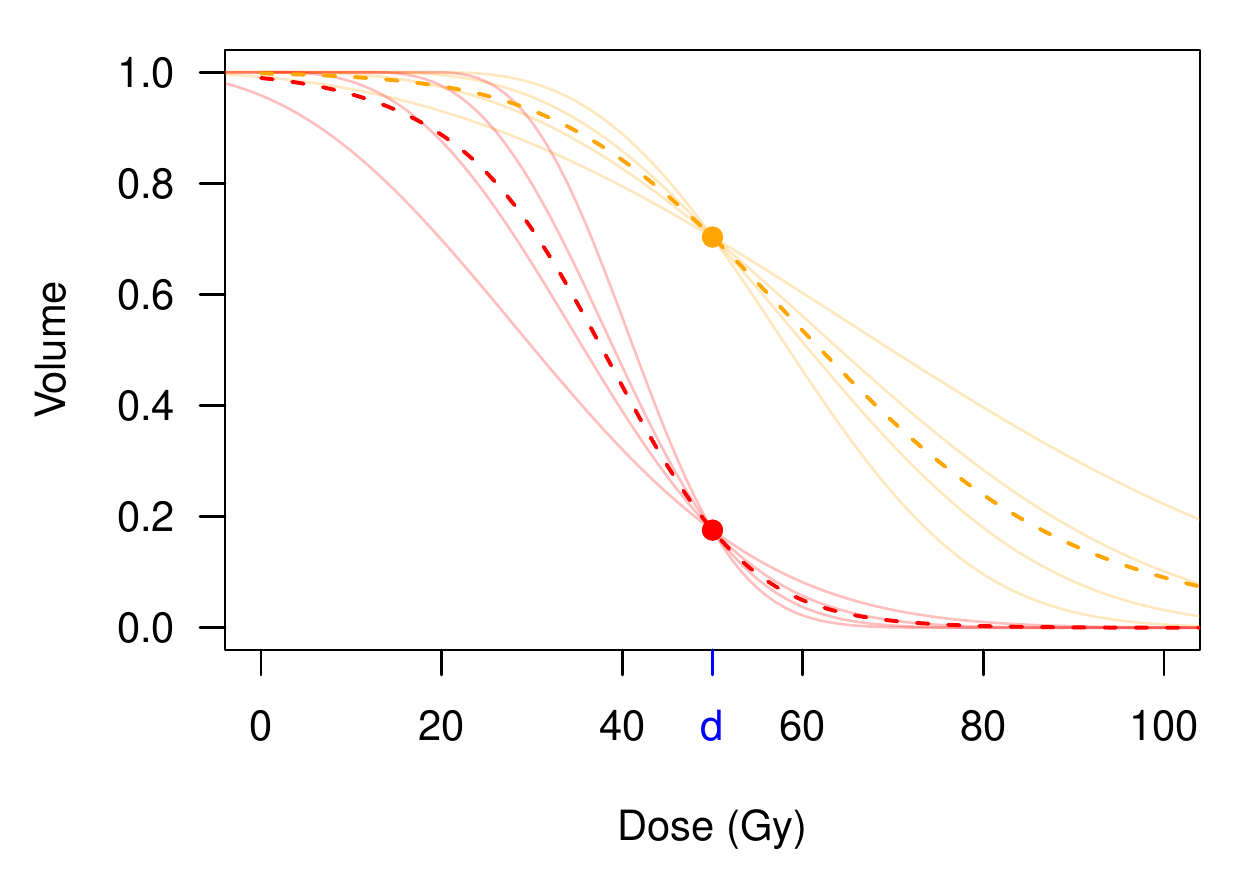}
        \label{fig:stochast-order-average-DVH}
    }
    \subfloat[Pointwise average  dDVHs]{\includegraphics[width = 0.5\textwidth]{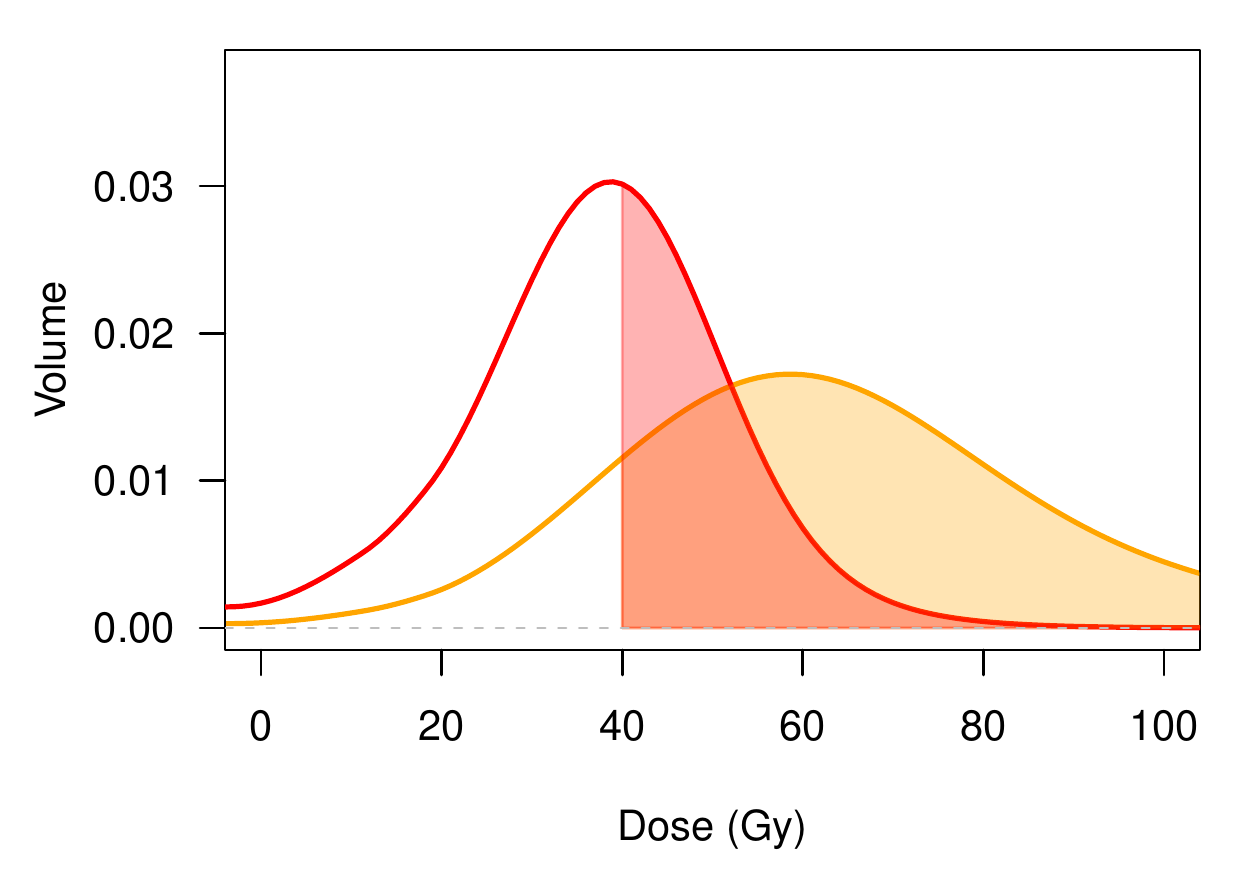}
        \label{fig:stochast-order-average-dDVH}
    }
    \captionsetup{labelfont=bf}
    \caption{Stochastic ordering of (a) pointwise average DVHs (dashed; represented as complementary CDFs), passing through ($d, g^*$) (red) and ($d, g$) (orange) and their (b) corresponding density functions. Shaded regions reflect the complementary CDF evaluated at 40 Gy.}
\end{figure}

\begin{figure}[H]
    \centering
    \scalebox{0.85}{\begin{tikzpicture}[
                > = {Latex[length=2mm, width=2mm]}, 
                shorten > = 1pt, 
                auto,
                scale = 1.3,
                node distance = 1cm, 
                semithick 
            ]
    
            \tikzstyle{state}=[
                circle,
                draw = black,
                thick,
                minimum size = 15mm,
                inner sep = 6.5pt,
                fill = white,
                align = center,
                font = \Large
            ]
    
            \node[state] (Z) at (-2, -1.25) {$G_{d^*}$};
            \node[state] (M) at (0, -1.25) {$\boldsymbol{G}_{-d^*}$};
            \node[state] (F) at (2, -1.25) {$Y$};
            \node[state] (S) at (-2, 1) {$\sigma$};
            \node[state] (X) at (-2, 3.25) {$\boldsymbol{X}$};
            
            \node[state] (X1) at (0, 1) {$\mu$};
            \node[state] (V) at (-4, -1.25) {$V_{d^*}$};
            \path[->, color = black] (Z) edge node[midway] {} (V);
            \path[->] (X) edge node[midway] {} (X1);
            \path[->, color = black] (X) edge [bend left=35, color = black] node {}  coordinate[pos=0.5] (curve) (F);
            \path[->] (S) edge node[midway] {} (Z);
            \path[->] (S) edge node[midway] {} (M);
            \path[->] (X1) edge node[midway] {} (F);
            \path[->] (X1) edge node[midway] {} (Z);
            \path[->] (X1) edge node[midway] {} (M);
            \path[->] (X) edge node {} (V);
\end{tikzpicture}}
    \captionsetup{labelfont=bf}
    \caption{A DAG of the simulated data-generating mechanism.}
    \label{figure:dag_sim}
\end{figure}

\begin{figure}[H]
    \centering
    \includegraphics[width = 0.95\textwidth]{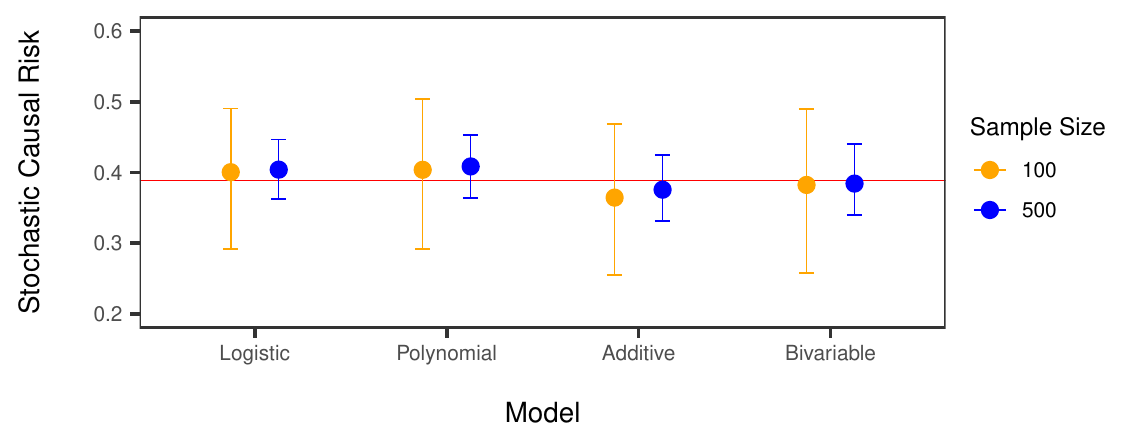}
    \captionsetup{labelfont=bf}
    \caption{Sampling distribution (Monte Carlo mean and 95\% quantile confidence intervals) of the model-estimated stochastic causal risk under the hypothetical truncated intervention, $E[Y^{(V_{d^*})}]$. The horizontal red line denotes the true stochastic causal risk.}
    \label{fig:stochast_samp}
\end{figure}

\begin{figure}[H]
    \centering
    \subfloat[Bladder DVHs stratified by $\geq$Grade 2 acute genitourinary toxicity]{\includegraphics[width = 0.45\textwidth]{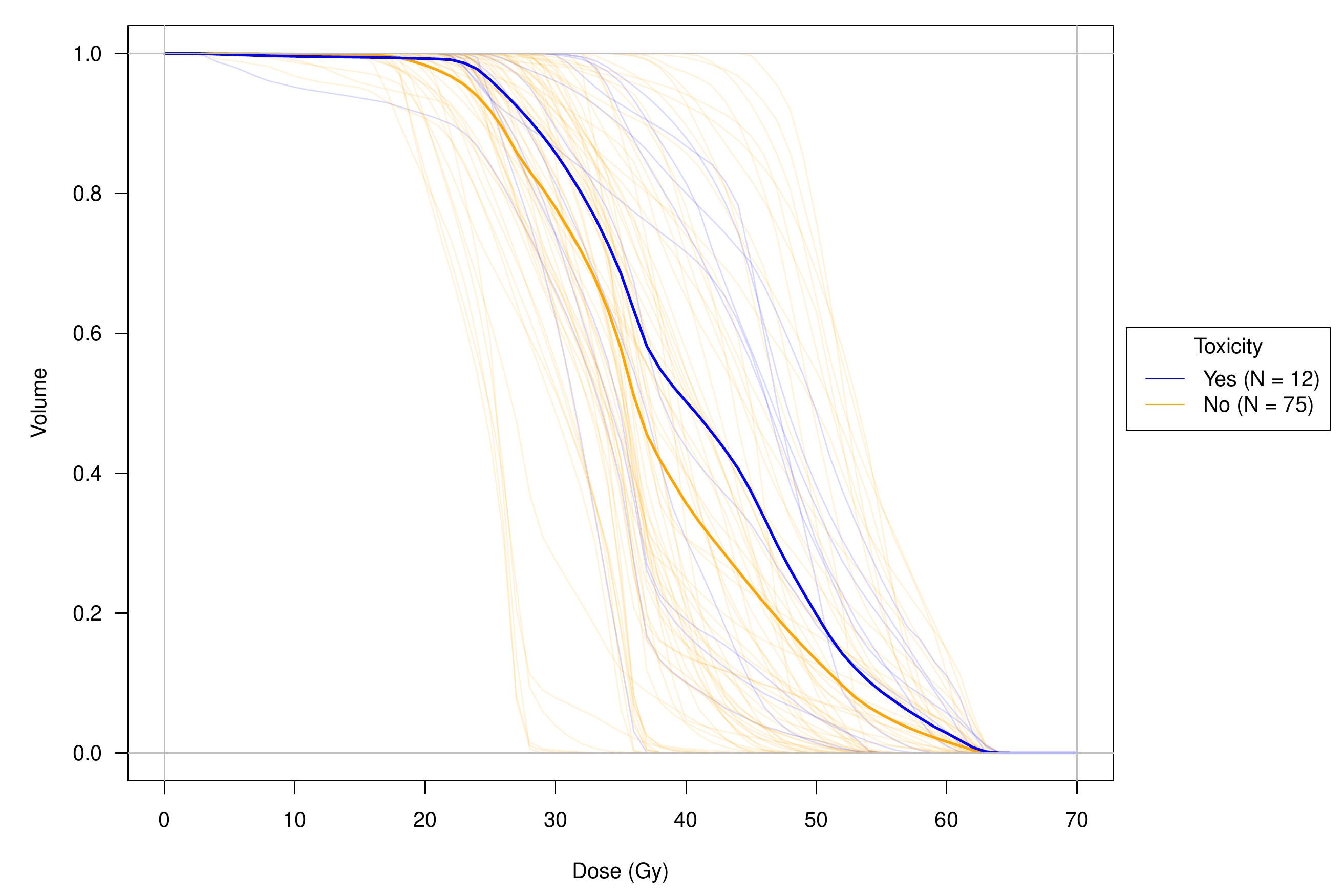}
    \label{fig:pmh-dvh-gutox}
    }\hspace{0.3cm}
    \subfloat[Skin DVHs stratified by $\geq$Grade 3 acute skin (perinanal, inguinal, or genital)  toxicity]{\includegraphics[width = 0.45\textwidth]{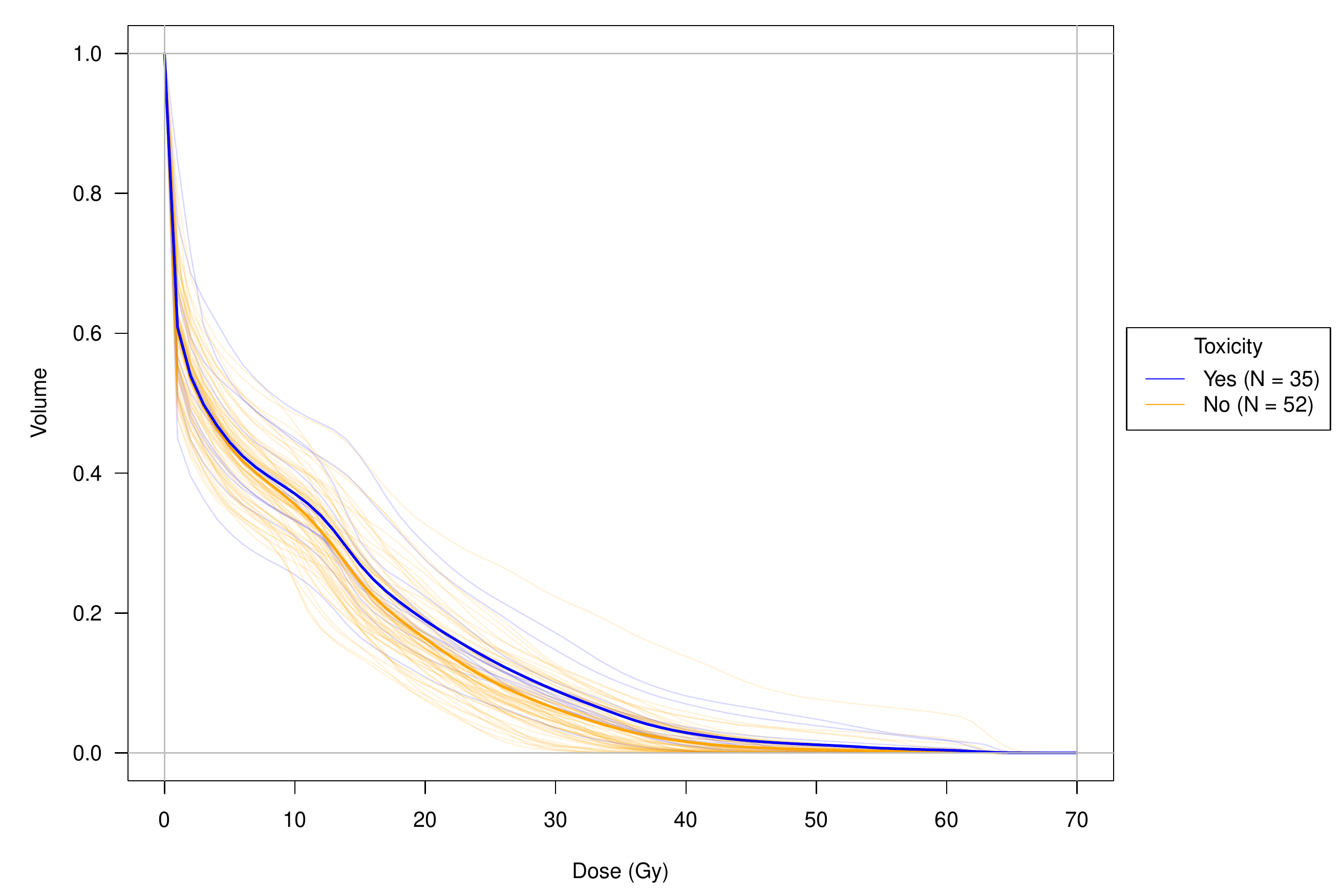}
    \label{fig:pmh-dvh-gitox}
    }
    \captionsetup{labelfont=bf}
    \caption{Dose-volume histograms (presented as complementary CDFs) of 87 anal canal cancer patients. Pointwise average DVHs are illustrated by darker, solid lines.}
\end{figure}

\begin{figure}[H]
    \centering
    \subfloat[Logistic]{\includegraphics[width = 0.45\textwidth]{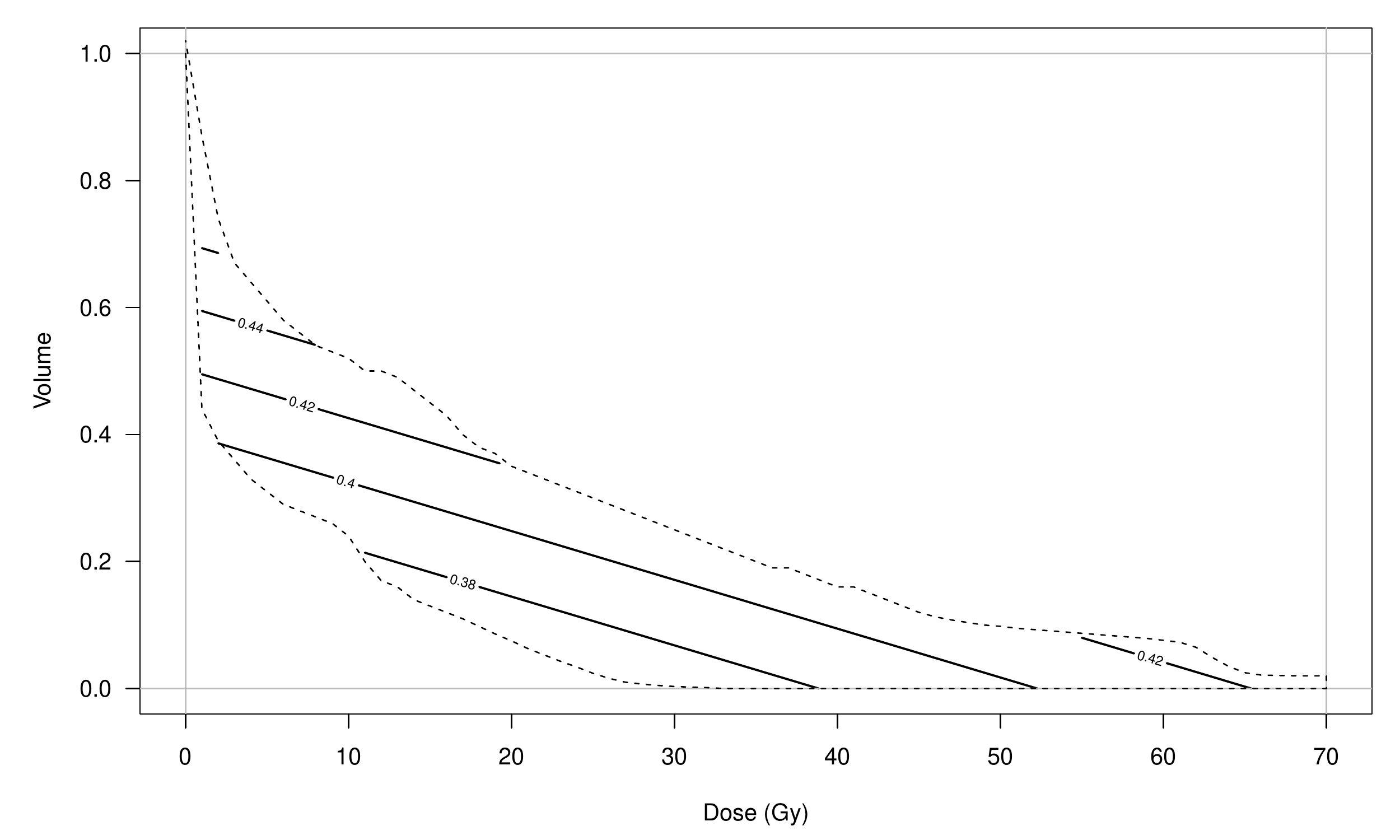}} \hspace{0.6cm}
     \subfloat[Polynomial logistic]{\includegraphics[width = 0.45\textwidth]{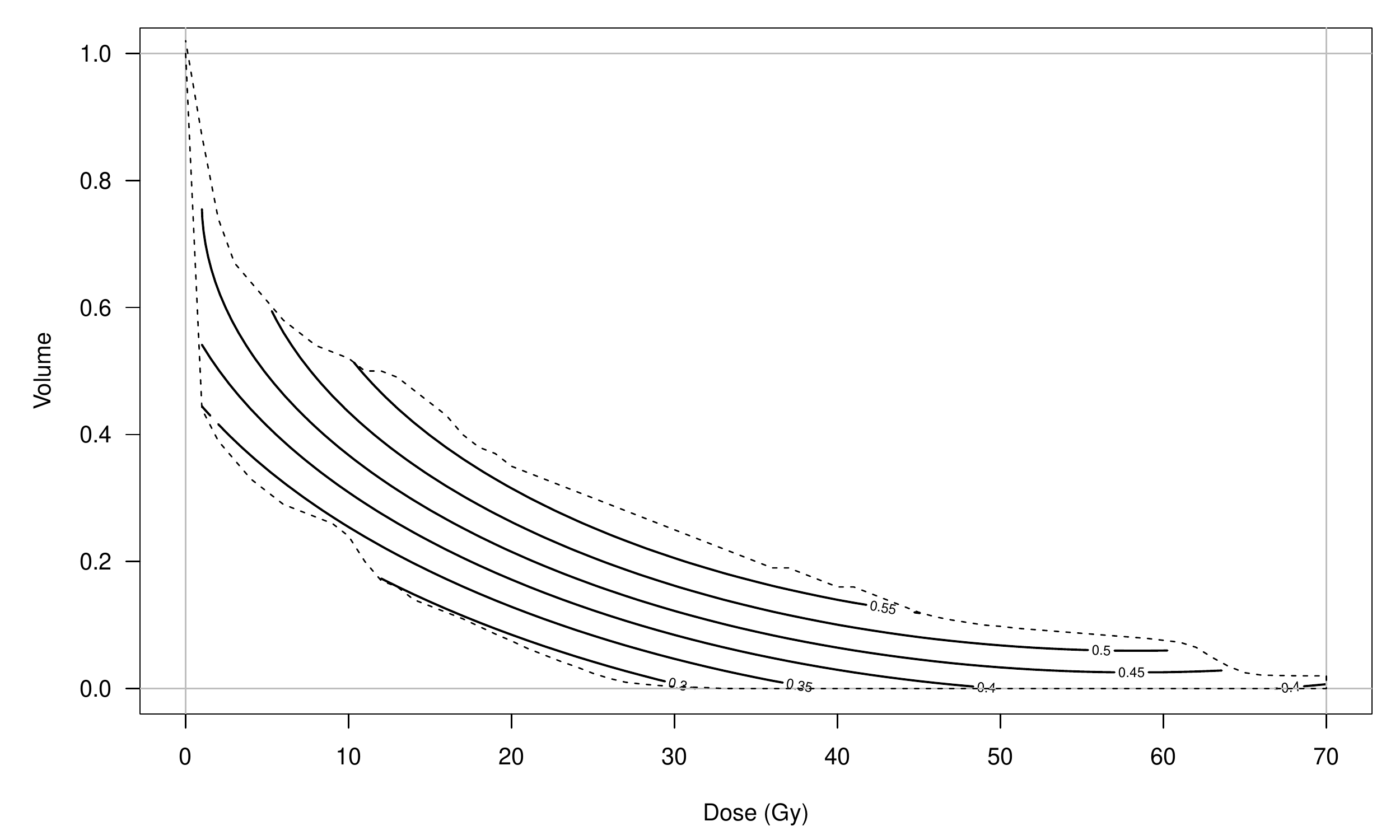}} \\[-0.5em]
    \subfloat[Additive]{\includegraphics[width = 0.45\textwidth]{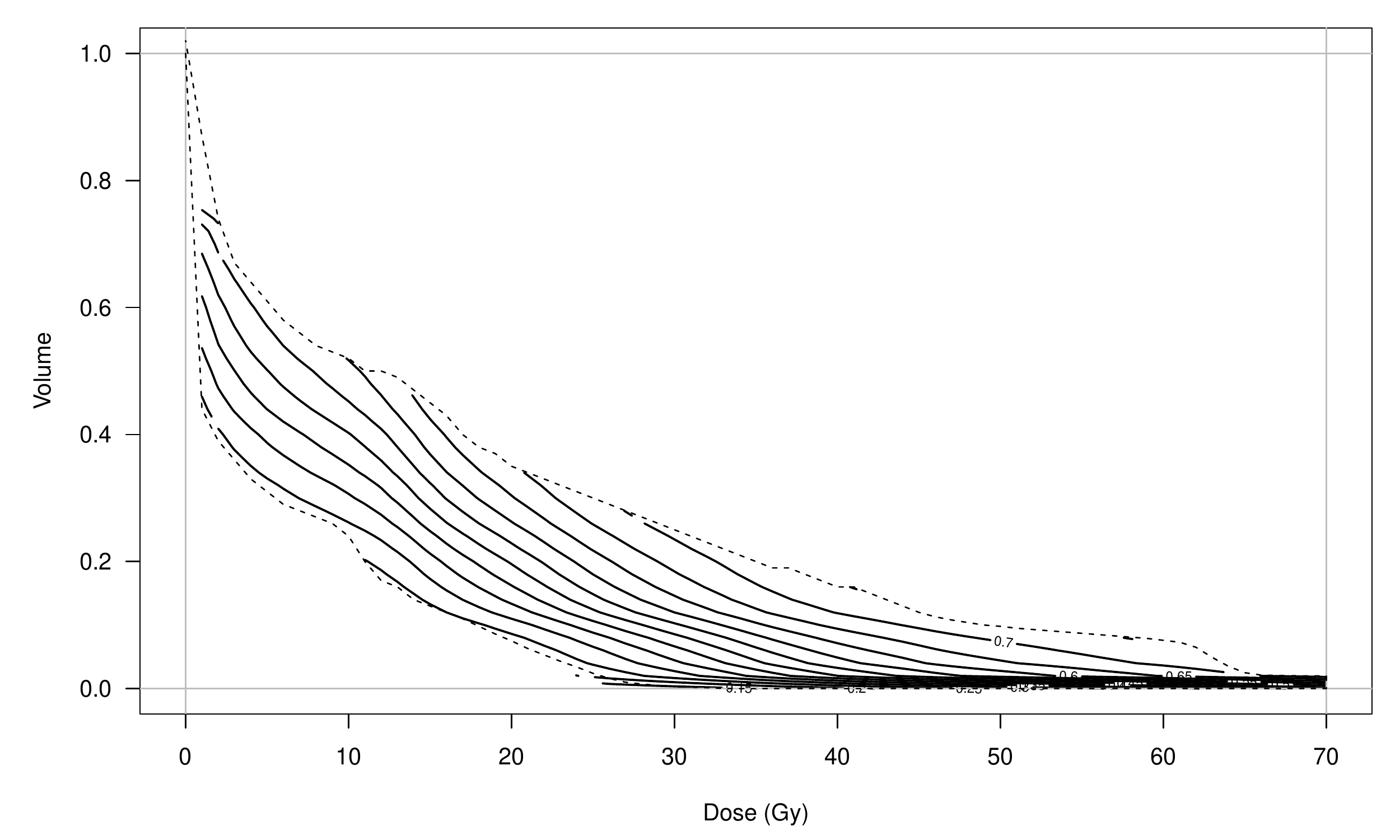}} \hspace{0.6cm}
     \subfloat[Bivariable monotone]{\includegraphics[width = 0.45\textwidth]{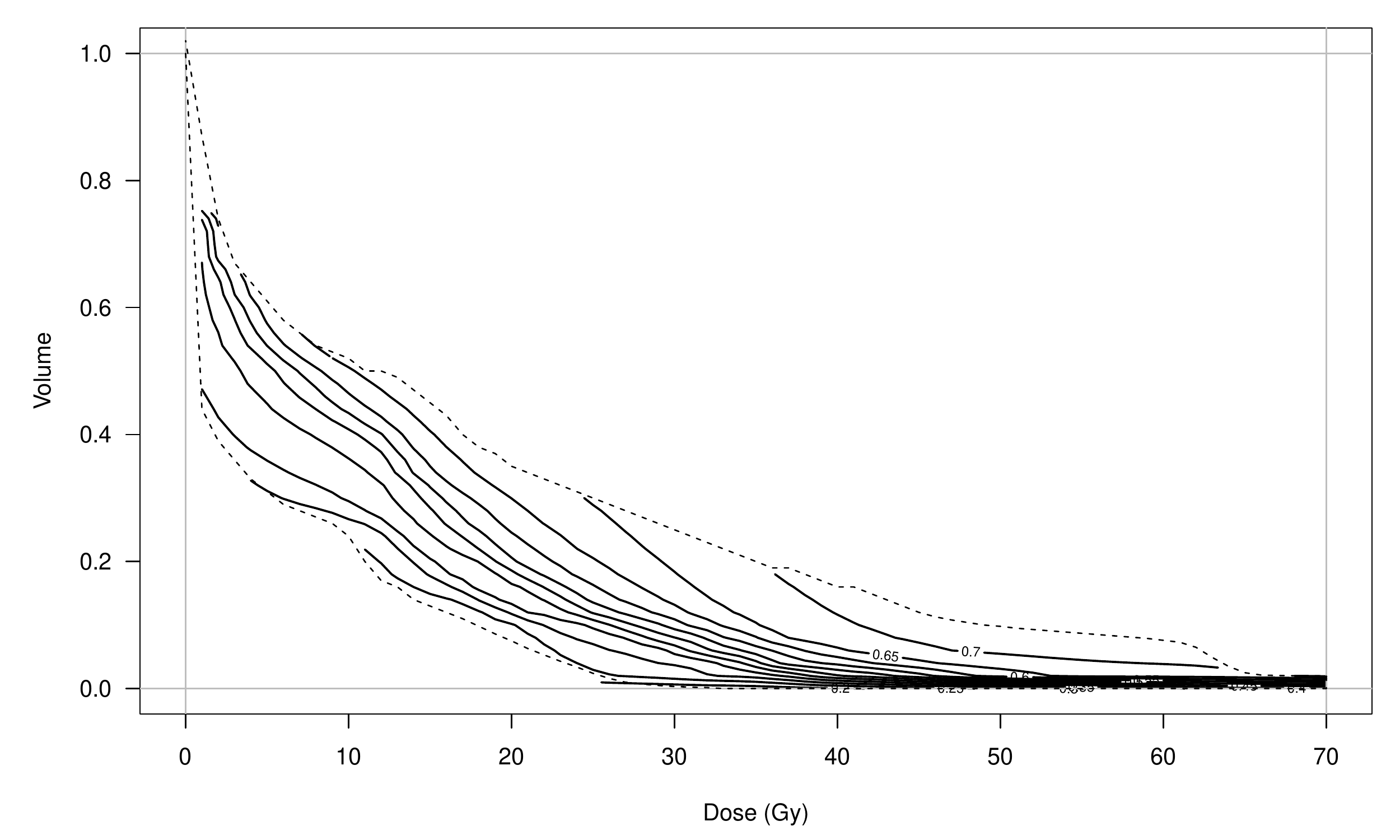}}
     \captionsetup{labelfont=bf}
    \caption{Contour plots for the model-estimated pointwise causal NTCP for skin toxicity at each dose-volume coordinate within the skin DVH domains (outlined by dotted lines) in anal canal cancer patients.}
    \label{figure:mean_contour_gitox}
\end{figure}

\newpage

\section{Additional simulation figures}\label{apdx:sims-sect}

\begin{figure}[H]
    \centering
    \vspace{-1em}
    \subfloat[Logistic]{\includegraphics[width = 0.45\textwidth]{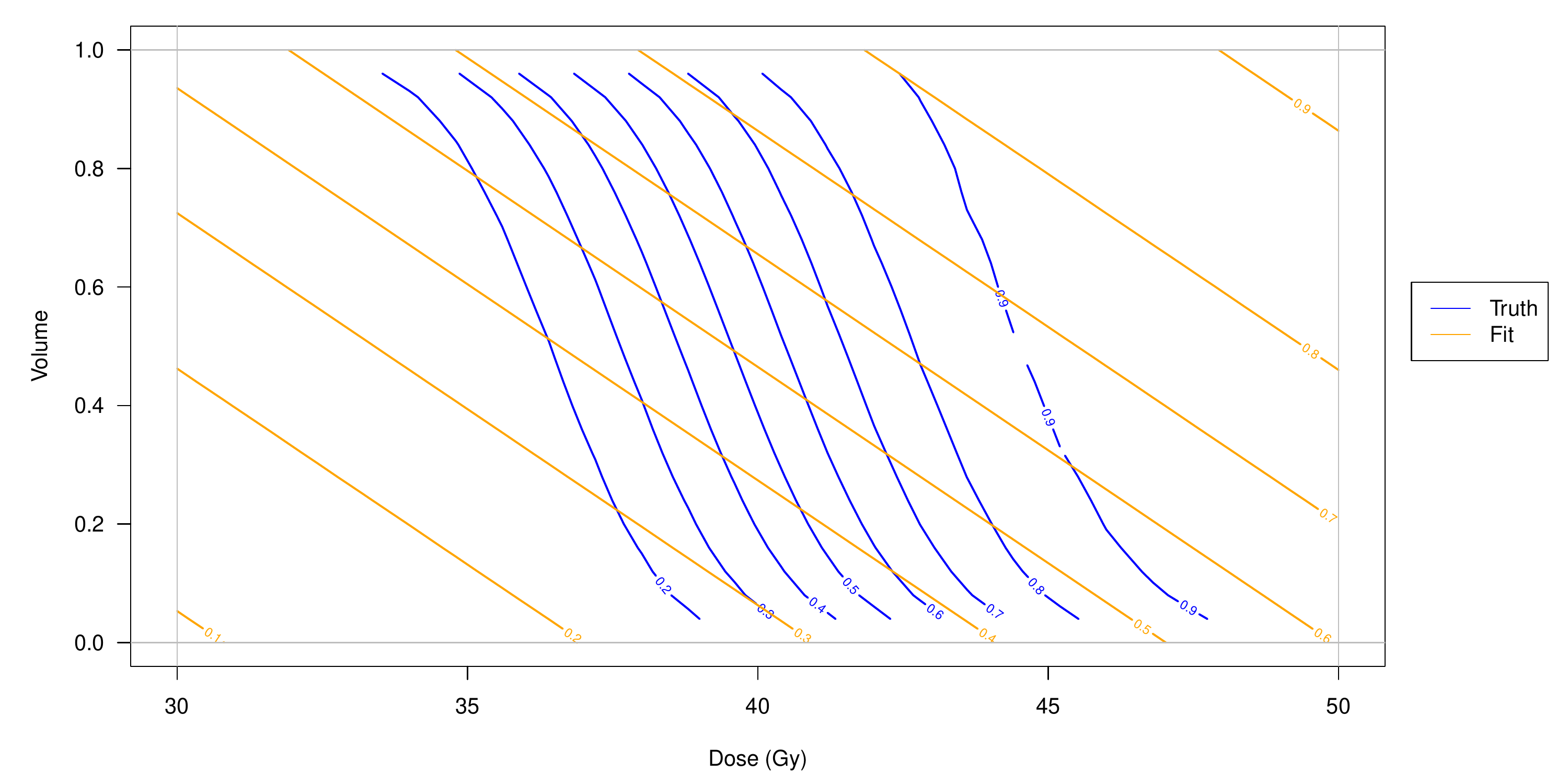}} \hspace{0.6cm}
     \subfloat[Polynomial logistic]{\includegraphics[width = 0.45\textwidth]{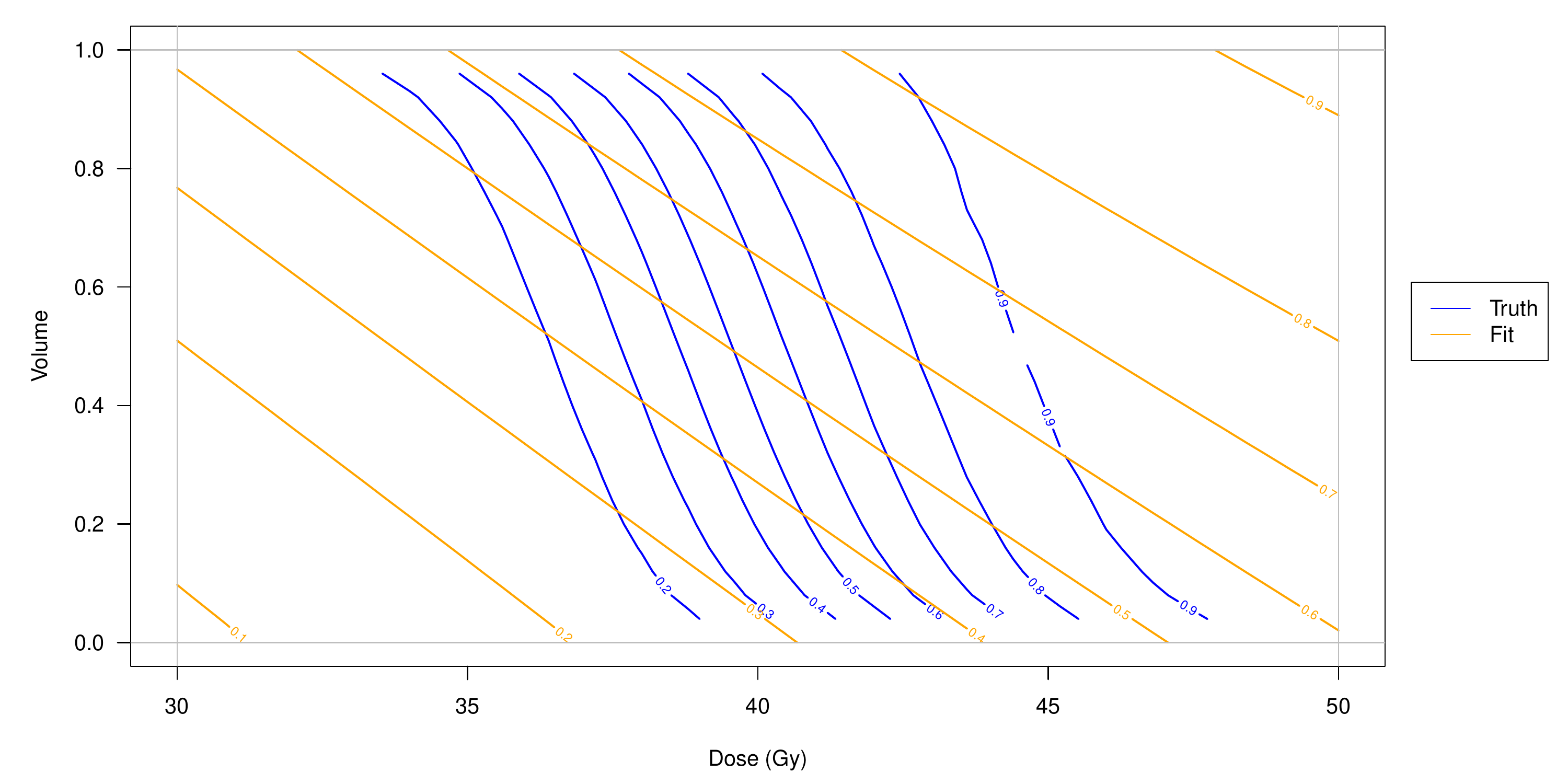}}  \\[-0.5em]
    \subfloat[Additive]{\includegraphics[width = 0.45\textwidth]{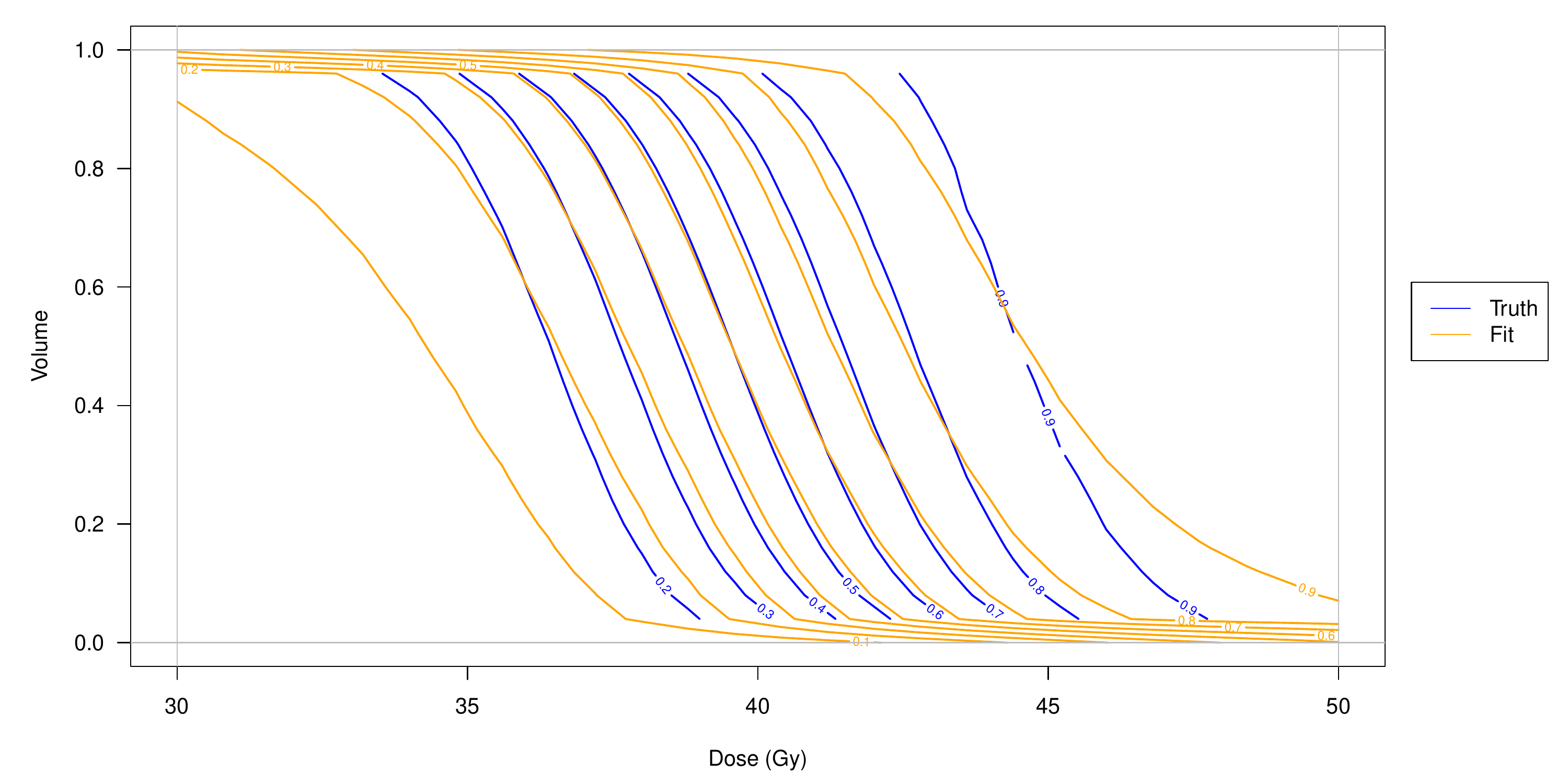}} \hspace{0.6cm}
     \subfloat[Bivariable monotone]{\includegraphics[width = 0.45\textwidth]{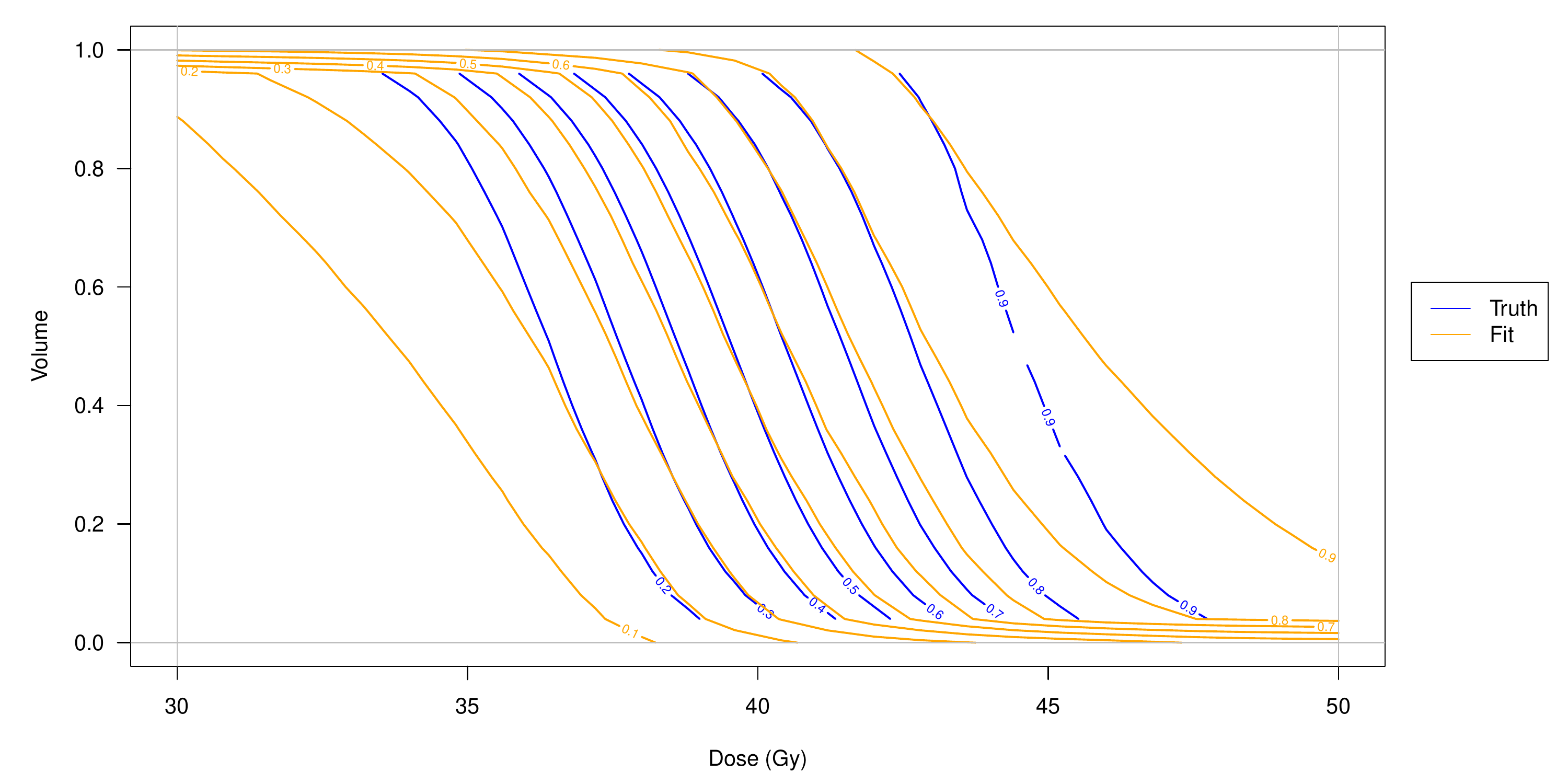}}
     \captionsetup{labelfont=bf}
    \caption{Contour plots for the model-estimated (orange) and true (blue) pointwise-causal risk by DVH volume and radiation dose for $n = 100$.}
    \label{figure:mean_contour_n=100}
\end{figure}
\begin{figure}[H]
    \centering
    \subfloat[Logistic]{\includegraphics[width = 0.45\textwidth]{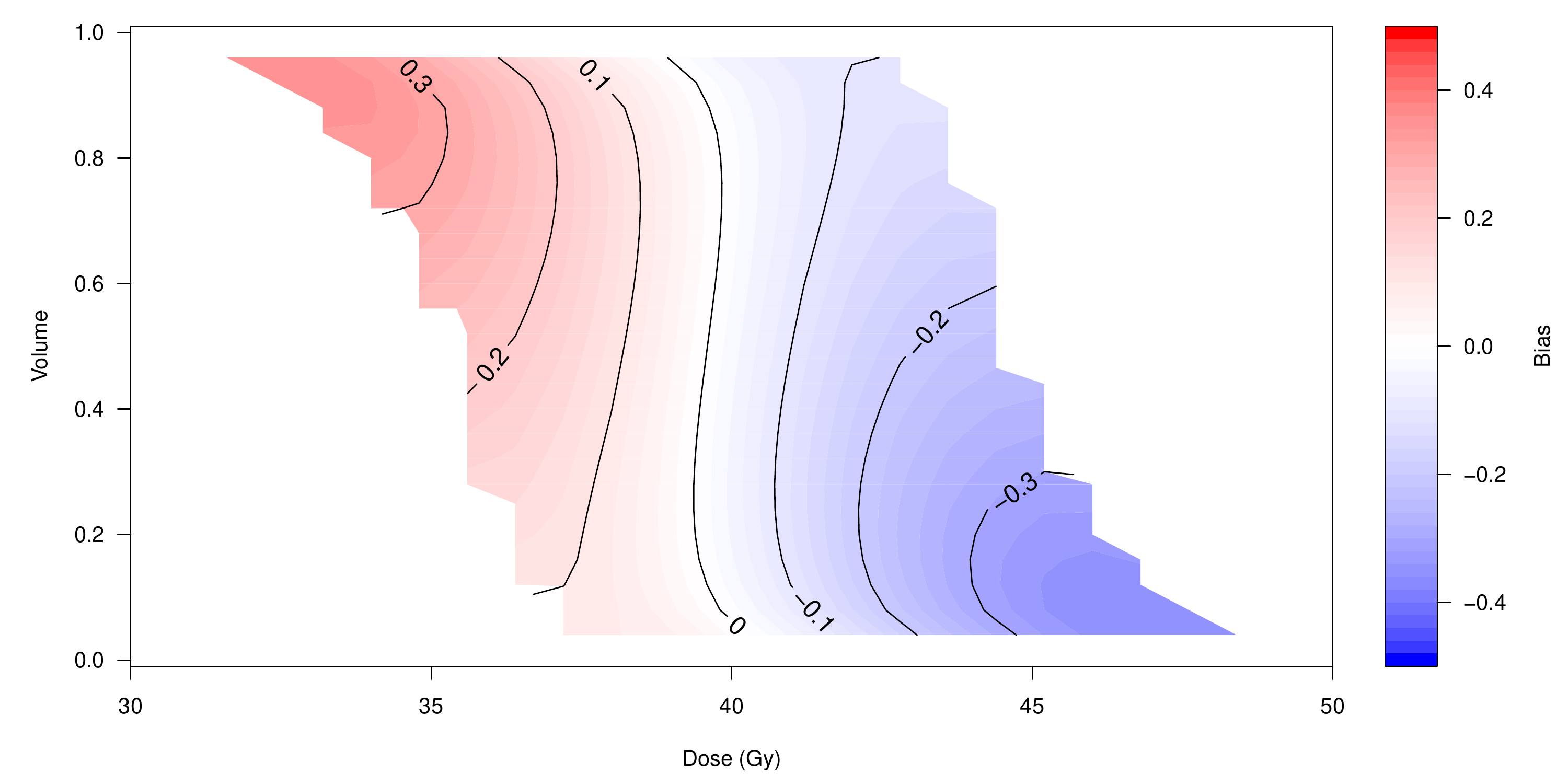}} \hspace{0.6cm}
     \subfloat[Polynomial logistic]{\includegraphics[width = 0.45\textwidth]{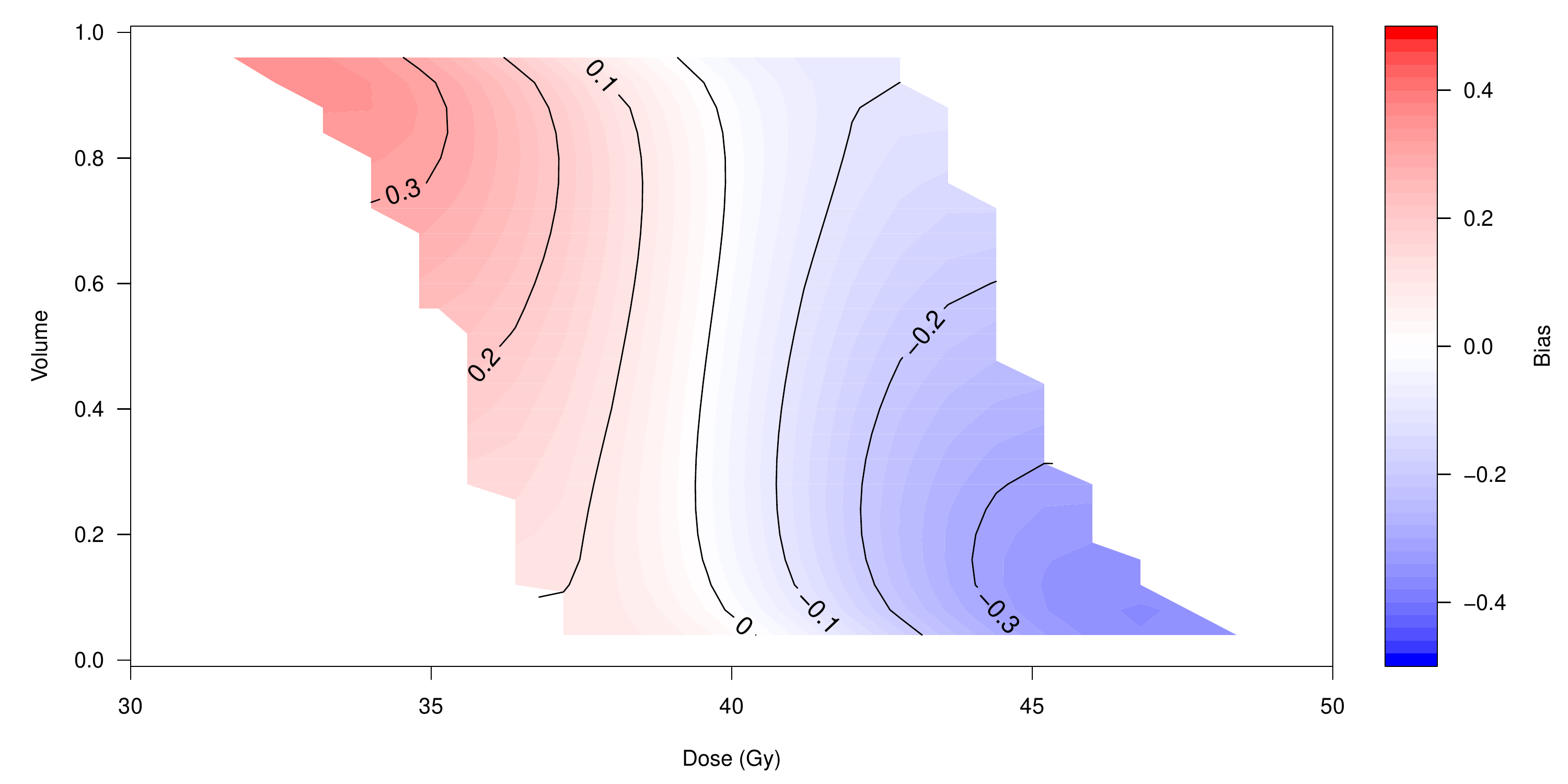}}  \\[-0.5em]
    \subfloat[Additive]{\includegraphics[width = 0.45\textwidth]{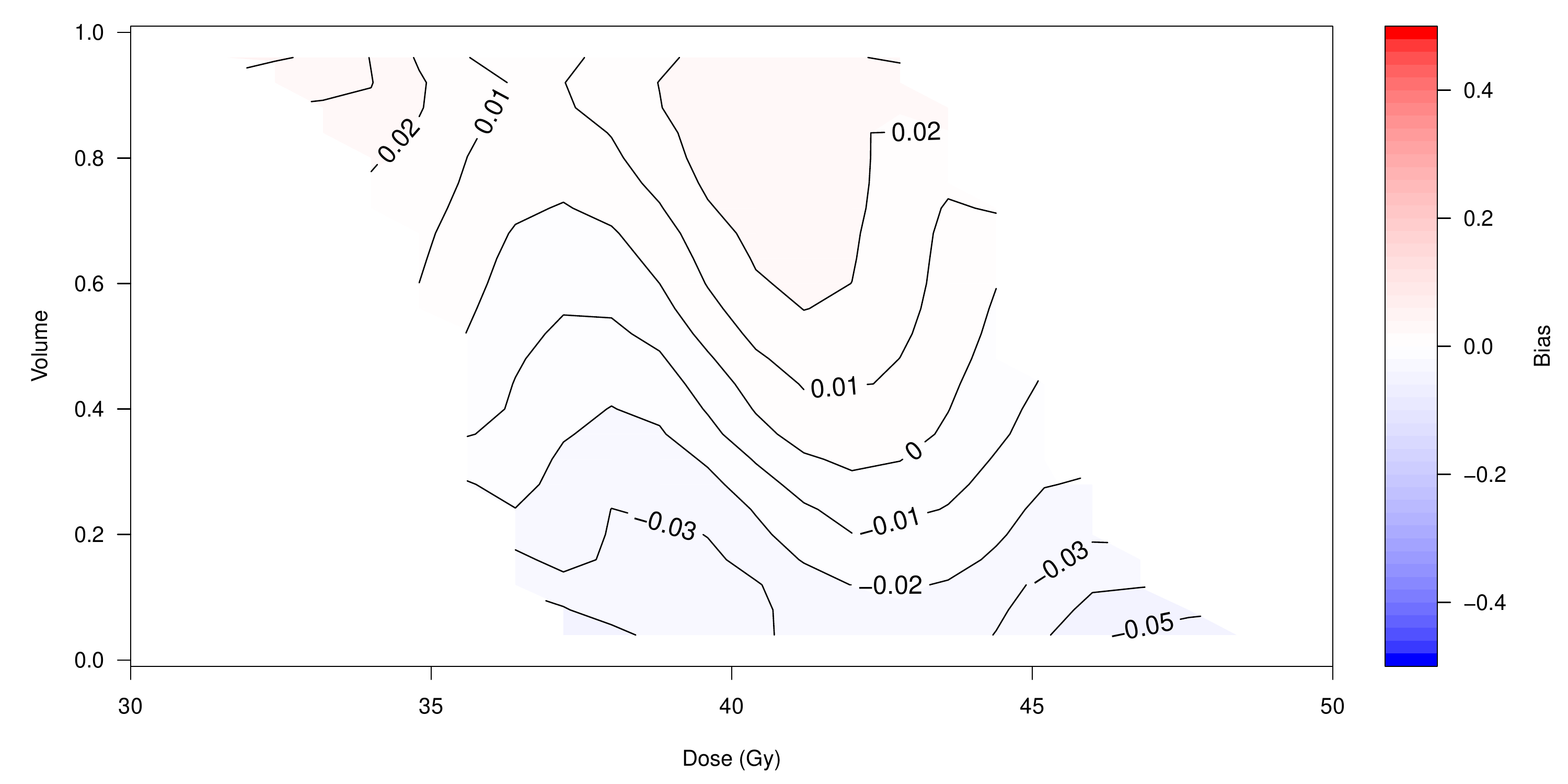}} \hspace{0.6cm}
     \subfloat[Bivariable monotone]{\includegraphics[width = 0.45\textwidth]{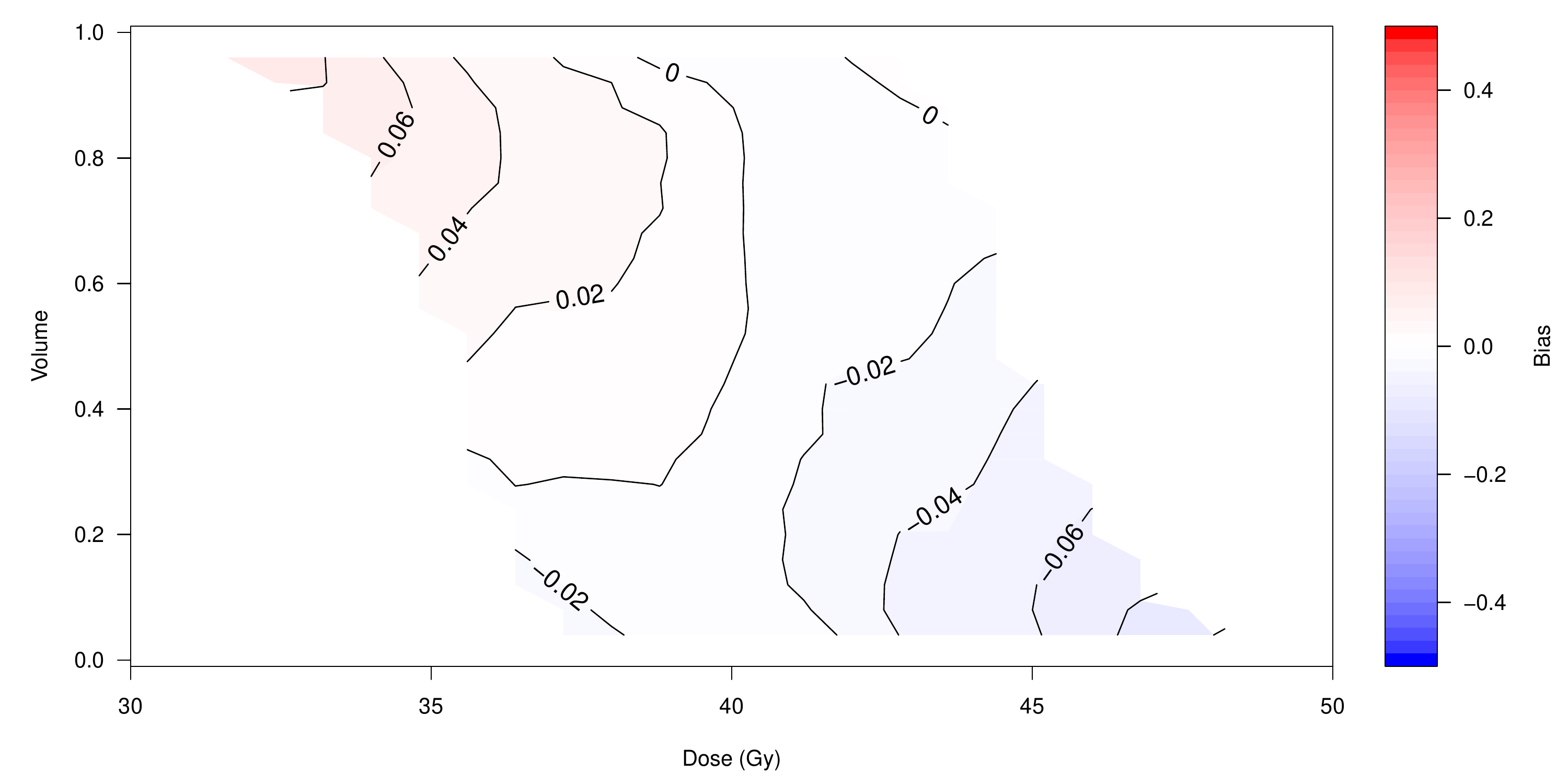}}
     \captionsetup{labelfont=bf}
    \caption{Contour plots for the bias of model-estimated pointwise-causal risk by DVH volume and radiation dose for $n = 100$.}
    \label{figure:bias_contour_n=100}
\end{figure}

\begin{figure}[H]
    \centering
    \subfloat[Logistic]{\includegraphics[width = 0.45\textwidth]{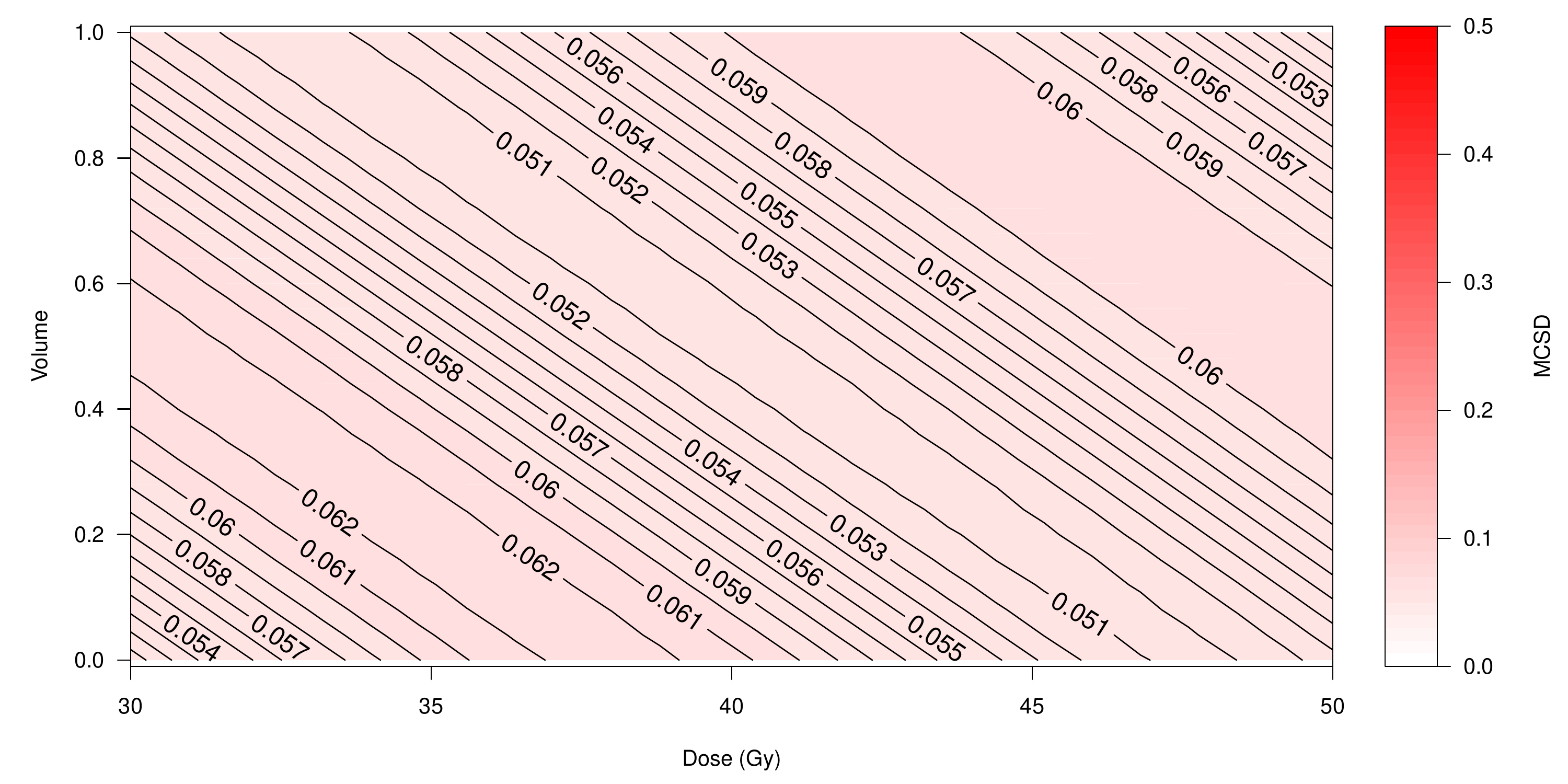}} \hspace{0.6cm}
     \subfloat[Polynomial logistic]{\includegraphics[width = 0.45\textwidth]{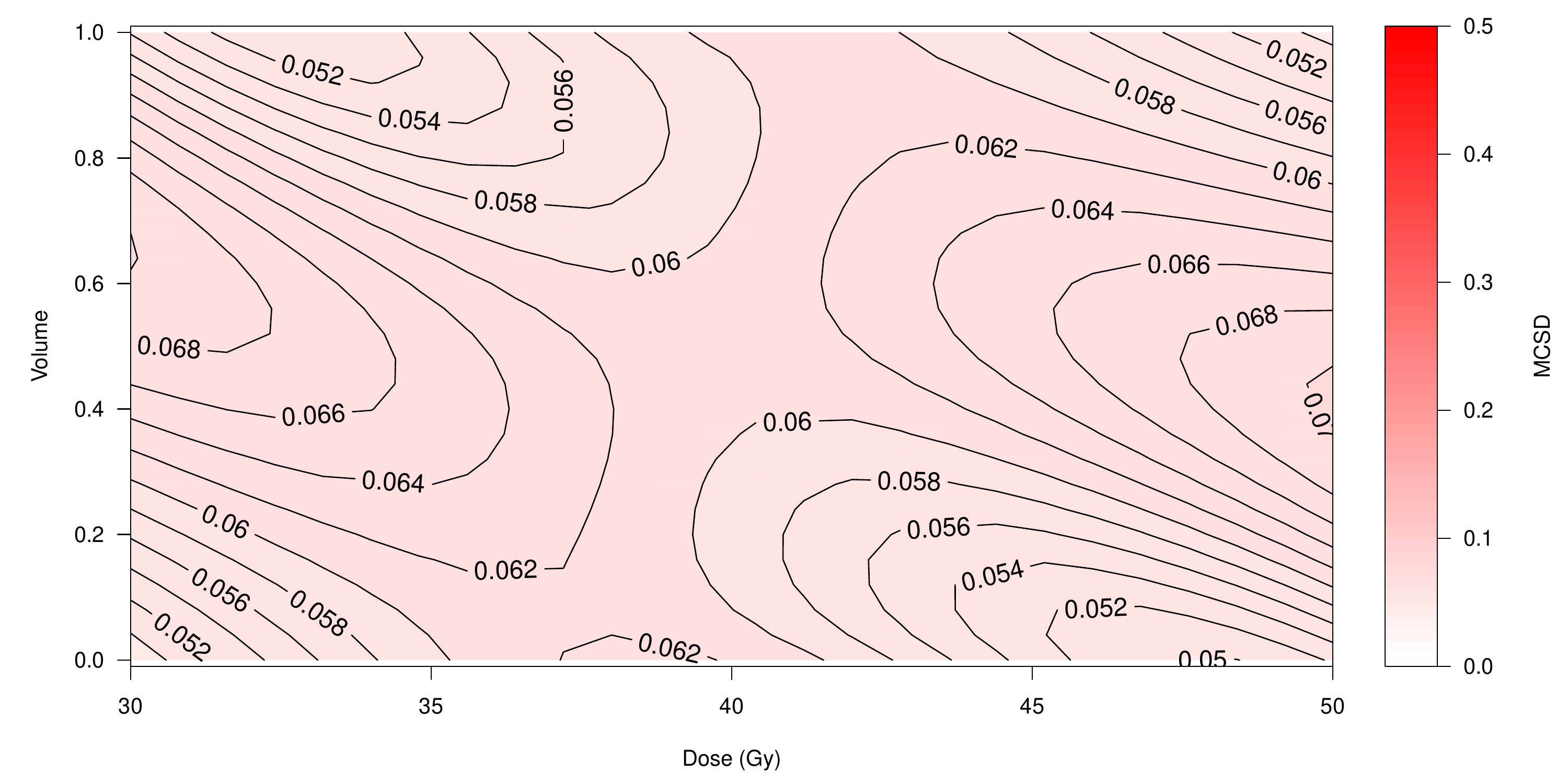}}  \\[-0.5em]
    \subfloat[Additive]{\includegraphics[width = 0.45\textwidth]{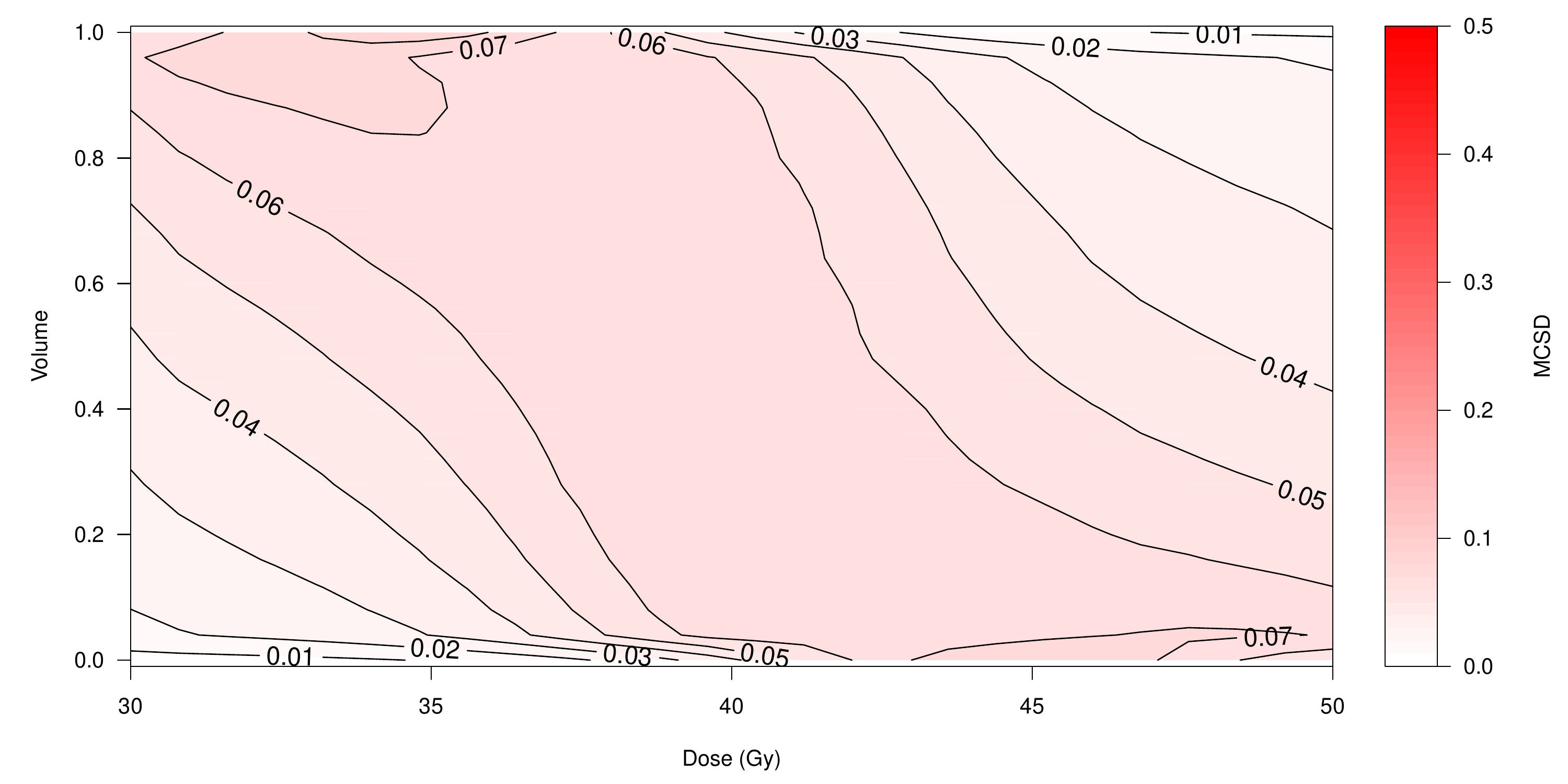}} \hspace{0.6cm}
     \subfloat[Bivariable monotone]{\includegraphics[width = 0.45\textwidth]{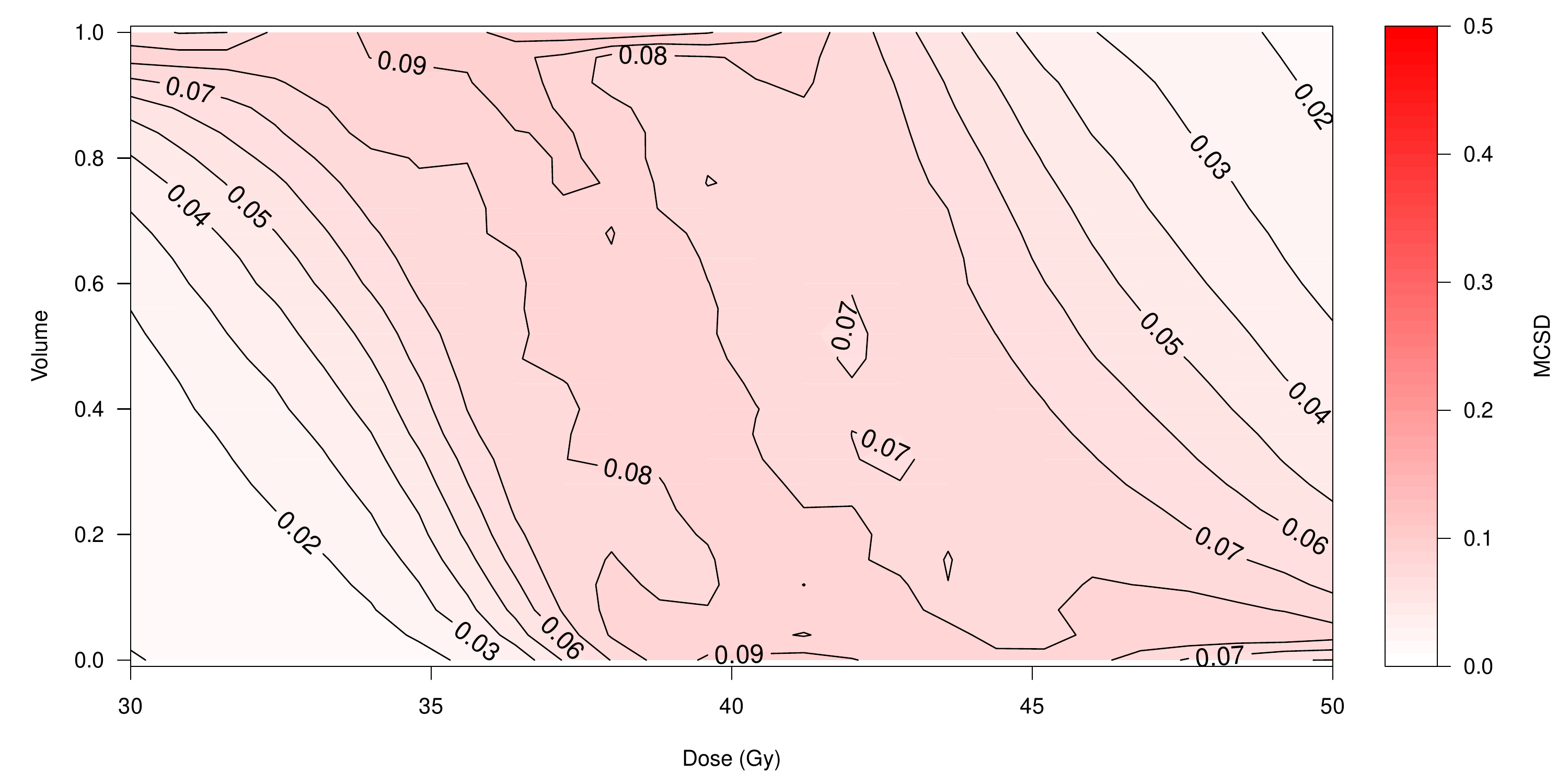}}
     \captionsetup{labelfont=bf}
    \caption{Contour plots for the Monte Carlo standard deviation of model-estimated pointwise-causal risk by DVH volume and radiation dose for $n = 100$.}
    \label{figure:mcsd_contour_n=100}
\end{figure}

\begin{figure}[H]
    \centering
    \subfloat[Logistic]{\includegraphics[width = 0.45\textwidth]{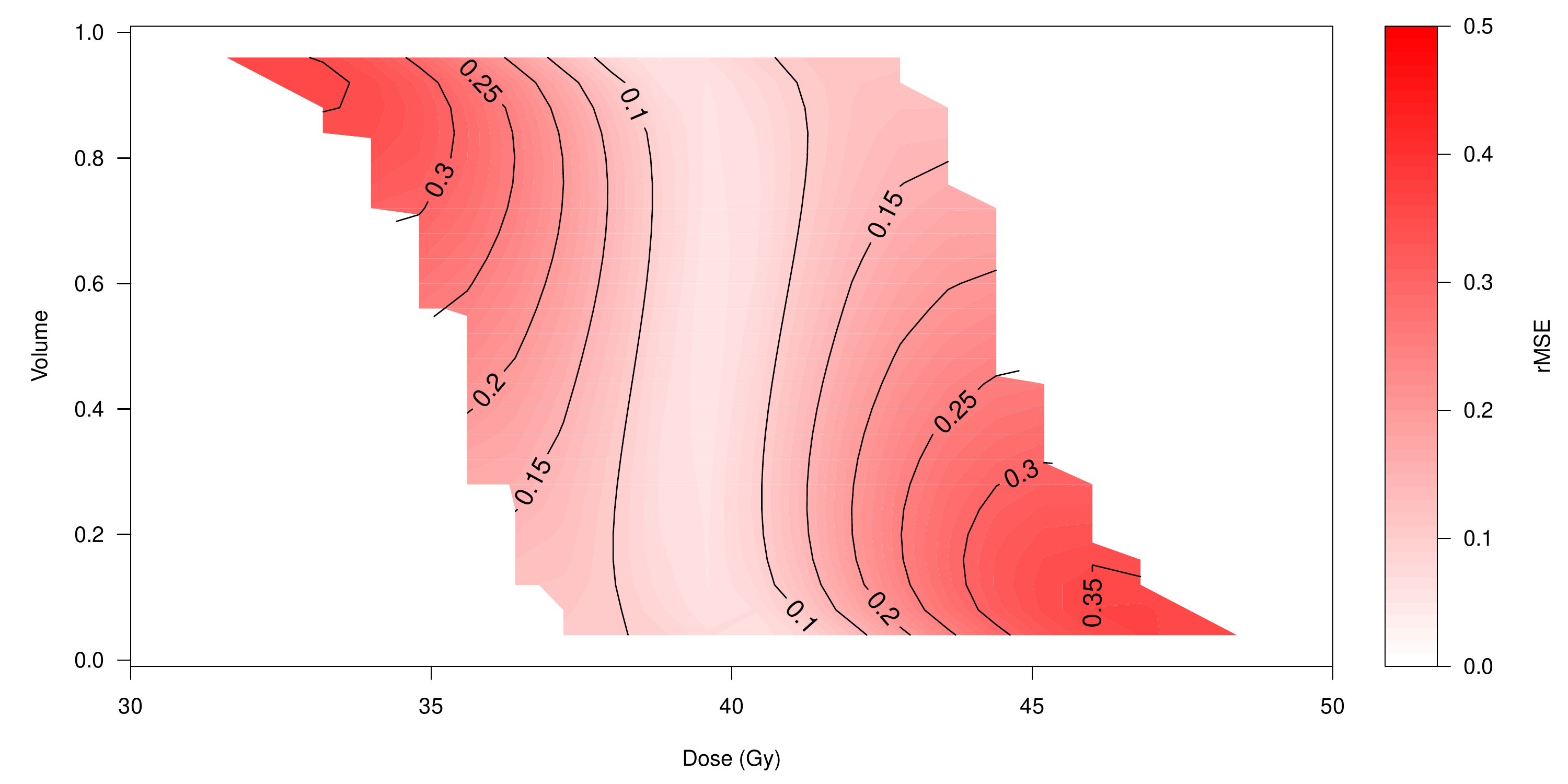}} \hspace{0.6cm}
     \subfloat[Polynomial logistic]{\includegraphics[width = 0.45\textwidth]{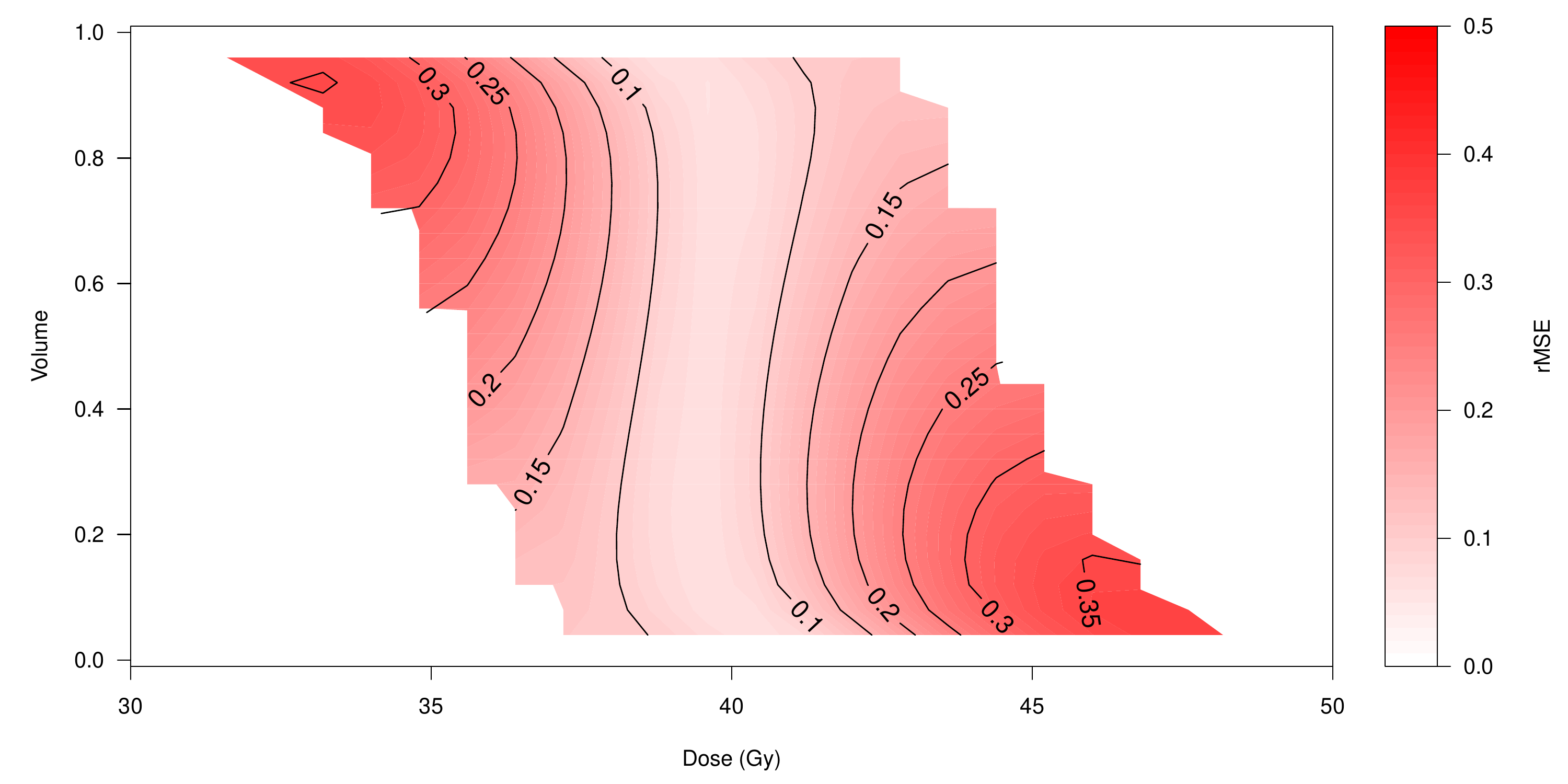}}  \\[-0.5em]
    \subfloat[Additive]{\includegraphics[width = 0.45\textwidth]{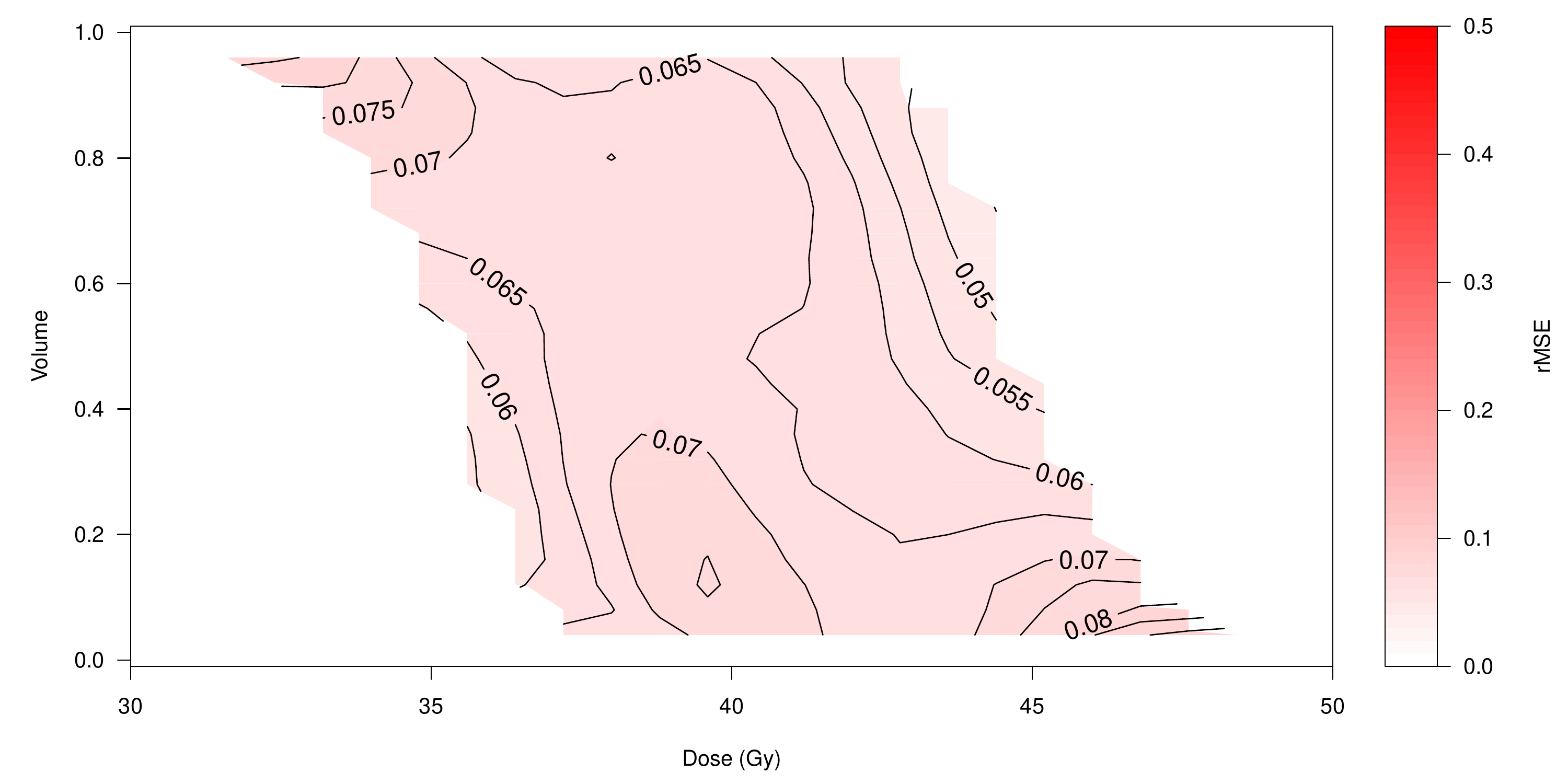}} \hspace{0.6cm}
     \subfloat[Bivariable monotone]{\includegraphics[width = 0.45\textwidth]{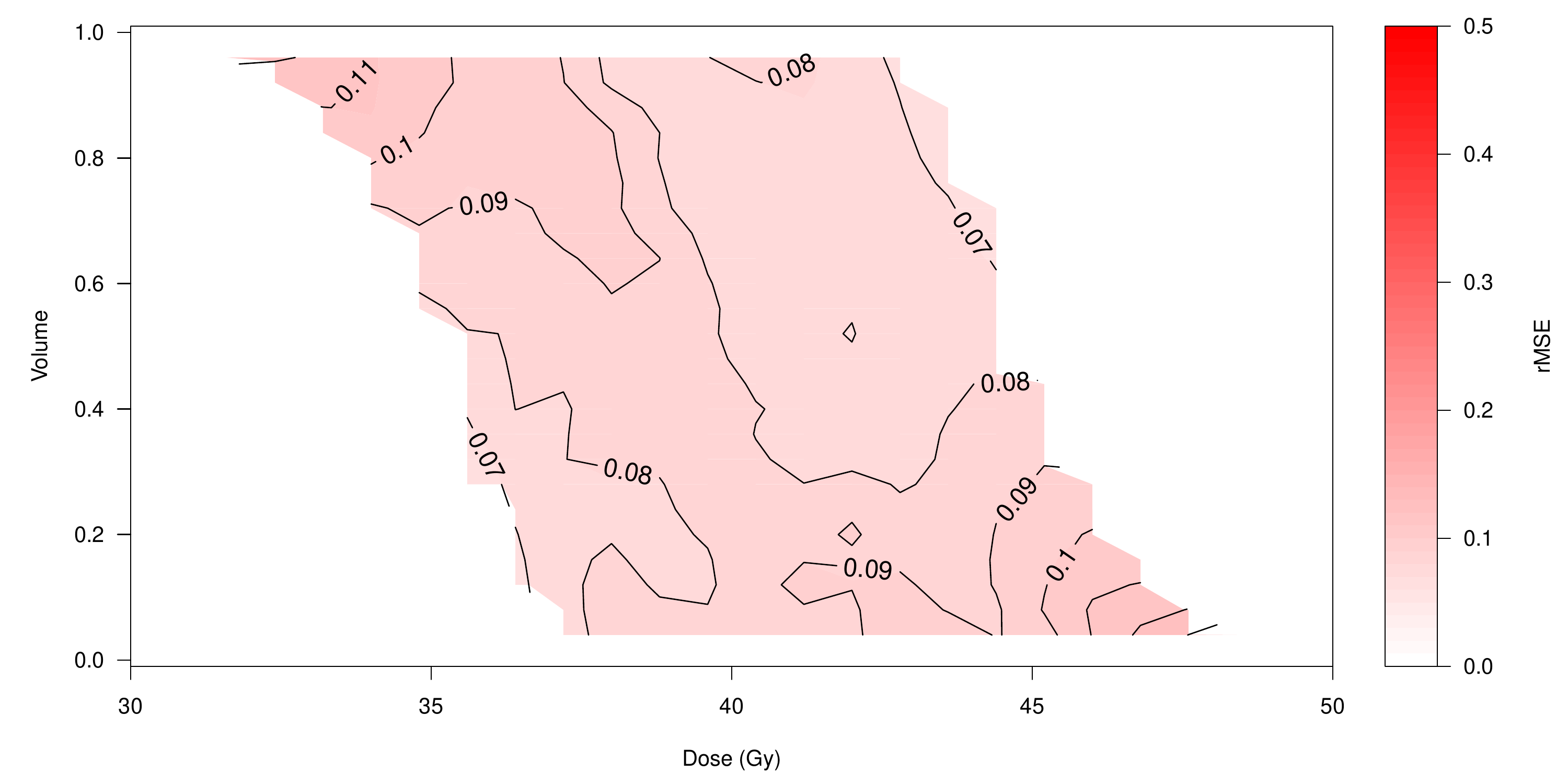}}
    \captionsetup{labelfont=bf}
    \caption{Contour plots for the root mean-square error of model-estimated pointwise-causal risk by DVH volume and radiation dose for $n = 100$.}
    \label{figure:rmse_contour_n=100}
\end{figure}

\begin{figure}[H]
    \centering
    \subfloat[Logistic]{\includegraphics[width = 0.45\textwidth]{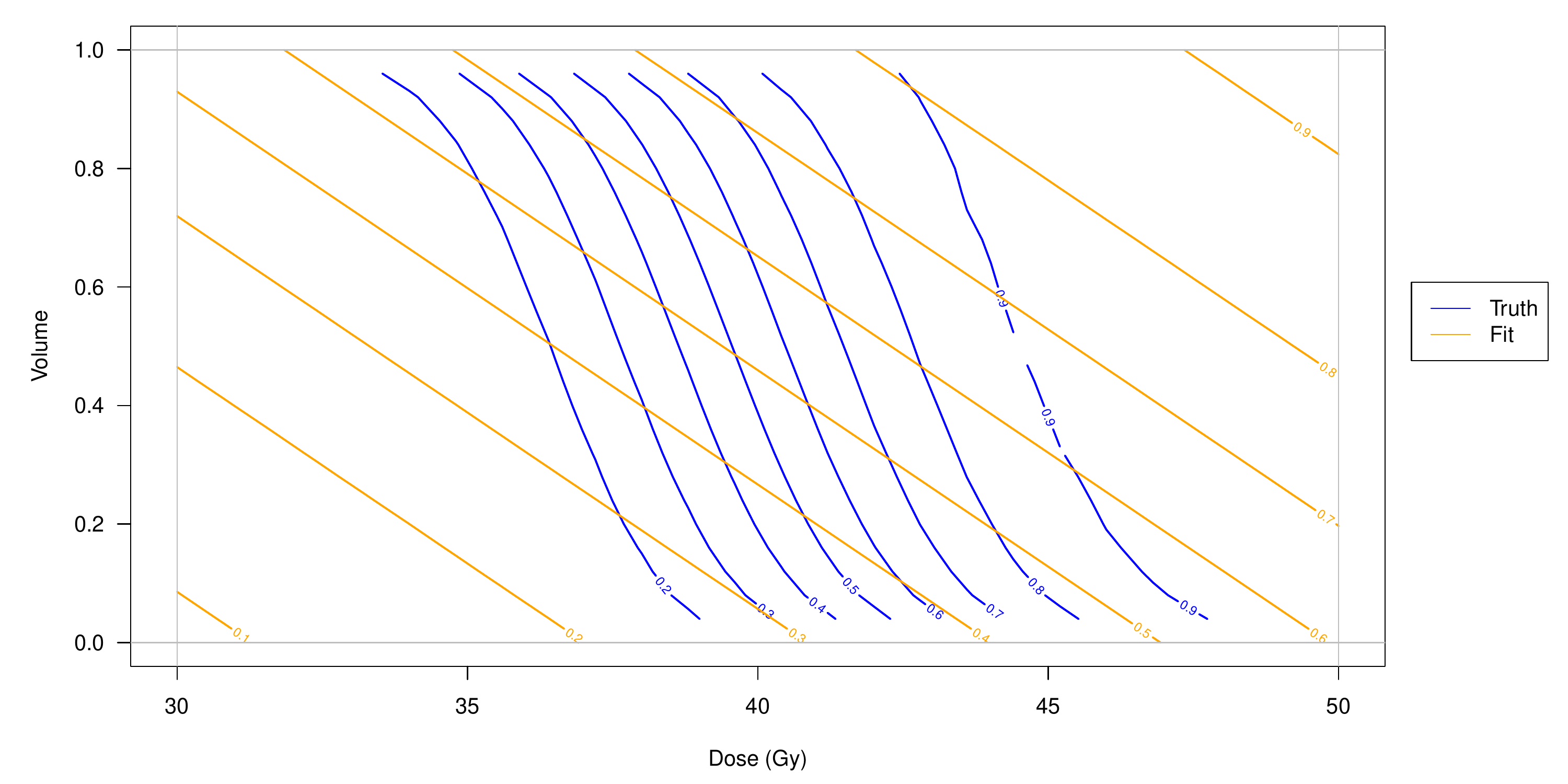}} \hspace{0.6cm}
     \subfloat[Polynomial logistic]{\includegraphics[width = 0.45\textwidth]{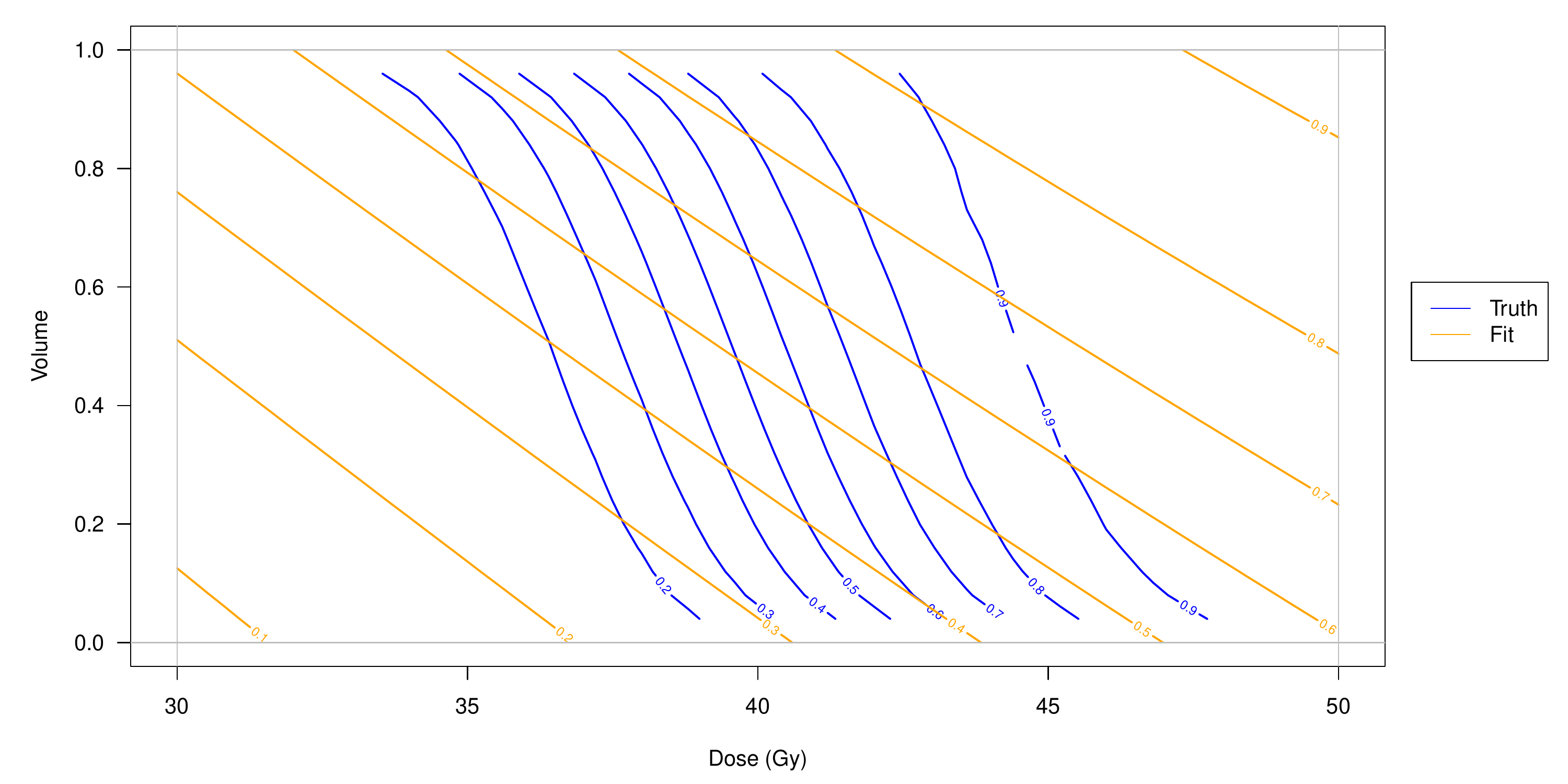}}  \\[-0.5em]
    \subfloat[Additive]{\includegraphics[width = 0.45\textwidth]{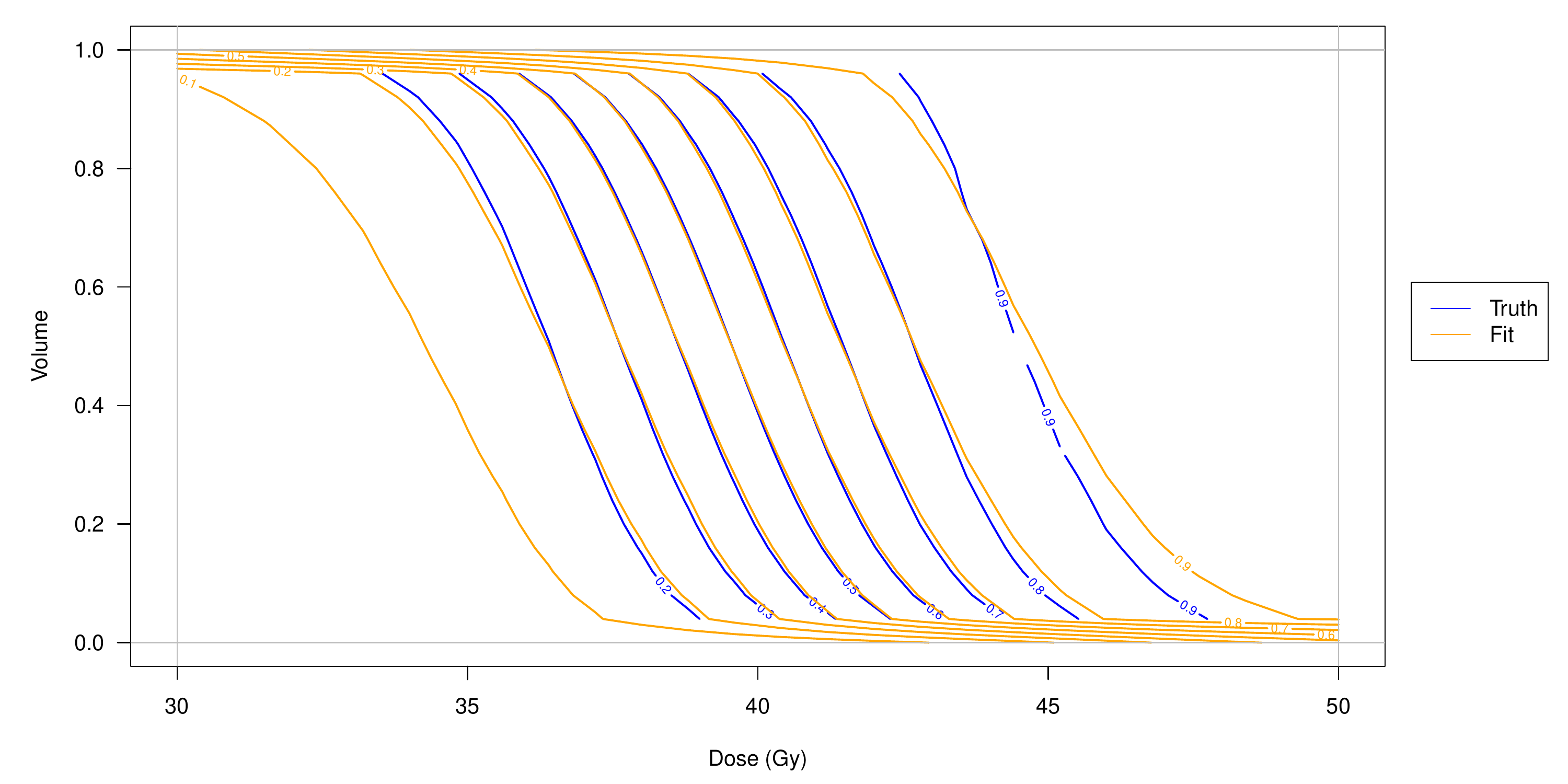}} \hspace{0.6cm}
     \subfloat[Bivariable monotone]{\includegraphics[width = 0.45\textwidth]{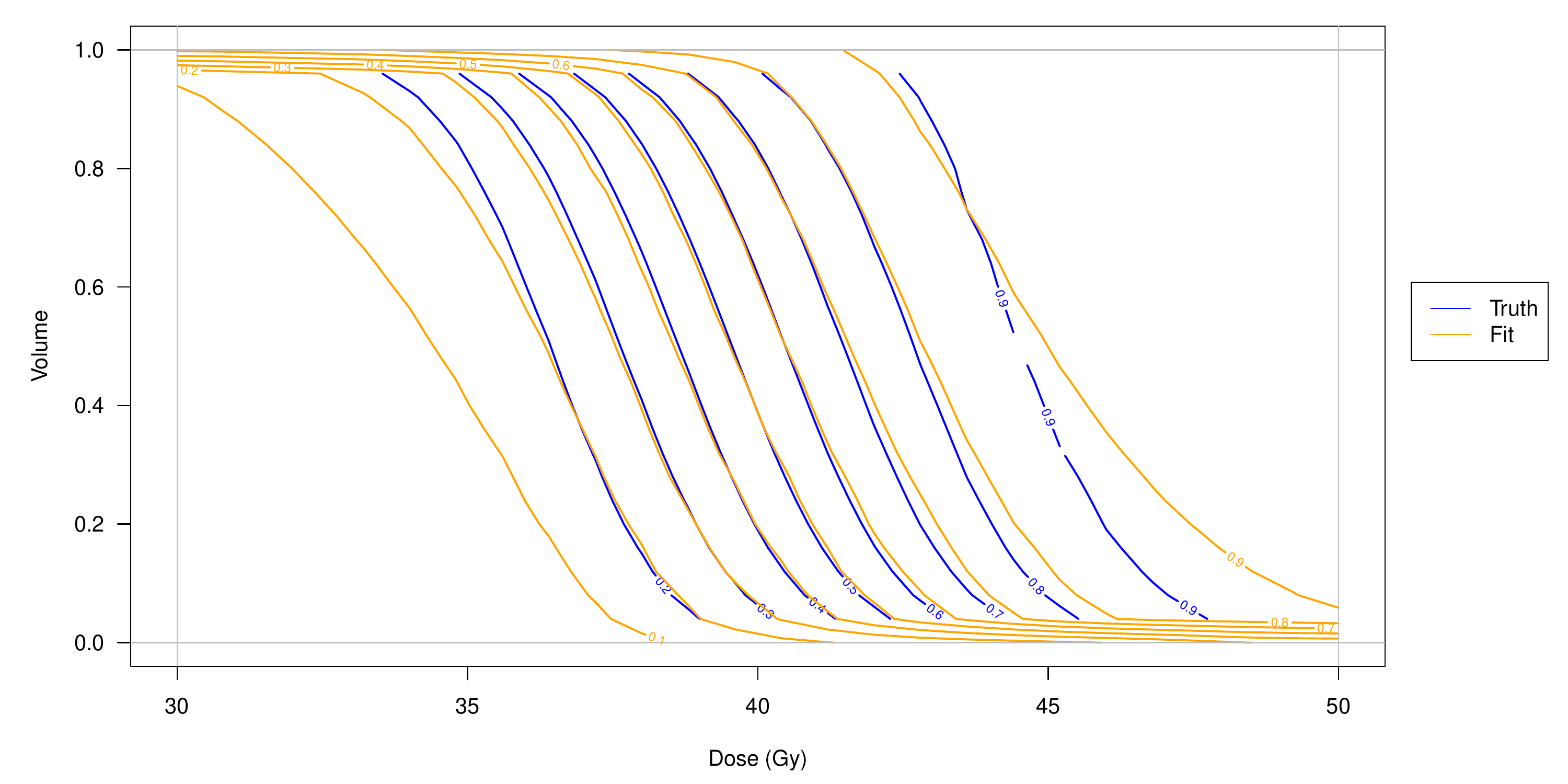}}
    \captionsetup{labelfont=bf}
    \caption{Contour plots for the model-estimated (orange) and true (blue) pointwise-causal risk by DVH volume and radiation dose for $n = 500$.}
    \label{figure:mean_contour_n=500}
\end{figure}

\begin{figure}[H]
    \centering
    \subfloat[Logistic]{\includegraphics[width = 0.45\textwidth]{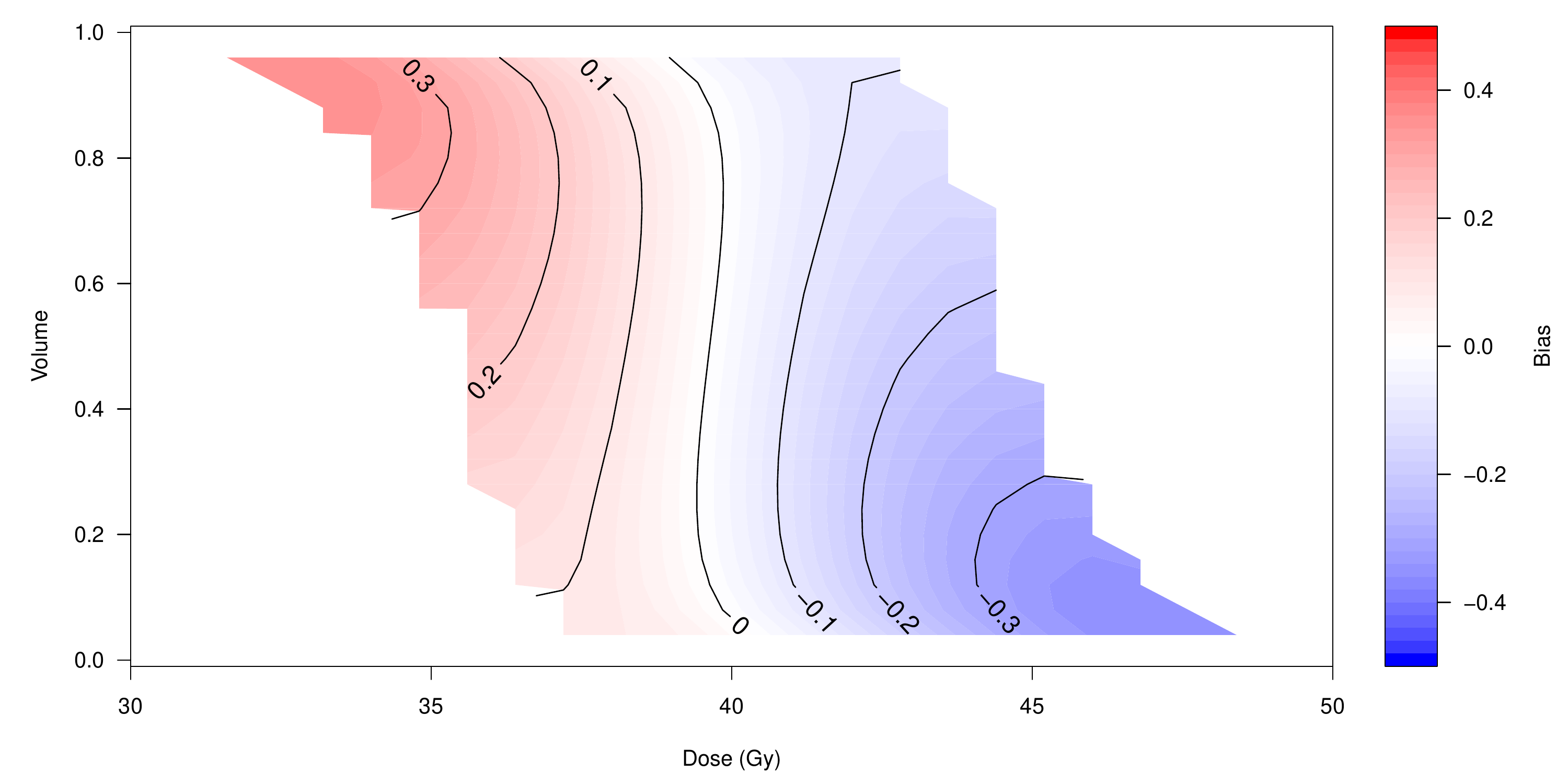}} \hspace{0.6cm}
     \subfloat[Polynomial logistic]{\includegraphics[width = 0.45\textwidth]{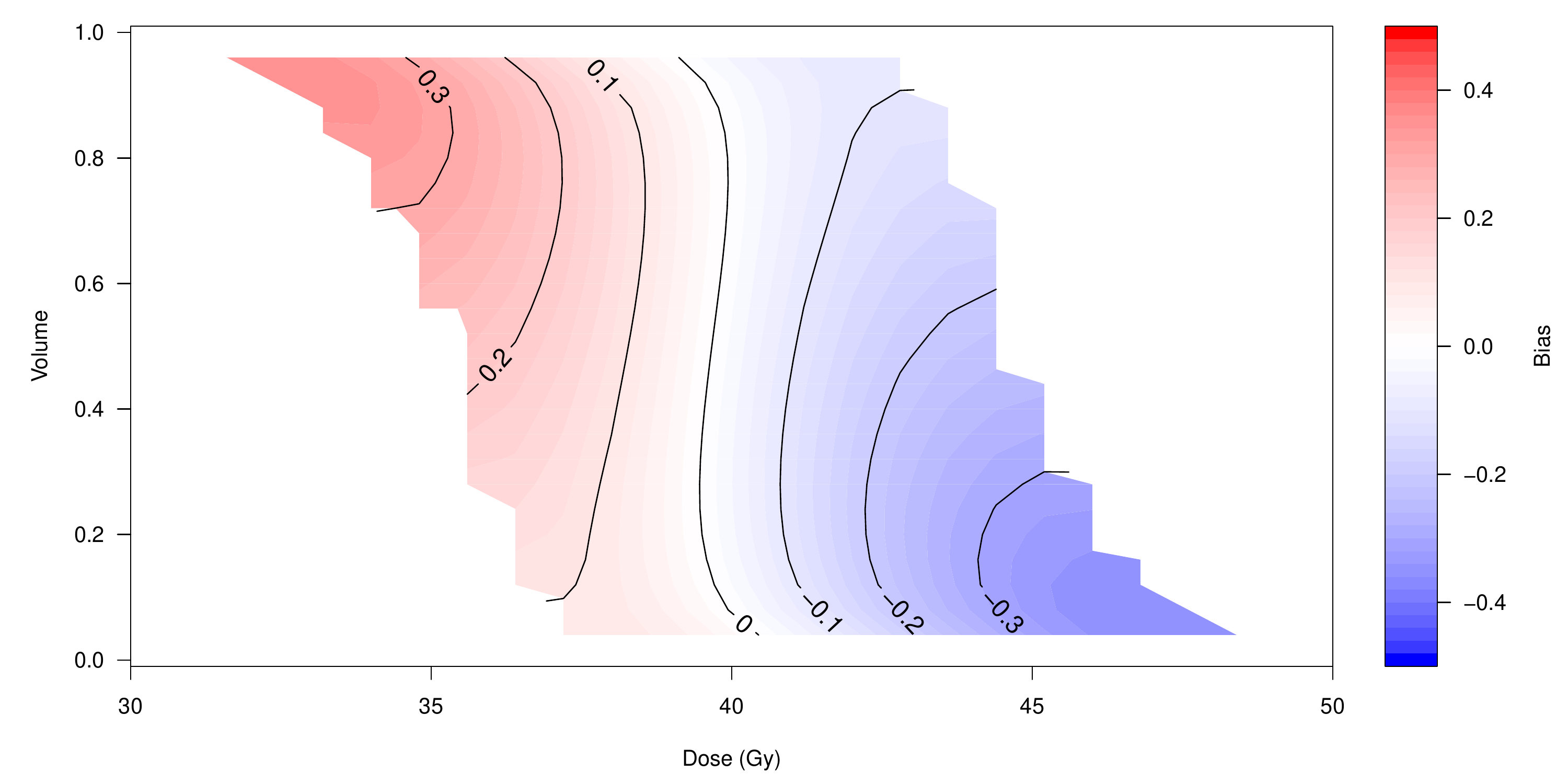}}  \\[-0.5em]
    \subfloat[Additive]{\includegraphics[width = 0.45\textwidth]{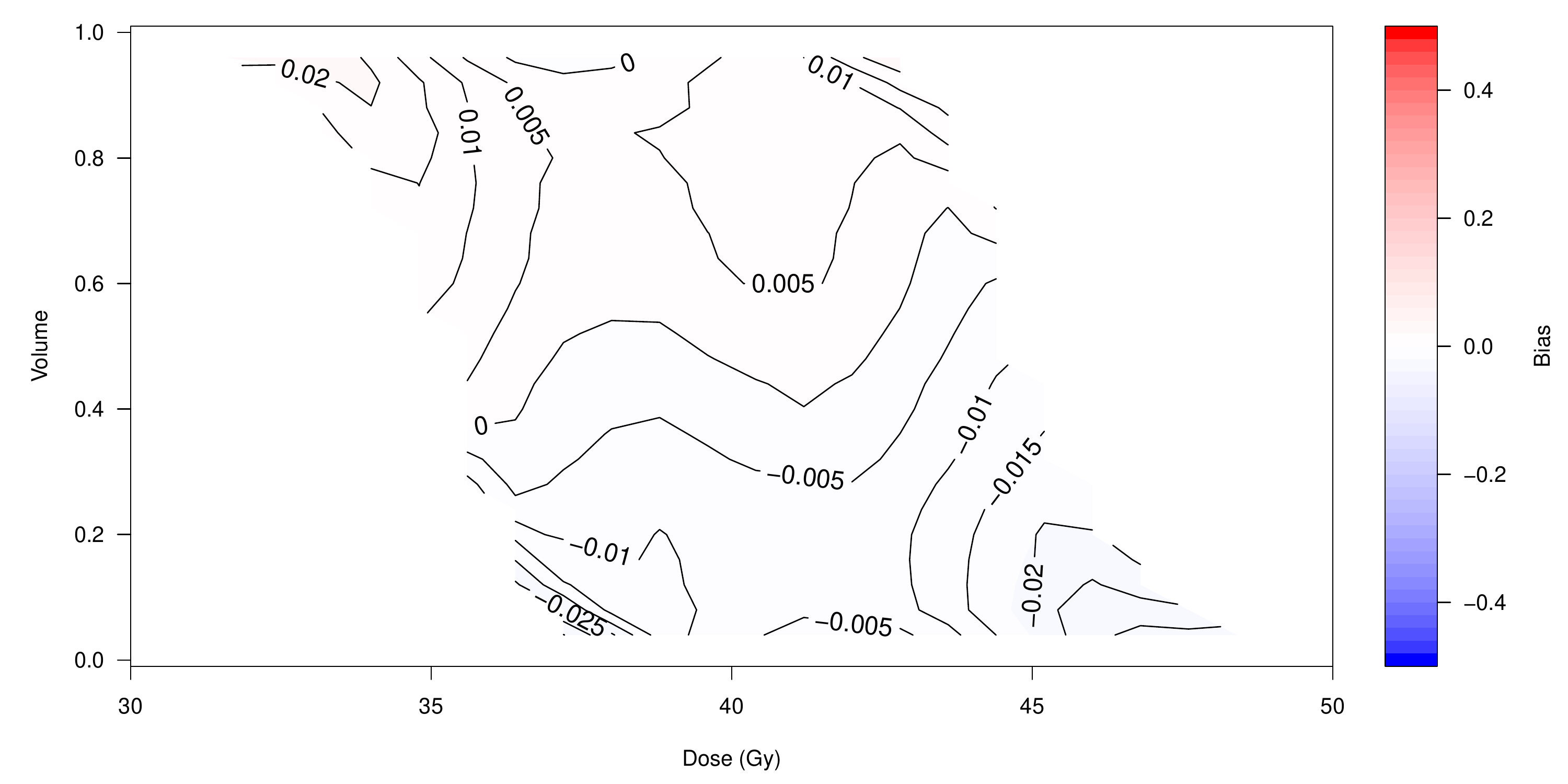}} \hspace{0.6cm}
     \subfloat[Bivariable monotone]{\includegraphics[width = 0.45\textwidth]{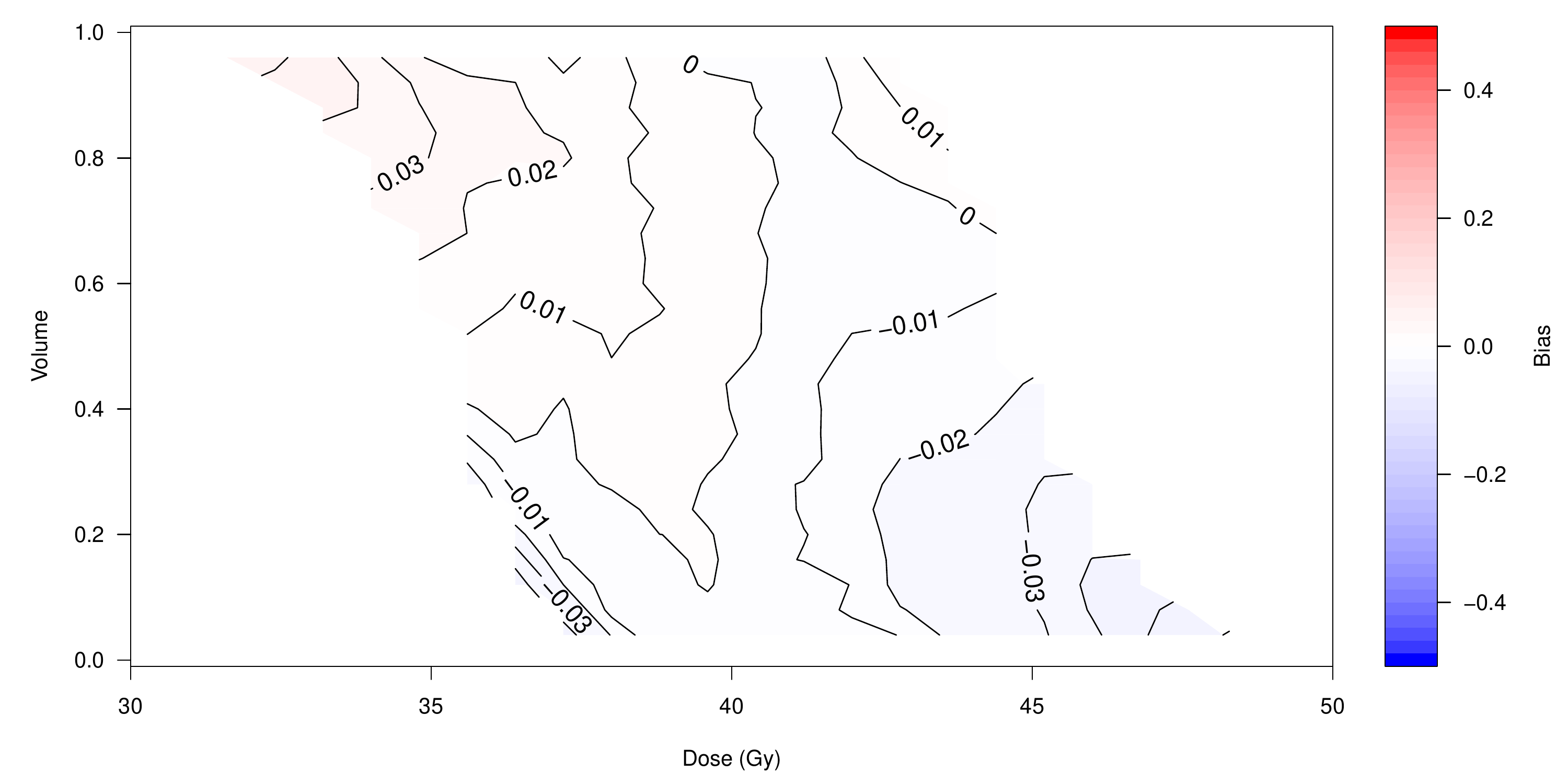}}
    \captionsetup{labelfont=bf}
    \caption{Contour plots for the bias of model-estimated pointwise-causal risk by DVH volume and radiation dose for $n = 500$.}
    \label{figure:bias_contour_n=500}
\end{figure}

\begin{figure}[H]
    \centering
    \subfloat[Logistic]{\includegraphics[width = 0.45\textwidth]{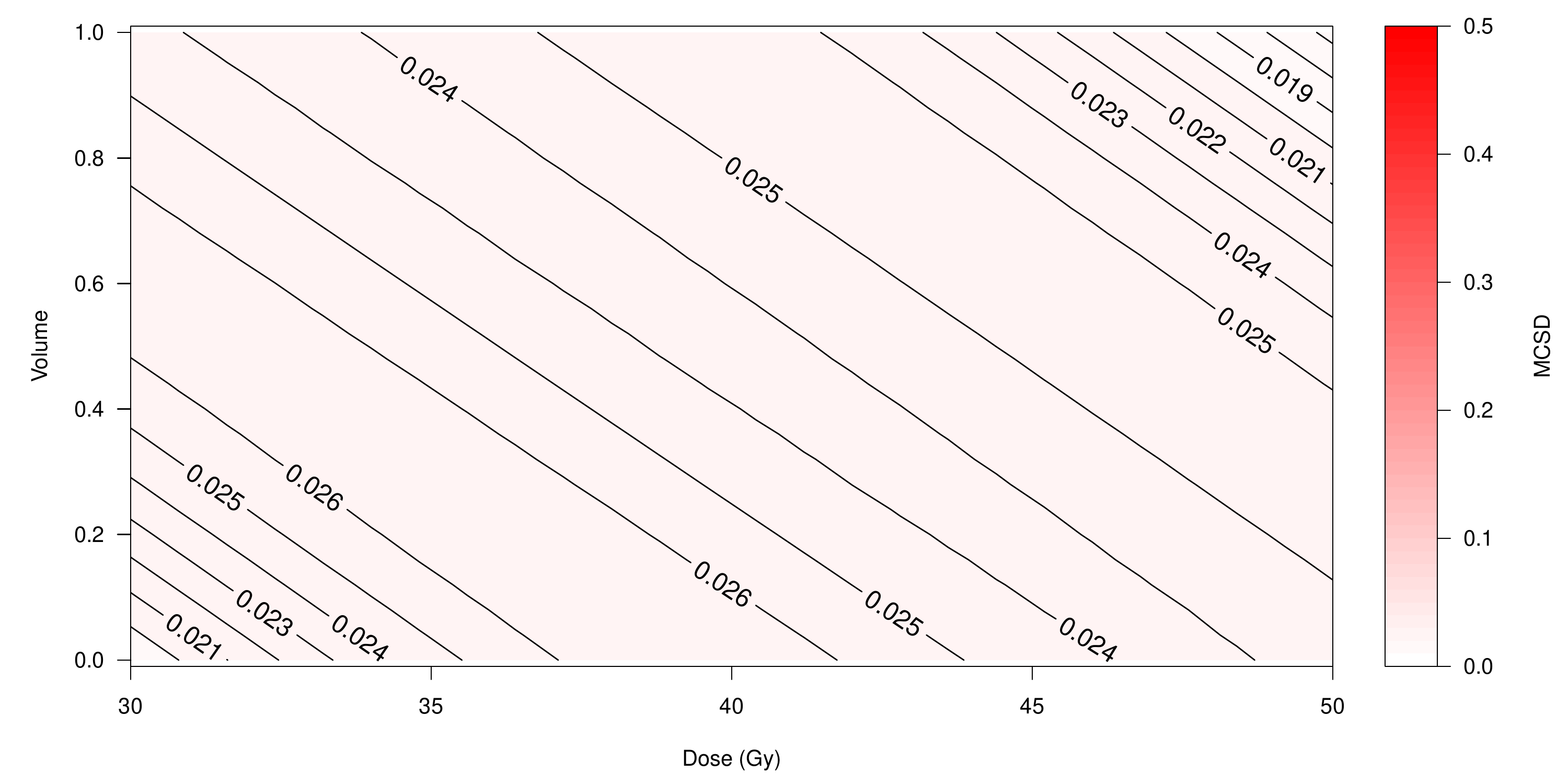}} \hspace{0.6cm}
     \subfloat[Polynomial logistic]{\includegraphics[width = 0.45\textwidth]{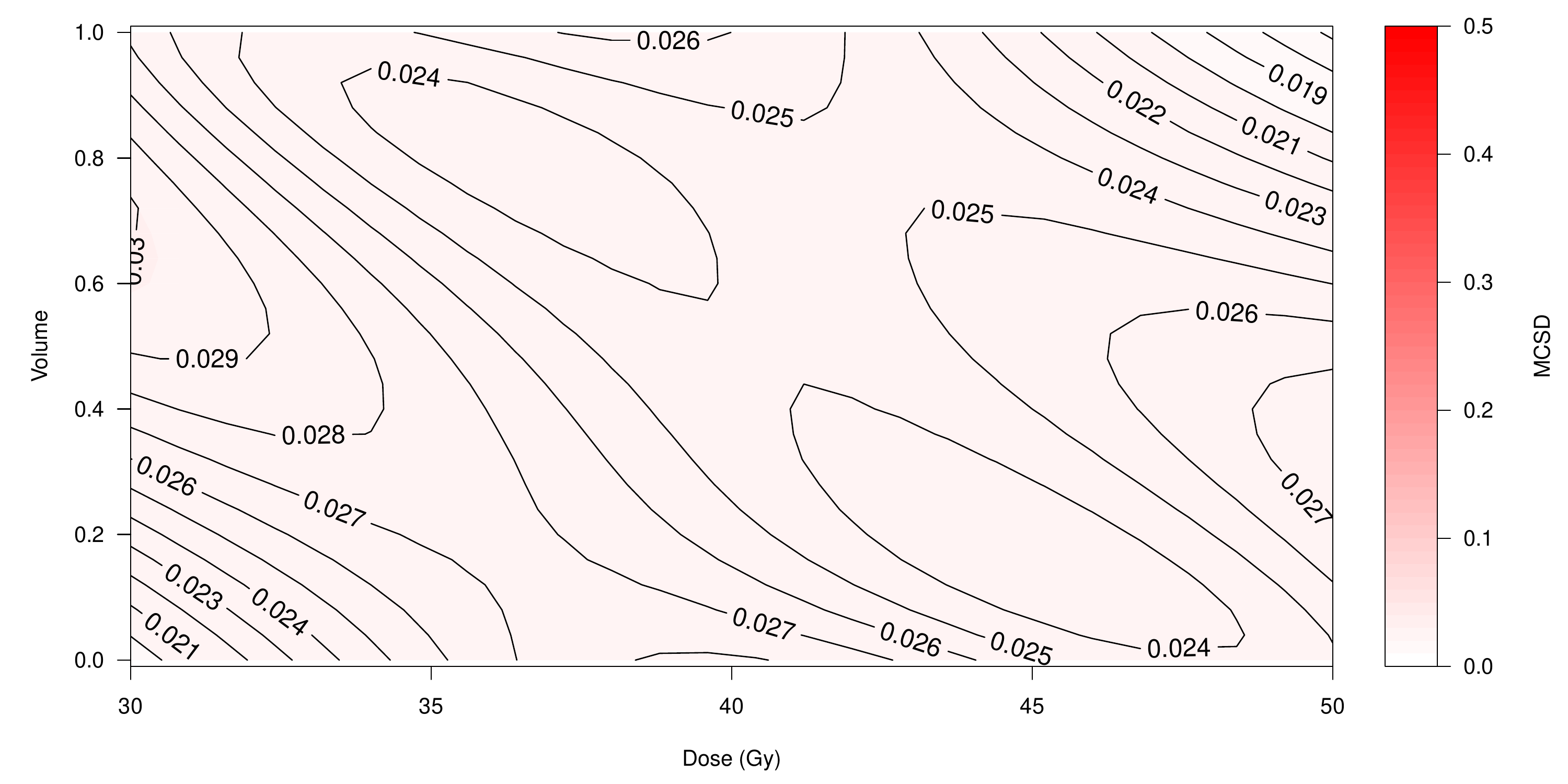}}  \\[-0.5em]
    \subfloat[Additive]{\includegraphics[width = 0.45\textwidth]{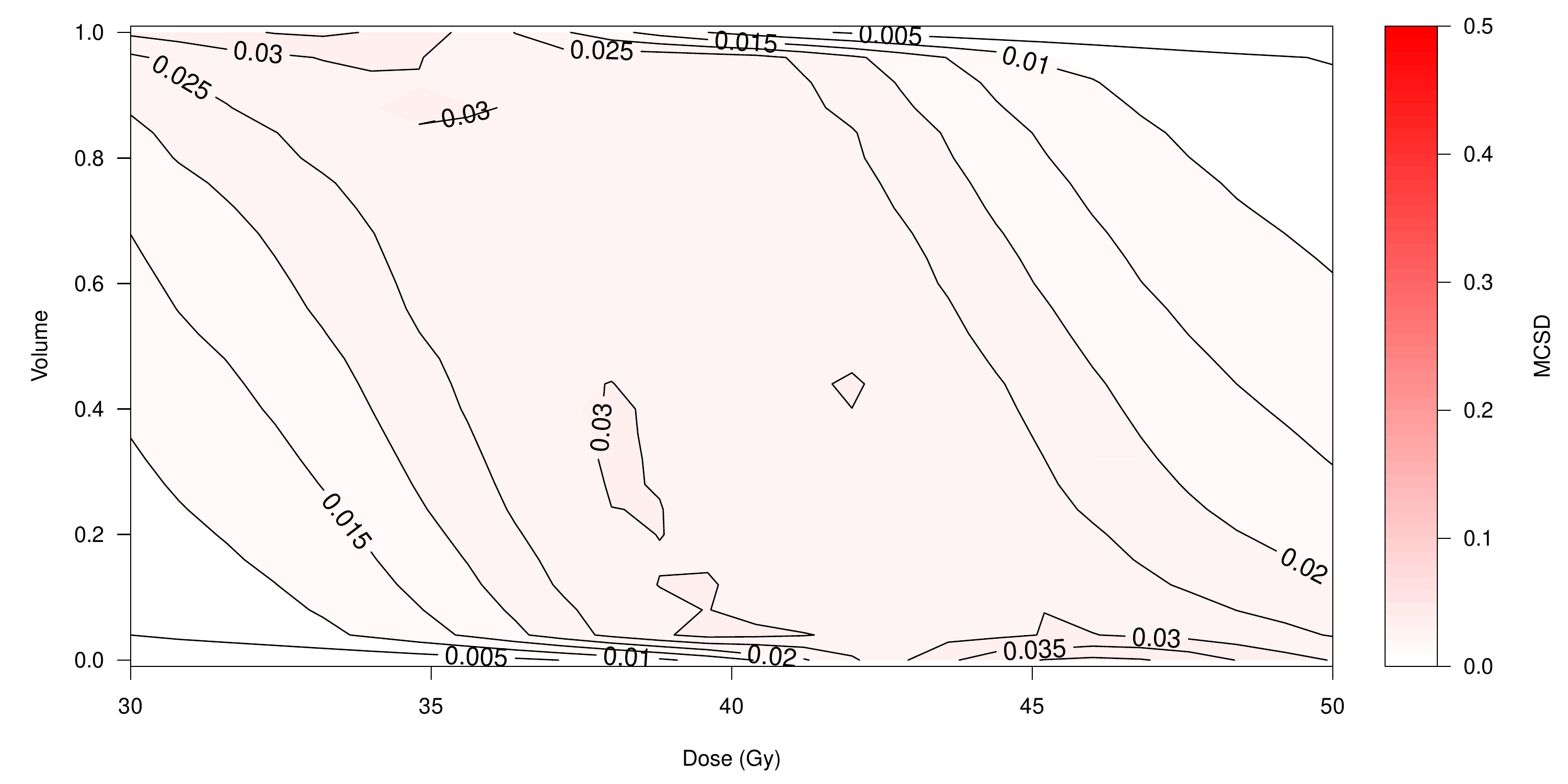}} \hspace{0.6cm}
     \subfloat[Bivariable monotone]{\includegraphics[width = 0.45\textwidth]{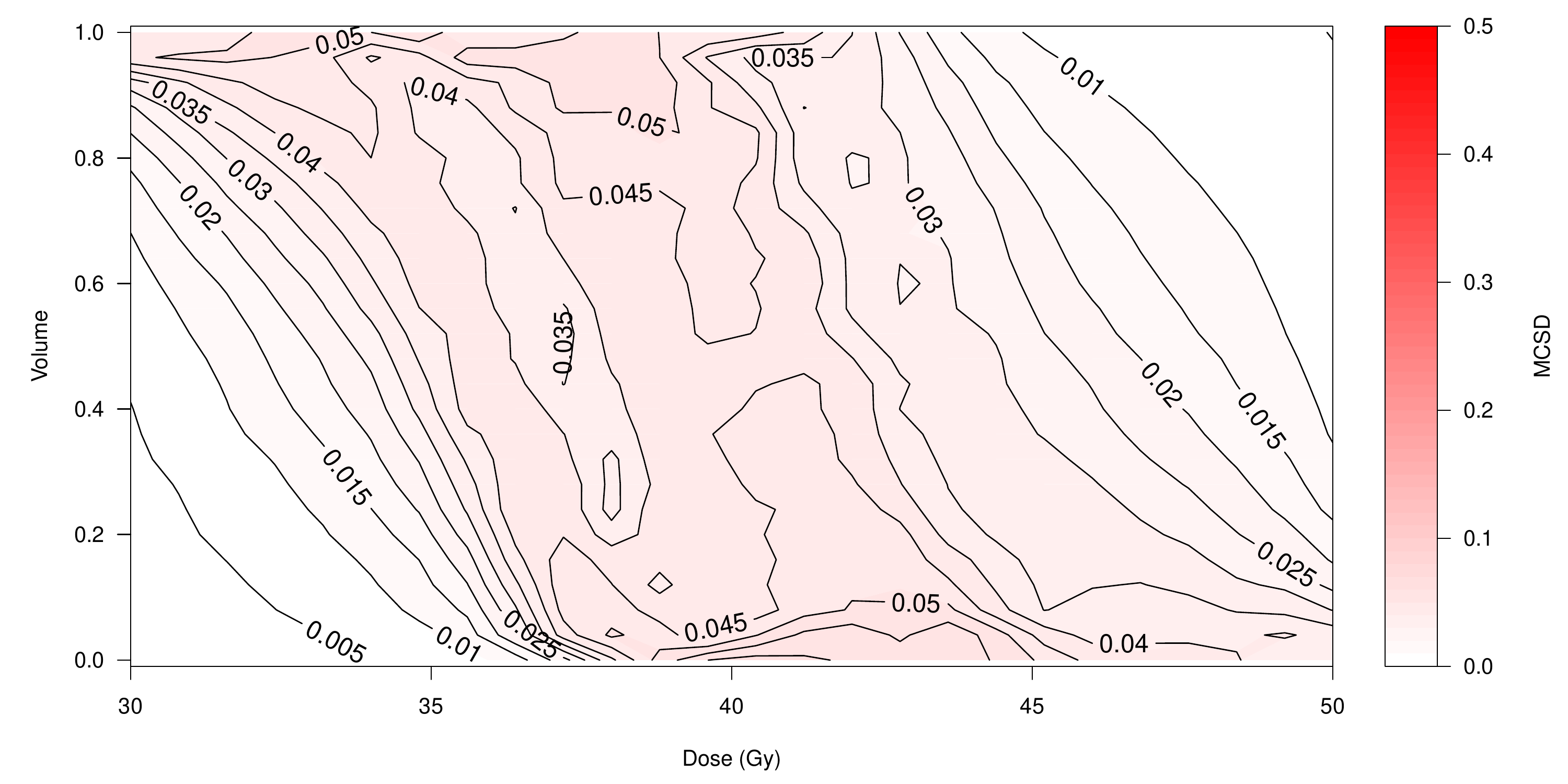}}
    \captionsetup{labelfont=bf}
    \caption{Contour plots for the Monte Carlo standard deviation of model-estimated pointwise-causal risk by DVH volume and radiation dose for $n = 500$.}
    \label{figure:mcsd_contour_n=500}
\end{figure}

\begin{figure}[H]
    \centering
    \subfloat[Logistic]{\includegraphics[width = 0.45\textwidth]{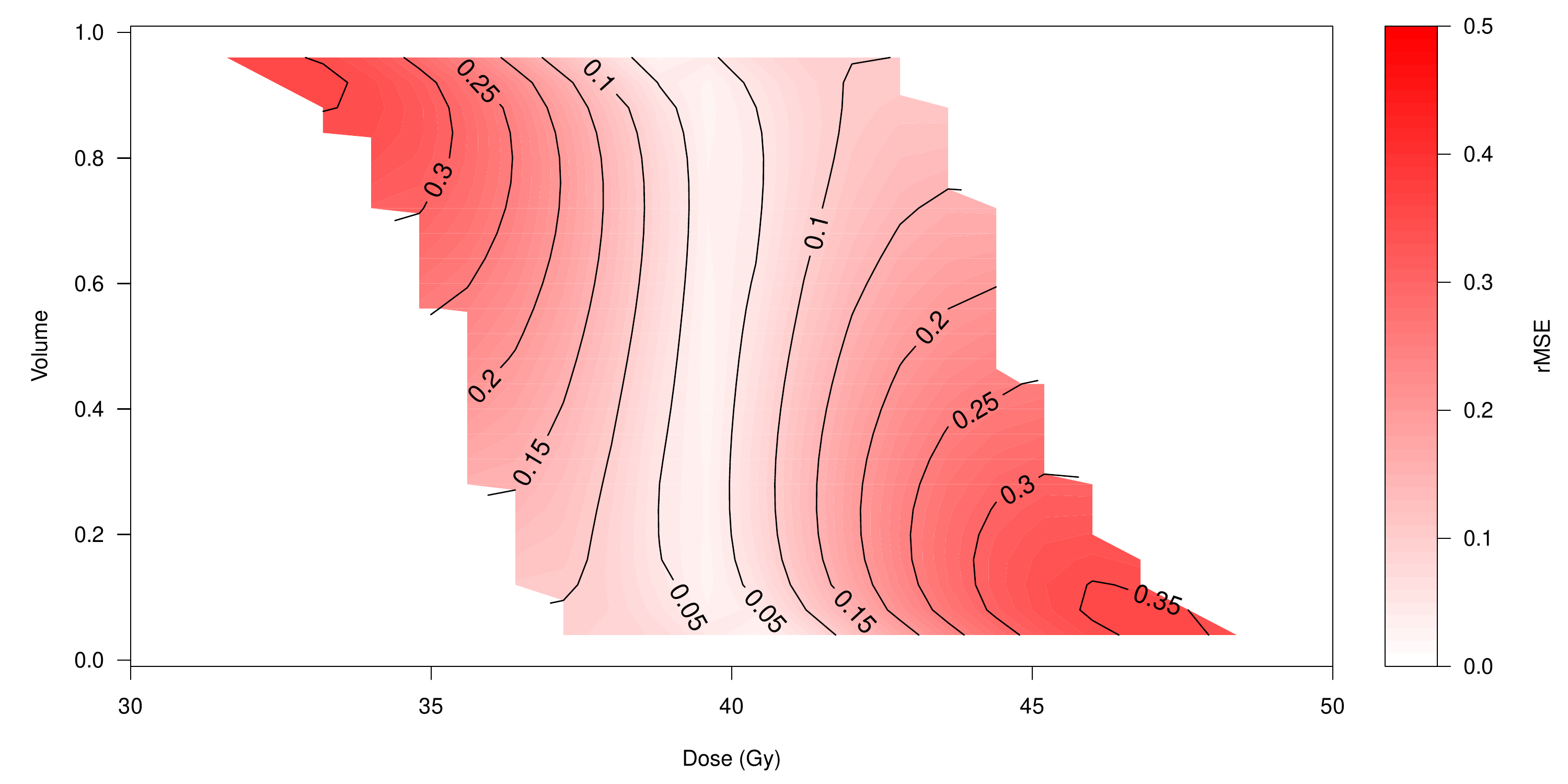}} \hspace{0.6cm}
     \subfloat[Polynomial logistic]{\includegraphics[width = 0.45\textwidth]{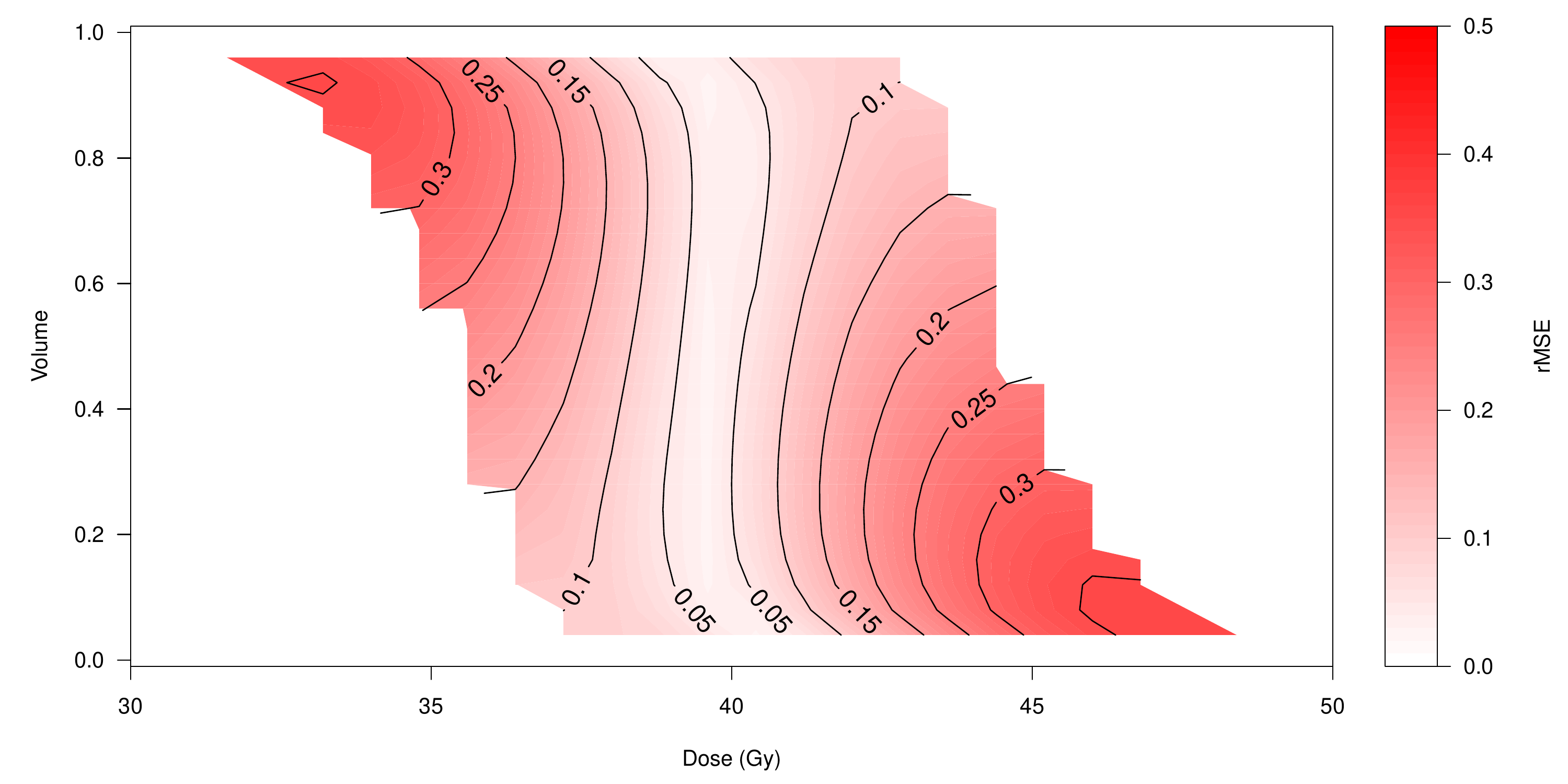}}  \\[-0.5em]
    \subfloat[Additive]{\includegraphics[width = 0.45\textwidth]{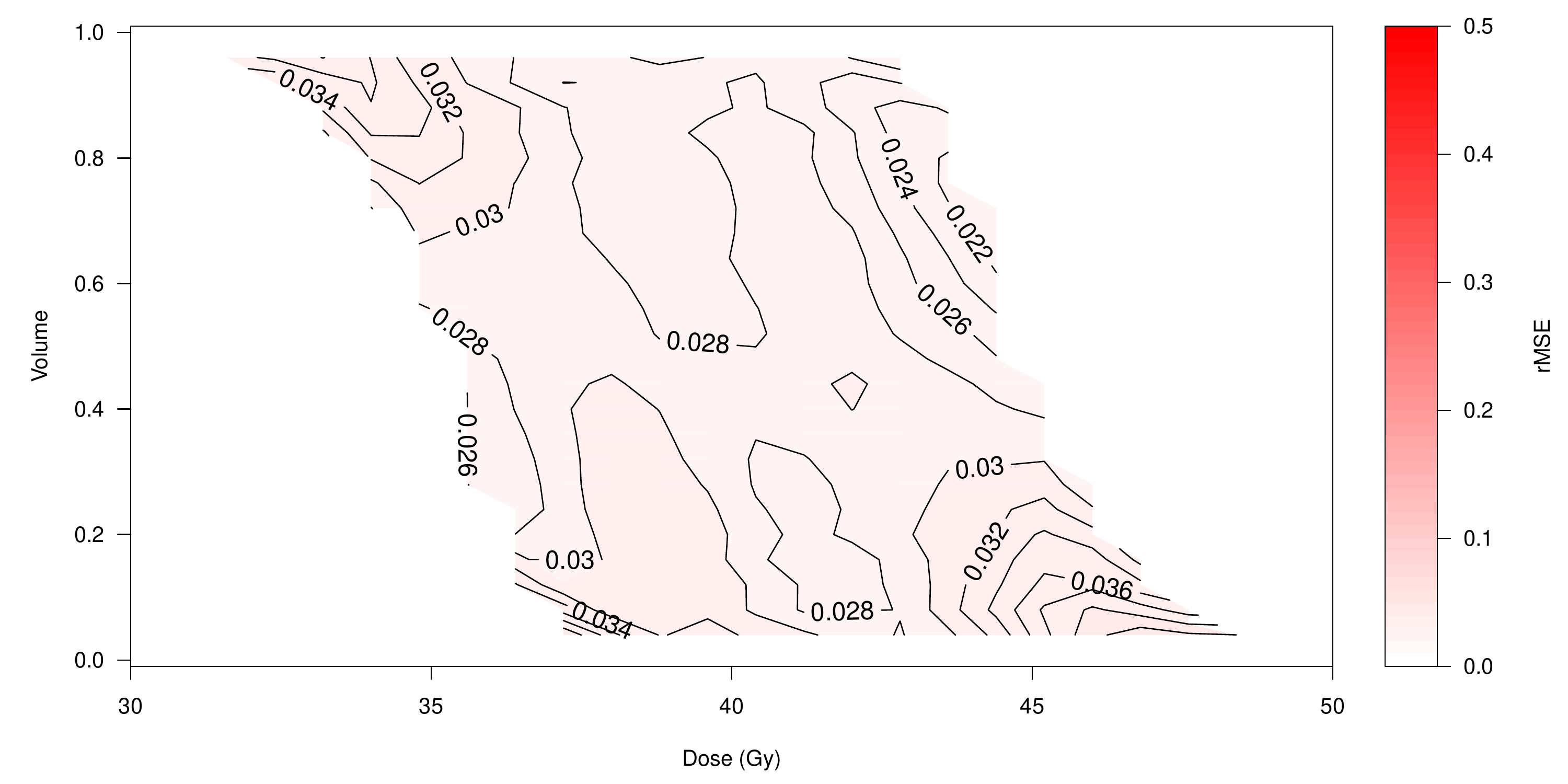}} \hspace{0.6cm}
     \subfloat[Bivariable monotone]{\includegraphics[width = 0.45\textwidth]{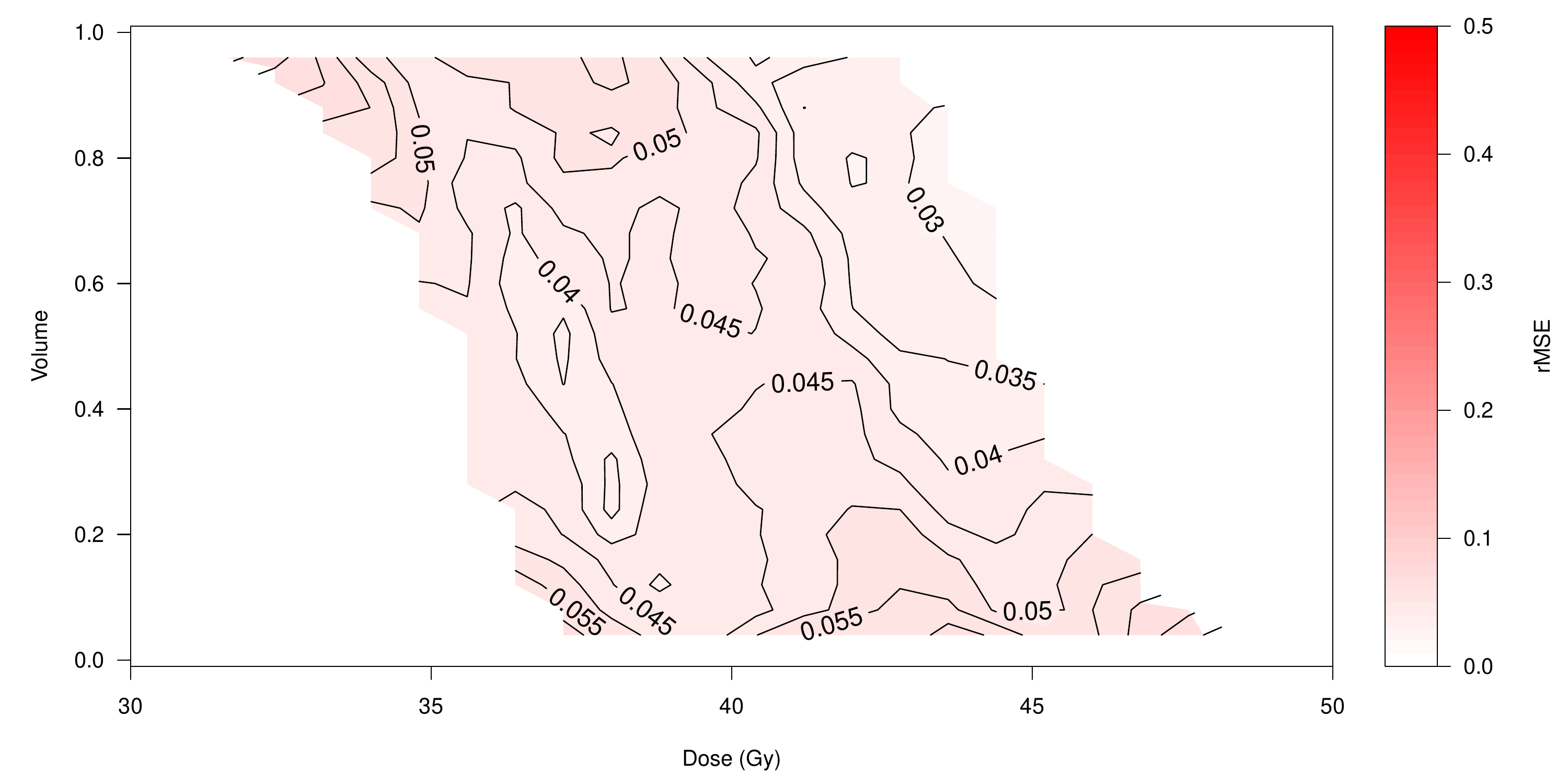}}
    \captionsetup{labelfont=bf}
    \caption{Contour plots for the root mean-square error of model-estimated pointwise-causal risk by DVH volume and radiation dose for $n = 500$.}
    \label{figure:rmse_contour_n=500}
\end{figure}

\section{Simulations assessing the performance of MSMs for the pointwise NTCP in the presence of binary and continuous confounders}\label{apdx:sims-scen}

\subsection{Simulation design}

In the second simulation, for $i = 1, \ldots, n= 500$ we denote $X_{i1} \sim \text{Bernoulli}(0.4)$ as a binary confounder and $X_{i2} \sim N(0, 1)$ as a continuous confounder. The dose distribution was discretized into $\mathcal{D} = 26$ histogram bins of size $\Delta d = \frac{20}{27}$ on the closed interval $[30, 50]$. The corresponding volumes were simulated as complementary CDFs from a Normal distribution, $G_{d, i} = 1 - \Phi(30 + d \Delta d \mid \mu_i(\boldsymbol{X}_i), \sigma_i), \; d \in \{1, \ldots, 26\}$, where $\Phi(\cdot; \mu_i, \sigma_i)$ is the CDF of a $N(\mu_i, \sigma_i^2)$ random variable, and $\boldsymbol{X}_i = (X_{i1}, X_{i2})$. The patient-specific mean doses $\mu_i$ were simulated from a domain-shifted Beta regression model, $\mu_i(\boldsymbol{X}_i) = a_{\text{min}} + (a_{\text{max}} - a_{\text{min}}) \times \text{Beta}(\alpha(\boldsymbol{X}_i), \beta(\boldsymbol{X}_i))$, with linear predictor, $\eta(\boldsymbol{X}_i) = 0.2 X_{i1} + 0.5 X_{i2}$, and shape parameters $\alpha(\boldsymbol{X}_i) = \frac{10}{3} \cdot \text{expit}(\eta(\boldsymbol{X}_i))$, $\beta(\boldsymbol{X}) = \frac{10}{3}(1 - \text{expit}(\eta(\boldsymbol{X}_i))$. $a_{\text{min}} = 35$ and $a_{\text{max}} = 45$ are the upper and lower limits of the domain of mean dose, while the standard deviations were simulated from $\sigma_i \sim U(1, 2)$. The binary normal tissue complication outcome was simulated using the mean dose and confounder, $Y_i \sim \mathrm{Bernoulli}(\text{expit}\{\gamma_0 + \gamma_1 \mu_i + \gamma_2 X_{i1} + \gamma_3 X_{i2}\})$. We consider three scenarios in generating the outcome, namely no confounding, $(\gamma_0, \gamma_1, \gamma_2, \gamma_3) = (-18, 0.45, 0, 0)$, weak confounding, $(\gamma_0, \gamma_1, \gamma_2, \gamma_3) = (-18, 0.45, 0.5, 0.5)$, and strong confounding, $(\gamma_0, \gamma_1, \gamma_2, \gamma_3) = (-21, 0.5, 1, 3)$. The causal pathway between the confounders and the outcome is eliminated under the no confounding scenario, but present in the confounding scenarios.

The true marginal pointwise causal NTCP at dose-volume point ($d, g$) in (\ref{eq:cond-risk}) can be extended to accommodate the binary and continuous covariates as,
\begin{align*}
    E&\left[Y^{\left(g, \; \boldsymbol{V}_{-d} \right)}\right] = \int_{\boldsymbol{x}} \left[ \int_{\mu_{\text{min}}(g)}^{\mu_{\text{max}}(g)} \text{expit}\{\gamma_0 + \gamma_1 \mu + \gamma_2 x_1 + \gamma_3 x_2\} \right. \\
    & \times \left. \dfrac{ (\mu - a_{\text{min}})^{\alpha(\boldsymbol{x}) - 1} (a_{\text{max}} - \mu)^{\beta(\boldsymbol{x}) - 1} |\text{\textbf{J}}|(\mu, g)}{\int_{\mu_{\text{min}}(g)}^{\mu_{\text{max}}(g)} (\mu - a_{\text{min}})^{\alpha(\boldsymbol{x}) - 1} (a_{\text{max}} - \mu)^{\beta(\boldsymbol{x}) - 1} |\text{\textbf{J}}|(\mu, g) \, \mathrm{d}\mu} \, \mathrm{d}\mu \right] f_{\boldsymbol{X}}(\boldsymbol{x}) \, \mathrm{d}\boldsymbol{x},
\end{align*}
which can be calculated over a $10 \times 10$ grid of doses on $[30, 50]$ (scaled to the interval $[0, 1]$ for modeling purposes) and relative volumes on $[0, 1]$.

We compare the performance of logistic MSMs where the functional forms are parametrized by linear, polynomial, two additive flexible monotonic functions, and a bivariable flexible monotonic function:
\begin{align*}
    \text{logit}\left(E\left[Y^{(g, \; \boldsymbol{V}_{-d})} \mathrel{\big|}  \boldsymbol{X}; \boldsymbol{\beta}\right]\right) &= \beta_0 + \beta_1 X_1 + \beta_2 X_2 + \beta_3 d + \beta_4 g  \\
    \text{logit}\left(E\left[Y^{(g, \; \boldsymbol{V}_{-d})} \mathrel{\big|}  \boldsymbol{X}; \boldsymbol{\beta}\right]\right) &= \beta_0 + \beta_1 X_1 + \beta_2 X_2 + \beta_3 d + \beta_4 g + \beta_5 d^2 + \beta_6 g^2 \\
    \text{logit}\left(E\left[Y^{(g, \; \boldsymbol{V}_{-d})} \mathrel{\big|}  \boldsymbol{X}; \boldsymbol{\theta}\right]\right) &= \beta_0 + \beta_1 X_1 + \beta_2 X_2 + \lambda_1(d) + \lambda_2(g) \\
    \text{logit}\left(E\left[Y^{(g, \; \boldsymbol{V}_{-d})} \mathrel{\big|}  \boldsymbol{X}; \boldsymbol{\theta}\right]\right) &= \beta_0 + \beta_1 X_1 + \beta_2 X_2 + \lambda(d, g)  ,
\end{align*}
adjusted for both confounders, as well as versions of these models without the confounder adjustment. All the models were fitted through the \texttt{monoreg} package developed by \citet{monoreg} using 2000 iterations with a burn-in of 1000 iterations. The same out-of-sample and in-sample performance metrics were used as described in Section \ref{section:simulation-design}. The pointwise causal NTCP estimator in (\ref{eq:msm-pointwise}) was used to calculate a $10 \times 10$ grid of doses on $[30, 50]$ (scaled to the interval $[0, 1]$) and relative volumes on $[0, 1]$.

\subsection{Simulation results}

\begin{TableNotes}[flushleft]\item \linespread{1}\scriptsize $n$ = sample size, $|\mathrm{Bias}|$ = absolute bias, MCSD = Monte Carlo standard deviation, rMSE = root mean-square error, MCE = Monte Carlo error, $k$ = effective number of parameters, DIC = deviance information criterion \end{TableNotes}
    \newcolumntype{L}[1]{>{\raggedright\arraybackslash}m{#1}}
    \newcolumntype{C}[1]{>{\centering\arraybackslash}m{#1}}
    \newcolumntype{R}[1]{>{\raggedleft\arraybackslash}m{#1}}
    \renewcommand\theadgape{\Gape[2pt]}
    \renewcommand\cellgape{\Gape[0pt]}
     \renewcommand\theadfont{}
    {\fontsize{8pt}{9.6pt}\selectfont
    \newcommand{\colsize}{1cm}
    \renewcommand{\arraystretch}{1.1}
    \begin{longtable}{C{1.5cm}  C{2.7cm}  C{0.8cm} C{0.8cm}  C{0.8cm} C{0.8cm} C{0.8cm} C{1cm} C{0.8cm} C{1cm} }
    \caption{Summarized performance metrics averaged over a dose-volume grid of pointwise-causal risk estimates.} \\
    \toprule \multirow{2}{*}[-0.5em]{\textbf{Model}} &  \multirow{2}{*}[-0.5em]{\textbf{Parametrization}} &  \multicolumn{4}{c}{\textbf{Out-of-Sample Metric}} & \multicolumn{4}{c}{\textbf{In-Sample Metric}} \\ \cmidrule(lr){3-6} \cmidrule(lr){7-10}
    & & $|\text{Bias}|$ & MCSD & rMSE & MCE & Brier & Deviance & $k$ & DIC  \\ 
    \midrule
    \multicolumn{10}{l}{\textit{Scenario: No Confounding}} \\ \midrule
   \multirow{4}{*}{Unadjusted} & Logistic & 0.147 & 0.022 & 0.151 & 0.002 & 22.48 & 16623.95 & 2.99 & 16626.95 \\
  & Polynomial & 0.147 & 0.024 & 0.151 & 0.002 & 22.48 & 16623.33 & 4.14 & 16627.48 \\
  & Additive & 0.007 & 0.026 & 0.027 & 0.003 & 19.07 & 14767.34 & 128.12 & 14895.45 \\  
  & Bivariable & 0.012 & 0.036 & 0.038 & 0.003 & 19.03 & 14786.09 & 267.76 & 15053.85 \\ 
  \midrule
\multirow{4}{*}{Adjusted} & Logistic & 0.153 & 0.024 & 0.157 & 0.002 & 21.84 & 16279.55 & 4.68 & 16284.23 \\
  & Polynomial & 0.153 & 0.025 & 0.157 & 0.002 & 21.84 & 16279.62 & 6.62 & 16286.24 \\
  & Additive & 0.007 & 0.029 & 0.030 & 0.003 & 18.96 & 14696.02 & 117.76 & 14813.77 \\ 
  & Bivariable & 0.013 & 0.037 & 0.041 & 0.004 & 18.91 & 14710.43 & 236.82 & 14947.25 \\ 
  \midrule
      \multicolumn{10}{l}{\textit{Scenario: Weak Confounding}} \\ \midrule
   \multirow{4}{*}{Unadjusted} & Logistic & 0.136 & 0.022 & 0.141 & 0.002 & 21.98 & 16328.91 & 2.98 & 16331.89 \\
  & Polynomial & 0.137 & 0.024 & 0.141 & 0.002 & 21.98 & 16329.60 & 4.45 & 16334.05 \\
  & Additive & 0.029 & 0.026 & 0.041 & 0.003 & 18.13 & 14175.81 & 145.56 & 14321.36 \\  
  & Bivariable & 0.026 & 0.038 & 0.048 & 0.004 & 18.12 & 14215.86 & 335.44 & 14551.30 \\ 
  \midrule
   \multirow{4}{*}{Adjusted} &  Logistic & 0.147 & 0.024 & 0.151 & 0.002 & 19.57 & 14909.90 & 4.80 & 14914.70 \\
  & Polynomial & 0.147 & 0.025 & 0.151 & 0.002 & 19.57 & 14910.84 & 6.39 & 14917.23 \\
  & Additive & 0.012 & 0.031 & 0.034 & 0.003 & 17.07 & 13446.77 & 93.74 & 13540.51 \\ 
  & Bivariable & 0.017 & 0.040 & 0.045 & 0.004 & 17.02 & 13459.41 & 228.19 & 13687.60 \\ 
    \midrule
      \multicolumn{10}{l}{\textit{Scenario: Strong Confounding}} \\ \midrule
    \multirow{4}{*}{Unadjusted} & Logistic & 0.108 & 0.023 & 0.112 & 0.002 & 22.23 & 16476.56 & 3.10 & 16479.66 \\
  & Polynomial & 0.108 & 0.024 & 0.113 & 0.002 & 22.22 & 16470.72 & 4.31 & 16475.04 \\
  & Additive & 0.119 & 0.026 & 0.124 & 0.003 & 18.74 & 14520.40 & 138.17 & 14658.56 \\  
  & Bivariable & 0.114 & 0.037 & 0.124 & 0.004 & 18.73 & 14560.86 & 254.24 & 14815.10 \\ \midrule
     \multirow{4}{*}{Adjusted} &  Logistic & 0.089 & 0.025 & 0.095 & 0.002 & 10.80 & 8791.13 & 4.74 & 8795.87 \\
  & Polynomial & 0.088 & 0.025 & 0.094 & 0.002 & 10.79 & 8788.24 & 6.51 & 8794.75 \\
  & Additive & 0.033 & 0.029 & 0.045 & 0.003 & 9.46 & 7856.64 & 75.65 & 7932.29 \\ 
  & Bivariable & 0.034 & 0.035 & 0.051 & 0.003 & 9.37 & 7849.67 & 164.61 & 8014.29 \\ 
    \bottomrule 
    \insertTableNotes
    \label{tab:perf_mets_apdx_sim}
    \end{longtable}
    }

\begin{figure}[H]
    \centering
    \begin{tabular}{r@{\hskip 2em}cc}
        & Unadjusted & Adjusted  \\[0.5em]
         \rotatebox{90}{Logistic} & \adjustbox{valign=m}{\includegraphics[width = 0.33\textwidth]{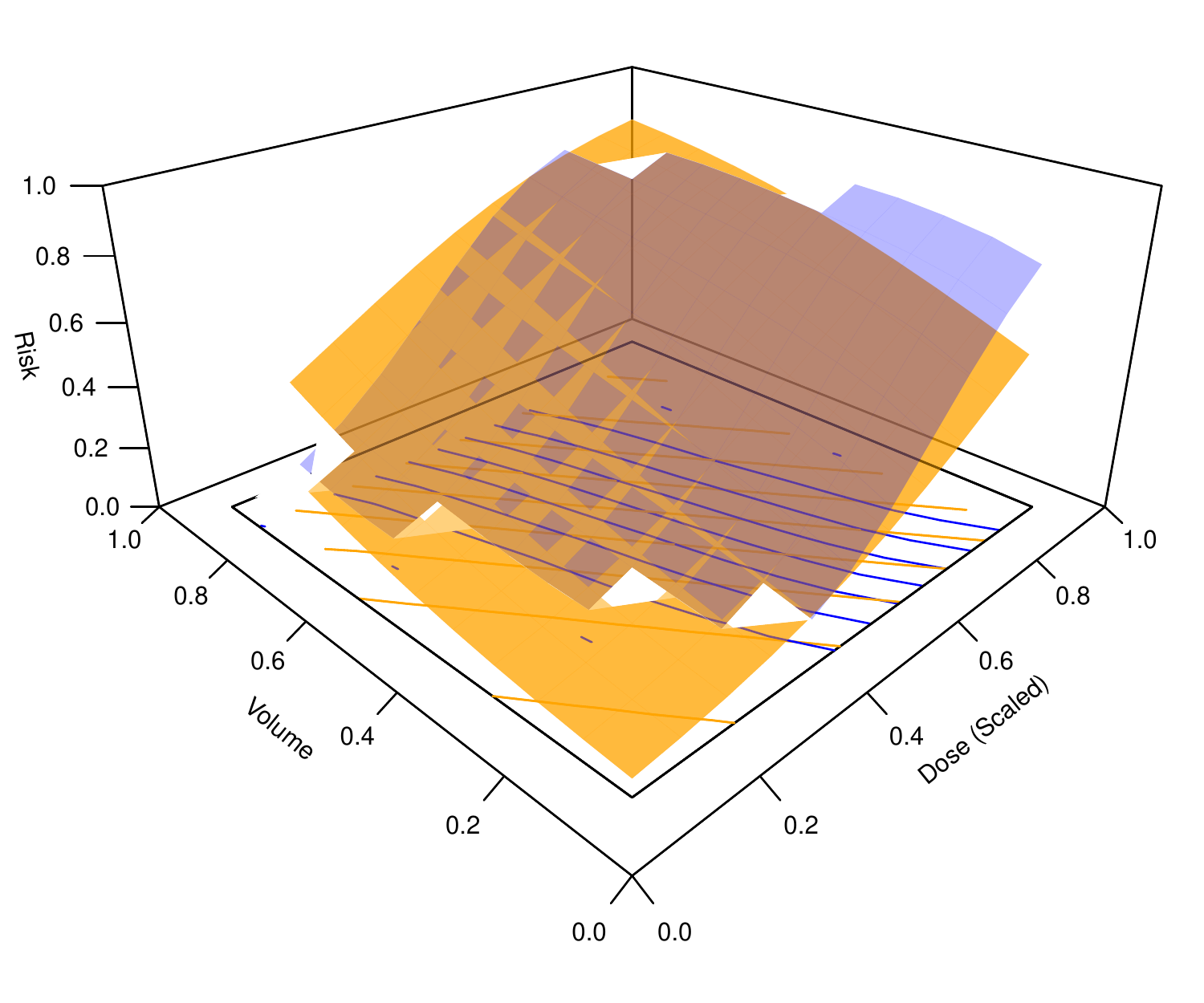}} & \adjustbox{valign=m}{\includegraphics[width = 0.33\textwidth]{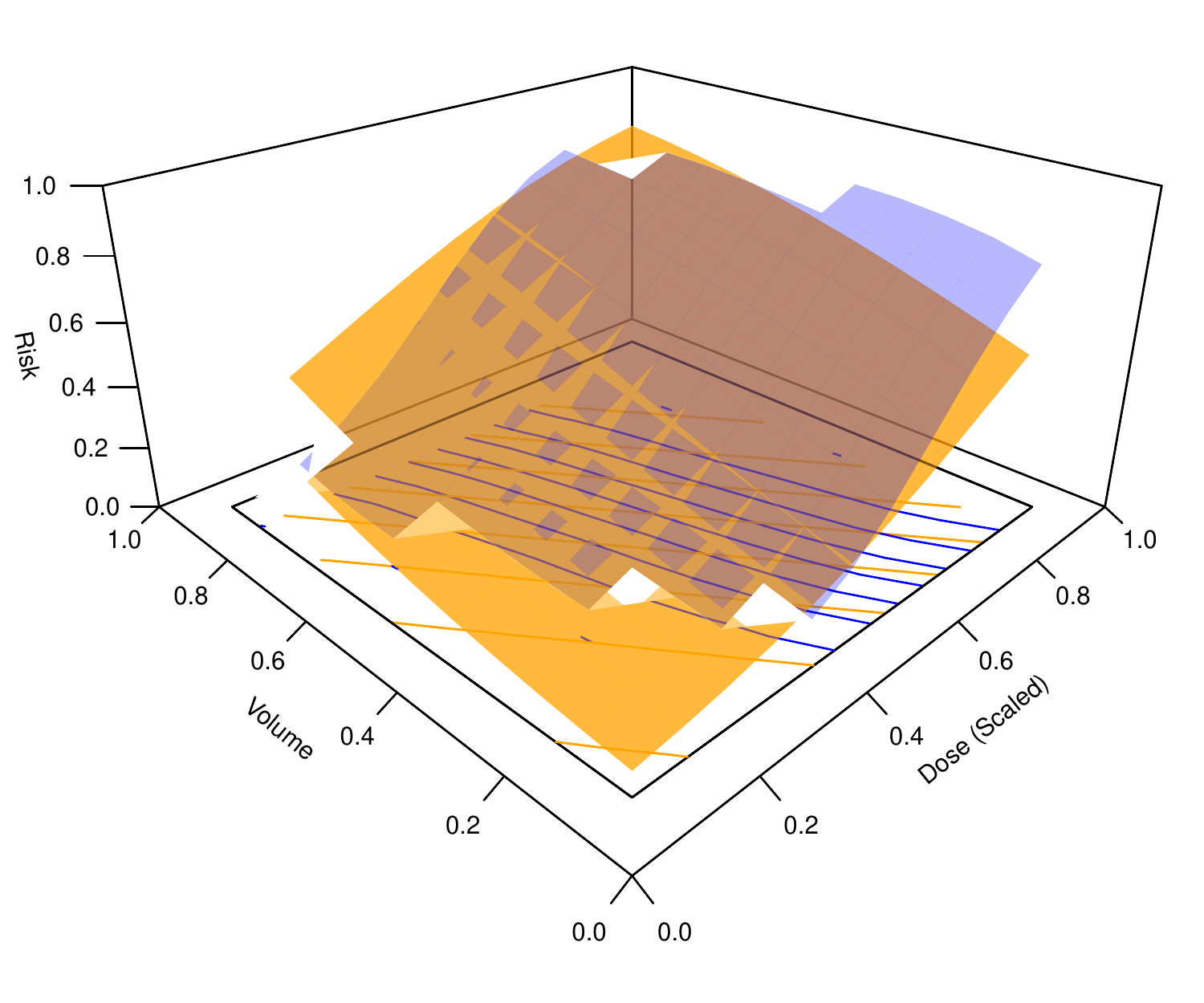}} \\ 
         \rotatebox{90}{\makecell{Polynomial \\[-0.15em] logistic}} & \adjustbox{valign=m}{\includegraphics[width = 0.33\textwidth]{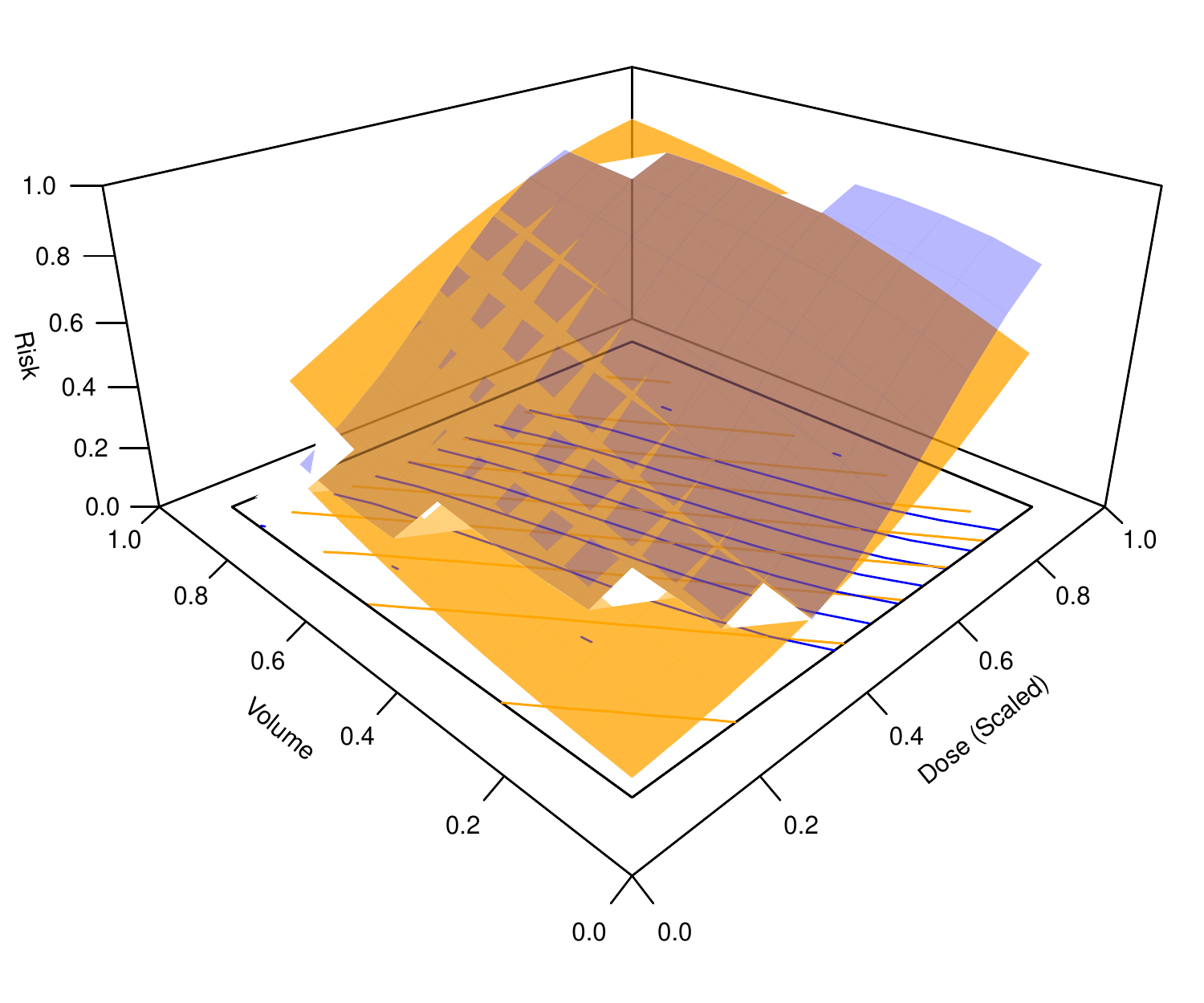}} & \adjustbox{valign=m}{\includegraphics[width = 0.33\textwidth]{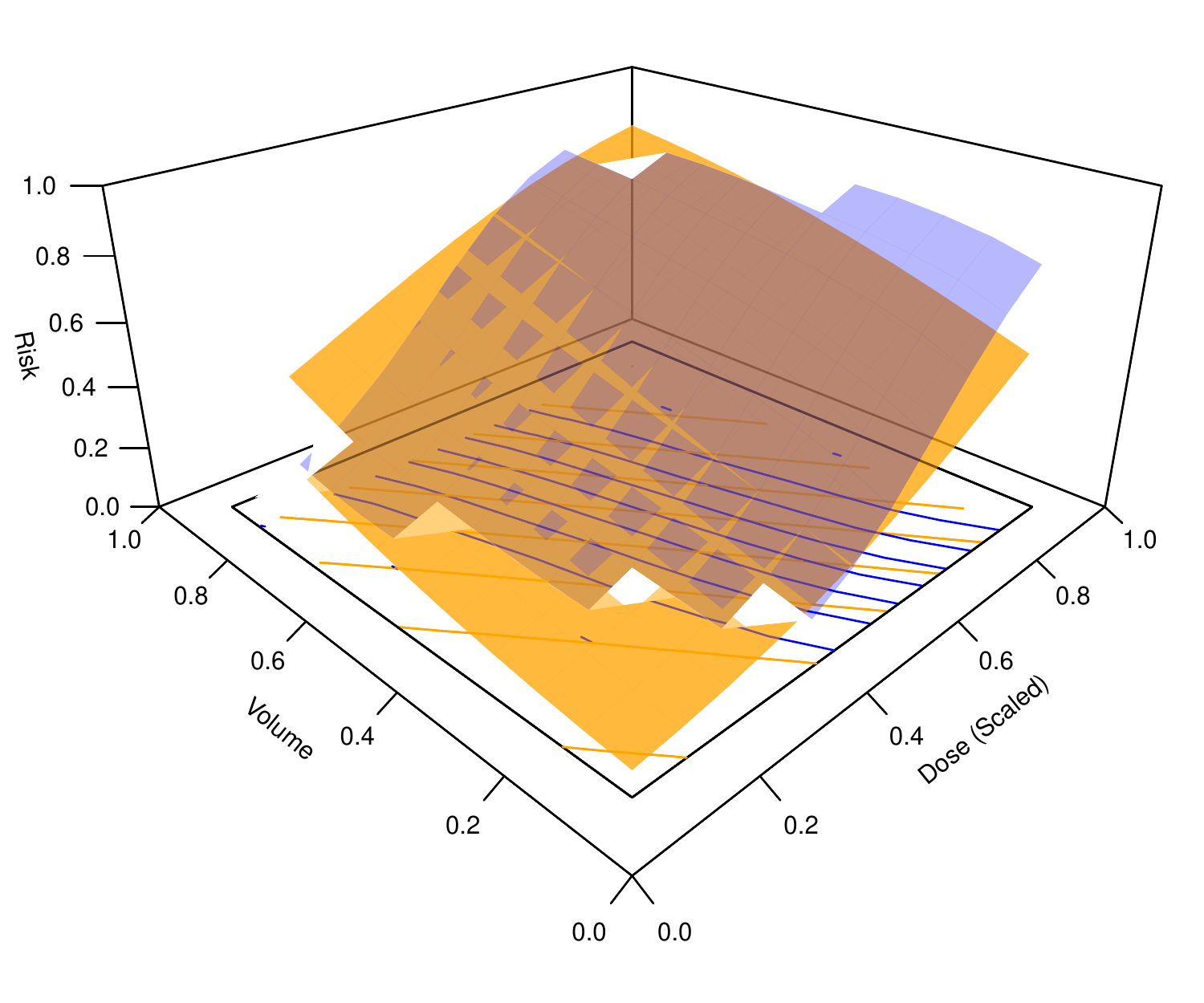}} \\ 
         \rotatebox{90}{Additive} & \adjustbox{valign=m}{\includegraphics[width = 0.33\textwidth]{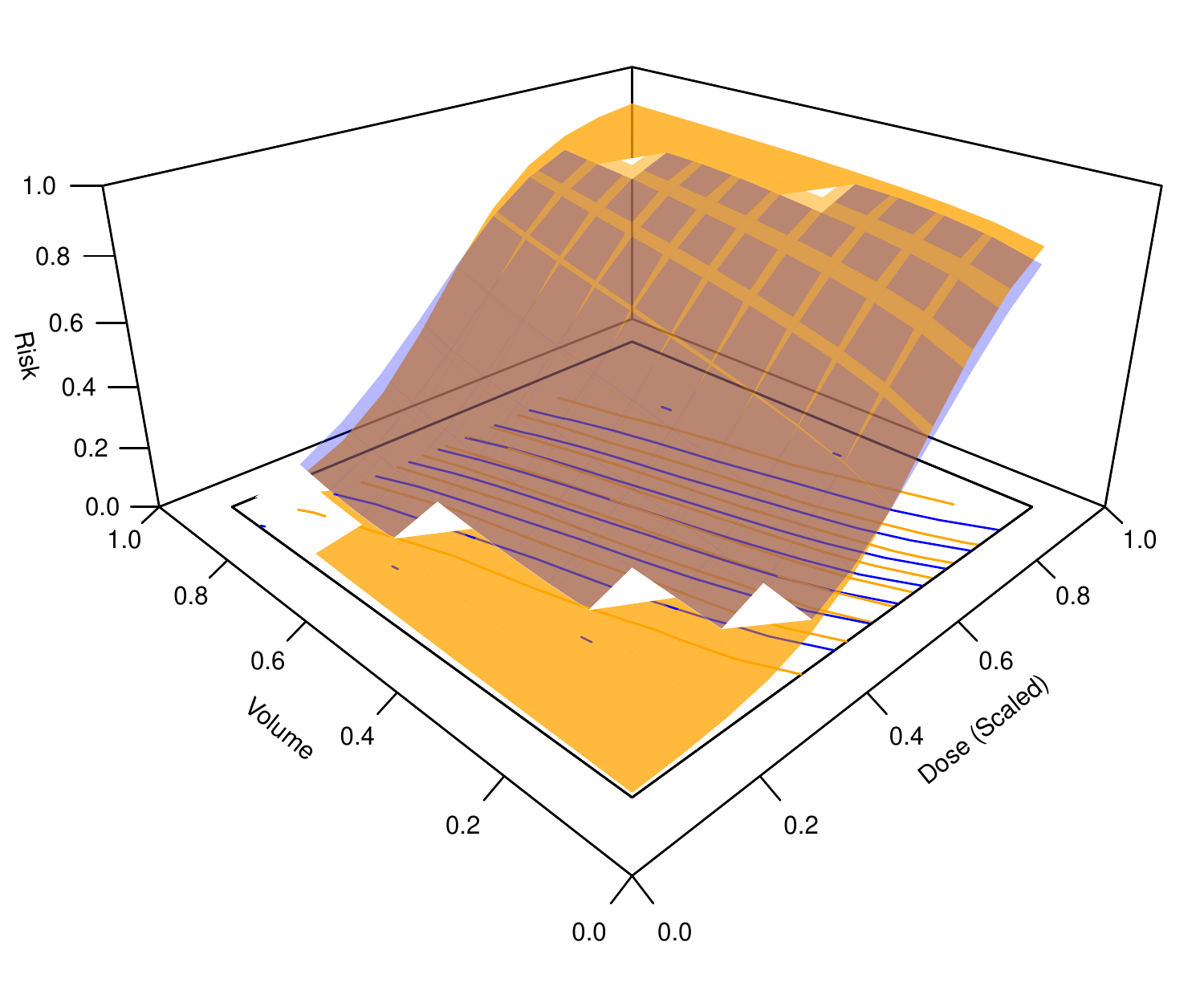}} & \adjustbox{valign=m}{\includegraphics[width = 0.33\textwidth]{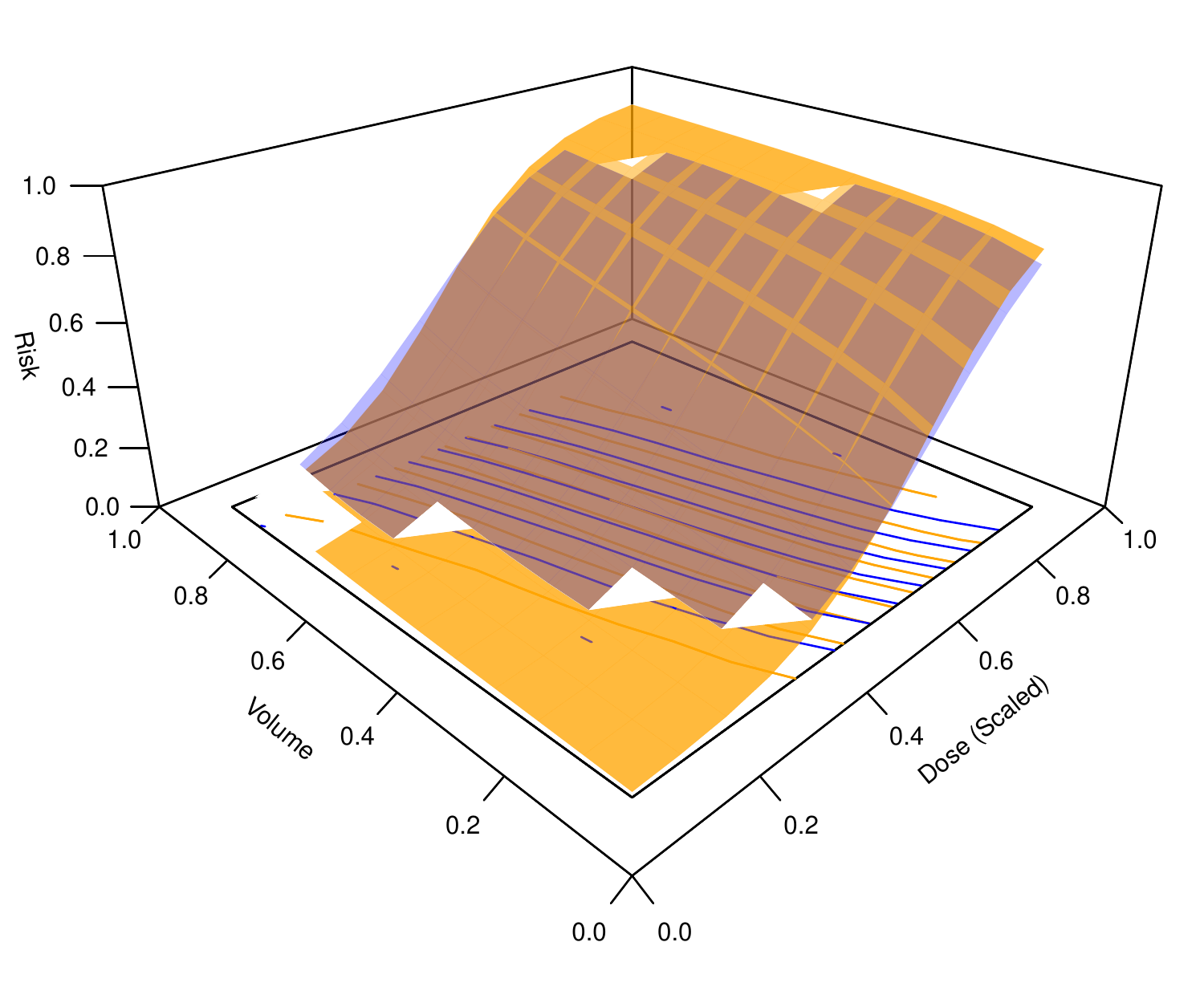}} \\ 
         \rotatebox{90}{\makecell{Bivariable \\[-0.15em] monotone}} & \adjustbox{valign=m}{\includegraphics[width = 0.33\textwidth]{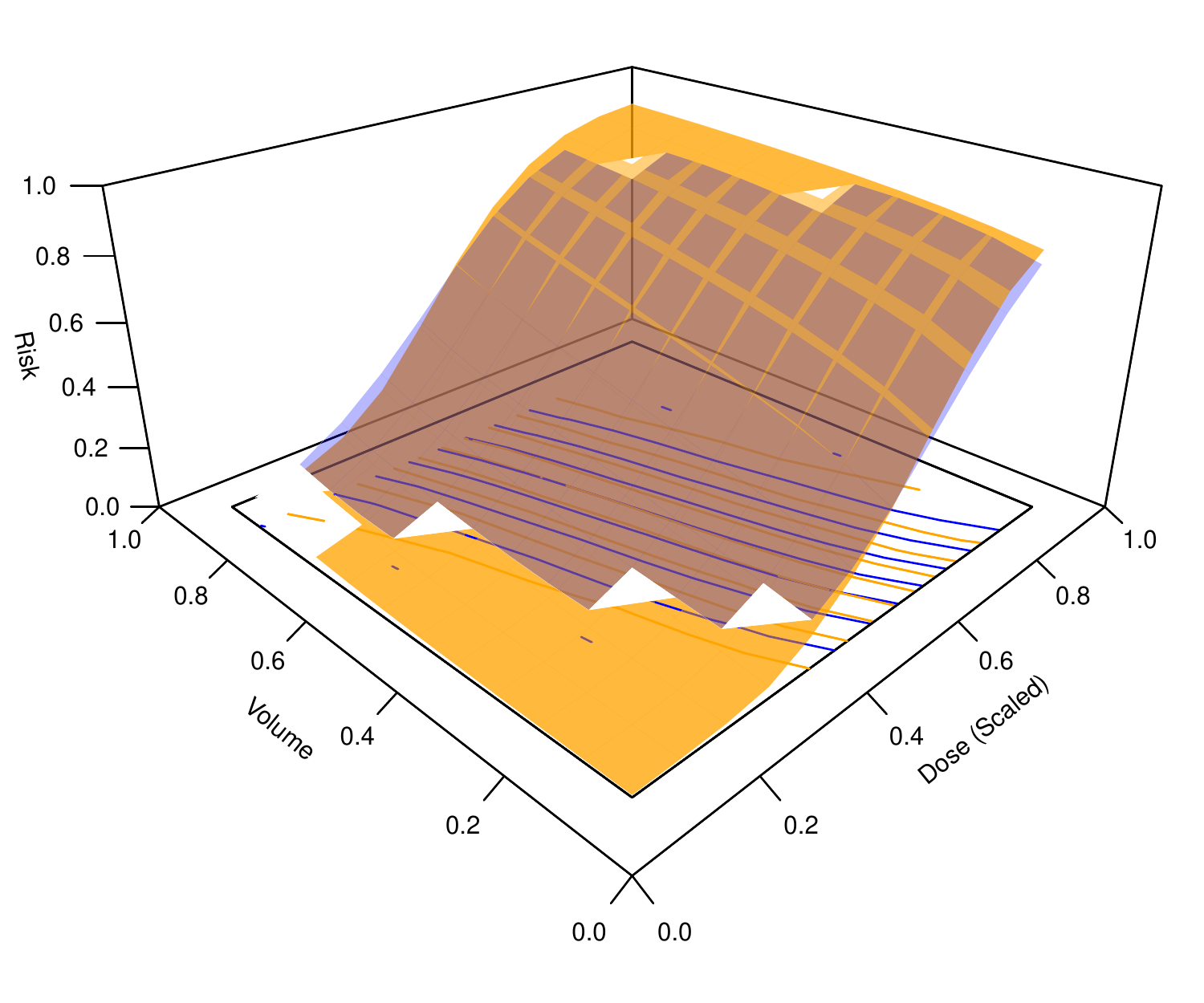}} & \adjustbox{valign=m}{\includegraphics[width = 0.33\textwidth]{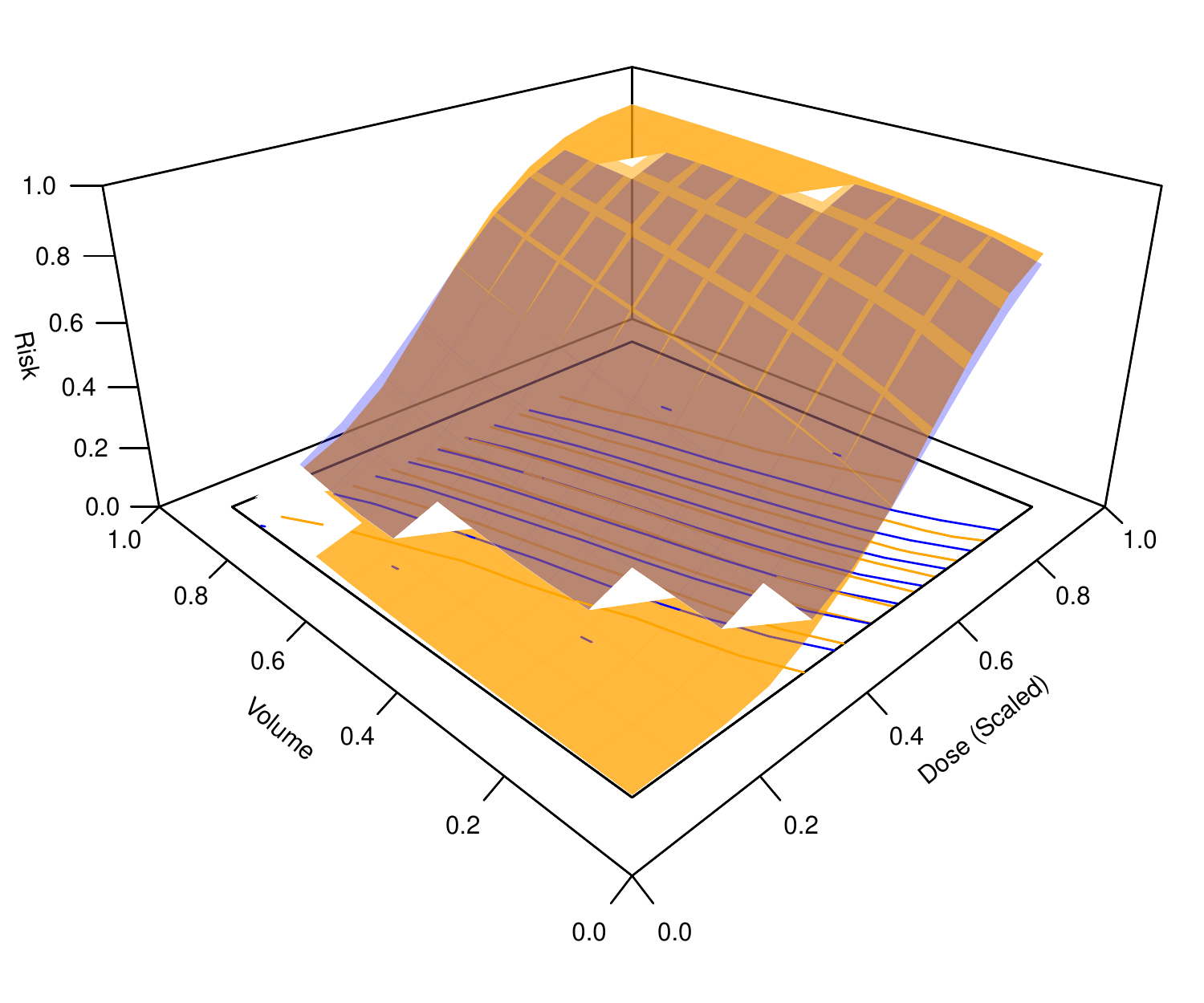}}
    \end{tabular}
    \caption{Perspective plots of the model-based estimated (orange) and true (blue) pointwise causal NTCP by dose/volume coordinate under the \textbf{no confounding scenario}. The rows list the functional form parametrizations while the columns list the adjustment for the binary and continuous confounders of the MSMs.}
    \label{fig:apdx-persp-noconf}
\end{figure}

\begin{figure}[H]
    \centering
    \begin{tabular}{r@{\hskip 2em}cc}
        & Unadjusted & Adjusted  \\[0.5em]
         \rotatebox{90}{Logistic} & \adjustbox{valign=m}{\includegraphics[width = 0.33\textwidth]{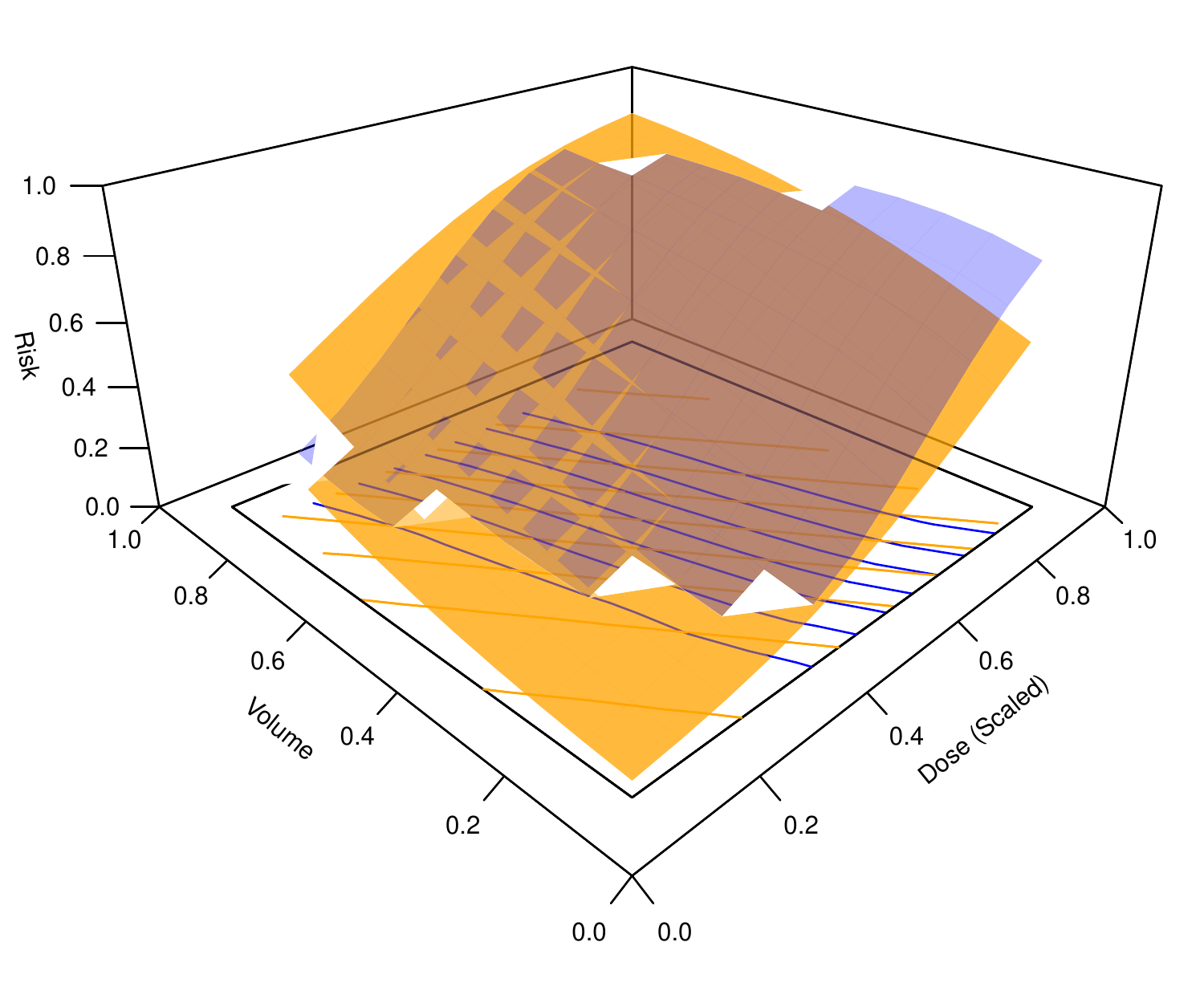}} & \adjustbox{valign=m}{\includegraphics[width = 0.33\textwidth]{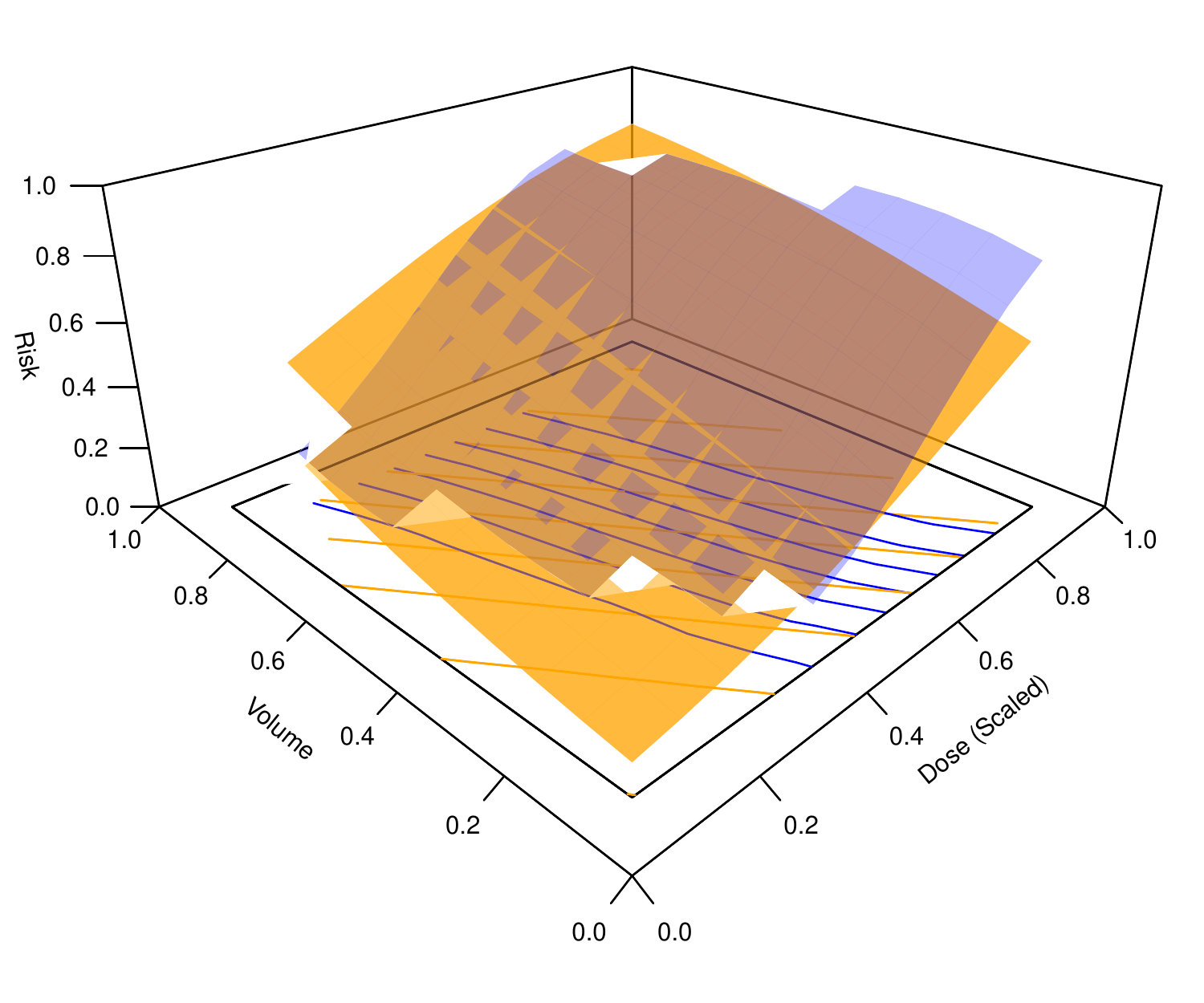}} \\ 
         \rotatebox{90}{\makecell{Polynomial \\[-0.15em] logistic}} & \adjustbox{valign=m}{\includegraphics[width = 0.33\textwidth]{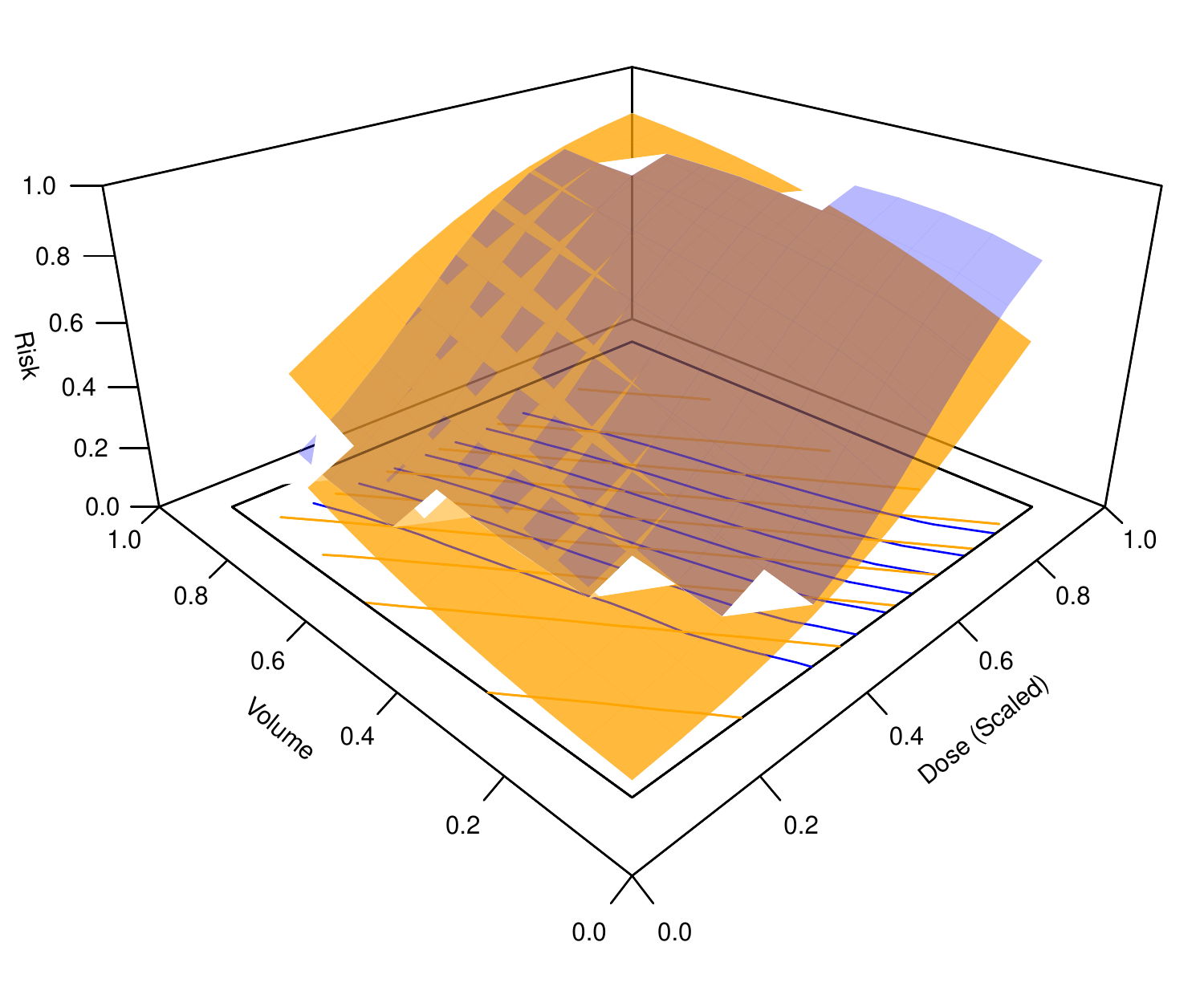}} & \adjustbox{valign=m}{\includegraphics[width = 0.33\textwidth]{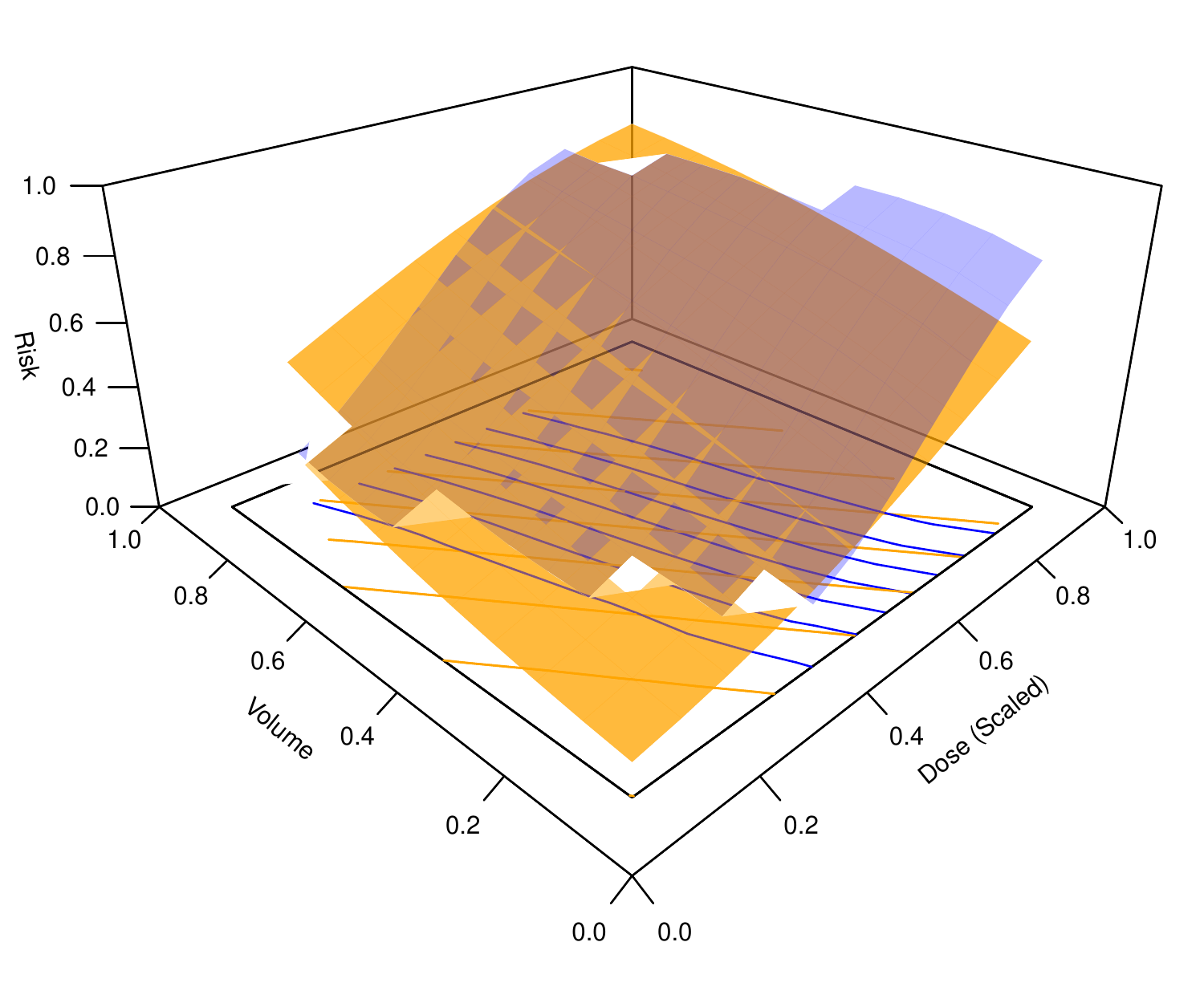}} \\ 
         \rotatebox{90}{Additive} & \adjustbox{valign=m}{\includegraphics[width = 0.33\textwidth]{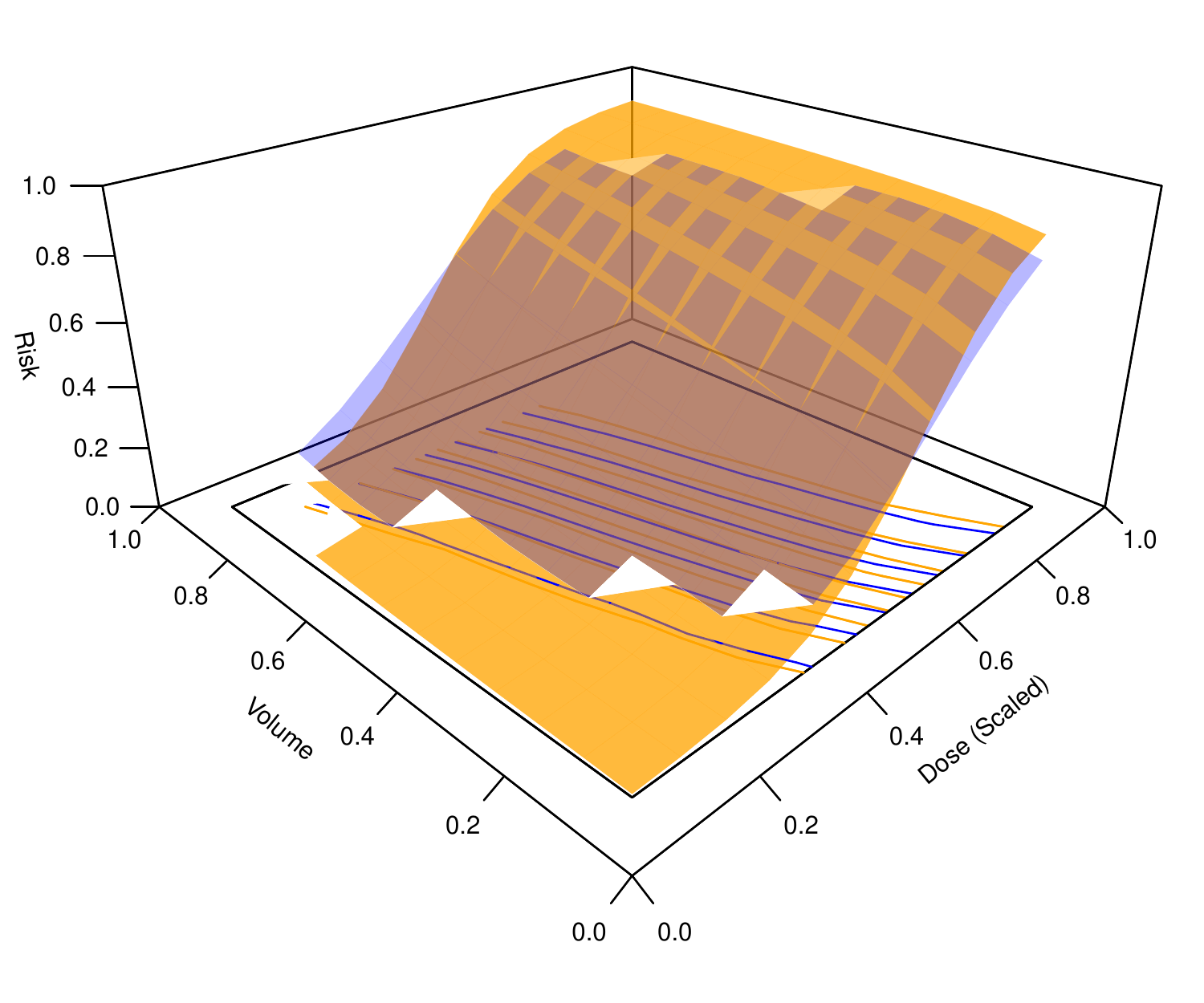}} & \adjustbox{valign=m}{\includegraphics[width = 0.33\textwidth]{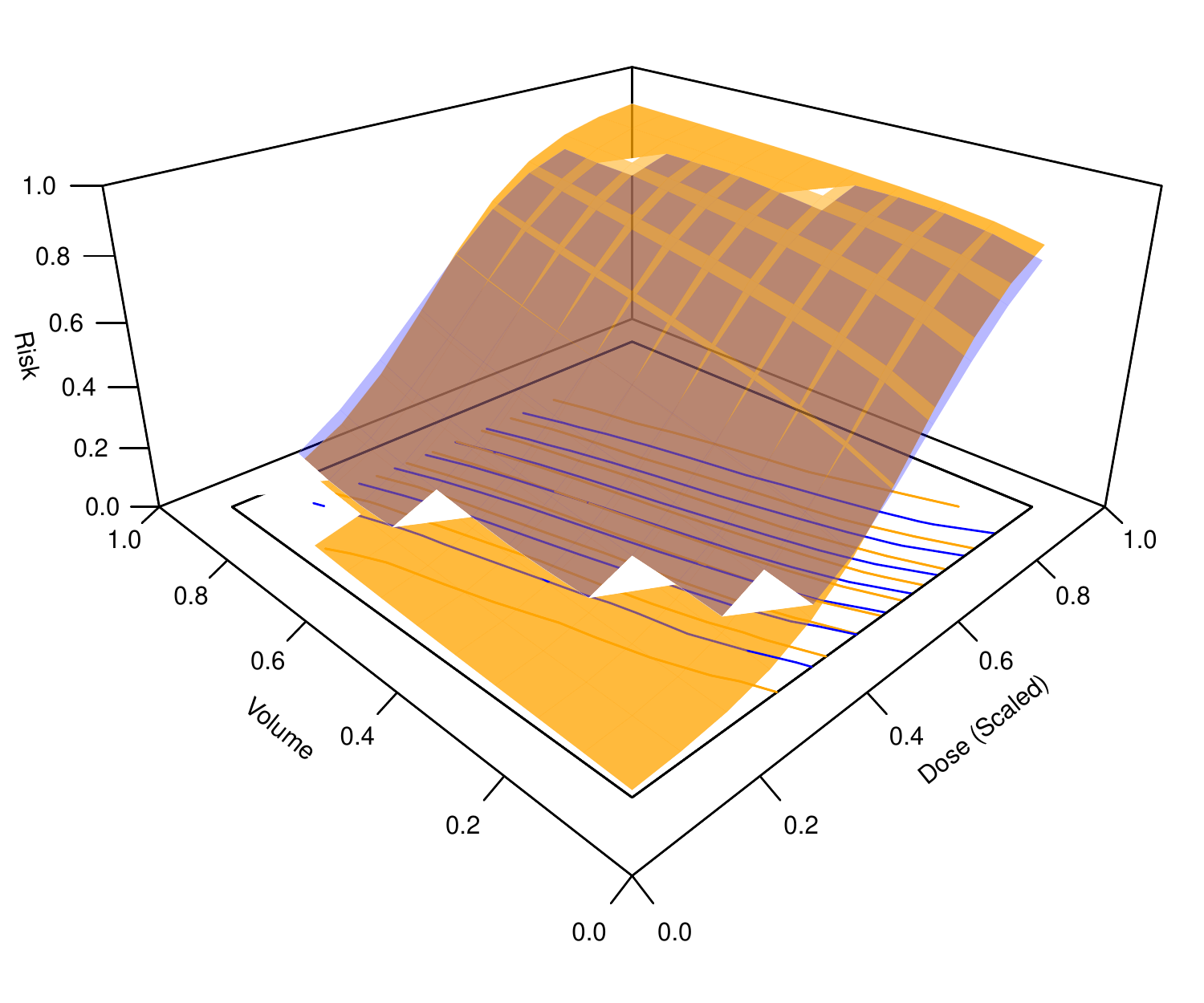}} \\ 
         \rotatebox{90}{\makecell{Bivariable \\[-0.15em] monotone}} & \adjustbox{valign=m}{\includegraphics[width = 0.33\textwidth]{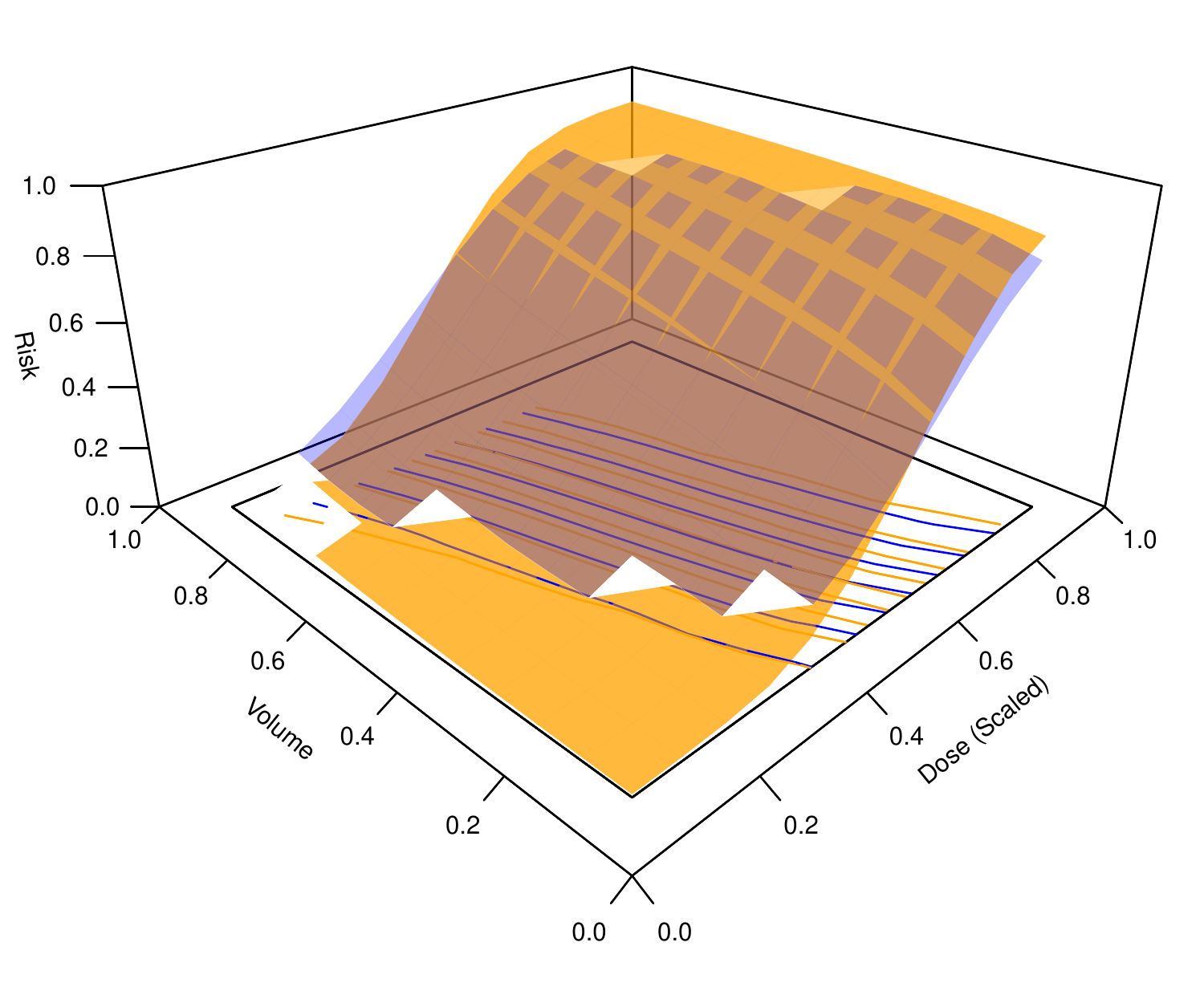}} & \adjustbox{valign=m}{\includegraphics[width = 0.33\textwidth]{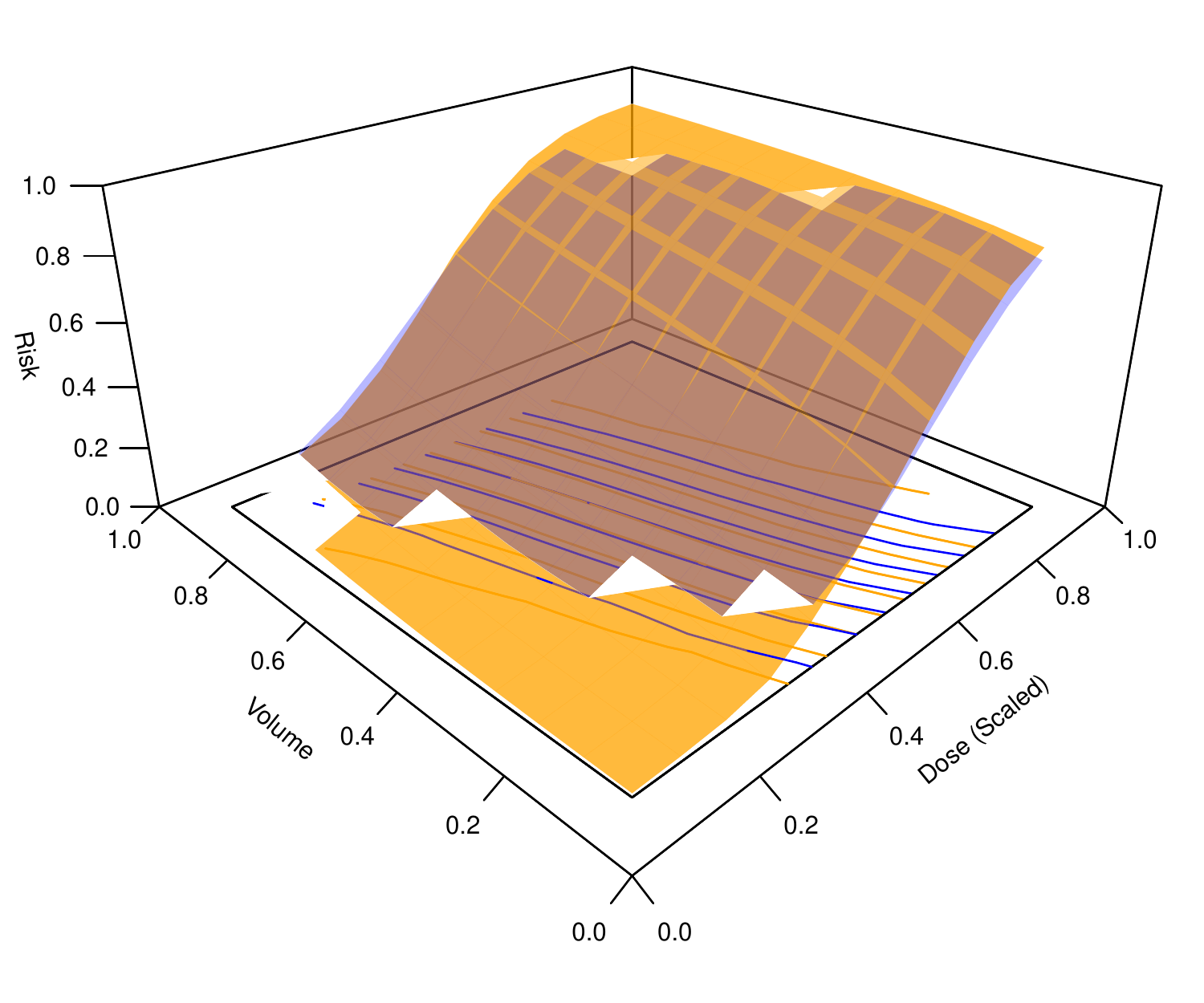}}
    \end{tabular}
    \caption{Perspective plots of the model-based estimated (orange) and true (blue) pointwise causal NTCP by dose/volume coordinate under the \textbf{weak confounding scenario}. The rows list the functional form parametrizations while the columns list the adjustment for the binary and continuous confounders of the MSMs.}
    \label{fig:apdx-persp-weakconf}
\end{figure}

\begin{figure}[H]
    \centering
    \begin{tabular}{r@{\hskip 2em}cc}
        & Unadjusted & Adjusted  \\[0.5em]
         \rotatebox{90}{Logistic} & \adjustbox{valign=m}{\includegraphics[width = 0.33\textwidth]{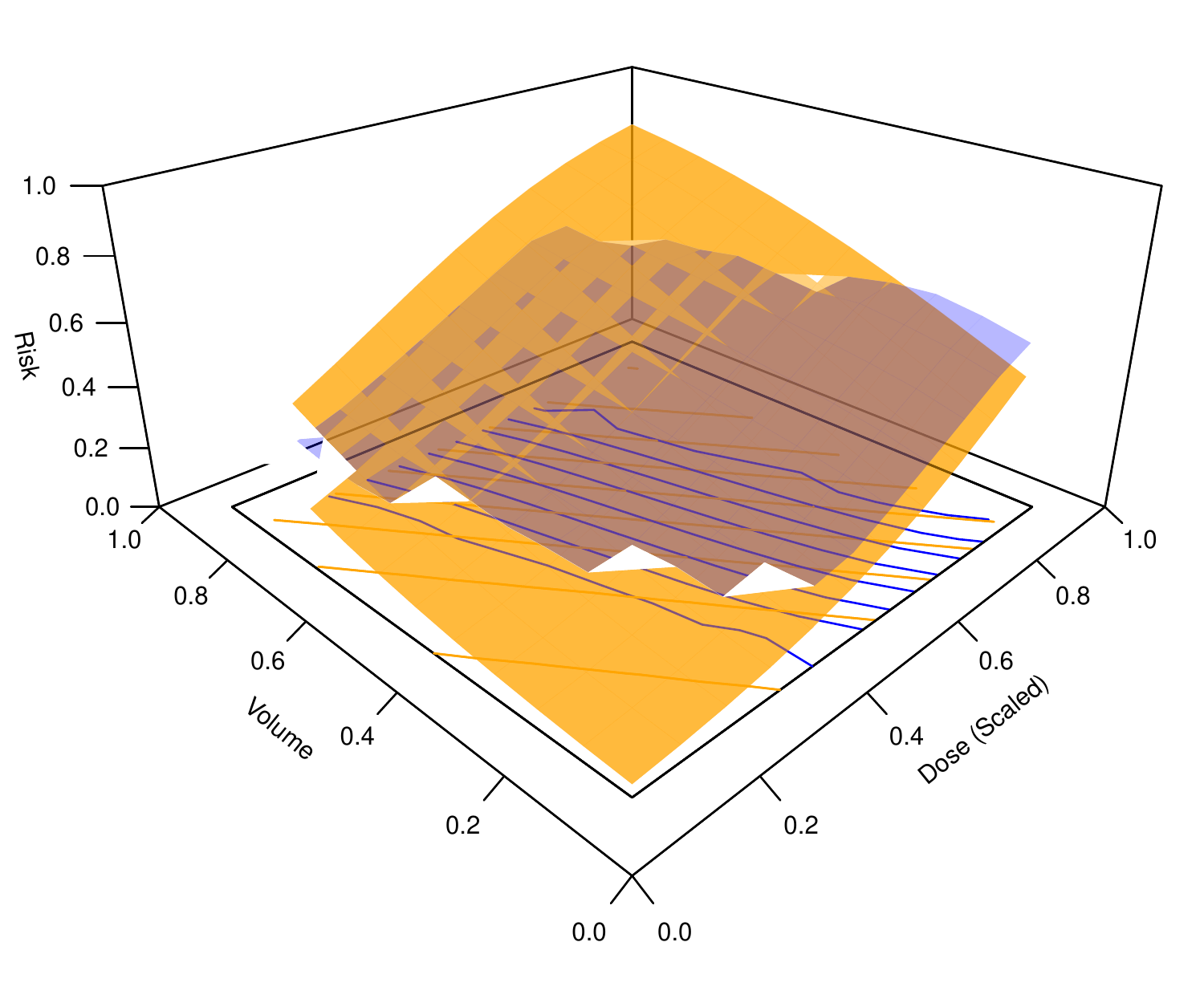}} & \adjustbox{valign=m}{\includegraphics[width = 0.33\textwidth]{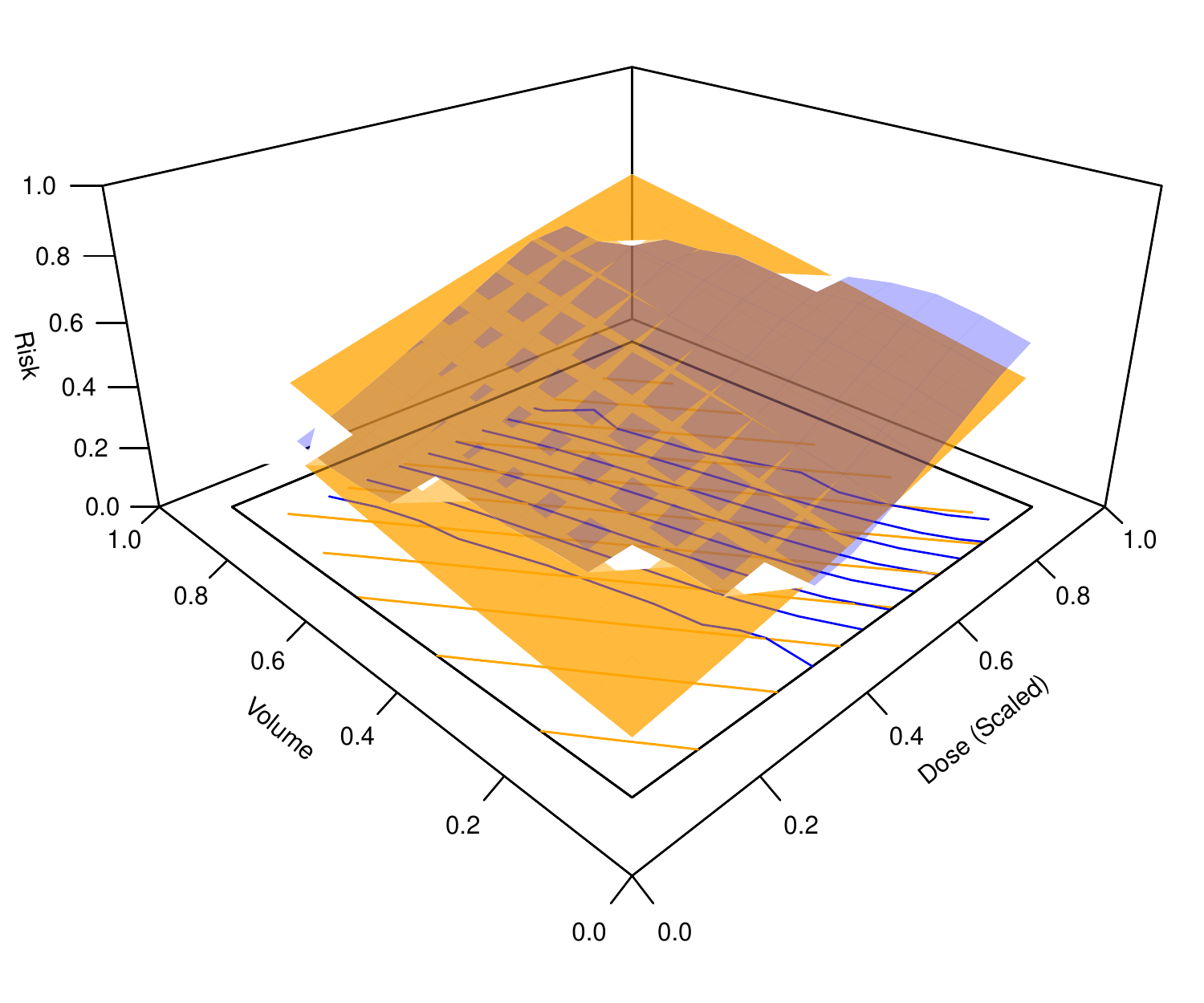}} \\ 
         \rotatebox{90}{\makecell{Polynomial \\[-0.15em] logistic}} & \adjustbox{valign=m}{\includegraphics[width = 0.33\textwidth]{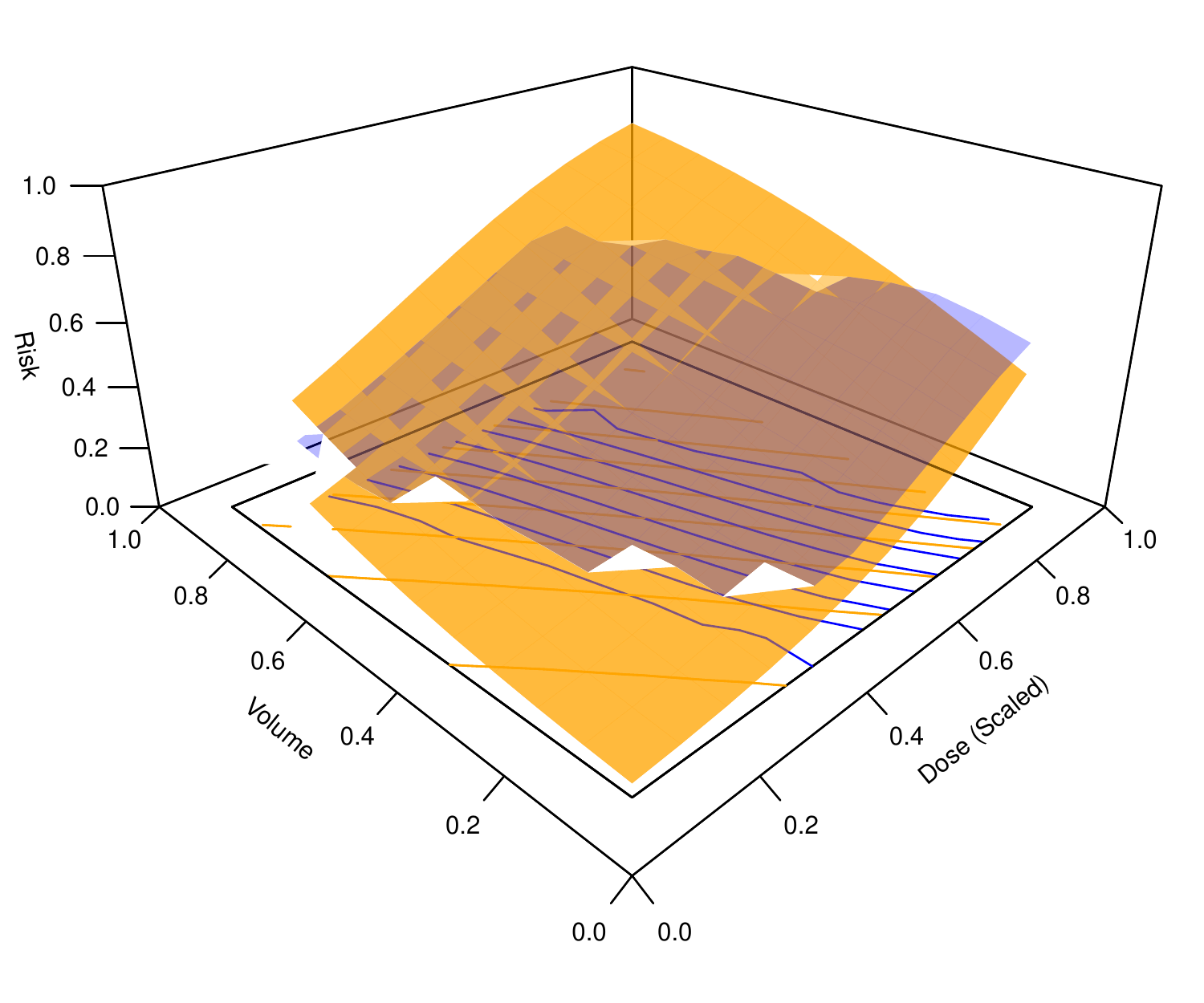}} & \adjustbox{valign=m}{\includegraphics[width = 0.33\textwidth]{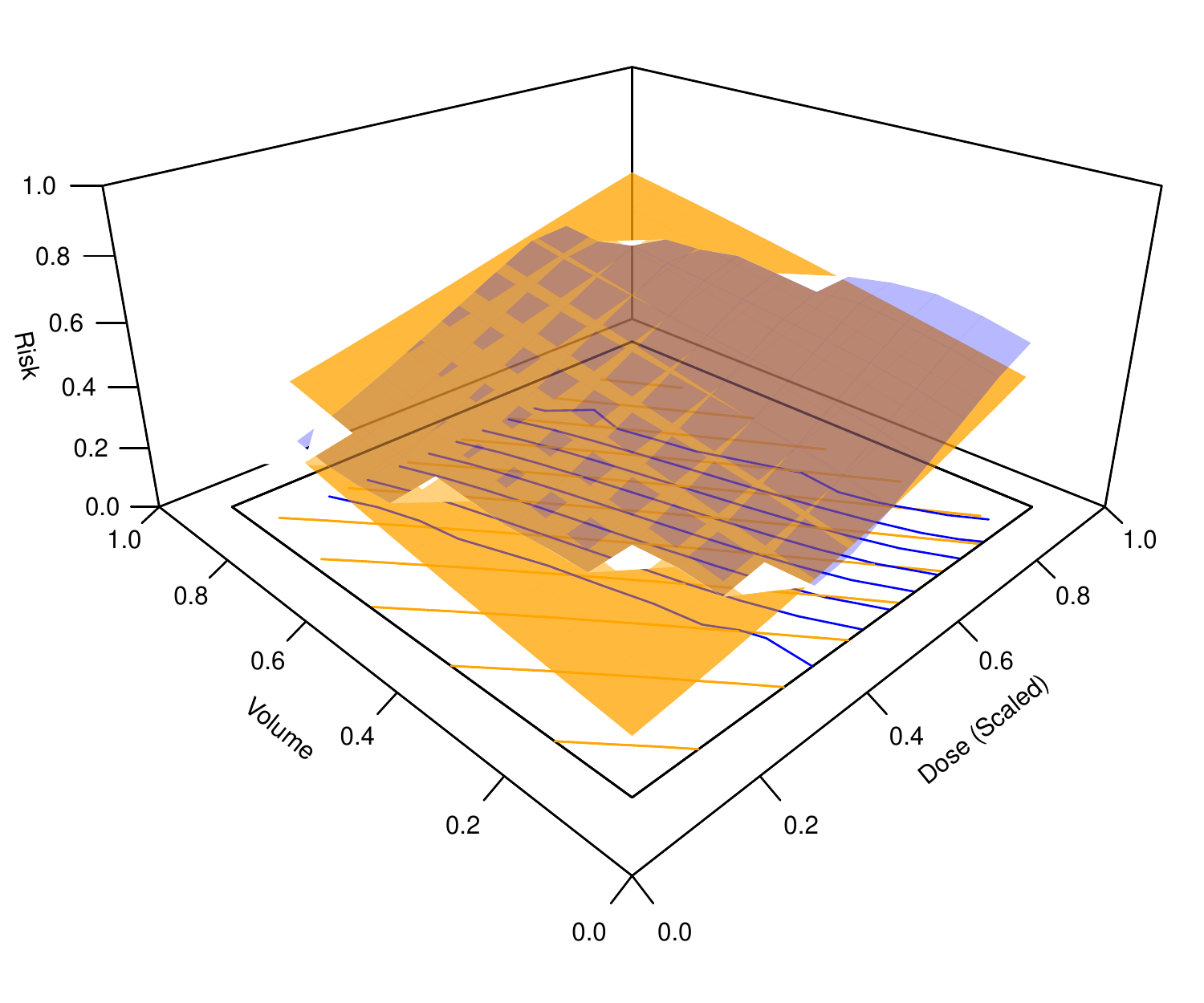}} \\ 
         \rotatebox{90}{Additive} & \adjustbox{valign=m}{\includegraphics[width = 0.33\textwidth]{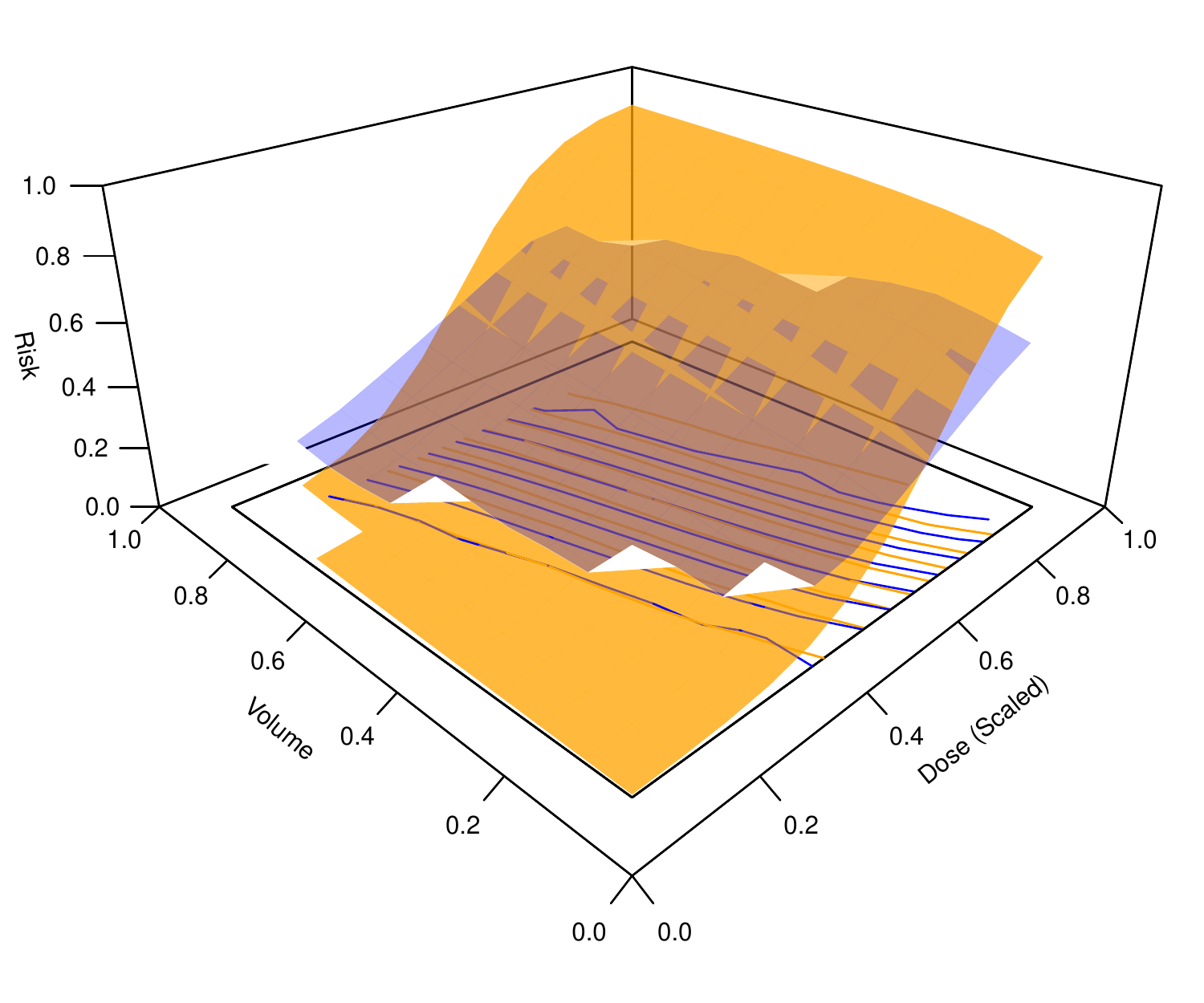}} & \adjustbox{valign=m}{\includegraphics[width = 0.33\textwidth]{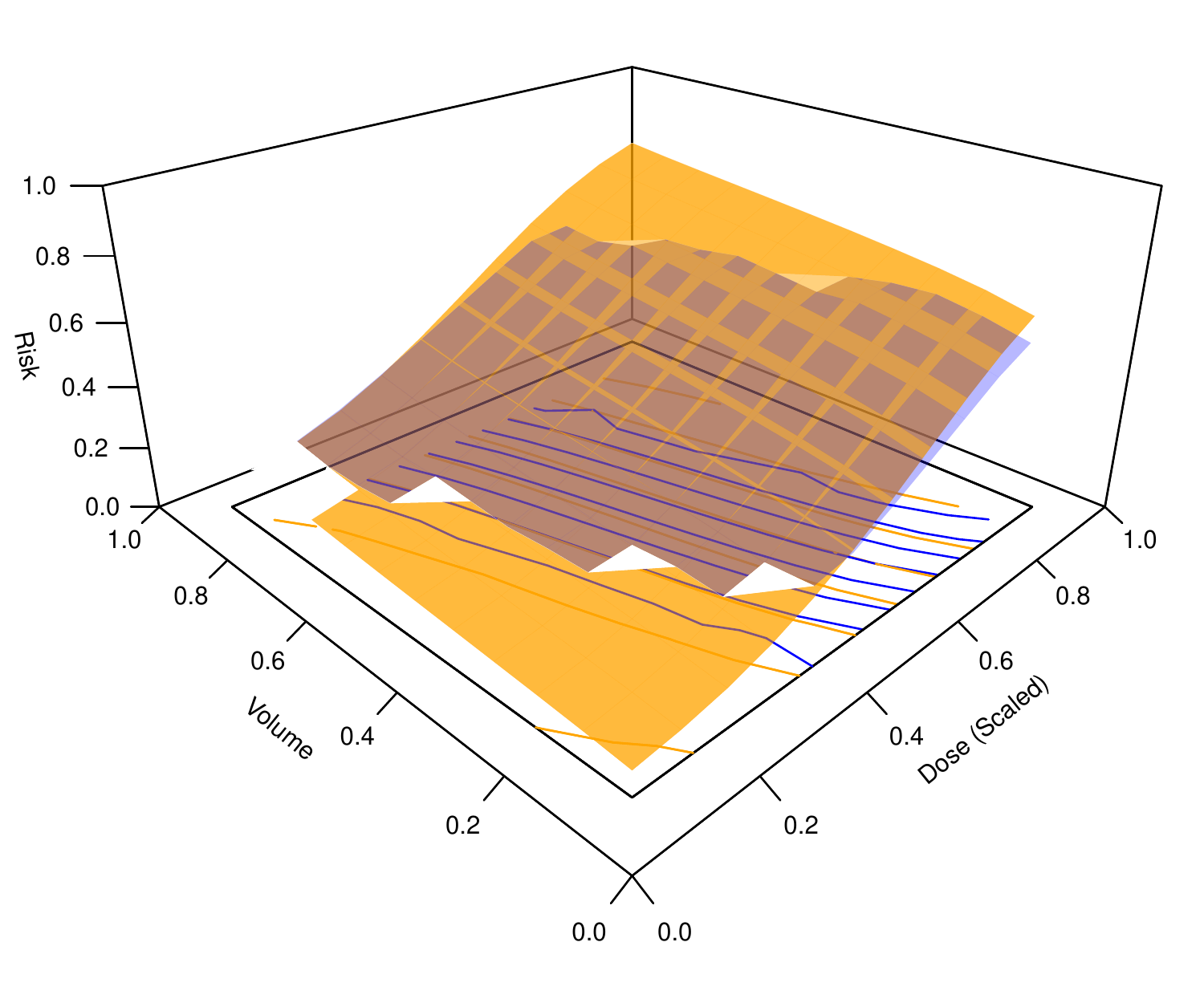}} \\ 
         \rotatebox{90}{\makecell{Bivariable \\[-0.15em] monotone}} & \adjustbox{valign=m}{\includegraphics[width = 0.33\textwidth]{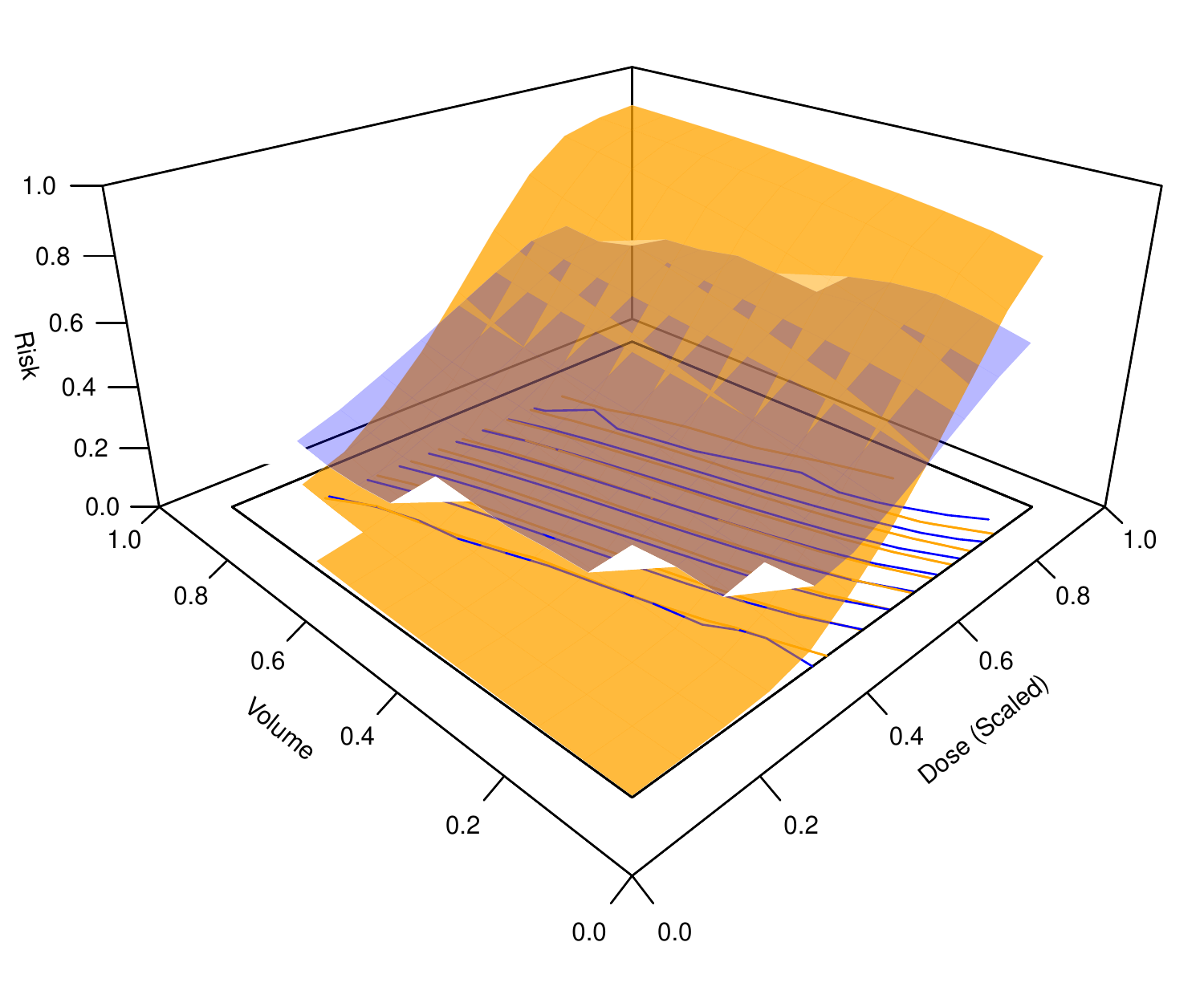}} & \adjustbox{valign=m}{\includegraphics[width = 0.33\textwidth]{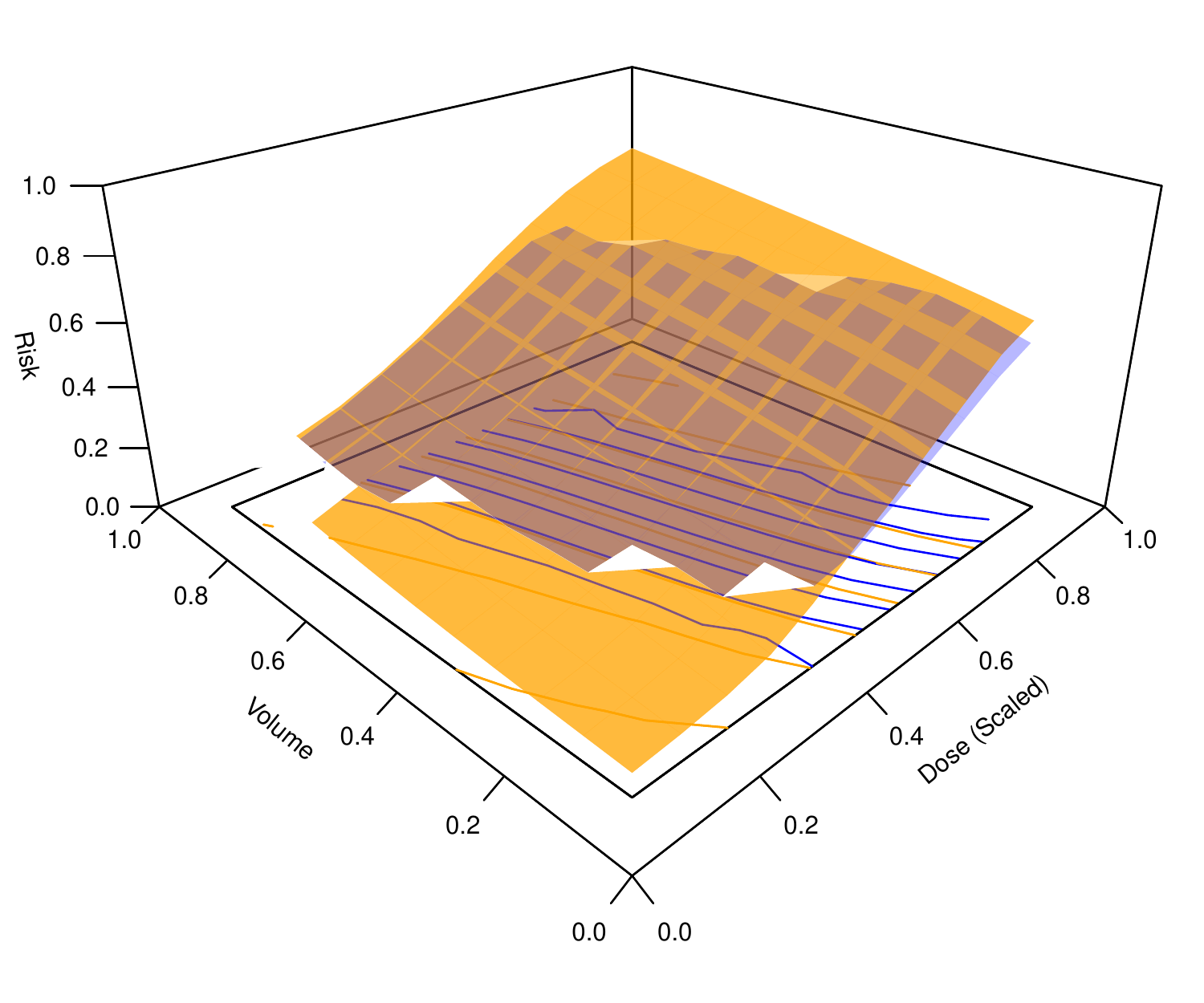}}
    \end{tabular}
    \caption{Perspective plots of the model-based estimated (orange) and true (blue) pointwise causal NTCP by dose/volume coordinate under the \textbf{strong confounding scenario}. The rows list the functional form parametrizations while the columns list the adjustment for the binary and continuous confounders of the MSMs.}
    \label{fig:apdx-persp-strongconf}
\end{figure}

\newpage

The true and average estimated pointwise causal NTCPs based on 104 replicates are illustrated by perspective plots and contour plots under the no confounding (Supplementary Figures \ref{fig:apdx-persp-noconf}), weak confounding (Supplementary Figures \ref{fig:apdx-persp-weakconf}), and strong confounding (Supplementary Figures \ref{fig:apdx-persp-strongconf}) scenarios, with summarized performance metrics are reported in Supplementary Table \ref{tab:perf_mets_apdx_sim}. Confounder-adjusted and unadjusted NTCP estimates were also reported.

All models under all scenarios yielded bivariable monotone increasing estimates of NTCP surfaces with respect to dose and volume, evident by the non-decreasing NTCP surfaces shown in the perspective plots (Supplementary Figures \ref{fig:apdx-persp-noconf}--\ref{fig:apdx-persp-strongconf}). Under the no confounding scenario, unadjusted and confounder-adjusted models produced similar NTCP estimates due to the absence of confounding bias. In addition, the parametric logistic and polynomial logistic models were unable to capture the true surface in contrast to the flexibly specified additive and bivariable monotone functions, evident by the larger mean absolute bias and small Monte Carlo standard deviations of the logistic and polynomial logistic models, as the assumptions contributed to more stable but biased estimates (Supplementary Table \ref{tab:perf_mets_apdx_sim}). 

Under the weak confounding scenario, the additive and bivariable monotone parametrizations again better captured the true surface while the adjusted versions of these models also resulted in less confounding bias (Supplementary Table \ref{tab:perf_mets_apdx_sim}). For the parametric models, bias due to misspecification was still dominating the results. Under the strong confounding scenario, adjustment for confounding produced less biased estimates across all models (Supplementary Table \ref{tab:perf_mets_apdx_sim}).

\end{document}